# Sn-modification of Pt$_7$/alumina model catalysts: Suppression of carbon deposition and enhanced thermal stability

Guangjing Li[a], Borna Zandkarimi[b], Ashley C. Cass[a], Timothy J. Gorey[a], Bradley J. Allen[a] and Anastassia N. Alexandrova[b,c]* and Scott L. Anderson[a]*

[a]Chemistry Department, University of Utah, Salt Lake City, UT 84112

[b]Chemistry and Biochemistry, University of California, Los Angeles, and [c]California NanoSystems Institute, Los Angeles, CA 90095

*Senior Authors, Corresponding Authors: Scott Anderson, (801) 585/7289, anderson@utah.edu, Anastassia Alexandrova, (310) 825/3769, ana@chem.ucla.edu

## ABSTRACT

An atomic layer deposition process is used to modify size-selected Pt$_7$/alumina model catalysts by Sn addition, both before and after Pt$_7$ cluster deposition. Surface science methods are used to probe the effects of Sn-modification on the electronic properties, reactivity, and morphology of the clusters. Sn addition, either before or after cluster deposition, is found to strongly affect the binding properties of a model alkene, ethylene, changing the number and type of binding sites, and suppressing decomposition leading to carbon deposition and poisoning of the catalyst. Density functional theory on a model system, Pt$_4$Sn$_3$/alumina, shows that the Sn and Pt atoms are mixed, forming alloy clusters with substantial electron transfer from Sn to Pt. The presence of Sn also makes all the thermally accessible structures closed shell, such that ethylene binds only by π-bonding to a single Pt atom. The Sn-modified catalysts are quite stable in repeated ethylene temperature programmed reaction experiments, suggesting that the presence of Sn also reduces the tendency of the sub-nano clusters to undergo thermal sintering.

## I.      INTRODUCTION

Catalyst deactivation by coking, i.e., carbon deposition, is a significant problem in many catalysis applications at high temperature under hydrocarbon-rich conditions, such as dehydrogenation of alkanes to alkenes[1-3] for petroleum processing, including applications based on Pt nanoparticle catalysts.[4-9] In industrial applications, it is possible to regenerate catalysts by periodically oxidizing the coke away at high temperatures; however, for nano-catalysts, the high regeneration temperatures can lead to sintering and loss of activity.  Sub-nano clusters have most or all of the catalytic metal in the reactant-accessible surface layer, thus improving the use of precious metals. However, such small clusters are less thermodynamically stable than larger particles, thus sintering is even more of a problem.  One application of interest is the use of alkane-to-alkene dehydrogenation in hot hydrocarbon fuels as an endothermic reaction to cool air vehicle fuel systems at high temperatures, the problem being that dehydrogenation of the alkene product creates coke precursors. Coking is a particular problem because it can lead to clogging of small channels or injectors, and regeneration may not be a practical option.  One strategy for sintering prevention is to stabilize the catalyst by forming porous silica or alumina coating over the nanoparticle catalysts;[10-13] however, despite achieving anti-sintering effects, this can also block the catalytic sites.[14]

Previous research on Pt bimetallic catalysts introduced elements such as tin, nickel, cobalt, copper, and germanium into the nanoparticle catalysts[15-22]. The role of these second elements is to alter the geometric and electronic properties of Pt particles to improve the stability of the catalyst and selectivity of catalytic reactions.  Among these elements, Sn has been widely investigated, and PtSn bimetallic catalysts showed favorable selectivity and low activity loss, especially toward light alkene production[1, 23-26].

For example, Koel and co-workers reported UHV studies of ordered PtSn surface alloys formed by depositing Sn on Pt (111) single crystals. In particular, they found that Sn-alloying suppressed dehydrogenation of adsorbed ethylene and other alkenes such as propylene and isobutylene, thus reducing formation of coke precursors.[27, 28] The temperature-programmed desorption (TPD) results showed that as the Sn coverage increased, the binding energies of the alkenes to the surface decreased substantially, resulting in desorption temperatures dropping by ~100 K in the case of ethylene. As a result, $H_2$ desorption, signaling ethylene decomposition and coking, was almost completely suppressed. These changes in binding energy and desorption behavior occurred despite the observation that Sn alloying had no effect on low temperature sticking probability, nor on the saturation alkene coverage. It was proposed that ethylene binds in a di-σ geometry on both the Pt and PtSn alloy surfaces.

Recently, Hook et al. used density functional theory (DFT) to examine ethane dehydrogenation over PtSn surface alloys, finding, consistent with the experiments of Koel and co-workers, that Sn depresses the alkene desorption energy below the barrier for dehydrogenation, due to a combination of geometric and electronic effects.[29] Ethylene was found to bind in a di-σ fashion on both the Pt and PtSn alloy surfaces. The results were in line with an earlier DFT study by Yang et al. of propane dehydrogenation, where formation of coke precursors was also predicted to be suppressed by Sn alloying due to lowering of the propylene desorption energy below the barrier for further dehydrogenation.[30]

Dumesic and co-workers used microcalorimetric and IR spectroscopy methods to study ethylene interactions with $Pt/SiO_2$ and several different stoichiometry $PtSn/SiO_2$ catalysts with metal particle sizes in the 2 to 5 nm range.[31, 32] On $Pt/SiO_2$, both di-σ and π-bonded ethylene were observed. Sn alloying reduced the heat of adsorption of ethylene on the catalysts, and with increasing Sn content, the fraction of π-bonded ethylene increased, thus differing from the results on extended PtSn surface alloys. As shown below, moving to sub-nanometer clusters introduces additional differences.

We have been using a combination of experiments and density functional theory (DFT) to probe methods to mitigate coking and sintering of sub-nano, size-selected model $Pt_n$/alumina catalysts. Ethylene temperature-programmed desorption (TPD), along with surface probes such as low energy ion scattering (ISS) and X-ray photoelectron spectroscopy (XPS) is used to test the propensity toward coking and sintering. Monitoring the branching between desorption and dehydrogenation, which deposits carbon, provides a signature of coking, and observing the evolution of the catalyst morphology and adsorption properties during multiple TPD runs, provides insight into thermal processes. Pure $Pt_n$/alumina catalysts were found to deactivate rapidly in sequential ethylene TPD experiments, primarily due to the deposition of carbon on the clusters.[33] We showed that diborane could be used to selectively borate size-selected $Pt_n$ seed clusters on the alumina support,[34] thus generating clusters anchored to the support by Pt-B-O bonds, and modifying the electronic properties such that dehydrogenation and carbon deposition were almost completely suppressed.[35] The success of coking prevention was attributed to boration substantially reducing the ethylene binding energy, so that desorption occurred in preference to dehydrogenation. DFT showed that the PtB clusters support only weak π-interaction with ethylene, favoring ethylene desorption, whereas pure Pt clusters can also bind ethylene in a strong di-σ fashion, activating it for dehydrogenation. Importantly, this property was observed for the entire thermally-accessible ensemble of cluster isomers, assessed via global optimization and Boltzmann statistics.[36, 37] A concern regarding boration, however, is that because it substantially weakens the alkene (i.e., ethylene) binding to the clusters, it might also increase the activation barriers for the desired alkane-to-alkene conversion.

Recently we developed an atomic layer deposition (ALD) approach that allows Sn to be selectively deposited on size-selected $Pt_n$ clusters on oxide thin film supports. The clusters act as seeds to initiate the ALD process, thus $Sn_mPt_n$ clusters of well-defined size and stoichiometry can be prepared.[38] For $Pt_n$ on $SiO_2$ the Sn deposition process was highly selective, depositing Sn on Pt cluster sites at least 50 times more efficiently than on the $SiO_2$ support sites. Here we apply this Sn deposition method to modifying model $Pt_n$/alumina catalysts and examining the effects on ethylene binding and suppression of carbon deposition. Alumina was chosen for the initial catalysis study of $Pt_nSn_m$ clusters, to allow comparisons with the $Pt_n$/alumina and $Pt_nB_m$/alumina systems, where detailed experimental and DFT results are available.

The resulting model Sn-Pt catalysts are studied using *in situ* surface chemistry methods to characterize their stoichiometries, electronic properties, morphology, and activity for ethylene binding and dehydrogenation. DFT is used to model the structures, energetics, electronic properties of both model SnPt supported clusters, and binding of ethylene to the clusters. Finally, we compare the effects of Sn-modifying the support, prior to deposition of the size-selected $Pt_7$ clusters, vs. modifying the samples after $Pt_7$ deposition. We focus the experimental work here on $Pt_7$ because this was identified as a particular active cluster size for $Pt_n$/alumina, and also coked/deactivated rapidly. The DFT work focused on a smaller, more tractable size: $Pt_4Sn_3$/alumina.

**METHODOLOGY**

**Experimental Methodology**

The instrument used in the experiments has been described previously.[33, 34, 39] Briefly, a laser vaporization/supersonic expansion source was used to produce $Pt_n^+$ clusters that were collected by a set of quadrupole ion guides, mass-selected by a quadrupole mass filter, and then guided by a final quadrupole into the ultrahigh vacuum (UHV) system ($1.5 \times 10^{-10}$ Torr), where they were deposited on a planar alumina support to form a model catalyst. The support is mounted via heater wires to a cryostat, and its temperature can be controlled between 120 K and 2100 K.

The alumina support was a thin film grown on a Ta (110) single crystal by the following procedure. The Ta single crystal was cooled below 130 K and then annealed to 2100 K for 5 minutes to desorb Pt and $Al_2O_3$ from the previous experiment, with the cleanliness checked by X-ray photoelectron spectroscopy (XPS). The Ta crystal was then transferred to a UHV antechamber, where Al was evaporated onto it in $5 \times 10^{-6}$ Torr of $O_2$, maintaining the crystal at 970 K. According to Chen and Goodman,[40] thin films of alumina grown on Ta(110) by this method have a distorted hexagonal lattice similar to the (0001) face of α-alumina or (111) face of γ-alumina. The film thickness was determined by modeling[41] the measured Al/Ta XPS ratio using photoemission cross sections and asymmetry parameters from Yeh and Lindau[42], and electron effective attenuation lengths (EALs) calculated with the database program of Powell and Jablonski.[43] Thicknesses were in the range from 4.7 to 5.7 nm, which we previously found to be thick enough to give thickness-independent chemistry for supported catalyst clusters, but thin enough to avoid charging during deposition or XPS.

For cluster deposition, the $Al_2O_3$ support was positioned just behind a 2 mm diameter exposure mask and then flashed to 700 K to remove adventitious adsorbates. As the sample cooled, $Pt_n^+$ deposition was started as the sample temperature reached 300 K, with the coverage monitored via the neutralization current. The deposition continued as the sample cooled to 120 K and was terminated when the coverage reached $1.5 \times 10^{14}$ Pt atoms/cm$^2$, corresponding to ~0.1 of a close-packed Pt monolayer (0.1 ML). In these experiments with $Pt_7$ clusters, $2.15 \times 10^{13}$ clusters were deposited *per* cm$^2$.

Sn was incorporated into the samples using an ALD process developed by Gorey *et al.*[38] to prepare $Pt_nSn_m$ alloy clusters on silica supports. Here, a freshly deposited $Pt_7/Al_2O_3$ sample was transferred into the UHV antechamber, then exposed to $H_2$ (6000 L), followed by $SnCl_4$ (24 L), followed by a final $H_2$ dose (6000 L), all at a sample temperature of 300 K. Gorey *et al.* showed that H selectively adsorbed on Pt atoms in the clusters, and that $SnCl_4$ preferentially reacted with hydrogenated Pt sites, losing $HCl_{(g)}$ and binding $SnCl_x$ to the clusters. During the second $H_2$ dose, additional HCl desorbed, leaving <10% of the Cl originally associated with the adsorbed Sn on the surface. This remaining Cl was desorbed as HCl by flashing the samples to 700 K. For consistency with the Gorey nomenclature, the sequence of gas exposures will be referred to as the "full treatment", i.e., a $Pt_7$/alumina sample thus treated will be denoted $FT/Pt_7/Al_2O_3$. After heating, such a sample would be denoted $heat/FT/Pt_7/Al_2O_3$. For comparison, samples were also prepared by first exposing a freshly grown alumina/Ta(110) substrate to the full $H_2/SnCl_4/H_2$ treatment, then heating the substrate, then depositing 0.1 ML of $Pt_7$. These samples will be referred to as $Pt_7/heat/FT/Al_2O_3$. Finally, control experiments were done on Pt-free alumina given the full treatment, and these are denoted $FT/Al_2O_3$. Mg Kα XPS was used to quantify the amount of Sn deposited on the samples. Samples were characterized as-deposited and after being flashed to 700 K to desorb residual H and Cl.

Low energy He$^+$ scattering (ISS) was used to probe the surface morphology of the samples. A 0.2 µA He$^+$ beam impinged on the surface at 1 keV with a 45° angle of incidence and scattered He$^+$ was detected along the surface normal. The He$^+$ beam caused some sputtering and other sample damage, and, to avoid damaging effects, ISS was performed either at the end of other experiments or on separately prepared samples. To compensate for any run-to-run variations in the He$^+$ flux, the Pt and Sn ISS peak intensities were normalized to the total scattered He$^+$ intensity.

Adsorbate binding and reactivity of the samples were probed by temperature-programmed desorption (TPD). In the TPD experiments, the sample temperature was held at 150 K while it was dosed with 5 L of $C_2D_4$. This dose temperature was chosen as being low enough to allow the cluster binding sites to be saturated while minimizing adsorption on the alumina support. After dosing, the sample was cooled to 130 K, and then ramped to 700 K at 3 K *per* second. During the heat ramp, molecules desorbing from the surface were monitored by a differentially pumped mass spectrometer that views the sample through the ~3 mm aperture in a skimmer cone. The ion signals are used to estimate the number of neutral molecules of each type desorbing from the sample, using a calibration procedure discussed elsewhere, and described in more detail in the SI. The absolute uncertainty of the resulting desorption numbers is estimated to be 50%.

**Computational Methodology**

Calculations were performed as discussed in previous work[34]. Projector-augmented wave method[44] and PBE functional[45] were used within the Vienna ab initio simulation package (VASP).[46-49] Plane-wave kinetic energy cut-off of 400.0 eV and convergence criteria of $10^{-6}$ eV for electronic (SCF) steps were employed. Note that the bottom half of the slab was kept fixed during the relaxation. Geometric relaxation was performed until forces on each atom were smaller than 0.01 eV/Å. In addition, Gaussian smearing with the sigma value of 0.1 eV was used. The hexagonal symmetry alumina thin film used in the experiments was modeled computationally as an α-alumina (0001) surface. The cell parameters of $a$ = 4.807 Å and $c$ = 13.126 Å were obtained by optimizing the bulk structure with stringent electronic and geometric convergence criteria, the details of which can be found here.[33] Moreover, the slab was modeled as a (3 × 3) unit cell with a vacuum gap of 15 Å. Since a fairly large super cell was used in this study, a k-point grid of 1 × 1 × 1 centered at the Γ-point was instituted.

In order to produce initial geometries for clusters on the $Al_2O_3$ surface, we used our in-house parallel global optimization and pathway toolkit (PGOPT), which automatically generates initial structures based on the bond length distribution algorithm (BLDA).[50] Each structure was optimized with DFT, and duplicates were filtered out thereafter. Finally, partial charges on each atom in the cluster were obtained using the Bader charge scheme.[51-54]

## RESULTS AND DISCUSSION

### A. XPS

Mg Kα XPS studies of these samples were done over a period when the spectrometer sensitivity varied significantly due to aging of the electron detector, thus the raw intensities must be corrected to allow sample-to-sample comparisons. If our alumina film supports were more than ~3 times thicker than the Al 2p electron effective attenuation length (EAL), the intensity of the Al 2p peak would be independent of film thickness, and the Pt and Sn intensities could be corrected by simply normalizing them to the Al 2p intensities. The Al 2p EAL in alumina, calculated using the database program of Powell and Jablonski,[43] is 2.1 nm, while our alumina films had thicknesses varying from 4.7 to 5.7 nm – not quite thick enough to give thickness-independent Al 2p signal.

Therefore, the sensitivity correction must be done in two steps, as summarized in Table S1: 1. The film thickness for each sample was determined from the Al/Ta XPS ratio as described above, and then the EAL was used to correct the measured Al 2p intensities to what they would have been if all the samples had films with identical thickness, chosen as 4.7 nm because several of the samples did have this thickness. 2. The thickness-corrected Al 2p intensities were then used to calculate the factors needed to correct for the declining sensitivity of the electron detector, and these factors (last column, Table S1) were then used to scale the Pt and Sn intensities in Fig. 1. The assumption in this correction process is that the Al 2p intensities are unaffected by the deposition of Pt and/or Sn on the alumina films. Given that the Pt and Sn coverages are small, on the order of 0.1 ML, any attenuation of Al 2p signal should be in the few percent range, and the resulting error is within the uncertainties due to low XPS signal level for Pt and Sn, which we estimate to be $\pm$ 15% for Pt, and $\pm$ 5% for Sn.

The background-subtracted and sensitivity-corrected XPS results for all the samples are shown in Fig. 1, along with the fits used to extract the integrated intensities. The integrated intensities and intensity ratios are given in Table 1, and the binding energies (BEs) for the Pt $4d_{5/2}$ and Sn $3d_{5/2}$ peaks are in Table S2.

The left column of Fig. 1 shows the Pt 4d XPS for all the samples. The first point to note is that the intensities are similar, as expected, because all samples had identical coverages of $Pt_7$. Some variation in the Pt intensity is expected because different samples have Sn deposited on top of the Pt, or have been heated which may change the cluster structure. Therefore, the attenuation of Pt photoelectrons should vary from sample to sample. Note that the Pt intensities are quite low, because the Pt coverage is low, and also because we monitored the Pt 4d XPS, rather than the more intense 4f since the 4f has substantial interference from Al 2p. As a result, the estimated uncertainties in the corrected integrated Pt intensities given in Table 1 are ~±15%.

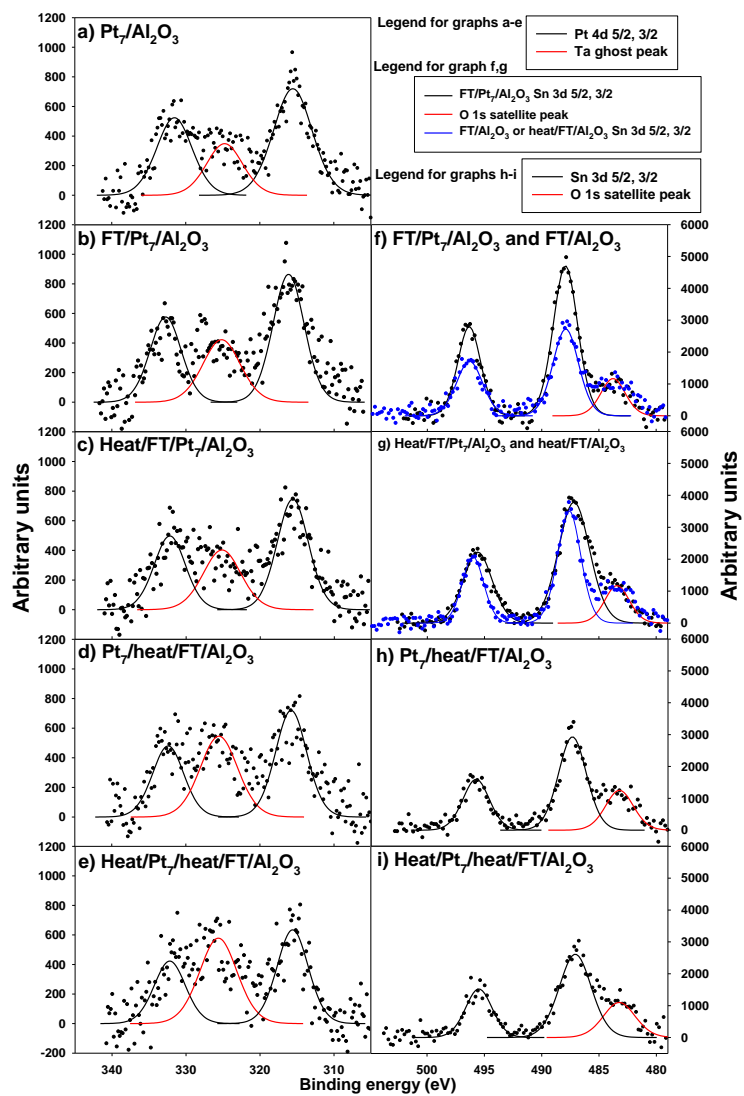

Figure 1. Left column: Pt 4d XPS region spectra. Peaks for Pt $4d_{5/2}$ and $4d_{3/2}$ are indicated with black fitted curves, the red curve indicates a Ta 4s ghost peak. Right column: Sn 3d XPS region. Peaks for Sn $3d_{5/2}$ and $3d_{3/2}$ are indicated with black or blue curves. The red curve indicates an Mg $K_\beta$ O 1s satellite peak. "FT" refers to the full $H_2/SnCl_4/H_2$ treatment.

As shown in Table S2, the Pt $4d_{5/2}$ BE for $Pt_7/Al_2O_3$ is 315.4 eV, compared to 314.6 eV for bulk Pt.[55] This 0.8 eV shift to higher BE is attributed to the effects of the small cluster size on stabilization of the photoemission final state.[56, 57] After the full Sn deposition treatment, and before heating ($FT/Pt_7/Al_2O_3$), the Pt 4d BE is shifted to 316.2 eV, presumably due to the combined effects of adsorbed Sn, Cl, and H, which may affect both the Pt oxidation state, and the final state stabilization.  After heating to drive off H and Cl, the BE dropped back to 315.5 eV.  When the alumina support was given the full Sn treatment and heated prior to $Pt_7$ deposition, the Pt 4d BE was 315.7 eV, and after heating to 700 K, the BE also dropped back to 315.5 eV.

The Sn XPS provides insight into both the amount of Sn present and its chemical state. First, consider the unheated samples. Fig. 1f compares the Sn XPS for the $FT/Al_2O_3$ and $FT/Pt_7/Al_2O_3$ samples. By the integrated Sn intensities (Table 1), a substantial Sn signal is observed in the absence of $Pt_7$, but that signal increases by a factor of ~1.6 when $Pt_7$ is present. Given that the $Pt_7$ is present at only 0.1 ML coverage, this implies that Sn deposition is roughly 6 times more efficient on Pt sites, compared to alumina sites.  In comparison to Sn deposition by the same process on $Pt_n/SiO_2$, the selectivity is lower here, largely because the $SiO_2$ was much more inert toward the $H_2/SnCl_4/H_2$ treatment process.  The intensities are analyzed below to extract numbers of Sn atoms deposited.

Table 1. Corrected intensities of Sn 3d, Pt 4d, Al 2s and values of their ratios. "FT" refers to the full $H_2/SnCl_4/H_2$ treatment. The bold font highlights the Sn and Pt to Al ratios.

| | corrected intensities | | | | | | |
|---|---|---|---|---|---|---|---|
| | $Al_2O_3$ thickness (nm) | Sn | Pt | Al | **Sn/Al** | **Pt/Al** | **Sn/Pt** |
| **$Pt_7$/ $Al_2O_3$** | 5.3 | | 8051 | 54972 | | **0.15** | |
| **FT/$Al_2O_3$** | 5.2 | 12771 | | 54972 | **0.23** | | |
| **Heat/FT/$Al_2O_3$** | 5.2 | 14074 | | 54972 | **0.26** | | |
| **FT/$Pt_7$/$Al_2O_3$** | 5.7 | 20006 | 7812 | 54972 | **0.36** | **0.14** | 2.56 |
| **Heat/FT/$Pt_7$/$Al_2O_3$** | 5.7 | 20413 | 6741 | 54972 | **0.37** | **0.12** | 3.03 |
| **$Pt_7$/heat/FT/$Al_2O_3$** | 4.7 | 13889 | 6505 | 54972 | **0.25** | **0.12** | 2.14 |
| **Heat/$Pt_7$/heat/FT/$Al_2O_3$** | 4.7 | 13865 | 5735 | 54972 | **0.25** | **0.10** | 2.42 |
| **$Pt_7$/heat/FT/$Al_2O_3$ post 6 TPDs** | 4.7 | 14841 | 5889 | 54972 | **0.27** | **0.11** | 2.52 |

After heating, the Sn intensity for FT/$Al_2O_3$ increased by ~10%, suggesting that the Cl and H present after the full treatment were attenuating the Sn signal, and demonstrating no significant Sn desorption occurring at 700 K.  For the FT/$Pt_7$/$Al_2O_3$ sample, the Sn intensity also increased, but only by ~2%.  The smaller increase suggests that the post-heating Sn binding geometry is such that there is still some attenuation of the Sn signal, as might occur if some Sn is alloyed with the Pt so that it is not in the surface layer.  The ISS and theory discussed below address this point.

For $Pt_7$/heat/FT/$Al_2O_3$ sample, we might expect that the Sn intensity should be similar to that of the Ft/$Al_2O_3$ sample since the Sn treatment was applied to Pt-free alumina in both cases. The intensity was actually ~10% higher.  Note, however, that the alumina film thicknesses were ~0.5 nm different in the two samples, which may influence Sn deposition.  After heating, the Sn intensity for the heat/$Pt_7$/heat/FT/$Al_2O_3$ sample was essentially unchanged.

The Sn 3d$_{5/2}$ peak for FT/$Pt_7$/$Al_2O_3$ is broadened and shifted to lower binding energies at 487.9 eV, when $Pt_7$ was present, and also at 487.9 eV for FT/$Al_2O_3$.  For the samples prepared by Sn deposition on alumina and heating, prior to $Pt_7$ deposition, the Sn BEs are shifted to lower energy (487.1 to 487.3 eV), but in all cases, the BEs are substantially higher compared that the BE reported for Sn bulk metal (~485 eV), and closer to the values reported for various oxygenated or halogenated Sn compounds (up to ~488 eV).[55] The high Sn binding energies, thus, suggest that Sn is in an oxidized state on the samples.  DFT results regarding this question are presented below.

The Sn/Pt stoichiometry of the samples can be determined from the corrected integrated intensity ratios in Table 1.   For samples containing 0.1 ML of Pt in the form of $Pt_7$, we can use the Sn/Pt ratio, along with Pt 4d and Sn 3d photoemission cross sections ($\sigma_{Sn}$, $\sigma_{Pt}$), obtained from the work of Yeh and Lindau.[42] Taking advantage of the fact that both Pt and Sn are present only in the surface of the sample, and assuming the attenuation of the associated photoelectrons is negligible, the Sn/Pt coverage ratio ($X_{Sn}/X_{Pt}$) is

Table 2. The ratio of the number of Sn atoms to Pt atoms and the number of Sn atoms per 10 nm$^2$. "FT" refers to the full $H_2$/SnCl$_4$/$H_2$ treatment.

| | Sn atoms/Pt atoms | Sn atoms/10 nm$^2$ |
|---|---|---|
| FT/Al$_2$O$_3$ | | 15.0 |
| Heat/FT/Al$_2$O$_3$ | | 16.5 |
| FT/Pt$_7$/Al$_2$O$_3$ | 1.62 | 24.3 |
| Heat/FT/Pt$_7$/Al$_2$O$_3$ | 1.65 | 24.8 |
| Sn bound to Pt from FT/Pt$_7$/Al$_2$O$_3$ and heat/FT/Pt$_7$/Al$_2$O$_3$ | 0.55 | 8.3 |
| Pt$_7$/Heat/FT/Al$_2$O$_3$ | 1.09 | 16.4 |
| Heat/Pt$_7$/Heat/FT/Al$_2$O$_3$ | 1.09 | 16.4 |

related to the XPS intensity ratio ($I_{Sn}/I_{Pt}$) by the following relation:

$$X_{Sn}/X_{Pt} = (I_{Sn}/\sigma_{Sn})/(I_{Pt}/\sigma_{Pt}) \qquad (1)$$

The coverage of Pt-free samples can be estimated by comparison of the Sn intensities for the Pt$_7$-containing and Pt-free samples, and the results are summarized in Table 2.

For the FT/Al$_2$O$_3$ sample, the Sn intensity before heating gives an estimated Sn coverage of 15.0 Sn atoms *per* 10 nm$^2$ of the Al$_2$O$_3$ substrate, or 1.5 x 10$^{14}$ Sn/cm$^2$, however, at this stage the Sn is still covered with adsorbed Cl (and possibly H), which attenuates the XPS.  After heating to desorb these species, the Sn XPS intensity increased by ~10%, leading to an estimate of the Sn coverage of 16.5 Sn/10 nm$^2$, or 1.65 x 10$^{14}$ Sn atoms/cm$^2$. This is comparable to the amount of Pt deposited in the Pt$_7$-containing samples, 1.5 x 10$^{14}$ Pt atoms/cm$^2$, i.e., the non-selective Sn deposition on the alumina film gives coverage comparable to the Pt coverage in the cluster-containing samples. Sn deposition by the same approach on SiO$_2$ substrates was found to deposit only 3.5 Sn atoms/10 nm$^2$, i.e., roughly 1/5$^{th}$ the number on the alumina support. We attribute the much larger amount of non-selective Sn deposition on alumina to a larger density of defects or other sites capable of supporting Sn deposition.

When Pt$_7$ is present on alumina, ~60% more Sn was deposited, and for the FT/Pt$_7$/Al$_2$O$_3$ sample, the Sn/Pt ratio was 1.62, corresponding to 24.3 Sn atoms and 15 Pt atoms/10 nm$^2$, and when that sample was heated, the Sn intensity increased ~2% (due to desorption of Cl and H), but the Pt decreased by ~10%. No Pt desorption is observed, and thus, we attribute the decrease to the effects of alloying the clusters with Sn, which may include both the Sn atoms initially deposited on the Pt clusters, as well as Sn initially deposited on the alumina, that diffuses and binds to the clusters when the samples were heated. The ISS experiments, below, provide additional insight into this point. The increase in the Sn XPS intensity after heating implies 24.8 Sn atoms/10 nm$^2$, and if we assume that the Pt coverage is unchanged, then the Sn/Pt ratio after heating is 1.65.

The question is how many of the 24.8 Sn/10 nm$^2$ were deposited directly on the clusters, and how many become associated with the clusters after heating, as opposed to binding at defects on the alumina support. We can estimate the number of Sn depositing on clusters by assuming that the presence of 0.1 ML of Pt$_7$ on the surface does not affect the deposition of Sn on the surrounding Al$_2$O$_3$. In that limit, we can estimate that 24.8 – 16.5 = 8.3 additional Sn atoms *per* 10 nm$^2$ deposited in association with the Pt$_7$ clusters, corresponding to 0.55 Sn/Pt atom, or ~4 Sn atoms/Pt$_7$ cluster. To the extent that the presence of Pt$_7$ blocks Sn deposition on the alumina, or captures Sn initially impinging on the alumina, then the number of Sn atoms associated with the $Pt_7$ would be higher, up to 11.6 Sn/$Pt_7$ if all the Sn diffused and bound to the clusters. Note that, for the $Pt_n$/$SiO_2$ system, where little non-selective Sn deposition is observed on the $SiO_2$, the estimated Sn/Pt stoichiometry was ~6 Sn/$Pt_7$.

If similar analysis is applied to the $Pt_7$/heat/FT/alumina and heat/$Pt_7$/heat/FT/alumina samples, the conclusion is that the number of Sn depositing non-selectively on the alumina support was ~16 Sn atoms/10 $nm^2$, with essentially no difference between calculations based on the two spectra. The coverage of $Pt_7$ deposited on top of the Sn-modified surface is 2.14 $Pt_7$ clusters/10 $nm^2$, corresponding to 15 Pt atoms.

The important questions regarding these model catalysts are:

1. What effect does Sn modification have on the activity, selectivity, and stability of the catalysts?

2. Does Sn initially deposited on the alumina support migrate and bind to the Pt clusters, either as-deposited or when heated?

3. Is there a significant difference between the effects of Sn deposited initially on the alumina vs. on the Pt clusters?

**TPD Results**

Figure 2 summarizes the results of TPD experiments in which the Sn-modified model catalysts were saturated with $C_2D_4$ at 150 K, and then the numbers of desorbing $C_2D_4$ and $D_2$ molecules were monitored during 3 K/sec heat ramps to 700 K. A series of 6 TPD experiments was done for each sample, each preceded by a 150 K $C_2D_4$ dose. The calculated numbers of each type of molecule desorbing *per* cluster are tabulated in Table S3. For comparison, TPD data for $Pt_7$/$Al_2O_3$ measured under identical conditions by Baxter *et al.*[33] is plotted in the top row. Also shown in each frame of the figure is the desorption observed from a Pt and Sn-free $Al_2O_3$ thin film sample.

As discussed in the Introduction, for our purposes, the interesting questions pertaining to the catalysis are: 1. How is the binding energy of $C_2D_4$ affected by the presence of Sn either on the clusters or the support? 2. How does Sn deposited on the clusters vs. on the support affect the branching to dehydrogenation? 3. In repeated TPD runs, how do the catalysts evolve, due to processes such as sintering or carbon deposition?

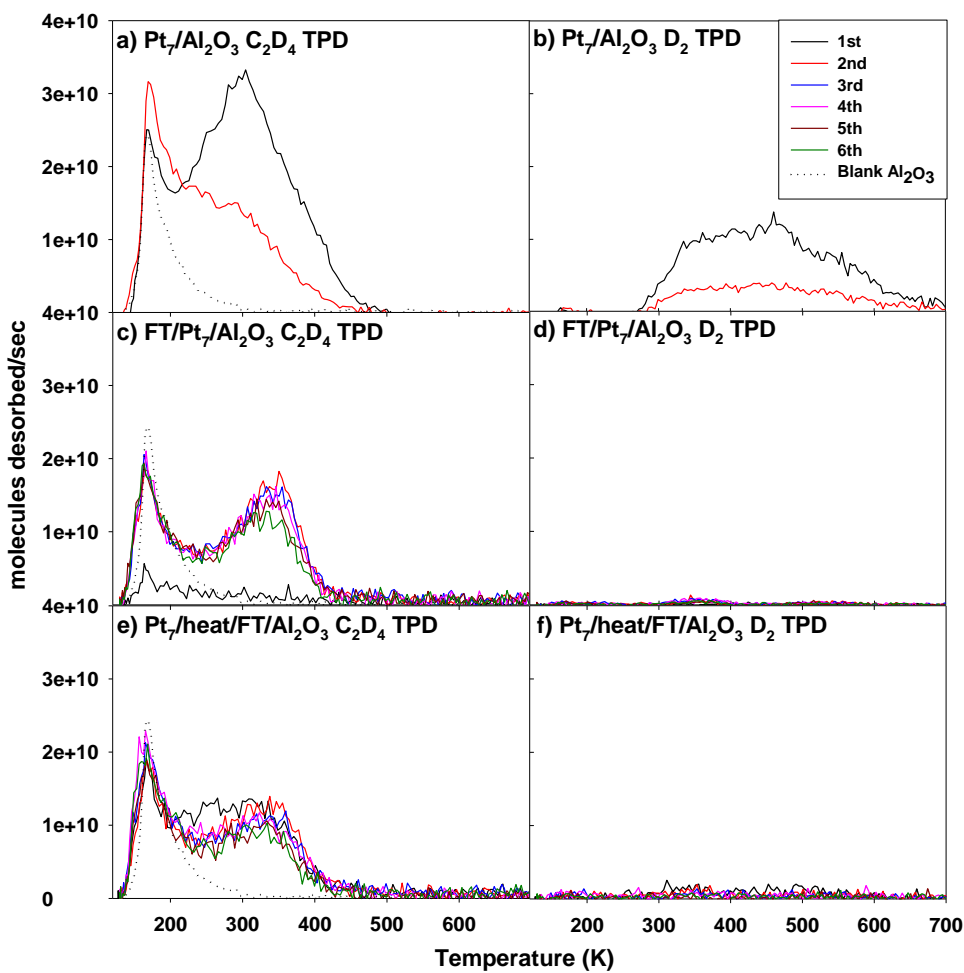

Figure 2. $C_2D_4$ and $D_2$ TPD data for the samples: a) $Pt_7/Al_2O_3$ $C_2D_4$ TPD, b) $Pt_7/Al_2O_3$ $D_2$ TPD, c) $FT/Pt_7/Al_2O_3$ $C_2D_4$ TPD, d) $FT/Pt_7/Al_2O_3$ $D_2$ TPD, e) $Pt_7/heat/FT/Al_2O_3$ $C_2D_4$ TPD, f) $Pt_7/heat/FT/Al_2O_3$ $D_2$ TPD. "FT" refers to the full $H_2/SnCl_4/H_2$ treatment.

It can be seen that for Sn-free $Pt_7/Al_2O_3$ (top row), the 1st TPD run has a low-temperature $C_2D_4$ desorption feature that matches that from the $Al_2O_3$ support, implying that this feature results from $C_2D_4$ desorbing from the support. Because the intensity of this peak is rather low, corresponding to only a small fraction of a monolayer of $C_2D_4$ desorbing, it is attributed to $C_2D_4$ bound at defects or other strong-binding sites on the surface.

There is also a higher temperature feature clearly associated with $C_2D_4$ desorbing from the $Pt_7$ clusters. No $D_2$ desorption is observed from the $Al_2O_3$ support, but in the 1st run on $Pt_7/Al_2O_3$, a large $D_2$ desorption feature is observed, implying that a significant fraction of the $C_2D_4$ undergoes dehydrogenation. We previously showed that no acetylene desorbs from this surface,[33] and thus, the $D_2$ signal is associated with carbon deposition – one C atom per $D_2$ molecule.

In the 2nd TPD run, the intensity of the Pt-associated $C_2D_4$ desorption decreased substantially, as did the amount of $D_2$ desorption. These changes were attributed mainly to the effects of the poisoning of cluster-associated binding sites by carbon deposition, observable by post-TPD C 1s XPS. Sintering probably also contributed to the deactivation process, although ISS and comparison of the effects of heating with and without $C_2D_4$,[33, 35] suggested that sintering during a single TPD run had only a modest effect for $Pt_7$ on the relatively defective $Al_2O_3$ thin film support. Only two TPD runs were reported in these experiments because the deactivation was so fast.

For the $FT/Pt_7/Al_2O_3$ sample, the 1st $C_2D_4$ TPD experiment was done on the unheated sample, and as shown in the 2nd row of the figure, little $C_2D_4$ desorption and no $D_2$ is observed. This implies that the Pt cluster sites were fully blocked by adsorbed H, Cl, and Sn and that many of the $Al_2O_3$ defect sites were also blocked, such that almost no $C_2D_4$ adsorbed during the 150 K dose. During the first TPD run, HCl desorption is observed, and there probably is also $H_2$ desorption above 300

K, although our mass spectrometer background at mass 2 is too large to allow direct observation of the relatively small signal for $H_2$ desorption from a coverage of clusters.

During the $2^{nd}$ and subsequent TPD runs, substantial signals are observed for $C_2D_4$ desorption, indicating that HCl (and $H_2$) desorption during the $1^{st}$ run exposed Pt sites. There is a low-temperature feature with about $1/3^{rd}$ the intensity of that seen for $Al_2O_3$, which we attribute to $C_2D_4$ desorbing from alumina defect sites that were unblocked during the $1^{st}$ TPD. The high temperature feature has about 46% the intensity of the high temperature feature observed for $Pt_7/Al_2O_3$ in the $1^{st}$ TPD, suggesting that a significant fraction of the exposed Pt binding sites are still blocked, presumably by Sn. Critically, however, little if any $D_2$ desorption is observed. Note also that the ethylene desorption is essentially unchanged in four additional TPD runs.

The bottom frame of the figure shows the results of depositing $Pt_7$ on pre-heated, Sn-modified $Al_2O_3$. As noted above, this results in the deposition of 16.4 Sn atoms/10 $nm^2$ – about $2/3^{rd}$ the amount present when Sn deposition is done after $Pt_7$ deposition. The $C_2D_4$ TPD results are actually quite similar to those observed when Sn is deposited after the clusters, with the exception that significant $C_2D_4$ desorption is observed in the $1^{st}$ TPD. That simply reflects the fact that the heat/FT/$Al_2O_3$ sample was heated to desorb H and Cl prior to $Pt_7$ deposition, so that binding sites on both the alumina support and the clusters were available to binding $C_2D_4$. Again, little $D_2$ desorption is observed, and the high-temperature $C_2D_4$ desorption is reasonably stable in repeated TPD runs.

There are several interesting points. Little $D_2$ is observed even in the $1^{st}$ TPD run, indicating that the $Pt_7$ cluster dehydrogenation properties were already strongly modified by the Sn, even though there was not Sn deposited directly onto the clusters. This suggests that some Sn may have diffused to the clusters, even though most of the clusters were deposited at cryogenic temperatures, and were never heated prior to the TPD run. It is also possible that the clusters deposited on or immediately adjacent to Sn on the surface, and obtained some Sn modification in this way. It can also be seen that there is a significant change in the $C_2D_4$ desorption between the 1st and 2nd TPD runs, with the amount desorbing between ~200 and 300 K dropping significantly. This suggests that during the heating accompanying the 1st TPD run, the clusters changed somehow, and either additional Sn diffused to the clusters, in addition, the cluster morphology is likely to have evolved during the TPD heating. ISS experiments to probe changes in morphology are discussed next.

It is interesting to compare the behavior shown here for sub-nano PtSn clusters/alumina to that seen for sub-nano borated $Pt_nB_m$ clusters/alumina,[35] and for ordered PtSn surface alloys.[27, 28] In all three cases, ethylene dehydrogenation is strongly suppressed, compared to pure $Pt_n$/alumina or Pt(111), however, there are several important differences. In the case of PtSn surface alloys, the ethylene desorption temperature dropped with increasing Sn coverage from ~285 K for pure Pt(111)[58] to 184 K for the $\sqrt{3}x\sqrt{3}$ R30° alloy. That alloy has a 1:2 Sn:Pt ratio in the surface layer, i.e., significantly lower than the Sn:Pt ratio in our clusters. For pure Pt clusters on alumina, the ethylene desorption occurs in a broad feature peaking near 300 K (Fig. 2a), and after boration, the desorption temperature drops substantially, peaking at cryogenic temperatures (overlapping the desorption from the alumina support). Thus, for both the extended PtSn alloy surfaces and $Pt_nB_m$/alumina, a substantial reduction in ethylene binding energy is at least partly responsible for the suppression of dehydrogenation – the ethylene desorbs before the onset of dehydrogenation. From the perspective of catalyzing selective alkane-to-alkene dehydrogenation, the reduction in alkene binding energy may be problematic, because that implies that the exothermicity of the $alkane_{adsorbed} \rightarrow alkene_{adsorbed}$ reaction is lower, which would tend to increase the activation energy.

In contrast, for the Sn-doped $Pt_7$/alumina clusters, the ethylene desorption feature remained above room temperature, with peak desorption temperature actually *increasing* relative to that for Sn-free Pt clusters. Nonetheless, dehydrogenation and the accompanying carbon deposition are suppressed, indicating that there must be another mechanism operating. A similar effect was predicted for supported PtGe, so far purely from theory.[59]

Another interesting difference is that in the PtSn surface alloys the saturation ethylene coverage was unaffected by Sn surface concentration, but in both borated and Sn-alloyed $Pt_n$/alumina, the saturation coverages decreased significantly, by on the order of 50% in both cases. This difference may simply reflect a lower Sn:Pt ratio in the surface layer of the extended alloys (Sn:Pt $\leq$ 1:2) compared to the levels of boron (~1:1 B:Pt) and tin in the cluster experiments. As noted, the additional Sn deposited when $Pt_7$ was present corresponded to only 4 $Sn/Pt_7$, but the experiments where $Pt_7$ was deposited after Sn-treating the alumina support suggest that Sn initially on the alumina support diffuses to the clusters, raising the Sn:Pt ratio. Campbell reviewed chemisorption of small molecules on numerous bimetallic surface alloys and overlayer systems,[60] and also discussed the weaker binding of small molecules (NO and CO) on the exothermic PtSn alloy that resulted in significant decrease (>100 K) of desorption temperatures.[61] For ethylene adsorption, a variety of different effects have been observed, ranging from simple site blocking to strong electronic effects on binding energies. In our clusters, Sn (and B) reduce the saturation ethylene coverage, presumably due to blocking of Pt sites. In addition, however, boration reduces the negative charge on the Pt clusters. Electronic effects of Sn addition are discussed below in the context of our DFT results.

The other important point from a catalysis perspective is that the high-temperature $C_2D_4$ desorption feature for the Sn-treated clusters was essentially unchanged in six TPD runs, whereas the analogous feature for $Pt_7$/alumina decreased rapidly.  This enhanced stability in repeated TPD cycling is not unexpected, since there is substantial carbon deposition for pure $Pt_n$/alumina model catalysts under these conditions,[33] however, it also indicates that the Sn-treated clusters must be reasonably stable with respect to sintering.  This conclusion is consistent with calorimetric experiments by Anres *et al.* which found that bulk PtSn has a highly negative enthalpy of formation,[62] which would slow the sintering kinetics in Campbell's modeling of sintering kinetics.[63, 64]  Similarly, Liu and Ascencio reported DFT and MD simulations for clusters of a few hundred atoms, finding that PtSn clusters were more energetically favorable than pure Pt clusters[65] Results of our DFT calculations for sub-nano PtSn clusters are presented below.

**Low energy He+ scattering (ISS)**

Example raw ISS spectra for select samples are shown in Fig. S1.  More than 99% of $He^+$ impinging on surfaces either neutralizes or implants in the surface, and thus the ISS signal corresponds to the small fraction of $He^+$ ions that survive and backscatter along the surface normal.[66]  Peaks in ISS result primarily from events in which an $He^+$ scatters from a single atom in the top-most layer of the surface.  Scattering from atoms deeper in the sample, or scattering involving multiple surface atoms is weak due to a combination of shadowing, blocking, and reduced ion survival probability (ISP), contributing mainly to the weak background observed at low $E/E_0$.  Thus, the intensities of ISS peaks are highly sensitive to sample structure, including cluster geometry and the presence of adsorbed species on top of cluster atoms. In the following, we present ISS data in the form of ISS peak intensities as a function of exposure to the $He^+$ beam, which slowly sputters the surface, providing some depth information. The intensities are normalized to the total scattered intensity in order to correct for any day-to-day changes in the $He^+$ beam flux.

Fig. 3 compares ISS sputter series results for $Pt_7/Al_2O_3$, $FT/Pt_7/Al_2O_3$, and $Pt_7/heat/FT/Al_2O_3$ samples. These measurements were done on a separate set of samples, but are directly comparable to the XPS and TPD data in Figs. 1 and 2. In each ISS experiment, the as-prepared sample was first exposed to two ISS scans at low (0.2 μA) $He^+$ flux. Then the sample was flashed to 700 K, driving the same processes that would have occurred in other 700 K heating experiments, including TPD. Finally, repeated ISS scans were used to measure changes in the peak intensities as the sample was slowly sputtered.

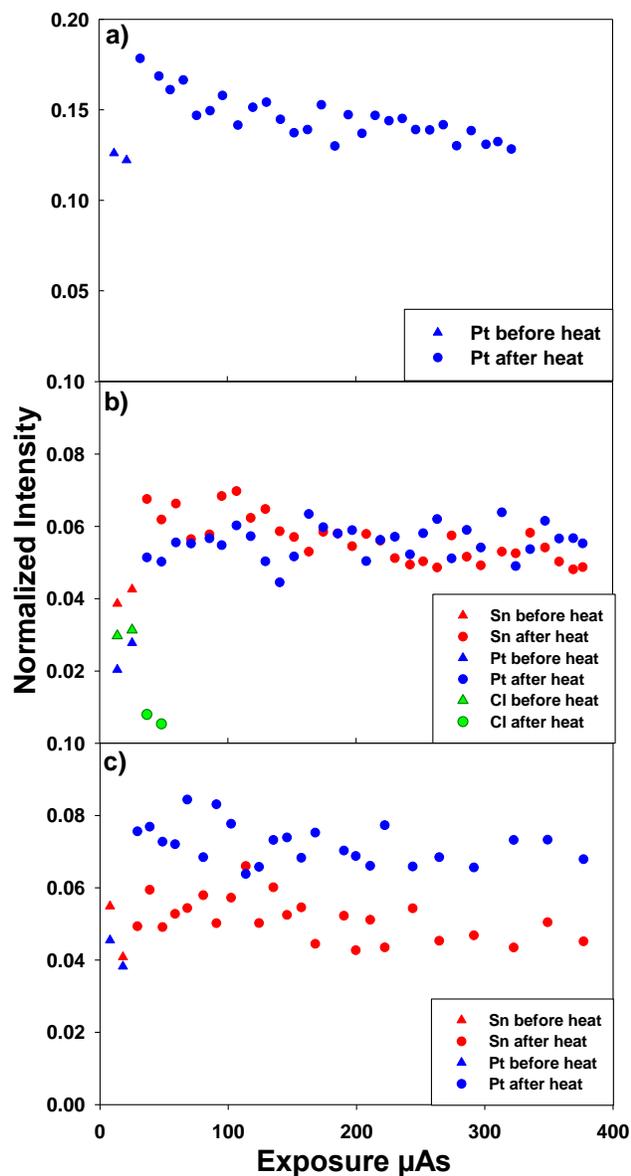

Figure 3. ISS intensities vs. $He^+$ exposure for: a) $Pt_7/Al_2O_3$, b) $FT/Pt_7/Al_2O_3$, c) $Pt_7/heat/FT/Al_2O_3$. In each experiment, 2 ISS scans were taken for the as-prepared sample, followed by a 700 K flash, and a series of ISS scans. "FT" refers to the full $H_2/SnCl_4/H_2$ treatment.

The top frame shows the Pt ISS intensity for a $Pt_7/Al_2O_3$ sample. The first two points are for the as-deposited sample, and the rest are for the sample after flashing to 700 K. The increase in Pt intensity after flashing is attributed to desorption of adventitious adsorbates, primarily CO, which adsorbs very efficiently on samples with sub-nanometer dispersed clusters due to substrate-mediated adsorption (reverse spillover). $Pt_7$ desorption was done as the samples cooled from room temperature to ~120 K. This is above the temperature for CO desorption from the alumina support,[67] however, the lifetime of CO on the alumina is sufficient to allow some diffusion, thus allowing CO initially landing on the alumina to bind at a nearby $Pt_7$ cluster. After heating, the Pt ISS intensity slowly decreased, consistent with the slow sputtering of Pt from the surface.

For the as-prepared $FT/Pt_7/Al_2O_3$ sample, both Sn and Pt ISS peaks are observed, but with much lower intensity compared to the Pt peak intensity in the $Pt_7/Al_2O_3$ sample. This low intensity reflects the presence of Sn bound to the Pt cluster and of Cl and H atoms adsorbed on Pt and Sn sites, attenuating ISS from the underlying atoms by a combination of shadowing, blocking, and reduced ISP. The presence of Cl is confirmed by the appearance of a weak peak for Cl ISS, as shown in the figure. After heating, the Cl intensity is attenuated nearly to the background level, and the Pt and Sn intensity both increase. Because ISP varies significantly for different elements, it is not possible to infer the ratio of Sn/Pt in the surface layer from the Sn/Pt intensity ratio. Note, however, that just after heating, the Sn intensity is higher than that for Pt, but it decays with time, while the Pt intensity is nearly constant (unlike the slow decline observed for $Pt_7/Al_2O_3$). This behavior indicates that the sputter rate for Sn is faster than that for Pt, which could reflect differences in the binding energy to the surface, but also would be expected if the heated clusters have Sn in the surface layer.

For the Pt$_7$/heat/FT/Al$_2$O$_3$ sample (bottom frame), the initial Sn ISS intensity is similar to that for the FT/Pt$_7$/Al$_2$O$_3$ sample, but the Pt intensity is twice as high as Pt in that sample. This is expected. The number of Sn atoms on the surface of both samples is high, but for FT/Pt$_7$/Al$_2$O$_3$, the Pt$_7$ has Sn, Cl, and H adsorbed on top, while for Pt$_7$/heat/FT/Al$_2$O$_3$, the Pt clusters are deposited on top after desorbing H and Cl, and thus should have less adsorbate coverage. Interestingly, however, the initial Pt intensity for Pt$_7$/heat/FT/Al$_2$O$_3$ is still only ~40% of that for the Pt$_7$/Al$_2$O$_3$ sample. This implies that even though the Pt clusters were deposited after the substrate was Sn-modified and heated, something is attenuating ISS from the Pt atoms. The most likely possibility is that some Sn atoms have aggregated to the clusters, which, given the ~0.1 ML coverage of Sn in these samples, is not surprising. If the Sn is initially present as isolated atoms, the average Sn-Sn separation is less than 1 nm, and there would be a high probability for Pt$_7$ clusters to deposit on top of, or very near to Sn atoms on the substrate. Again, cluster isomerizing to flatter shapes that favor Pt-Sn binding cannot be ruled out. Both suppositions that Sn already is binding to the clusters as-deposited would also explain why no significant D$_2$ desorption is observed in the 1$^{st}$ TPD on the Pt$_7$/heat/FT/Al$_2$O$_3$ sample – the clusters are already Sn-modified.

After heating, the Sn intensity in the Pt$_7$/heat/FT/Al$_2$O$_3$ sample was essentially unchanged, but the Pt intensity increased by nearly a factor of two. Some of this increase can be attributed to the desorption of adventitious CO (as in the Pt$_7$/Al$_2$O$_3$ data in the top frame), but there may also be some restructuring of the clusters.

## DFT Calculations

Because the Sn-modified Pt clusters are quite large, and the number of Sn atoms present is not well defined, it was not feasible to try to compute the structures. We previously reported on the structures of Pt$_7$/Al$_2$O$_3$ with and without 1-3 ethylene molecules adsorbed.[33] Here we report DFT calculations of structures, atomic charges, and ethylene binding behavior for a small model system, consisting of $Pt_4Sn_3$ clusters on an alumina support.

The global optimization was carried out, with the purpose of finding the low-lying minima that could be thermally populated at temperatures of the TPD. The three lowest energy structures found for $Pt_4Sn_3$/alumina are shown in Fig. 4, with the atomic charges, energies relative to the global minimum structure, and the Boltzmann populations estimated at 700 K. The $Pt_4Sn_3$ clusters were all singlets with a net negative charge indicating net electron transfer from the alumina support to the clusters (see Table S4). The structures also tend to have Pt binding to underlying Al atoms, while the Sn binds to both Al and O atoms on the surface.

The global minimum structure is nearly planar with all atoms exposed in the surface layer. The Sn atoms are all positively

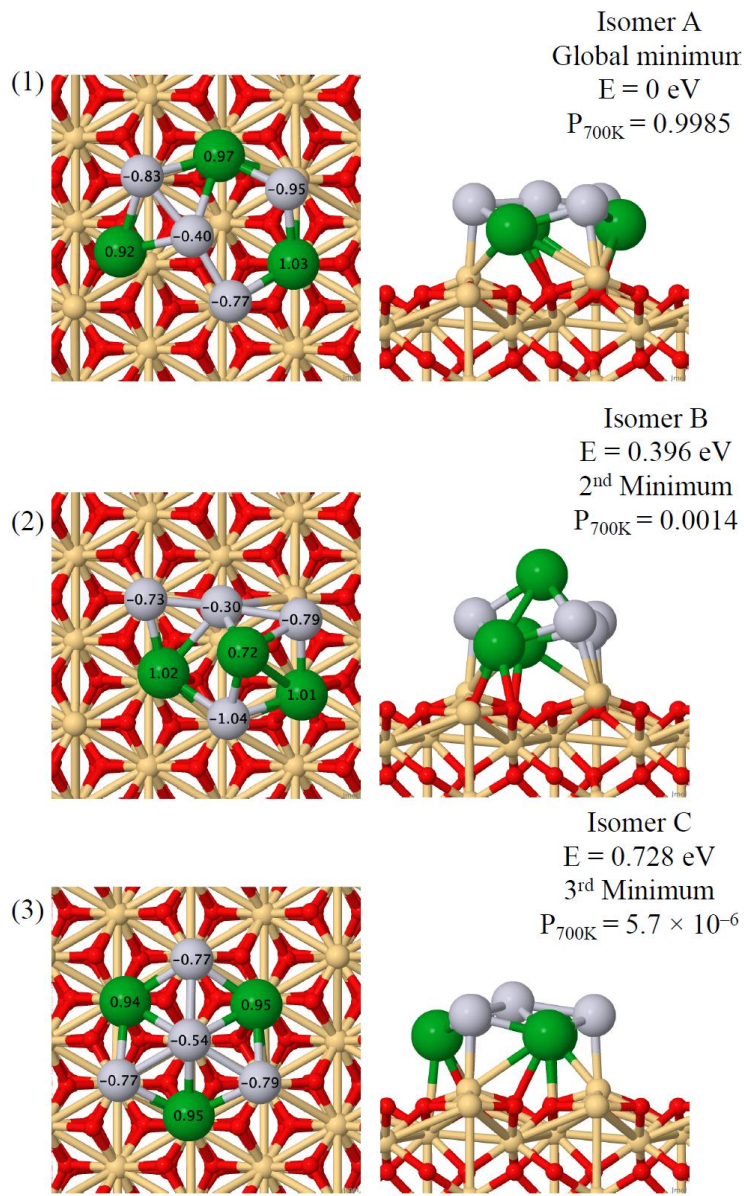

(1) Isomer A
Global minimum
E = 0 eV
$P_{700K}$ = 0.9985

(2) Isomer B
E = 0.396 eV
$2^{nd}$ Minimum
$P_{700K}$ = 0.0014

(3) Isomer C
E = 0.728 eV
$3^{rd}$ Minimum
$P_{700K}$ = 5.7 × $10^{-6}$

Figure 4. (1) Isomer A, the global minimum structure for $Pt_4Sn_3$/alumina. (2) Isomer B, the $2^{nd}$ minimum structure for $Pt_4Sn_3$/alumina. (3) Isomer C, the $3^{rd}$ minimum. E indicates the energy relative to the global minimum structure. $P_{700K}$ is the Boltzmann population at 700 K.

charged by nearly 1 e, and the Pt atoms are negative, with charges ranging from -0.4 e to -0.95 e, thus, there is substantial electron transfer from Sn to Pt.  The charge distribution is consistent with the high Sn 3d BEs measured by XPS, which indicates oxidized Sn.  The Sn and Pt atoms are mixed, rather than phase-separated.

The second minimum structure is substantially (0.396 eV) higher in energy than the global minimum, thus even at 700 K, its predicted population remains below 1 %.  Again, the Sn atoms are all positively charged by ~1e, and the Pt atoms are negative, with charges between -0.3 and -1.04 e.  This structure is three dimensional, with a Sn atom in the top layer.  The third isomer is a roughly hexagonal, quasi-planar structure with alternative Pt and Sn atoms around the periphery, but it is >0.7 eV above the global minimum and has near-zero population at 700 K.  Again, there is substantial Sn-to-Pt electron transfer. One important observation can be made about the global minimum that both Sn and Pt atoms are in contact with the support. This kind of structure could form either if Sn is deposited on alumina-supported Pt clusters and the cluster restructures upon heating, or if Sn is on the surface before the Pt cluster arrives and it mixes into Pt upon heating. Hence both or our aforementioned hypotheses of the effect of heating on the ethylene binding remain valid, and in fact they might be indistinguishable.

DFT calculations were also performed for ethylene binding to the three Pt$_4$Sn$_3$/alumina structures. A total of 10 structures were found to be within thermal reach, as shown in Fig. 5. Again, the Sn atoms are positively charged by roughly 1 e each, and the Pt atoms are negatively charged by -0.4 e to -1 e. There are several interesting points. Note that the two lowest energy structures for $C_2H_4$/$Pt_4Sn_3$/alumina are based on ethylene binding to Isomer C of $Pt_4Sn_3$/alumina, which is ~0.4 eV above the global minimum for the bare cluster. These two nearly isoenergetic structures both have $C_2H_4$ π-bonded to a single Pt atom around the edge of the nearly planar Isomer C structure. The 3rd and 4th isomers are built on Isomer A, i.e., the global minimum for $Pt_4Sn_3$/alumina, and again are nearly isoenergetic structures with $C_2H_4$ π-bonded to single Pt atoms at the edge of the cluster. The 5th isomer is again built on Isomer C of $Pt_4Sn_3$/alumina, but in this case has $C_2H_4$ bound on top of the cluster, with substantial distortion of the cluster. This 5th isomer is already 0.265 eV above the global minimum, and thus has <1% population weighting at 700 K. The other 5 structures are all build on Isomer B of $Pt_4Sn_3$/alumina, and have energies between 0.313 and 0.444 eV above the global minimum, hence having very small population weighting.

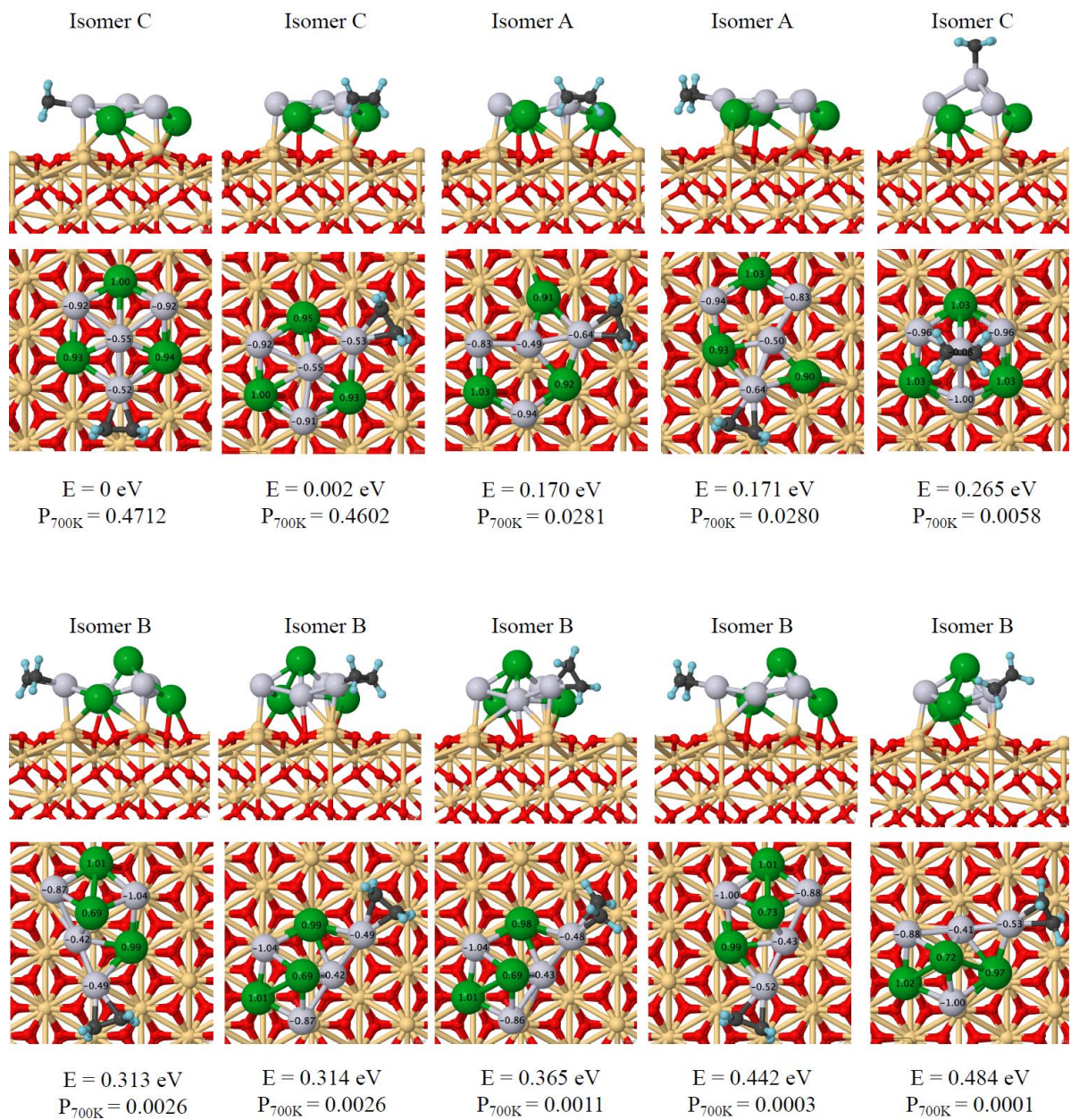

Figure 5. Structures of Pt$_4$Sn$_3$/alumina with one ethylene bound, with energies relative to the global minimum (GM), and thermal populations at 700 K. Note that the GM with ethylene is based on Isomer C of the bare cluster.

It is interesting to compare these structures to those found for ethylene bound to $Pt_n$/alumina previously. For the Sn-free Pt clusters, ethylene binds in both π and di-σ geometries (i.e., with each C atom bound to a different Pt atom, and $sp^2$ hybridization broken). Di-σ bonding has been associated with a propensity for dehydrogenation, thus the DFT structures are consistent with the observation of significant $D_2$ desorption from the Sn-free $Pt_7$/alumina sample. In contrast, all the structures found for $Pt_4Sn_3$/alumina have only π-bonding of ethylene, which is presumably a factor contributing to the observation that little $D_2$ is generated for Sn-treated $Pt_4Sn_3$/alumina. We confirmed the strong correlation between the mode of binding of ethylene and the barrier to dehydrogenation for other co-alloying elements, Si and Ge[59, 68]. The barriers associated with ethylene undergoing dehydrogenation from the π-mode is typically several times larger than that for the di-σ mode. In addition, upon dehydrogenation, a detached H atom needs to bind to Pt, and, when most of Pt atoms are separated by Sn, the sites for H atom arrival becomes unavailable. In contrast, for the preceding step of ethane dehydrogenation, there is no need for two adjacent Pt sites, as the reaction occurs on a single Pt atom[59]. Hence, in agreement with the experiment, theory shows that PtSn clusters should prevent dehydrogenation of alkanes beyond olefins.

**CONCLUSIONS**

We have shown that an ALD-like process can be used to modify $Pt_7$/alumina samples with Sn, both before and after Pt cluster deposition. Sn deposition is roughly 5 to 6 times more efficient on Pt sites than on the alumina support; however, it is far less selective for Pt sites than the $Pt_n/SiO_2$, studied previously. Modification of the alumina (e.g. use of thicker, less defective thin films) would presumably reduce the number of alumina sites that support Sn deposition, making the process more selective for Pt sites.

For these model catalysts, Sn modification almost completely suppresses carbon deposition, i.e. coking, and unlike the B-induced suppression of coking, it does it without lowering the ethylene binding energy. The effects of adding Sn to the support before depositing clusters, vs. depositing Sn after cluster deposition, are quite similar, and both the TPD and ISS results indicating that Sn present on the support prior to $Pt_7$ deposition must be binding to, and modifying the chemical properties of the Pt clusters, even at the low deposition temperatures.

The DFT results show that Sn and Pt mix intimately, with substantial Pt-to-Sn electron transfer. The clusters have exclusively spin-singlet isomers, and ethylene binds to Pt in PtSn in an exclusively π-mode. In contrast, Sn-free Pt clusters have both closed and open-shell structures, allowing ethylene to bind strongly in di-σ geometries that tend to undergo dehydrogenation, rather than desorption. The electronic, structural, and mechanistic effects of Sn closely resemble those for Si and Ge recently predicted theoretically.

Sn-modification clearly makes two highly desirable changes to sub-nano $Pt_n$/alumina catalysts for potential applications in fuel-rich, high-temperature applications. First, it almost completely suppresses dehydrogenation to deposit carbon on the surface. In addition, repeated TPD experiments show very little run-to-run change in the $C_2D_4$ desorption intensity or temperature dependence, indicating that the catalysts are also more thermally stable than Sn-free $Pt_n$ clusters on alumina. Both factors would be highly desirable in applications where the catalysts must be stable against coking and sintering. One such application is alkane-to-alkene dehydrogenation as an endothermic reaction for air vehicle cooling. The gas-phase Sn deposition process used could, presumably, be adapted to modifying small Pt clusters supported on alumina-coated lines and other fuel system parts.

**SUPPLEMENTARY MATERIAL**

Details of the XPS intensity correction process, example raw ISS spectra, and coordinates of the DFT structures can be found in the supplementary information.

**Acknowledgments**

This work was supported by the U.S. Air Force Office of Scientific Research under AFOSR Grants FA9550-19-1-0261 and FA9550-16-1-0141.

## Supporting Information

Table S1. XPS data calibration: The thickness of the alumina film was different for different samples, so the total number of electrons passing out of the film were at different values. We can assume that the thickness was all same at 4.7 nm, by calculating the number of electrons passing out of all the different films, then find the relative percentage decrease that is required for the Al intensities of original XPS data thicker films and that brings us the values for "Al peak intensities after correction to same thickness". We then assume that all these Al intensities should be of the same value as that of the $Pt_7$-$Al_2O_3$ sample and hence all the obtained XPS data were scaled up by percentages listed in the last column of table S1.

| | $Al_2O_3$ thickness (nm) | Al 2s peak intensities of original XPS data | Al 2s peak intensity after correction to same thickness | Percentage increased for XPS data normalization |
|---|---|---|---|---|
| $Pt_7/Al_2O_3$ | 5.3 | 56605 | 54972 | 0% |
| $FT/Al_2O_3$ | 5.2 | 43209 | 42141 | 30% |
| $Heat/FT/Al_2O_3$ | 5.2 | 45678 | 44549 | 23% |
| $FT/Pt_7/Al_2O_3$ | 5.7 | 47755 | 45688 | 20% |
| $Heat/FT/Pt_7/Al_2O_3$ | 5.7 | 54398 | 52043 | 6% |
| $Pt_7/heat/FT/Al_2O_3$ | 4.7 | 33667 | 33667 | 63% |
| $Heat/Pt_7/heat/FT/Al_2O_3$ | 4.7 | 32246 | 32246 | 70% |
| $Pt_7/heat/FT/Al_2O_3$ post 6 TPDs | 4.7 | 46671 | 46671 | 18% |

Table S2. Binding energies of Pt 4d 5/2 and Sn 3d 5/2 peaks taken from figure 1. Results were calibrated by using the binding energy, from O 1s in $Al_2O_3$, of 531.7 eV.

| Binding energy in XPS | Pt 4d 5/2 | Sn 3d 5/2 |
|---|---|---|
| $Pt_7/Al_2O_3$ | 315.4 eV | |
| $FT/Al_2O_3$ | | 487.9 eV |
| $Heat/FT/Al_2O_3$ | | 487.6 eV |
| $FT/Pt_7/Al_2O_3$ | 316.2 eV | 487.9 eV |
| $Heat/FT/Pt_7/Al_2O_3$ | 315.5 eV | 487.1 eV |
| $Pt_7/heat/FT/Al_2O_3$ | 315.7 eV | 487.3 eV |
| $Heat/Pt_7/heat/FT/Al_2O_3$ | 315.5 eV | 487.1 eV |

Table S3 Molecules desorbed per cluster for $C_2D_4$ and $D_2$ from experimental samples quantified from figure 2, assuming all molecules desorbed were from the clusters. $C_2D_4$ and $D_2$ molecules desorbed were quantified by taking the integral of the desorption peaks (range was 130 K to 500 K), then were divided by the number of Pt clusters in the area of analysis. Corresponding to 0.1 ML of Pt coverage and 0.0314 $cm^2$ of cluster spot size on the sample, $6.73 \times 10^{11}$ of clusters in the area of analysis were assumed for the TPD calculations.

| | 1st TPD | 2nd TPD | 3rd TPD | 4th TPD | 5th TPD | 6th TPD |
|---|---|---|---|---|---|---|
| $Pt_7/Al_2O_3$ | | | | | | |
| $C_2D_4$ | 8.64 | 5.44 | | | | |
| $D_2$ | 4.30 | 1.44 | | | | |
| $FT/Pt_7/Al_2O_3$ | | | | | | |
| $C_2D_4$ | 0.70 | 4.52 | 4.43 | 4.32 | 4.12 | 3.88 |
| $D_2$ | 0.03 | 0.13 | 0.15 | 0.15 | 0.06 | 0.10 |
| $Pt_7/heat/FT/Al_2O_3$ | | | | | | |
| $C_2D_4$ | 4.53 | 4.28 | 4.19 | 4.33 | 3.58 | 3.67 |
| $D_2$ | 0.55 | 0.40 | 0.24 | 0.35 | 0.10 | 0.21 |
| Blank $Al_2O_3$ | | | | | | |
| $C_2D_4$ | 1.63 | | | | | |

Molecules desorbed conversion

We start by leaking in $2 \times 10^{-8}$ Torr of ethylene into the main chamber and the electronics recorded the corresponding number of counts per second.

We then correct for the pressure from the $2 \times 10^{-8}$ Torr of ethylene that was leaked in,

$$Corrected\ pressure\ =\ 2\ \times 10^{-8}\ Torr\ \times\ ion\ gauge\ sensitivity\ factor \qquad (1)$$

Then calculate the molecular density from the corrected pressure by:

$$\rho = corrected\ pressure\ \times\ N_A/RT \qquad (2)$$

We now calculate velocity of the gas molecule,

$$V_{rms} = 158\sqrt{T/m(atomic\ wt\ of\ molecule)} \qquad (3)$$

And the flux of molecules into the mass spectrometer,

$$Z_{wall} = 0.25 \times \rho \times V_{rms} \qquad (4)$$

Which leads to molecules entering the mass spectrometer per second,

$$molecules\ entering\ mass\ spec\ per\ second = Z_{wall} \times$$
$$aperture\ area\ of\ the\ skimmer\ cone\ on\ the\ mass\ spec \qquad (5)$$

Finally, we obtain molecules per count from molecules entering mass spec per second divided by the number of counts per second as recorded in electronics from leaking in ethylene in the first step of this conversion,

$$Molecules\ per\ count\ = \frac{molecules\ entering\ mass\ spec/second}{number\ of\ counts/second} \qquad (6)$$

The calibration is not yet finished as in any raw TPD data, a total of 8 different masses were observed, the first mass is scanned for 0.05 seconds followed by a wait time then another mass is scanned, this is repeated for all 8 masses, therefore it takes around 1.2 seconds for the electronics to return to scanning the first mass again. This cycle is defined as a duty cycle and the count time of 0.05 seconds divided by the duty cycle is defined as the duty factor.

The TPD results collected were initially in counts, this is divided by the duty factor to obtain the true counts, then multiplied by the molecules per count obtained in (6) to produce the final results in figure 2.

Figure S1. Raw ISS spectrum of Heat/$Pt_7$/$Al_2O_3$ (blue) and Heat/FT/$Pt_7$/$Al_2O_3$ (red). O, Al, Sn and Pt peaks are labeled. Cl is labeled around the area where it would exist if present on surface of the samples.

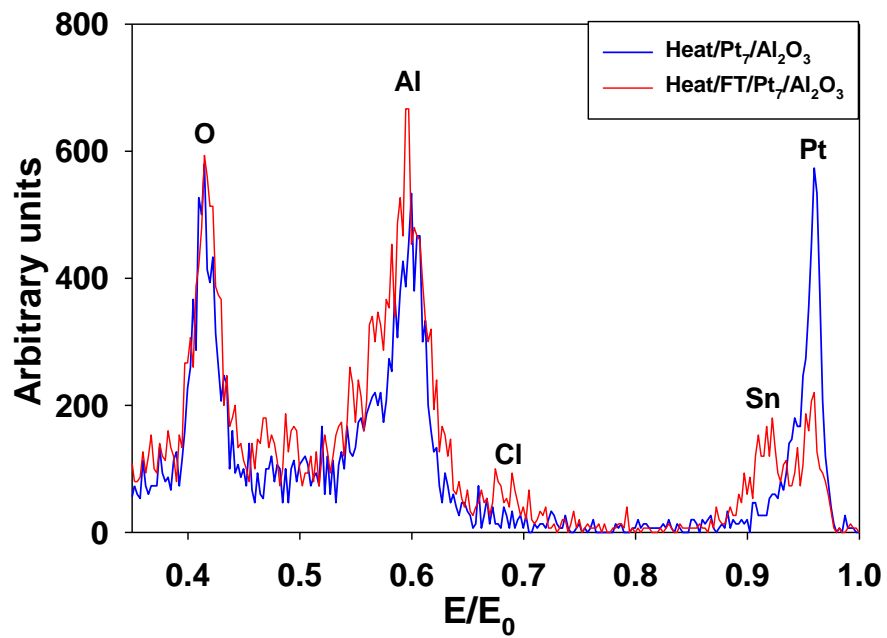

Table S4. DFT calculations for $Pt_4Sn_3$ on alumina. E is the energy relative to global minimum, 2S+1 is the spin multiplicity and $Q_{cluster}$ is the net charge of the cluster.

| Isomer | E(eV) | 2S+1 | $Q_{cluster}$(e) |
|--------|-------|------|------------------|
| A | 0 | 1 | -0.04 |
| B | 0.396 | 1 | -0.1 |
| C | 0.728 | 1 | -0.03 |
| D | 0.756 | 1 | -0.07 |

Table S5 DFT calculations for ethylene binding $Pt_4Sn_3$ on alumina. E is the energy relative to global minimum, 2S+1 is the spin multiplicity and $Q_{cluster}$ is the net charge of the cluster. Mode means the type of interaction between ethylene and cluster. C-C is the distance between the two carbon on ethylene.

| Isomer | E(eV) | 2S+1 | $Q_{cluster}$(e) | Mode | C-C(A) |
|---|---|---|---|---|---|
| C | 0 | 1 | -0.03 | pi | 1.41 |
| C | 0.001 | 1 | -0.03 | pi | 1.41 |
| A | 0.170 | 1 | -0.04 | pi | 1.40 |
| A | 0.171 | 1 | -0.04 | pi | 1.40 |
| C | 0.265 | 1 | 0.08 | pi | 1.43 |
| B | 0.313 | 1 | -0.13 | pi | 1.41 |
| B | 0.314 | 1 | -0.12 | pi | 1.41 |
| B | 0.365 | 1 | -0.12 | pi | 1.41 |

Figure 4

Coordinates for Pt$_4$Sn$_3$, the format is

Number of atom

Comment line: cluster-support-isomer

symbol    x    y    z    bader_charge

| | | | | |
|---|---|---|---|---|
| 277 | | | | |
| pt4sn3-al2o3-A | | | | |
| Al | 9.69003000 | 2.79452000 | 0.00000000 | 2.16903200 |
| Al | 4.84978000 | 11.17808000 | 0.00000000 | 2.16893610 |
| Al | 0.00953000 | 2.79452000 | 0.00000000 | 2.16899470 |
| Al | 7.26991000 | 6.98630000 | 0.00000000 | 2.16889230 |
| Al | -4.83072000 | 11.17808000 | 0.00000000 | 2.16890010 |
| Al | -2.41059000 | 6.98630000 | 0.00000000 | 2.16899870 |
| Al | 4.84978000 | 2.79452000 | 0.00000000 | 2.16886050 |
| Al | 0.00953000 | 11.17808000 | 0.00000000 | 2.16895800 |
| Al | 2.42966000 | 6.98630000 | 0.00000000 | 2.16901000 |
| O | 10.62741000 | 1.39726000 | 0.85241000 | -1.55246510 |
| O | 12.85153000 | 0.11316000 | 0.85241000 | -1.56066860 |
| O | 12.85153000 | 2.68136000 | 0.85241000 | -1.55848090 |
| O | 5.78716000 | 9.78082000 | 0.85241000 | -1.55217950 |
| O | 8.01128000 | 8.49672000 | 0.85241000 | -1.56064900 |
| O | 8.01128000 | 11.06492000 | 0.85241000 | -1.55830160 |
| O | 0.94691000 | 1.39726000 | 0.85241000 | -1.55238230 |
| O | 8.20728000 | 5.58904000 | 0.85241000 | -1.55242460 |
| O | 10.43141000 | 4.30494000 | 0.85241000 | -1.56100600 |
| O | 3.17103000 | 0.11317000 | 0.85241000 | -1.56066520 |
| O | 10.43140000 | 6.87314000 | 0.85241000 | -1.55839870 |
| O | 3.17103000 | 2.68136000 | 0.85241000 | -1.55831920 |
| O | -3.89334000 | 9.78082000 | 0.85241000 | -1.55245030 |
| O | -1.66922000 | 8.49672000 | 0.85241000 | -1.56080070 |
| O | -1.66922000 | 11.06492000 | 0.85241000 | -1.55842250 |
| O | -1.47322000 | 5.58904000 | 0.85241000 | -1.55253340 |
| O | 5.78716000 | 1.39726000 | 0.85241000 | -1.55228330 |
| O | 8.01128000 | 0.11317000 | 0.85241000 | -1.56068690 |
| O | 0.75091000 | 4.30494000 | 0.85241000 | -1.56061740 |
| O | 0.75090000 | 6.87314000 | 0.85241000 | -1.55846130 |
| O | 8.01128000 | 2.68136000 | 0.85241000 | -1.55840140 |
| O | 0.94691000 | 9.78082000 | 0.85241000 | -1.55240040 |
| O | 3.17103000 | 8.49672000 | 0.85241000 | -1.56048260 |
| O | 3.17103000 | 11.06492000 | 0.85241000 | -1.55834490 |

| | | | |
|---|---|---|---|
| O | 3.36703000 | 5.58904000 | 0.85241000 | -1.55220470 |
| O | 5.59116000 | 4.30494000 | 0.85241000 | -1.56062530 |
| O | 5.59116000 | 6.87314000 | 0.85241000 | -1.55825850 |
| Al | 2.42965000 | 12.57534000 | 1.70482000 | 2.46326480 |
| Al | 4.84978000 | 8.38356000 | 1.70482000 | 2.46327080 |
| Al | -7.25084000 | 12.57534000 | 1.70482000 | 2.46327130 |
| Al | 7.26991000 | 4.19178000 | 1.70482000 | 2.46335990 |
| Al | -4.83072000 | 8.38356000 | 1.70482000 | 2.46337230 |
| Al | -2.41059000 | 4.19178000 | 1.70482000 | 2.46349420 |
| Al | -2.41060000 | 12.57534000 | 1.70482000 | 2.46340340 |
| Al | 0.00953000 | 8.38356000 | 1.70482000 | 2.46336390 |
| Al | 2.42966000 | 4.19178000 | 1.70482000 | 2.46323490 |
| Al | 12.11016000 | 1.39726000 | 2.20274000 | 2.48269380 |
| Al | 7.26991000 | 9.78082000 | 2.20274000 | 2.48268280 |
| Al | 9.69003000 | 5.58904000 | 2.20274000 | 2.48279730 |
| Al | 2.42966000 | 1.39726000 | 2.20274000 | 2.48277460 |
| Al | -2.41059000 | 9.78082000 | 2.20274000 | 2.48282750 |
| Al | 7.26991000 | 1.39726000 | 2.20274000 | 2.48275630 |
| Al | 0.00953000 | 5.58904000 | 2.20274000 | 2.48269450 |
| Al | 2.42966000 | 9.78082000 | 2.20274000 | 2.48280940 |
| Al | 4.84978000 | 5.58904000 | 2.20274000 | 2.48271940 |
| O | 8.94865000 | 1.28410000 | 3.05515000 | -1.64220590 |
| O | 11.17279000 | 0.00001000 | 3.05515000 | -1.62408920 |
| O | 4.10840000 | 9.66766000 | 3.05515000 | -1.64211630 |
| O | 11.36879000 | 2.90768000 | 3.05515000 | -1.64999060 |
| O | 6.33254000 | 8.38356000 | 3.05515000 | -1.62409530 |
| O | 6.52853000 | 5.47588000 | 3.05515000 | -1.64192820 |
| O | -0.73185000 | 1.28410000 | 3.05515000 | -1.64217340 |
| O | 6.52853000 | 11.29124000 | 3.05515000 | -1.64987110 |
| O | 1.49229000 | 0.00001000 | 3.05515000 | -1.62406150 |
| O | 8.75266000 | 4.19179000 | 3.05515000 | -1.62430730 |
| O | -5.57210000 | 9.66766000 | 3.05515000 | -1.64198580 |
| O | 8.94866000 | 7.09946000 | 3.05515000 | -1.64997500 |
| O | 1.68828000 | 2.90768000 | 3.05515000 | -1.65005510 |
| O | -3.34797000 | 8.38357000 | 3.05515000 | -1.62431670 |
| O | 4.10840000 | 1.28410000 | 3.05515000 | -1.64198310 |
| O | -3.15197000 | 5.47588000 | 3.05515000 | -1.64233580 |
| O | -3.15197000 | 11.29124000 | 3.05515000 | -1.65019130 |
| O | -0.92784000 | 4.19178000 | 3.05515000 | -1.62435330 |
| O | 6.33254000 | 0.00001000 | 3.05515000 | -1.62426730 |
| O | 6.52854000 | 2.90768000 | 3.05515000 | -1.64989760 |
| O | -0.73185000 | 9.66766000 | 3.05515000 | -1.64219040 |
| O | -0.73184000 | 7.09946000 | 3.05515000 | -1.65024200 |

| | | | |
|---|---|---|---|
| O | 1.49228000 | 8.38356000 | 3.05515000 | -1.62423200 |
| O | 1.68829000 | 11.29124000 | 3.05515000 | -1.64999700 |
| O | 1.68828000 | 5.47588000 | 3.05515000 | -1.64215390 |
| O | 3.91241000 | 4.19178000 | 3.05515000 | -1.62409670 |
| O | 4.10841000 | 7.09946000 | 3.05515000 | -1.64980420 |
| Al | 9.69003000 | 2.79452000 | 3.90756000 | 2.47804340 |
| Al | 4.84978000 | 11.17808000 | 3.90756000 | 2.47790610 |
| Al | 0.00953000 | 2.79452000 | 3.90756000 | 2.47792840 |
| Al | 7.26991000 | 6.98630000 | 3.90756000 | 2.47778690 |
| Al | -4.83072000 | 11.17808000 | 3.90756000 | 2.47782460 |
| Al | -2.41059000 | 6.98630000 | 3.90756000 | 2.47804610 |
| Al | 4.84978000 | 2.79452000 | 3.90756000 | 2.47779910 |
| Al | 0.00953000 | 11.17808000 | 3.90756000 | 2.47788640 |
| Al | 2.42966000 | 6.98630000 | 3.90756000 | 2.47799910 |
| Al | 2.42965000 | 12.57534000 | 4.40548000 | 2.47224150 |
| Al | 4.84978000 | 8.38356000 | 4.40548000 | 2.47240090 |
| Al | -7.25084000 | 12.57534000 | 4.40548000 | 2.47235490 |
| Al | 7.26991000 | 4.19178000 | 4.40548000 | 2.47221690 |
| Al | -4.83072000 | 8.38356000 | 4.40548000 | 2.47227490 |
| Al | -2.41059000 | 4.19178000 | 4.40548000 | 2.47227870 |
| Al | -2.41060000 | 12.57534000 | 4.40548000 | 2.47230300 |
| Al | 0.00953000 | 8.38356000 | 4.40548000 | 2.47241280 |
| Al | 2.42966000 | 4.19178000 | 4.40548000 | 2.47247260 |
| O | 8.20728000 | 2.79452000 | 5.25789000 | -1.64328320 |
| O | 10.43141000 | 4.07863000 | 5.25789000 | -1.65412220 |
| O | 10.43142000 | 1.51042000 | 5.25789000 | -1.64891880 |
| O | 3.36702000 | 11.17808000 | 5.25789000 | -1.64355920 |
| O | 5.59116000 | 12.46219000 | 5.25789000 | -1.65405600 |
| O | 5.59116000 | 9.89398000 | 5.25789000 | -1.64903520 |
| O | -1.47322000 | 2.79452000 | 5.25789000 | -1.64355830 |
| O | 5.78715000 | 6.98630000 | 5.25789000 | -1.64320310 |
| O | 8.01128000 | 8.27041000 | 5.25789000 | -1.65410360 |
| O | 0.75091000 | 4.07863000 | 5.25789000 | -1.65393910 |
| O | 0.75091000 | 1.51042000 | 5.25789000 | -1.64897050 |
| O | 8.01129000 | 5.70220000 | 5.25789000 | -1.64905180 |
| O | -6.31347000 | 11.17808000 | 5.25789000 | -1.64334670 |
| O | -4.08934000 | 12.46219000 | 5.25789000 | -1.65405740 |
| O | -4.08934000 | 9.89398000 | 5.25789000 | -1.64918400 |
| O | -3.89335000 | 6.98630000 | 5.25789000 | -1.64344260 |
| O | 3.36703000 | 2.79452000 | 5.25789000 | -1.64357320 |
| O | 5.59116000 | 4.07863000 | 5.25789000 | -1.65403930 |
| O | -1.66922000 | 8.27041000 | 5.25789000 | -1.65378550 |
| O | 5.59116000 | 1.51042000 | 5.25789000 | -1.64908870 |

| | | | |
|---|---|---|---|
| O | -1.66921000 | 5.70220000 | 5.25789000 | -1.64910600 |
| O | -1.47323000 | 11.17808000 | 5.25789000 | -1.64348960 |
| O | 0.75091000 | 12.46219000 | 5.25789000 | -1.65395160 |
| O | 0.75091000 | 9.89398000 | 5.25789000 | -1.64895080 |
| O | 0.94690000 | 6.98630000 | 5.25789000 | -1.64362280 |
| O | 3.17103000 | 8.27041000 | 5.25789000 | -1.65379320 |
| O | 3.17104000 | 5.70220000 | 5.25789000 | -1.64897440 |
| Al | 12.11016000 | 1.39726000 | 6.11030000 | 2.47609940 |
| Al | 7.26991000 | 9.78082000 | 6.11030000 | 2.47618780 |
| Al | 9.69003000 | 5.58904000 | 6.11030000 | 2.47609450 |
| Al | 2.42966000 | 1.39726000 | 6.11030000 | 2.47616750 |
| Al | -2.41059000 | 9.78082000 | 6.11030000 | 2.47602880 |
| Al | 7.26991000 | 1.39726000 | 6.11030000 | 2.47600380 |
| Al | 0.00953000 | 5.58904000 | 6.11030000 | 2.47582370 |
| Al | 2.42966000 | 9.78082000 | 6.11030000 | 2.47611450 |
| Al | 4.84978000 | 5.58904000 | 6.11030000 | 2.47622030 |
| Al | 9.69003000 | 2.79452000 | 6.60910000 | 2.47675900 |
| Al | 4.84978000 | 11.17808000 | 6.60910000 | 2.47694100 |
| Al | 0.00953000 | 2.79452000 | 6.60910000 | 2.47670790 |
| Al | 7.26991000 | 6.98630000 | 6.60910000 | 2.47693160 |
| Al | -4.83072000 | 11.17808000 | 6.60910000 | 2.47684900 |
| Al | -2.41059000 | 6.98630000 | 6.60910000 | 2.47697910 |
| Al | 4.84978000 | 2.79452000 | 6.60910000 | 2.47663380 |
| Al | 0.00953000 | 11.17808000 | 6.60910000 | 2.47664420 |
| Al | 2.42966000 | 6.98630000 | 6.60910000 | 2.47695180 |
| O | 11.36395000 | 0.11351000 | 7.46570000 | -1.64888900 |
| O | 11.37150000 | 2.68537000 | 7.46570000 | -1.65015580 |
| O | 13.59502000 | 1.39291000 | 7.46570000 | -1.65377190 |
| O | 6.52370000 | 8.49707000 | 7.46570000 | -1.64827560 |
| O | 6.53125000 | 11.06893000 | 7.46570000 | -1.64957150 |
| O | 8.75477000 | 9.77647000 | 7.46570000 | -1.65366090 |
| O | 1.68345000 | 0.11351000 | 7.46570000 | -1.64882940 |
| O | 8.94383000 | 4.30529000 | 7.46570000 | -1.64966560 |
| O | 8.95137000 | 6.87715000 | 7.46570000 | -1.64964460 |
| O | 1.69100000 | 2.68537000 | 7.46570000 | -1.65026870 |
| O | 11.17490000 | 5.58469000 | 7.46570000 | -1.65479370 |
| O | 3.91452000 | 1.39291000 | 7.46570000 | -1.65384470 |
| O | -3.15680000 | 8.49707000 | 7.46570000 | -1.64965720 |
| O | -3.14925000 | 11.06893000 | 7.46570000 | -1.64938120 |
| O | -0.92573000 | 9.77647000 | 7.46570000 | -1.65402560 |
| O | -0.73667000 | 4.30528000 | 7.46570000 | -1.64944350 |
| O | 6.52370000 | 0.11351000 | 7.46570000 | -1.64905820 |
| O | -0.72913000 | 6.87715000 | 7.46570000 | -1.65006750 |
| O | 6.53125000 | 2.68537000 | 7.46570000 | -1.64926540 |
| O | 8.75477000 | 1.39291000 | 7.46570000 | -1.65456970 |
| O | 1.49440000 | 5.58468000 | 7.46570000 | -1.65347940 |
| O | 1.68345000 | 8.49707000 | 7.46570000 | -1.64900620 |
| O | 1.69100000 | 11.06893000 | 7.46570000 | -1.64917780 |
| O | 3.91452000 | 9.77646000 | 7.46570000 | -1.65380000 |
| O | 4.10358000 | 4.30529000 | 7.46570000 | -1.64896320 |
| O | 4.11112000 | 6.87715000 | 7.46570000 | -1.64892060 |
| O | 6.33465000 | 5.58469000 | 7.46570000 | -1.65389130 |
| Al | 2.42965000 | 12.57534000 | 8.32833000 | 2.47127630 |
| Al | 4.84978000 | 8.38356000 | 8.32833000 | 2.47059540 |
| Al | -7.25084000 | 12.57534000 | 8.32833000 | 2.47118140 |
| Al | 7.26991000 | 4.19178000 | 8.32833000 | 2.47181180 |
| Al | -4.83072000 | 8.38356000 | 8.32833000 | 2.47227660 |
| Al | -2.41059000 | 4.19178000 | 8.32833000 | 2.47178100 |
| Al | -2.41060000 | 12.57534000 | 8.32833000 | 2.47172950 |
| Al | 0.00953000 | 8.38356000 | 8.32833000 | 2.47087930 |
| Al | 2.42966000 | 4.19178000 | 8.32833000 | 2.47056480 |
| Al | 12.11016000 | 1.39726000 | 8.78548000 | 2.47180270 |
| Al | 7.26991000 | 9.78082000 | 8.78548000 | 2.47711810 |
| Al | 9.69003000 | 5.58904000 | 8.78548000 | 2.47107470 |
| Al | 2.42966000 | 1.39726000 | 8.78548000 | 2.47268720 |
| Al | -2.41059000 | 9.78082000 | 8.78548000 | 2.47201400 |
| Al | 7.26991000 | 1.39726000 | 8.78548000 | 2.47212220 |
| Al | 0.00953000 | 5.58904000 | 8.78548000 | 2.47259920 |
| Al | 2.42966000 | 9.78082000 | 8.78548000 | 2.47541800 |
| Al | 4.84978000 | 5.58904000 | 8.78548000 | 2.47643050 |
| O | 8.01661000 | 2.88194000 | 9.68989000 | -1.64587920 |
| O | 3.17636000 | 11.26550000 | 9.68989000 | -1.63798770 |
| O | 10.45104000 | 1.30159000 | 9.68989000 | -1.65748230 |
| O | 12.85686000 | 2.88194000 | 9.68989000 | -1.64465220 |
| O | 5.59648000 | 7.07372000 | 9.68989000 | -1.63494700 |
| O | 5.61079000 | 9.68514000 | 9.68989000 | -1.64476920 |
| O | 13.02257000 | 0.00826000 | 9.68989000 | -1.66087070 |
| O | 8.01661000 | 11.26550000 | 9.68989000 | -1.64296620 |
| O | 0.77054000 | 1.30159000 | 9.68989000 | -1.65628380 |
| O | 8.03091000 | 5.49337000 | 9.68989000 | -1.66096020 |
| O | 8.18232000 | 8.39182000 | 9.68989000 | -1.65696500 |
| O | 3.17636000 | 2.88194000 | 9.68989000 | -1.64352350 |
| O | 10.43673000 | 7.07372000 | 9.68989000 | -1.64490930 |
| O | -4.06971000 | 9.68514000 | 9.68989000 | -1.66021580 |
| O | 10.60245000 | 4.20004000 | 9.68989000 | -1.67087590 |
| O | 3.34207000 | 0.00826000 | 9.68989000 | -1.66705370 |

| | | | |
|---|---|---|---|
| O | -1.66389000 | 11.26550000 | 9.68989000 | -1.64234570 |
| O | -1.64958000 | 5.49336000 | 9.68989000 | -1.65929780 |
| O | 5.61079000 | 1.30159000 | 9.68989000 | -1.66003730 |
| O | -1.49818000 | 8.39182000 | 9.68989000 | -1.66860380 |
| O | 0.75623000 | 7.07372000 | 9.68989000 | -1.64216290 |
| O | 0.77054000 | 9.68514000 | 9.68989000 | -1.64700540 |
| O | 0.92195000 | 4.20004000 | 9.68989000 | -1.66640710 |
| O | 8.18232000 | 0.00827000 | 9.68989000 | -1.66495270 |
| O | 3.19067000 | 5.49337000 | 9.68989000 | -1.64641660 |
| O | 3.34207000 | 8.39182000 | 9.68989000 | -1.65482900 |
| O | 5.76220000 | 4.20004000 | 9.68989000 | -1.65246940 |
| Al | 9.69003000 | 2.79452000 | 10.71830000 | 2.47610410 |
| Al | 4.84978000 | 11.17808000 | 10.71830000 | 2.47885530 |
| Al | 0.00953000 | 2.79452000 | 10.71830000 | 2.47800240 |
| Al | 7.26991000 | 6.98630000 | 10.71830000 | 2.47709610 |
| Al | -4.83072000 | 11.17808000 | 10.71830000 | 2.48000960 |
| Al | -2.41059000 | 6.98630000 | 10.71830000 | 2.47485340 |
| Al | 4.84978000 | 2.79452000 | 10.71830000 | 2.47649360 |
| Al | 0.00953000 | 11.17808000 | 10.71830000 | 2.47973630 |
| Al | 2.42966000 | 6.98630000 | 10.71830000 | 2.47968600 |
| Al | 2.42965000 | 12.57534000 | 10.98601000 | 2.47027630 |
| Al | 4.84978000 | 8.38356000 | 10.98601000 | 2.46984090 |
| Al | -7.25084000 | 12.57534000 | 10.98601000 | 2.47314380 |
| Al | 7.26991000 | 4.19178000 | 10.98601000 | 2.47140400 |
| Al | -4.83072000 | 8.38356000 | 10.98601000 | 2.47060410 |
| Al | -2.41059000 | 4.19178000 | 10.98601000 | 2.46781730 |
| Al | -2.41060000 | 12.57534000 | 10.98601000 | 2.47138480 |
| Al | 0.00953000 | 8.38356000 | 10.98601000 | 2.46962720 |
| Al | 2.42966000 | 4.19178000 | 10.98601000 | 2.46968450 |
| O | 8.86951000 | 4.08969000 | 11.85298000 | -1.62583740 |
| O | 8.98287000 | 1.43849000 | 11.85300000 | -1.62265090 |
| O | 11.21492000 | 2.86251000 | 11.84856000 | -1.62268000 |
| O | 4.03712000 | 12.47075000 | 11.87802000 | -1.62741060 |
| O | 4.11759000 | 9.83431000 | 11.95278000 | -1.56465280 |
| O | 6.37582000 | 11.20797000 | 11.85751000 | -1.61310190 |
| O | -0.82383000 | 4.06927000 | 11.86088000 | -1.61728520 |
| O | 6.48660000 | 8.28064000 | 11.90423000 | -1.57679530 |
| O | -0.70624000 | 1.43453000 | 11.85225000 | -1.62544450 |
| O | 6.53714000 | 5.65577000 | 11.87659000 | -1.58451030 |
| O | 8.80422000 | 7.02430000 | 11.87663000 | -1.63114800 |
| O | 1.53229000 | 2.86347000 | 11.85474000 | -1.62428760 |
| O | -5.65294000 | 12.47976000 | 11.86148000 | -1.62192880 |
| O | -5.54745000 | 9.81096000 | 11.84329000 | -1.61084220 |
| O | -3.29818000 | 11.24067000 | 11.85247000 | -1.62501710 |
| O | -3.22827000 | 8.27039000 | 11.85142000 | -1.61649280 |
| O | 4.03026000 | 4.11179000 | 11.84634000 | -1.61404130 |
| O | -3.10294000 | 5.63182000 | 11.85263000 | -1.62863190 |
| O | 4.13826000 | 1.43708000 | 11.85253000 | -1.62233740 |
| O | 6.38995000 | 2.85778000 | 11.86166000 | -1.62038930 |
| O | -0.87049000 | 7.04737000 | 11.85449000 | -1.62847190 |
| O | -0.80922000 | 12.46976000 | 11.85340000 | -1.62017710 |
| O | -0.70494000 | 9.80729000 | 11.86826000 | -1.62823960 |
| O | 1.54085000 | 11.23798000 | 11.82347000 | -1.60965110 |
| O | 1.60633000 | 8.28630000 | 11.94575000 | -1.56750540 |
| O | 1.69521000 | 5.64796000 | 11.87755000 | -1.62850160 |
| O | 3.95986000 | 7.02777000 | 11.82093000 | -1.61090910 |
| Al | 12.11014000 | 1.40897000 | 11.85728000 | 2.41395720 |
| Al | 7.32273000 | 9.81356000 | 12.39393000 | 2.37616300 |
| Al | 9.71794000 | 5.56400000 | 11.81685000 | 2.41366610 |
| Al | 2.43022000 | 1.40457000 | 11.89446000 | 2.41591150 |
| Al | -2.41900000 | 9.77964000 | 11.88240000 | 2.41562830 |
| Al | 7.27140000 | 1.39386000 | 11.88335000 | 2.41511970 |
| Al | -0.02062000 | 5.56448000 | 11.90605000 | 2.40415860 |
| Al | 2.37287000 | 9.84123000 | 12.40011000 | 2.38017150 |
| Al | 4.80328000 | 5.58314000 | 12.39372000 | 2.37615320 |
| Pt | 3.86236000 | 7.93009000 | 14.91642000 | -0.40218810 |
| Pt | 5.12104000 | 5.76434000 | 14.82597000 | -0.77300650 |
| Pt | 2.20799000 | 9.91803000 | 14.82750000 | -0.83114230 |
| Pt | 6.98661000 | 9.49436000 | 14.79336000 | -0.95405530 |
| Sn | 4.74507000 | 10.57814000 | 14.15397000 | 0.97173270 |
| Sn | 7.20575000 | 7.00771000 | 13.99509000 | 1.03112890 |
| Sn | 1.48192000 | 7.41734000 | 14.12845000 | 0.92054150 |

277
pt4sn3-al2o3-B

| | | | | |
|---|---|---|---|---|
| Al | 9.69003000 | 2.79452000 | 0.00000000 | 2.17018520 |
| Al | 4.84978000 | 11.17808000 | 0.00000000 | 2.16969020 |
| Al | 0.00953000 | 2.79452000 | 0.00000000 | 2.17008610 |
| Al | 7.26991000 | 6.98630000 | 0.00000000 | 2.16996350 |
| Al | -4.83072000 | 11.17808000 | 0.00000000 | 2.16995190 |
| Al | -2.41059000 | 6.98630000 | 0.00000000 | 2.17014850 |
| Al | 4.84978000 | 2.79452000 | 0.00000000 | 2.17001870 |
| Al | 0.00953000 | 11.17808000 | 0.00000000 | 2.16987570 |
| Al | 2.42966000 | 6.98630000 | 0.00000000 | 2.16976020 |
| O | 10.62741000 | 1.39726000 | 0.85241000 | -1.55154860 |
| O | 12.85153000 | 0.11316000 | 0.85241000 | -1.55999370 |
| O | 12.85153000 | 2.68136000 | 0.85241000 | -1.55795250 |

| | | | |
|---|---|---|---|
| O | 5.78716000 | 9.78082000 | 0.85241000 | -1.55178430 |
| O | 8.01128000 | 8.49672000 | 0.85241000 | -1.55994830 |
| O | 8.01128000 | 11.06492000 | 0.85241000 | -1.55776710 |
| O | 0.94691000 | 1.39726000 | 0.85241000 | -1.55173240 |
| O | 8.20728000 | 5.58904000 | 0.85241000 | -1.55190460 |
| O | 10.43141000 | 4.30494000 | 0.85241000 | -1.56018880 |
| O | 3.17103000 | 0.11317000 | 0.85241000 | -1.55999850 |
| O | 10.43140000 | 6.87314000 | 0.85241000 | -1.55789270 |
| O | 3.17103000 | 2.68136000 | 0.85241000 | -1.55787640 |
| O | -3.89334000 | 9.78082000 | 0.85241000 | -1.55181880 |
| O | -1.66922000 | 8.49672000 | 0.85241000 | -1.56001370 |
| O | -1.66922000 | 11.06492000 | 0.85241000 | -1.55789310 |
| O | -1.47322000 | 5.58904000 | 0.85241000 | -1.55186860 |
| O | 5.78716000 | 1.39726000 | 0.85241000 | -1.55181370 |
| O | 8.01128000 | 0.11317000 | 0.85241000 | -1.56000560 |
| O | 0.75091000 | 4.30494000 | 0.85241000 | -1.56003650 |
| O | 0.75090000 | 6.87314000 | 0.85241000 | -1.55801060 |
| O | 8.01128000 | 2.68136000 | 0.85241000 | -1.55789860 |
| O | 0.94691000 | 9.78082000 | 0.85241000 | -1.55183000 |
| O | 3.17103000 | 8.49672000 | 0.85241000 | -1.55987310 |
| O | 3.17103000 | 11.06492000 | 0.85241000 | -1.55784580 |
| O | 3.36703000 | 5.58904000 | 0.85241000 | -1.55188520 |
| O | 5.59116000 | 4.30494000 | 0.85241000 | -1.56002010 |
| O | 5.59116000 | 6.87314000 | 0.85241000 | -1.55776980 |
| Al | 9.69003000 | 0.00000000 | 1.70482000 | 2.46329390 |
| Al | 4.84978000 | 8.38356000 | 1.70482000 | 2.46339140 |
| Al | 0.00954000 | 0.00000000 | 1.70482000 | 2.46338140 |
| Al | 7.26991000 | 4.19178000 | 1.70482000 | 2.46323660 |
| Al | -4.83072000 | 8.38356000 | 1.70482000 | 2.46328860 |
| Al | -2.41059000 | 4.19178000 | 1.70482000 | 2.46323780 |
| Al | 4.84978000 | 0.00000000 | 1.70482000 | 2.46327790 |
| Al | 0.00953000 | 8.38356000 | 1.70482000 | 2.46351720 |
| Al | 2.42966000 | 4.19178000 | 1.70482000 | 2.46338120 |
| Al | 12.11016000 | 1.39726000 | 2.20274000 | 2.48263480 |
| Al | 7.26991000 | 9.78082000 | 2.20274000 | 2.48285460 |
| Al | 9.69003000 | 5.58904000 | 2.20274000 | 2.48269790 |
| Al | 2.42966000 | 1.39726000 | 2.20274000 | 2.48261900 |
| Al | -2.41059000 | 9.78082000 | 2.20274000 | 2.48279930 |
| Al | 7.26991000 | 1.39726000 | 2.20274000 | 2.48269400 |
| Al | 0.00953000 | 5.58904000 | 2.20274000 | 2.48275260 |
| Al | 2.42966000 | 9.78082000 | 2.20274000 | 2.48283030 |
| Al | 4.84978000 | 5.58904000 | 2.20274000 | 2.48274940 |
| O | 8.94865000 | 1.28410000 | 3.05515000 | -1.64157110 |

| | | | |
|---|---|---|---|
| O | 11.17279000 | 0.00001000 | 3.05515000 | -1.62396950 |
| O | 4.10840000 | 9.66766000 | 3.05515000 | -1.64135720 |
| O | 11.36879000 | 2.90768000 | 3.05515000 | -1.64974270 |
| O | 6.33254000 | 8.38356000 | 3.05515000 | -1.62387210 |
| O | 6.52853000 | 5.47588000 | 3.05515000 | -1.64154090 |
| O | -0.73185000 | 1.28410000 | 3.05515000 | -1.64150130 |
| O | 6.52853000 | 11.29124000 | 3.05515000 | -1.64939480 |
| O | 1.49229000 | 0.00001000 | 3.05515000 | -1.62394860 |
| O | 8.75266000 | 4.19179000 | 3.05515000 | -1.62403720 |
| O | -5.57210000 | 9.66766000 | 3.05515000 | -1.64167530 |
| O | 8.94866000 | 7.09946000 | 3.05515000 | -1.64943470 |
| O | 1.68828000 | 2.90768000 | 3.05515000 | -1.64949390 |
| O | -3.34797000 | 8.38357000 | 3.05515000 | -1.62406450 |
| O | 4.10840000 | 1.28410000 | 3.05515000 | -1.64148320 |
| O | -3.15197000 | 5.47588000 | 3.05515000 | -1.64183200 |
| O | -3.15197000 | 11.29124000 | 3.05515000 | -1.64953900 |
| O | -0.92784000 | 4.19178000 | 3.05515000 | -1.62398950 |
| O | 6.33254000 | 0.00001000 | 3.05515000 | -1.62378490 |
| O | 6.52854000 | 2.90768000 | 3.05515000 | -1.64957140 |
| O | -0.73185000 | 9.66766000 | 3.05515000 | -1.64163400 |
| O | -0.73184000 | 7.09946000 | 3.05515000 | -1.64969440 |
| O | 1.49228000 | 8.38356000 | 3.05515000 | -1.62373960 |
| O | 1.68829000 | 11.29124000 | 3.05515000 | -1.64947050 |
| O | 1.68828000 | 5.47588000 | 3.05515000 | -1.64155250 |
| O | 3.91241000 | 4.19178000 | 3.05515000 | -1.62386040 |
| O | 4.10841000 | 7.09946000 | 3.05515000 | -1.64932500 |
| Al | 9.69003000 | 2.79452000 | 3.90756000 | 2.47770000 |
| Al | 4.84978000 | 11.17808000 | 3.90756000 | 2.47754660 |
| Al | 0.00953000 | 2.79452000 | 3.90756000 | 2.47777790 |
| Al | 7.26991000 | 6.98630000 | 3.90756000 | 2.47767120 |
| Al | -4.83072000 | 11.17808000 | 3.90756000 | 2.47777610 |
| Al | -2.41059000 | 6.98630000 | 3.90756000 | 2.47780020 |
| Al | 4.84978000 | 2.79452000 | 3.90756000 | 2.47778720 |
| Al | 0.00953000 | 11.17808000 | 3.90756000 | 2.47757130 |
| Al | 2.42966000 | 6.98630000 | 3.90756000 | 2.47752010 |
| Al | 9.69003000 | 0.00000000 | 4.40548000 | 2.47226230 |
| Al | 4.84978000 | 8.38356000 | 4.40548000 | 2.47241820 |
| Al | 0.00954000 | 0.00000000 | 4.40548000 | 2.47238870 |
| Al | 7.26991000 | 4.19178000 | 4.40548000 | 2.47213170 |
| Al | -4.83072000 | 8.38356000 | 4.40548000 | 2.47244720 |
| Al | -2.41059000 | 4.19178000 | 4.40548000 | 2.47246140 |
| Al | 4.84978000 | 0.00000000 | 4.40548000 | 2.47224090 |
| Al | 0.00953000 | 8.38356000 | 4.40548000 | 2.47227570 |

| | | | |
|---|---|---|---|
| Al | 2.42966000 | 4.19178000 | 4.40548000 | 2.47235500 |
| O | 8.20728000 | 2.79452000 | 5.25789000 | -1.64331510 |
| O | 10.43141000 | 4.07863000 | 5.25789000 | -1.65419730 |
| O | 10.43142000 | 1.51042000 | 5.25789000 | -1.64882820 |
| O | 3.36702000 | 11.17808000 | 5.25789000 | -1.64330780 |
| O | 5.59116000 | 12.46219000 | 5.25789000 | -1.65423630 |
| O | 5.59116000 | 9.89398000 | 5.25789000 | -1.64888220 |
| O | -1.47322000 | 2.79452000 | 5.25789000 | -1.64360130 |
| O | 5.78715000 | 6.98630000 | 5.25789000 | -1.64296020 |
| O | 8.01128000 | 8.27041000 | 5.25789000 | -1.65410850 |
| O | 0.75091000 | 4.07863000 | 5.25789000 | -1.65394660 |
| O | 0.75091000 | 1.51042000 | 5.25789000 | -1.64904380 |
| O | 8.01129000 | 5.70220000 | 5.25789000 | -1.64917860 |
| O | -6.31347000 | 11.17808000 | 5.25789000 | -1.64299720 |
| O | -4.08934000 | 12.46219000 | 5.25789000 | -1.65419140 |
| O | -4.08934000 | 9.89398000 | 5.25789000 | -1.64946910 |
| O | -3.89335000 | 6.98630000 | 5.25789000 | -1.64362210 |
| O | 3.36703000 | 2.79452000 | 5.25789000 | -1.64332330 |
| O | 5.59116000 | 4.07863000 | 5.25789000 | -1.65414310 |
| O | -1.66922000 | 8.27041000 | 5.25789000 | -1.65372800 |
| O | 5.59116000 | 1.51042000 | 5.25789000 | -1.64907480 |
| O | -1.66921000 | 5.70220000 | 5.25789000 | -1.64903260 |
| O | -1.47323000 | 11.17808000 | 5.25789000 | -1.64352390 |
| O | 0.75091000 | 12.46219000 | 5.25789000 | -1.65403520 |
| O | 0.75091000 | 9.89398000 | 5.25789000 | -1.64896030 |
| O | 0.94690000 | 6.98630000 | 5.25789000 | -1.64328930 |
| O | 3.17103000 | 8.27041000 | 5.25789000 | -1.65375800 |
| O | 3.17104000 | 5.70220000 | 5.25789000 | -1.64910730 |
| Al | 12.11016000 | 1.39726000 | 6.11030000 | 2.47603510 |
| Al | 7.26991000 | 9.78082000 | 6.11030000 | 2.47610130 |
| Al | 9.69003000 | 5.58904000 | 6.11030000 | 2.47618360 |
| Al | 2.42966000 | 1.39726000 | 6.11030000 | 2.47613770 |
| Al | -2.41059000 | 9.78082000 | 6.11030000 | 2.47604080 |
| Al | 7.26991000 | 1.39726000 | 6.11030000 | 2.47611560 |
| Al | 0.00953000 | 5.58904000 | 6.11030000 | 2.47588150 |
| Al | 2.42966000 | 9.78082000 | 6.11030000 | 2.47651390 |
| Al | 4.84978000 | 5.58904000 | 6.11030000 | 2.47609270 |
| Al | 9.69003000 | 2.79452000 | 6.60910000 | 2.47694160 |
| Al | 4.84978000 | 11.17808000 | 6.60910000 | 2.47745150 |
| Al | 0.00953000 | 2.79452000 | 6.60910000 | 2.47708830 |
| Al | 7.26991000 | 6.98630000 | 6.60910000 | 2.47680810 |
| Al | -4.83072000 | 11.17808000 | 6.60910000 | 2.47709660 |
| Al | -2.41059000 | 6.98630000 | 6.60910000 | 2.47691820 |
| Al | 4.84978000 | 2.79452000 | 6.60910000 | 2.47681250 |
| Al | 0.00953000 | 11.17808000 | 6.60910000 | 2.47701490 |
| Al | 2.42966000 | 6.98630000 | 6.60910000 | 2.47698120 |
| O | 11.36395000 | 0.11351000 | 7.46570000 | -1.64816570 |
| O | 11.37150000 | 2.68537000 | 7.46570000 | -1.65058090 |
| O | 13.59502000 | 1.39291000 | 7.46570000 | -1.65387700 |
| O | 6.52370000 | 8.49707000 | 7.46570000 | -1.64885870 |
| O | 6.53125000 | 11.06893000 | 7.46570000 | -1.64868500 |
| O | 8.75477000 | 9.77647000 | 7.46570000 | -1.65393410 |
| O | 1.68345000 | 0.11351000 | 7.46570000 | -1.64866000 |
| O | 8.94383000 | 4.30529000 | 7.46570000 | -1.64920900 |
| O | 8.95137000 | 6.87715000 | 7.46570000 | -1.65002870 |
| O | 1.69100000 | 2.68537000 | 7.46570000 | -1.65010710 |
| O | 11.17490000 | 5.58469000 | 7.46570000 | -1.65439890 |
| O | 3.91452000 | 1.39291000 | 7.46570000 | -1.65409770 |
| O | -3.15680000 | 8.49707000 | 7.46570000 | -1.64937680 |
| O | -3.14925000 | 11.06893000 | 7.46570000 | -1.64939790 |
| O | -0.92573000 | 9.77647000 | 7.46570000 | -1.65365480 |
| O | -0.73667000 | 4.30528000 | 7.46570000 | -1.64969740 |
| O | 6.52370000 | 0.11351000 | 7.46570000 | -1.64877600 |
| O | -0.72913000 | 6.87715000 | 7.46570000 | -1.65002200 |
| O | 6.53125000 | 2.68537000 | 7.46570000 | -1.64990120 |
| O | 8.75477000 | 1.39291000 | 7.46570000 | -1.65384050 |
| O | 1.49440000 | 5.58468000 | 7.46570000 | -1.65395130 |
| O | 1.68345000 | 8.49707000 | 7.46570000 | -1.64937990 |
| O | 1.69100000 | 11.06893000 | 7.46570000 | -1.64910540 |
| O | 3.91452000 | 9.77646000 | 7.46570000 | -1.65382710 |
| O | 4.10358000 | 4.30529000 | 7.46570000 | -1.64899030 |
| O | 4.11112000 | 6.87715000 | 7.46570000 | -1.64892700 |
| O | 6.33465000 | 5.58469000 | 7.46570000 | -1.65386240 |
| Al | 9.69003000 | 0.00000000 | 8.32833000 | 2.47028690 |
| Al | 4.84978000 | 8.38356000 | 8.32833000 | 2.47058470 |
| Al | 0.00954000 | 0.00000000 | 8.32833000 | 2.47048230 |
| Al | 7.26991000 | 4.19178000 | 8.32833000 | 2.47146440 |
| Al | -4.83072000 | 8.38356000 | 8.32833000 | 2.47221030 |
| Al | -2.41059000 | 4.19178000 | 8.32833000 | 2.47086540 |
| Al | 4.84978000 | 0.00000000 | 8.32833000 | 2.47152180 |
| Al | 0.00953000 | 8.38356000 | 8.32833000 | 2.47079530 |
| Al | 2.42966000 | 4.19178000 | 8.32833000 | 2.47119190 |
| Al | 12.11016000 | 1.39726000 | 8.78548000 | 2.47241810 |
| Al | 7.26991000 | 9.78082000 | 8.78548000 | 2.47612140 |
| Al | 9.69003000 | 5.58904000 | 8.78548000 | 2.46989550 |
| Al | 2.42966000 | 1.39726000 | 8.78548000 | 2.47283660 |

| | | | |
|---|---|---|---|
| Al | -2.41059000 | 9.78082000 | 8.78548000 | 2.47257870 |
| Al | 7.26991000 | 1.39726000 | 8.78548000 | 2.47206760 |
| Al | 0.00953000 | 5.58904000 | 8.78548000 | 2.47177860 |
| Al | 2.42966000 | 9.78082000 | 8.78548000 | 2.47657780 |
| Al | 4.84978000 | 5.58904000 | 8.78548000 | 2.47587930 |
| O | 8.01661000 | 2.88194000 | 9.68989000 | -1.64548650 |
| O | 3.17636000 | 11.26550000 | 9.68989000 | -1.63946240 |
| O | 10.45104000 | 1.30159000 | 9.68989000 | -1.65664550 |
| O | 12.85686000 | 2.88194000 | 9.68989000 | -1.64964290 |
| O | 5.59648000 | 7.07372000 | 9.68989000 | -1.63223220 |
| O | 5.61079000 | 9.68514000 | 9.68989000 | -1.64670820 |
| O | 13.02257000 | 0.00826000 | 9.68989000 | -1.66370360 |
| O | 8.01661000 | 11.26550000 | 9.68989000 | -1.63826050 |
| O | 0.77054000 | 1.30159000 | 9.68989000 | -1.65732480 |
| O | 8.03091000 | 5.49337000 | 9.68989000 | -1.66158890 |
| O | 8.18232000 | 8.39182000 | 9.68989000 | -1.65580500 |
| O | 3.17636000 | 2.88194000 | 9.68989000 | -1.64188090 |
| O | 10.43673000 | 7.07372000 | 9.68989000 | -1.64692400 |
| O | -4.06971000 | 9.68514000 | 9.68989000 | -1.66002150 |
| O | 10.60245000 | 4.20004000 | 9.68989000 | -1.66803450 |
| O | 3.34207000 | 0.00826000 | 9.68989000 | -1.66708200 |
| O | -1.66389000 | 11.26550000 | 9.68989000 | -1.64345320 |
| O | -1.64958000 | 5.49336000 | 9.68989000 | -1.65883550 |
| O | 5.61079000 | 1.30159000 | 9.68989000 | -1.65967230 |
| O | -1.49818000 | 8.39182000 | 9.68989000 | -1.66919850 |
| O | 0.75623000 | 7.07372000 | 9.68989000 | -1.64442810 |
| O | 0.77054000 | 9.68514000 | 9.68989000 | -1.64486630 |
| O | 0.92195000 | 4.20004000 | 9.68989000 | -1.66813610 |
| O | 8.18232000 | 0.00827000 | 9.68989000 | -1.66263400 |
| O | 3.19067000 | 5.49337000 | 9.68989000 | -1.64864460 |
| O | 3.34207000 | 8.39182000 | 9.68989000 | -1.65780180 |
| O | 5.76220000 | 4.20004000 | 9.68989000 | -1.65389420 |
| Al | 9.69003000 | 2.79452000 | 10.71830000 | 2.47483110 |
| Al | 4.84978000 | 11.17808000 | 10.71830000 | 2.48232030 |
| Al | 0.00953000 | 2.79452000 | 10.71830000 | 2.47525980 |
| Al | 7.26991000 | 6.98630000 | 10.71830000 | 2.47214740 |
| Al | -4.83072000 | 11.17808000 | 10.71830000 | 2.48123320 |
| Al | -2.41059000 | 6.98630000 | 10.71830000 | 2.47551170 |
| Al | 4.84978000 | 2.79452000 | 10.71830000 | 2.47662770 |
| Al | 0.00953000 | 11.17808000 | 10.71830000 | 2.48035180 |
| Al | 2.42966000 | 6.98630000 | 10.71830000 | 2.47411790 |
| Al | 9.69003000 | 0.00000000 | 10.98601000 | 2.47086950 |
| Al | 4.84978000 | 8.38356000 | 10.98601000 | 2.47166380 |

| | | | |
|---|---|---|---|
| Al | 0.00954000 | 0.00000000 | 10.98601000 | 2.47435270 |
| Al | 7.26991000 | 4.19178000 | 10.98601000 | 2.47047770 |
| Al | -4.83072000 | 8.38356000 | 10.98601000 | 2.46845230 |
| Al | -2.41059000 | 4.19178000 | 10.98601000 | 2.46875830 |
| Al | 4.84978000 | 0.00000000 | 10.98601000 | 2.47331450 |
| Al | 0.00953000 | 8.38356000 | 10.98601000 | 2.46935960 |
| Al | 2.42966000 | 4.19178000 | 10.98601000 | 2.46918010 |
| O | 8.86652000 | 4.09550000 | 11.85000000 | -1.63024260 |
| O | 8.97776000 | 1.45029000 | 11.84937000 | -1.61852320 |
| O | 11.22026000 | 2.86264000 | 11.84855000 | -1.61953480 |
| O | 4.02854000 | 12.47544000 | 11.86317000 | -1.62266050 |
| O | 4.11108000 | 9.80597000 | 11.86254000 | -1.58123270 |
| O | 6.39076000 | 11.21935000 | 11.84910000 | -1.62021560 |
| O | -0.81054000 | 4.08722000 | 11.84866000 | -1.61970980 |
| O | 6.47008000 | 8.24776000 | 11.85616000 | -1.59509680 |
| O | -0.70378000 | 1.44405000 | 11.85056000 | -1.62358840 |
| O | 6.57116000 | 5.64450000 | 11.91212000 | -1.57409620 |
| O | 8.80259000 | 7.00237000 | 11.86964000 | -1.61550810 |
| O | 1.54197000 | 2.86517000 | 11.85462000 | -1.61956530 |
| O | -5.64618000 | 12.47799000 | 11.85893000 | -1.62141210 |
| O | -5.58753000 | 9.79492000 | 11.86052000 | -1.60020210 |
| O | -3.30420000 | 11.23434000 | 11.85075000 | -1.62316870 |
| O | -3.23765000 | 8.27110000 | 11.85222000 | -1.62160400 |
| O | 4.04113000 | 4.11402000 | 11.84665000 | -1.61337460 |
| O | -3.09396000 | 5.63363000 | 11.84860000 | -1.62760500 |
| O | 4.13785000 | 1.44309000 | 11.85090000 | -1.62422360 |
| O | 6.38936000 | 2.86370000 | 11.86401000 | -1.62146910 |
| O | -0.87582000 | 7.05010000 | 11.85337000 | -1.61926800 |
| O | -0.81946000 | 12.47471000 | 11.85661000 | -1.62301600 |
| O | -0.70903000 | 9.81119000 | 11.86297000 | -1.62580100 |
| O | 1.52904000 | 11.25022000 | 11.82651000 | -1.60989510 |
| O | 1.65082000 | 8.29511000 | 11.86254000 | -1.59271560 |
| O | 1.71573000 | 5.64161000 | 11.88061000 | -1.62891850 |
| O | 3.95878000 | 7.02253000 | 11.94148000 | -1.56805090 |
| Al | 12.10920000 | 1.39569000 | 11.90056000 | 2.41691370 |
| Al | 7.23784000 | 9.77592000 | 12.33353000 | 2.38327320 |
| Al | 9.73762000 | 5.55334000 | 11.74937000 | 2.41071100 |
| Al | 2.42883000 | 1.40120000 | 11.89288000 | 2.41593570 |
| Al | -2.41630000 | 9.77630000 | 11.88779000 | 2.41470630 |
| Al | 7.27064000 | 1.39135000 | 11.89075000 | 2.41568110 |
| Al | -0.00178000 | 5.58539000 | 11.84950000 | 2.41540230 |
| Al | 2.40441000 | 9.84991000 | 12.35186000 | 2.38379500 |
| Al | 4.85768000 | 5.53714000 | 12.39424000 | 2.37564080 |
| Pt | 2.32997000 | 9.72856000 | 14.77189000 | -0.72846700 |
| Pt | 5.06107000 | 5.77596000 | 14.87949000 | -1.03619020 |
| Pt | 4.82737000 | 9.48232000 | 14.49827000 | -0.29675850 |
| Pt | 7.44135000 | 9.06656000 | 14.80091000 | -0.79234910 |
| Sn | 7.47894000 | 6.56467000 | 14.00179000 | 1.00640380 |
| Sn | 2.91786000 | 7.28327000 | 14.08013000 | 1.02096930 |
| Sn | 5.54642000 | 7.97372000 | 16.34329000 | 0.72218390 |

277

pt4sn3-al2o3-C

| | | | | |
|---|---|---|---|---|
| Al | 9.69003000 | 2.79452000 | 0.00000000 | 2.16892710 |
| Al | 4.84978000 | 11.17808000 | 0.00000000 | 2.16904140 |
| Al | 0.00953000 | 2.79452000 | 0.00000000 | 2.16898530 |
| Al | 7.26991000 | 6.98630000 | 0.00000000 | 2.16904850 |
| Al | -4.83072000 | 11.17808000 | 0.00000000 | 2.16910640 |
| Al | -2.41059000 | 6.98630000 | 0.00000000 | 2.16903440 |
| Al | 4.84978000 | 2.79452000 | 0.00000000 | 2.16903470 |
| Al | 0.00953000 | 11.17808000 | 0.00000000 | 2.16902820 |
| Al | 2.42966000 | 6.98630000 | 0.00000000 | 2.16906510 |
| O | 10.62741000 | 1.39726000 | 0.85241000 | -1.55200900 |
| O | 12.85153000 | 0.11316000 | 0.85241000 | -1.56035940 |
| O | 12.85153000 | 2.68136000 | 0.85241000 | -1.55810960 |
| O | 5.78716000 | 9.78082000 | 0.85241000 | -1.55204450 |
| O | 8.01128000 | 8.49672000 | 0.85241000 | -1.56038410 |
| O | 8.01128000 | 11.06492000 | 0.85241000 | -1.55814030 |
| O | 0.94691000 | 1.39726000 | 0.85241000 | -1.55210440 |
| O | 8.20728000 | 5.58904000 | 0.85241000 | -1.55202110 |
| O | 10.43141000 | 4.30494000 | 0.85241000 | -1.56032170 |
| O | 3.17103000 | 0.11317000 | 0.85241000 | -1.56042080 |
| O | 10.43140000 | 6.87314000 | 0.85241000 | -1.55816010 |
| O | 3.17103000 | 2.68136000 | 0.85241000 | -1.55813560 |
| O | -3.89334000 | 9.78082000 | 0.85241000 | -1.55204880 |
| O | -1.66922000 | 8.49672000 | 0.85241000 | -1.56038710 |
| O | -1.66922000 | 11.06492000 | 0.85241000 | -1.55818580 |
| O | -1.47322000 | 5.58904000 | 0.85241000 | -1.55199070 |
| O | 5.78716000 | 1.39726000 | 0.85241000 | -1.55205120 |
| O | 8.01128000 | 0.11317000 | 0.85241000 | -1.56036570 |
| O | 0.75091000 | 4.30494000 | 0.85241000 | -1.56027950 |
| O | 0.75090000 | 6.87314000 | 0.85241000 | -1.55810480 |
| O | 8.01128000 | 2.68136000 | 0.85241000 | -1.55811830 |
| O | 0.94691000 | 9.78082000 | 0.85241000 | -1.55197280 |
| O | 3.17103000 | 8.49672000 | 0.85241000 | -1.56033700 |
| O | 3.17103000 | 11.06492000 | 0.85241000 | -1.55810350 |
| O | 3.36703000 | 5.58904000 | 0.85241000 | -1.55203970 |
| O | 5.59116000 | 4.30494000 | 0.85241000 | -1.56029510 |
| O | 5.59116000 | 6.87314000 | 0.85241000 | -1.55807100 |
| Al | 9.69003000 | 0.00000000 | 1.70482000 | 2.46324260 |
| Al | 4.84978000 | 8.38356000 | 1.70482000 | 2.46327450 |
| Al | 0.00954000 | 0.00000000 | 1.70482000 | 2.46327530 |
| Al | 7.26991000 | 4.19178000 | 1.70482000 | 2.46324150 |
| Al | -4.83072000 | 8.38356000 | 1.70482000 | 2.46320220 |
| Al | -2.41059000 | 4.19178000 | 1.70482000 | 2.46324170 |
| Al | 4.84978000 | 0.00000000 | 1.70482000 | 2.46320300 |
| Al | 0.00953000 | 8.38356000 | 1.70482000 | 2.46324250 |
| Al | 2.42966000 | 4.19178000 | 1.70482000 | 2.46321220 |
| Al | 12.11016000 | 1.39726000 | 2.20274000 | 2.48268260 |
| Al | 7.26991000 | 9.78082000 | 2.20274000 | 2.48264390 |
| Al | 9.69003000 | 5.58904000 | 2.20274000 | 2.48265480 |
| Al | 2.42966000 | 1.39726000 | 2.20274000 | 2.48267660 |
| Al | -2.41059000 | 9.78082000 | 2.20274000 | 2.48270490 |
| Al | 7.26991000 | 1.39726000 | 2.20274000 | 2.48274620 |
| Al | 0.00953000 | 5.58904000 | 2.20274000 | 2.48272180 |
| Al | 2.42966000 | 9.78082000 | 2.20274000 | 2.48269680 |
| Al | 4.84978000 | 5.58904000 | 2.20274000 | 2.48272840 |
| O | 8.94865000 | 1.28410000 | 3.05515000 | -1.64184340 |
| O | 11.17279000 | 0.00001000 | 3.05515000 | -1.62407480 |
| O | 4.10840000 | 9.66766000 | 3.05515000 | -1.64165250 |
| O | 11.36879000 | 2.90768000 | 3.05515000 | -1.64967860 |
| O | 6.33254000 | 8.38356000 | 3.05515000 | -1.62413800 |
| O | 6.52853000 | 5.47588000 | 3.05515000 | -1.64179850 |
| O | -0.73185000 | 1.28410000 | 3.05515000 | -1.64170140 |
| O | 6.52853000 | 11.29124000 | 3.05515000 | -1.64987190 |
| O | 1.49229000 | 0.00001000 | 3.05515000 | -1.62424850 |
| O | 8.75266000 | 4.19179000 | 3.05515000 | -1.62399320 |
| O | -5.57210000 | 9.66766000 | 3.05515000 | -1.64212240 |
| O | 8.94866000 | 7.09946000 | 3.05515000 | -1.64983080 |
| O | 1.68828000 | 2.90768000 | 3.05515000 | -1.64985710 |
| O | -3.34797000 | 8.38357000 | 3.05515000 | -1.62403170 |
| O | 4.10840000 | 1.28410000 | 3.05515000 | -1.64188370 |
| O | -3.15197000 | 5.47588000 | 3.05515000 | -1.64202540 |
| O | -3.15197000 | 11.29124000 | 3.05515000 | -1.65008100 |
| O | -0.92784000 | 4.19178000 | 3.05515000 | -1.62400620 |
| O | 6.33254000 | 0.00001000 | 3.05515000 | -1.62415850 |
| O | 6.52854000 | 2.90768000 | 3.05515000 | -1.64969620 |
| O | -0.73185000 | 9.66766000 | 3.05515000 | -1.64207250 |
| O | -0.73184000 | 7.09946000 | 3.05515000 | -1.64980540 |
| O | 1.49228000 | 8.38356000 | 3.05515000 | -1.62400640 |
| O | 1.68829000 | 11.29124000 | 3.05515000 | -1.64972710 |
| O | 1.68828000 | 5.47588000 | 3.05515000 | -1.64167420 |
| O | 3.91241000 | 4.19178000 | 3.05515000 | -1.62409430 |
| O | 4.10841000 | 7.09946000 | 3.05515000 | -1.64960630 |
| Al | 9.69003000 | 2.79452000 | 3.90756000 | 2.47778360 |
| Al | 4.84978000 | 11.17808000 | 3.90756000 | 2.47773510 |
| Al | 0.00953000 | 2.79452000 | 3.90756000 | 2.47778430 |
| Al | 7.26991000 | 6.98630000 | 3.90756000 | 2.47773920 |
| Al | -4.83072000 | 11.17808000 | 3.90756000 | 2.47793100 |
| Al | -2.41059000 | 6.98630000 | 3.90756000 | 2.47791220 |
| Al | 4.84978000 | 2.79452000 | 3.90756000 | 2.47777650 |
| Al | 0.00953000 | 11.17808000 | 3.90756000 | 2.47786300 |
| Al | 2.42966000 | 6.98630000 | 3.90756000 | 2.47778940 |
| Al | 9.69003000 | 0.00000000 | 4.40548000 | 2.47245800 |
| Al | 4.84978000 | 8.38356000 | 4.40548000 | 2.47225480 |
| Al | 0.00954000 | 0.00000000 | 4.40548000 | 2.47224550 |
| Al | 7.26991000 | 4.19178000 | 4.40548000 | 2.47233140 |
| Al | -4.83072000 | 8.38356000 | 4.40548000 | 2.47233150 |
| Al | -2.41059000 | 4.19178000 | 4.40548000 | 2.47232840 |
| Al | 4.84978000 | 0.00000000 | 4.40548000 | 2.47242480 |
| Al | 0.00953000 | 8.38356000 | 4.40548000 | 2.47247280 |
| Al | 2.42966000 | 4.19178000 | 4.40548000 | 2.47245480 |
| O | 8.20728000 | 2.79452000 | 5.25789000 | -1.64351670 |
| O | 10.43141000 | 4.07863000 | 5.25789000 | -1.65395890 |
| O | 10.43142000 | 1.51042000 | 5.25789000 | -1.64914570 |
| O | 3.36702000 | 11.17808000 | 5.25789000 | -1.64331100 |
| O | 5.59116000 | 12.46219000 | 5.25789000 | -1.65406630 |
| O | 5.59116000 | 9.89398000 | 5.25789000 | -1.64912200 |
| O | -1.47322000 | 2.79452000 | 5.25789000 | -1.64337300 |
| O | 5.78715000 | 6.98630000 | 5.25789000 | -1.64324680 |
| O | 8.01128000 | 8.27041000 | 5.25789000 | -1.65415430 |
| O | 0.75091000 | 4.07863000 | 5.25789000 | -1.65379910 |
| O | 0.75091000 | 1.51042000 | 5.25789000 | -1.64909690 |
| O | 8.01129000 | 5.70220000 | 5.25789000 | -1.64913480 |
| O | -6.31347000 | 11.17808000 | 5.25789000 | -1.64325210 |
| O | -4.08934000 | 12.46219000 | 5.25789000 | -1.65377730 |
| O | -4.08934000 | 9.89398000 | 5.25789000 | -1.64875190 |
| O | -3.89335000 | 6.98630000 | 5.25789000 | -1.64353970 |
| O | 3.36703000 | 2.79452000 | 5.25789000 | -1.64347320 |
| O | 5.59116000 | 4.07863000 | 5.25789000 | -1.65382250 |
| O | -1.66922000 | 8.27041000 | 5.25789000 | -1.65373650 |
| O | 5.59116000 | 1.51042000 | 5.25789000 | -1.64900820 |
| O | -1.66921000 | 5.70220000 | 5.25789000 | -1.64898560 |

| | | | |
|---|---|---|---|
| O | -1.47323000 | 11.17808000 | 5.25789000 | -1.64364950 |
| O | 0.75091000 | 12.46219000 | 5.25789000 | -1.65386420 |
| O | 0.75091000 | 9.89398000 | 5.25789000 | -1.64889850 |
| O | 0.94690000 | 6.98630000 | 5.25789000 | -1.64353420 |
| O | 3.17103000 | 8.27041000 | 5.25789000 | -1.65402860 |
| O | 3.17104000 | 5.70220000 | 5.25789000 | -1.64933860 |
| Al | 12.11016000 | 1.39726000 | 6.11030000 | 2.47611880 |
| Al | 7.26991000 | 9.78082000 | 6.11030000 | 2.47606250 |
| Al | 9.69003000 | 5.58904000 | 6.11030000 | 2.47617030 |
| Al | 2.42966000 | 1.39726000 | 6.11030000 | 2.47597000 |
| Al | -2.41059000 | 9.78082000 | 6.11030000 | 2.47594620 |
| Al | 7.26991000 | 1.39726000 | 6.11030000 | 2.47608250 |
| Al | 0.00953000 | 5.58904000 | 6.11030000 | 2.47627680 |
| Al | 2.42966000 | 9.78082000 | 6.11030000 | 2.47617880 |
| Al | 4.84978000 | 5.58904000 | 6.11030000 | 2.47612470 |
| Al | 9.69003000 | 2.79452000 | 6.60910000 | 2.47690410 |
| Al | 4.84978000 | 11.17808000 | 6.60910000 | 2.47674650 |
| Al | 0.00953000 | 2.79452000 | 6.60910000 | 2.47669410 |
| Al | 7.26991000 | 6.98630000 | 6.60910000 | 2.47672950 |
| Al | -4.83072000 | 11.17808000 | 6.60910000 | 2.47675080 |
| Al | -2.41059000 | 6.98630000 | 6.60910000 | 2.47679750 |
| Al | 4.84978000 | 2.79452000 | 6.60910000 | 2.47696300 |
| Al | 0.00953000 | 11.17808000 | 6.60910000 | 2.47682650 |
| Al | 2.42966000 | 6.98630000 | 6.60910000 | 2.47703580 |
| O | 11.36395000 | 0.11351000 | 7.46570000 | -1.64865280 |
| O | 11.37150000 | 2.68537000 | 7.46570000 | -1.64974220 |
| O | 13.59502000 | 1.39291000 | 7.46570000 | -1.65431530 |
| O | 6.52370000 | 8.49707000 | 7.46570000 | -1.64875360 |
| O | 6.53125000 | 11.06893000 | 7.46570000 | -1.64982600 |
| O | 8.75477000 | 9.77647000 | 7.46570000 | -1.65449540 |
| O | 1.68345000 | 0.11351000 | 7.46570000 | -1.64980580 |
| O | 8.94383000 | 4.30529000 | 7.46570000 | -1.64934400 |
| O | 8.95137000 | 6.87715000 | 7.46570000 | -1.65001470 |
| O | 1.69100000 | 2.68537000 | 7.46570000 | -1.64951230 |
| O | 11.17490000 | 5.58469000 | 7.46570000 | -1.65417660 |
| O | 3.91452000 | 1.39291000 | 7.46570000 | -1.65422480 |
| O | -3.15680000 | 8.49707000 | 7.46570000 | -1.64945760 |
| O | -3.14925000 | 11.06893000 | 7.46570000 | -1.65020620 |
| O | -0.92573000 | 9.77647000 | 7.46570000 | -1.65374170 |
| O | -0.73667000 | 4.30528000 | 7.46570000 | -1.64925540 |
| O | 6.52370000 | 0.11351000 | 7.46570000 | -1.64922980 |
| O | -0.72913000 | 6.87715000 | 7.46570000 | -1.64922080 |
| O | 6.53125000 | 2.68537000 | 7.46570000 | -1.64959500 |
| O | 8.75477000 | 1.39291000 | 7.46570000 | -1.65410410 |
| O | 1.49440000 | 5.58468000 | 7.46570000 | -1.65351190 |
| O | 1.68345000 | 8.49707000 | 7.46570000 | -1.64901790 |
| O | 1.69100000 | 11.06893000 | 7.46570000 | -1.64940210 |
| O | 3.91452000 | 9.77646000 | 7.46570000 | -1.65356820 |
| O | 4.10358000 | 4.30529000 | 7.46570000 | -1.64880320 |
| O | 4.11112000 | 6.87715000 | 7.46570000 | -1.64895320 |
| O | 6.33465000 | 5.58469000 | 7.46570000 | -1.65347940 |
| Al | 9.69003000 | 0.00000000 | 8.32833000 | 2.47084740 |
| Al | 4.84978000 | 8.38356000 | 8.32833000 | 2.47148540 |
| Al | 0.00954000 | 0.00000000 | 8.32833000 | 2.47178070 |
| Al | 7.26991000 | 4.19178000 | 8.32833000 | 2.47168010 |
| Al | -4.83072000 | 8.38356000 | 8.32833000 | 2.47173100 |
| Al | -2.41059000 | 4.19178000 | 8.32833000 | 2.47100450 |
| Al | 4.84978000 | 0.00000000 | 8.32833000 | 2.47205370 |
| Al | 0.00953000 | 8.38356000 | 8.32833000 | 2.47024460 |
| Al | 2.42966000 | 4.19178000 | 8.32833000 | 2.47166280 |
| Al | 12.11016000 | 1.39726000 | 8.78548000 | 2.47260980 |
| Al | 7.26991000 | 9.78082000 | 8.78548000 | 2.47229220 |
| Al | 9.69003000 | 5.58904000 | 8.78548000 | 2.47239950 |
| Al | 2.42966000 | 1.39726000 | 8.78548000 | 2.47221050 |
| Al | -2.41059000 | 9.78082000 | 8.78548000 | 2.47204380 |
| Al | 7.26991000 | 1.39726000 | 8.78548000 | 2.47234750 |
| Al | 0.00953000 | 5.58904000 | 8.78548000 | 2.47592430 |
| Al | 2.42966000 | 9.78082000 | 8.78548000 | 2.47664790 |
| Al | 4.84978000 | 5.58904000 | 8.78548000 | 2.47619430 |
| O | 8.01661000 | 2.88194000 | 9.68989000 | -1.64430060 |
| O | 3.17636000 | 11.26550000 | 9.68989000 | -1.63713700 |
| O | 10.45104000 | 1.30159000 | 9.68989000 | -1.65680810 |
| O | 12.85686000 | 2.88194000 | 9.68989000 | -1.64165200 |
| O | 5.59648000 | 7.07372000 | 9.68989000 | -1.63710270 |
| O | 5.61079000 | 9.68514000 | 9.68989000 | -1.65688200 |
| O | 13.02257000 | 0.00826000 | 9.68989000 | -1.66765540 |
| O | 8.01661000 | 11.26550000 | 9.68989000 | -1.64435430 |
| O | 0.77054000 | 1.30159000 | 9.68989000 | -1.66021740 |
| O | 8.03091000 | 5.49337000 | 9.68989000 | -1.65873180 |
| O | 8.18232000 | 8.39182000 | 9.68989000 | -1.66767430 |
| O | 3.17636000 | 2.88194000 | 9.68989000 | -1.64384600 |
| O | 10.43673000 | 7.07372000 | 9.68989000 | -1.64425130 |
| O | -4.06971000 | 9.68514000 | 9.68989000 | -1.66023660 |
| O | 10.60245000 | 4.20004000 | 9.68989000 | -1.66786180 |
| O | 3.34207000 | 0.00826000 | 9.68989000 | -1.66954370 |
| O | -1.66389000 | 11.26550000 | 9.68989000 | -1.64248330 |

| | | | |
|---|---|---|---|
| O | -1.64958000 | 5.49336000 | 9.68989000 | -1.64769690 |
| O | 5.61079000 | 1.30159000 | 9.68989000 | -1.65962620 |
| O | -1.49818000 | 8.39182000 | 9.68989000 | -1.66639010 |
| O | 0.75623000 | 7.07372000 | 9.68989000 | -1.62974760 |
| O | 0.77054000 | 9.68514000 | 9.68989000 | -1.64799750 |
| O | 0.92195000 | 4.20004000 | 9.68989000 | -1.65799030 |
| O | 8.18232000 | 0.00827000 | 9.68989000 | -1.66377900 |
| O | 3.19067000 | 5.49337000 | 9.68989000 | -1.64499710 |
| O | 3.34207000 | 8.39182000 | 9.68989000 | -1.65603370 |
| O | 5.76220000 | 4.20004000 | 9.68989000 | -1.65543760 |
| Al | 9.69003000 | 2.79452000 | 10.71830000 | 2.47819320 |
| Al | 4.84978000 | 11.17808000 | 10.71830000 | 2.47551020 |
| Al | 0.00953000 | 2.79452000 | 10.71830000 | 2.47496700 |
| Al | 7.26991000 | 6.98630000 | 10.71830000 | 2.47991410 |
| Al | -4.83072000 | 11.17808000 | 10.71830000 | 2.47571910 |
| Al | -2.41059000 | 6.98630000 | 10.71830000 | 2.47453940 |
| Al | 4.84978000 | 2.79452000 | 10.71830000 | 2.47873460 |
| Al | 0.00953000 | 11.17808000 | 10.71830000 | 2.47954600 |
| Al | 2.42966000 | 6.98630000 | 10.71830000 | 3.00000000 |
| Al | 9.69003000 | 0.00000000 | 10.98601000 | 2.47130850 |
| Al | 4.84978000 | 8.38356000 | 10.98601000 | 2.46927120 |
| Al | 0.00954000 | 0.00000000 | 10.98601000 | 2.46993420 |
| Al | 7.26991000 | 4.19178000 | 10.98601000 | 2.47274220 |
| Al | -4.83072000 | 8.38356000 | 10.98601000 | 2.46906800 |
| Al | -2.41059000 | 4.19178000 | 10.98601000 | 2.47194180 |
| Al | 4.84978000 | 0.00000000 | 10.98601000 | 2.47093230 |
| Al | 0.00953000 | 8.38356000 | 10.98601000 | 2.47155930 |
| Al | 2.42966000 | 4.19178000 | 10.98601000 | 2.47090480 |
| O | 8.87185000 | 4.08919000 | 11.85246000 | -1.62252030 |
| O | 8.97815000 | 1.43855000 | 11.85286000 | -1.62312530 |
| O | 11.22060000 | 2.85590000 | 11.85265000 | -1.62612410 |
| O | 4.03515000 | 12.47052000 | 11.86107000 | -1.62019980 |
| O | 4.11165000 | 9.79568000 | 11.86778000 | -1.60113620 |
| O | 6.37688000 | 11.23835000 | 11.85301000 | -1.62800600 |
| O | -0.80424000 | 4.09614000 | 11.84558000 | -1.60592940 |
| O | 6.44971000 | 8.29571000 | 11.87028000 | -1.63239660 |
| O | -0.69787000 | 1.43613000 | 11.85291000 | -1.62591400 |
| O | 6.54927000 | 5.63123000 | 11.84621000 | -1.60434740 |
| O | 8.80023000 | 7.05321000 | 11.85311000 | -1.62429030 |
| O | 1.55368000 | 2.85025000 | 11.86987000 | -1.62759470 |
| O | -5.65337000 | 12.46407000 | 11.85180000 | -1.61960420 |
| O | -5.53323000 | 9.82227000 | 11.85150000 | -1.62394250 |
| O | -3.30511000 | 11.24790000 | 11.85127000 | -1.62125040 |
| O | -3.22715000 | 8.27885000 | 11.85277000 | -1.62059870 |
| O | 4.02200000 | 4.12451000 | 11.86716000 | -1.60812270 |
| O | -3.12283000 | 5.63423000 | 11.86095000 | -1.62850940 |
| O | 4.13840000 | 1.44144000 | 11.85319000 | -1.62526710 |
| O | 6.37652000 | 2.85357000 | 11.86063000 | -1.62444230 |
| O | -0.84656000 | 7.04031000 | 11.86797000 | -1.59932170 |
| O | -0.81369000 | 12.46974000 | 11.85298000 | -1.62132230 |
| O | -0.71445000 | 9.81333000 | 11.86950000 | -1.63089520 |
| O | 1.54269000 | 11.23281000 | 11.84538000 | -1.60857390 |
| O | 1.58140000 | 8.30753000 | 11.94203000 | -1.57806190 |
| O | 1.70776000 | 5.58892000 | 11.95055000 | -1.57463480 |
| O | 3.99756000 | 7.06003000 | 11.94655000 | -1.57048290 |
| Al | 12.11154000 | 1.39918000 | 11.88304000 | 2.41520240 |
| Al | 7.28307000 | 9.79332000 | 11.86084000 | 2.41615930 |
| Al | 9.68720000 | 5.58938000 | 11.88369000 | 2.41518900 |
| Al | 2.43408000 | 1.37943000 | 11.86062000 | 2.41597990 |
| Al | -2.42863000 | 9.78591000 | 11.85854000 | 2.41293910 |
| Al | 7.27088000 | 1.39475000 | 11.88337000 | 2.41345420 |
| Al | -0.04260000 | 5.56210000 | 12.37601000 | 2.37609720 |
| Al | 2.42651000 | 9.83511000 | 12.37061000 | 2.37769140 |
| Al | 4.89913000 | 5.56067000 | 12.37924000 | 2.37421080 |
| Pt | 2.38607000 | 7.01776000 | 15.27519000 | -0.53625490 |
| Pt | 0.05392000 | 5.68307000 | 14.86081000 | -0.77456260 |
| Pt | 2.46667000 | 9.69709000 | 14.85473000 | -0.77223970 |
| Pt | 4.70976000 | 5.61070000 | 14.86362000 | -0.78947210 |
| Sn | 2.36336000 | 4.51417000 | 14.19239000 | 0.95029750 |
| Sn | 0.27096000 | 8.29011000 | 14.20862000 | 0.94423340 |
| Sn | 4.59414000 | 8.19084000 | 14.19744000 | 0.94814440 |

277

pt4sn3-al2o3-D

| | | | | |
|---|---|---|---|---|
| Al | 9.69003000 | 2.79452000 | 0.00000000 | 2.17039470 |
| Al | 4.84978000 | 11.17808000 | 0.00000000 | 2.17048340 |
| Al | 0.00953000 | 2.79452000 | 0.00000000 | 2.17036490 |
| Al | 7.26991000 | 6.98630000 | 0.00000000 | 2.17054780 |
| Al | -4.83072000 | 11.17808000 | 0.00000000 | 2.17061210 |
| Al | -2.41059000 | 6.98630000 | 0.00000000 | 2.17040460 |
| Al | 4.84978000 | 2.79452000 | 0.00000000 | 2.17039470 |
| Al | 0.00953000 | 11.17808000 | 0.00000000 | 2.17040810 |
| Al | 2.42966000 | 6.98630000 | 0.00000000 | 2.17035400 |
| O | 10.62741000 | 1.39726000 | 0.85241000 | -1.55161990 |
| O | 12.85153000 | 0.11316000 | 0.85241000 | -1.55993110 |
| O | 12.85153000 | 2.68136000 | 0.85241000 | -1.55767630 |
| O | 5.78716000 | 9.78082000 | 0.85241000 | -1.55172600 |
| O | 8.01128000 | 8.49672000 | 0.85241000 | -1.56008270 |
| O | 8.01128000 | 11.06492000 | 0.85241000 | -1.55772870 |
| O | 0.94691000 | 1.39726000 | 0.85241000 | -1.55179960 |
| O | 8.20728000 | 5.58904000 | 0.85241000 | -1.55171000 |
| O | 10.43141000 | 4.30494000 | 0.85241000 | -1.55995190 |
| O | 3.17103000 | 0.11317000 | 0.85241000 | -1.55999740 |
| O | 10.43140000 | 6.87314000 | 0.85241000 | -1.55780340 |
| O | 3.17103000 | 2.68136000 | 0.85241000 | -1.55767490 |
| O | -3.89334000 | 9.78082000 | 0.85241000 | -1.55181480 |
| O | -1.66922000 | 8.49672000 | 0.85241000 | -1.55995650 |
| O | -1.66922000 | 11.06492000 | 0.85241000 | -1.55794180 |
| O | -1.47322000 | 5.58904000 | 0.85241000 | -1.55157880 |
| O | 5.78716000 | 1.39726000 | 0.85241000 | -1.55188730 |
| O | 8.01128000 | 0.11317000 | 0.85241000 | -1.55996000 |
| O | 0.75091000 | 4.30494000 | 0.85241000 | -1.55985630 |
| O | 0.75090000 | 6.87314000 | 0.85241000 | -1.55773890 |
| O | 8.01128000 | 2.68136000 | 0.85241000 | -1.55774680 |
| O | 0.94691000 | 9.78082000 | 0.85241000 | -1.55143900 |
| O | 3.17103000 | 8.49672000 | 0.85241000 | -1.55987650 |
| O | 3.17103000 | 11.06492000 | 0.85241000 | -1.55766370 |
| O | 3.36703000 | 5.58904000 | 0.85241000 | -1.55177620 |
| O | 5.59116000 | 4.30494000 | 0.85241000 | -1.55988090 |
| O | 5.59116000 | 6.87314000 | 0.85241000 | -1.55769270 |
| Al | 9.69003000 | 0.00000000 | 1.70482000 | 2.46327640 |
| Al | 4.84978000 | 8.38356000 | 1.70482000 | 2.46338220 |
| Al | 0.00954000 | 0.00000000 | 1.70482000 | 2.46329000 |
| Al | 7.26991000 | 4.19178000 | 1.70482000 | 2.46330650 |
| Al | -4.83072000 | 8.38356000 | 1.70482000 | 2.46327870 |
| Al | -2.41059000 | 4.19178000 | 1.70482000 | 2.46327520 |
| Al | 4.84978000 | 0.00000000 | 1.70482000 | 2.46344640 |
| Al | 0.00953000 | 8.38356000 | 1.70482000 | 2.46328930 |
| Al | 2.42966000 | 4.19178000 | 1.70482000 | 2.46345880 |
| Al | 12.11016000 | 1.39726000 | 2.20274000 | 2.48273730 |
| Al | 7.26991000 | 9.78082000 | 2.20274000 | 2.48265030 |
| Al | 9.69003000 | 5.58904000 | 2.20274000 | 2.48269180 |
| Al | 2.42966000 | 1.39726000 | 2.20274000 | 2.48270370 |
| Al | -2.41059000 | 9.78082000 | 2.20274000 | 2.48274730 |
| Al | 7.26991000 | 1.39726000 | 2.20274000 | 2.48280790 |
| Al | 0.00953000 | 5.58904000 | 2.20274000 | 2.48272660 |
| Al | 2.42966000 | 9.78082000 | 2.20274000 | 2.48275830 |
| Al | 4.84978000 | 5.58904000 | 2.20274000 | 2.48283140 |
| O | 8.94865000 | 1.28410000 | 3.05515000 | -1.64149890 |
| O | 11.17279000 | 0.00001000 | 3.05515000 | -1.62390890 |
| O | 4.10840000 | 9.66766000 | 3.05515000 | -1.64148320 |
| O | 11.36879000 | 2.90768000 | 3.05515000 | -1.64951900 |
| O | 6.33254000 | 8.38356000 | 3.05515000 | -1.62404900 |
| O | 6.52853000 | 5.47588000 | 3.05515000 | -1.64152990 |
| O | -0.73185000 | 1.28410000 | 3.05515000 | -1.64148070 |
| O | 6.52853000 | 11.29124000 | 3.05515000 | -1.64947670 |
| O | 1.49229000 | 0.00001000 | 3.05515000 | -1.62393910 |
| O | 8.75266000 | 4.19179000 | 3.05515000 | -1.62387940 |
| O | -5.57210000 | 9.66766000 | 3.05515000 | -1.64178730 |
| O | 8.94866000 | 7.09946000 | 3.05515000 | -1.64953700 |
| O | 1.68828000 | 2.90768000 | 3.05515000 | -1.64933680 |
| O | -3.34797000 | 8.38357000 | 3.05515000 | -1.62393180 |
| O | 4.10840000 | 1.28410000 | 3.05515000 | -1.64164120 |
| O | -3.15197000 | 5.47588000 | 3.05515000 | -1.64150980 |
| O | -3.15197000 | 11.29124000 | 3.05515000 | -1.64962270 |
| O | -0.92784000 | 4.19178000 | 3.05515000 | -1.62368540 |
| O | 6.33254000 | 0.00001000 | 3.05515000 | -1.62382780 |
| O | 6.52854000 | 2.90768000 | 3.05515000 | -1.64942610 |
| O | -0.73185000 | 9.66766000 | 3.05515000 | -1.64145870 |
| O | -0.73184000 | 7.09946000 | 3.05515000 | -1.64944890 |
| O | 1.49228000 | 8.38356000 | 3.05515000 | -1.62389810 |
| O | 1.68829000 | 11.29124000 | 3.05515000 | -1.64946560 |
| O | 1.68828000 | 5.47588000 | 3.05515000 | -1.64150370 |
| O | 3.91241000 | 4.19178000 | 3.05515000 | -1.62378220 |
| O | 4.10841000 | 7.09946000 | 3.05515000 | -1.64927850 |
| Al | 9.69003000 | 2.79452000 | 3.90756000 | 2.47766480 |
| Al | 4.84978000 | 11.17808000 | 3.90756000 | 2.47767700 |
| Al | 0.00953000 | 2.79452000 | 3.90756000 | 2.47763170 |
| Al | 7.26991000 | 6.98630000 | 3.90756000 | 2.47767530 |
| Al | -4.83072000 | 11.17808000 | 3.90756000 | 2.47780750 |
| Al | -2.41059000 | 6.98630000 | 3.90756000 | 2.47759290 |
| Al | 4.84978000 | 2.79452000 | 3.90756000 | 2.47770820 |
| Al | 0.00953000 | 11.17808000 | 3.90756000 | 2.47754490 |
| Al | 2.42966000 | 6.98630000 | 3.90756000 | 2.47753440 |
| Al | 9.69003000 | 0.00000000 | 4.40548000 | 2.47239610 |
| Al | 4.84978000 | 8.38356000 | 4.40548000 | 2.47243820 |
| Al | 0.00954000 | 0.00000000 | 4.40548000 | 2.47242790 |
| Al | 7.26991000 | 4.19178000 | 4.40548000 | 2.47227380 |
| Al | -4.83072000 | 8.38356000 | 4.40548000 | 2.47221770 |
| Al | -2.41059000 | 4.19178000 | 4.40548000 | 2.47233530 |
| Al | 4.84978000 | 0.00000000 | 4.40548000 | 2.47238610 |
| Al | 0.00953000 | 8.38356000 | 4.40548000 | 2.47223140 |
| Al | 2.42966000 | 4.19178000 | 4.40548000 | 2.47238240 |
| O | 8.20728000 | 2.79452000 | 5.25789000 | -1.64339840 |
| O | 10.43141000 | 4.07863000 | 5.25789000 | -1.65394570 |
| O | 10.43142000 | 1.51042000 | 5.25789000 | -1.64913650 |
| O | 3.36702000 | 11.17808000 | 5.25789000 | -1.64309360 |
| O | 5.59116000 | 12.46219000 | 5.25789000 | -1.65432650 |
| O | 5.59116000 | 9.89398000 | 5.25789000 | -1.64907040 |
| O | -1.47322000 | 2.79452000 | 5.25789000 | -1.64331470 |
| O | 5.78715000 | 6.98630000 | 5.25789000 | -1.64306800 |
| O | 8.01128000 | 8.27041000 | 5.25789000 | -1.65425860 |
| O | 0.75091000 | 4.07863000 | 5.25789000 | -1.65393300 |
| O | 0.75091000 | 1.51042000 | 5.25789000 | -1.64928750 |
| O | 8.01129000 | 5.70220000 | 5.25789000 | -1.64906700 |
| O | -6.31347000 | 11.17808000 | 5.25789000 | -1.64337090 |
| O | -4.08934000 | 12.46219000 | 5.25789000 | -1.65388640 |
| O | -4.08934000 | 9.89398000 | 5.25789000 | -1.64884530 |
| O | -3.89335000 | 6.98630000 | 5.25789000 | -1.64363420 |
| O | 3.36703000 | 2.79452000 | 5.25789000 | -1.64321530 |
| O | 5.59116000 | 4.07863000 | 5.25789000 | -1.65405540 |
| O | -1.66922000 | 8.27041000 | 5.25789000 | -1.65386890 |
| O | 5.59116000 | 1.51042000 | 5.25789000 | -1.64918410 |
| O | -1.66921000 | 5.70220000 | 5.25789000 | -1.64896900 |
| O | -1.47323000 | 11.17808000 | 5.25789000 | -1.64359460 |
| O | 0.75091000 | 12.46219000 | 5.25789000 | -1.65379430 |
| O | 0.75091000 | 9.89398000 | 5.25789000 | -1.64887670 |
| O | 0.94690000 | 6.98630000 | 5.25789000 | -1.64321490 |
| O | 3.17103000 | 8.27041000 | 5.25789000 | -1.65413700 |
| O | 3.17104000 | 5.70220000 | 5.25789000 | -1.64926520 |
| Al | 12.11016000 | 1.39726000 | 6.11030000 | 2.47603510 |
| Al | 7.26991000 | 9.78082000 | 6.11030000 | 2.47601970 |
| Al | 9.69003000 | 5.58904000 | 6.11030000 | 2.47633190 |
| Al | 2.42966000 | 1.39726000 | 6.11030000 | 2.47608500 |
| Al | -2.41059000 | 9.78082000 | 6.11030000 | 2.47600210 |
| Al | 7.26991000 | 1.39726000 | 6.11030000 | 2.47593240 |
| Al | 0.00953000 | 5.58904000 | 6.11030000 | 2.47644060 |
| Al | 2.42966000 | 9.78082000 | 6.11030000 | 2.47624200 |
| Al | 4.84978000 | 5.58904000 | 6.11030000 | 2.47620460 |
| Al | 9.69003000 | 2.79452000 | 6.60910000 | 2.47683180 |
| Al | 4.84978000 | 11.17808000 | 6.60910000 | 2.47699450 |
| Al | 0.00953000 | 2.79452000 | 6.60910000 | 2.47670070 |
| Al | 7.26991000 | 6.98630000 | 6.60910000 | 2.47695580 |
| Al | -4.83072000 | 11.17808000 | 6.60910000 | 2.47681610 |
| Al | -2.41059000 | 6.98630000 | 6.60910000 | 2.47686320 |
| Al | 4.84978000 | 2.79452000 | 6.60910000 | 2.47688580 |

| | | | |
|---|---|---|---|
| Al | 0.00953000 | 11.17808000 | 6.60910000 | 2.47697300 |
| Al | 2.42966000 | 6.98630000 | 6.60910000 | 2.47750040 |
| O | 11.36395000 | 0.11351000 | 7.46570000 | -1.64864940 |
| O | 11.37150000 | 2.68537000 | 7.46570000 | -1.64961010 |
| O | 13.59502000 | 1.39291000 | 7.46570000 | -1.65417570 |
| O | 6.52370000 | 8.49707000 | 7.46570000 | -1.64923410 |
| O | 6.53125000 | 11.06893000 | 7.46570000 | -1.64937120 |
| O | 8.75477000 | 9.77647000 | 7.46570000 | -1.65453810 |
| O | 1.68345000 | 0.11351000 | 7.46570000 | -1.64983070 |
| O | 8.94383000 | 4.30529000 | 7.46570000 | -1.64894450 |
| O | 8.95137000 | 6.87715000 | 7.46570000 | -1.65009200 |
| O | 1.69100000 | 2.68537000 | 7.46570000 | -1.64927960 |
| O | 11.17490000 | 5.58469000 | 7.46570000 | -1.65403520 |
| O | 3.91452000 | 1.39291000 | 7.46570000 | -1.65398960 |
| O | -3.15680000 | 8.49707000 | 7.46570000 | -1.64919400 |
| O | -3.14925000 | 11.06893000 | 7.46570000 | -1.65025020 |
| O | -0.92573000 | 9.77647000 | 7.46570000 | -1.65341050 |
| O | -0.73667000 | 4.30528000 | 7.46570000 | -1.64898820 |
| O | 6.52370000 | 0.11351000 | 7.46570000 | -1.64916780 |
| O | -0.72913000 | 6.87715000 | 7.46570000 | -1.64916660 |
| O | 6.53125000 | 2.68537000 | 7.46570000 | -1.64942710 |
| O | 8.75477000 | 1.39291000 | 7.46570000 | -1.65389880 |
| O | 1.49440000 | 5.58468000 | 7.46570000 | -1.65392870 |
| O | 1.68345000 | 8.49707000 | 7.46570000 | -1.64837120 |
| O | 1.69100000 | 11.06893000 | 7.46570000 | -1.64923840 |
| O | 3.91452000 | 9.77646000 | 7.46570000 | -1.65369760 |
| O | 4.10358000 | 4.30529000 | 7.46570000 | -1.64849550 |
| O | 4.11112000 | 6.87715000 | 7.46570000 | -1.64867920 |
| O | 6.33465000 | 5.58469000 | 7.46570000 | -1.65371960 |
| Al | 9.69003000 | 0.00000000 | 8.32833000 | 2.47039620 |
| Al | 4.84978000 | 8.38356000 | 8.32833000 | 2.47133560 |
| Al | 0.00954000 | 0.00000000 | 8.32833000 | 2.47207990 |
| Al | 7.26991000 | 4.19178000 | 8.32833000 | 2.47165350 |
| Al | -4.83072000 | 8.38356000 | 8.32833000 | 2.47190400 |
| Al | -2.41059000 | 4.19178000 | 8.32833000 | 2.47067970 |
| Al | 4.84978000 | 0.00000000 | 8.32833000 | 2.47243600 |
| Al | 0.00953000 | 8.38356000 | 8.32833000 | 2.46984300 |
| Al | 2.42966000 | 4.19178000 | 8.32833000 | 2.47101860 |
| Al | 12.11016000 | 1.39726000 | 8.78548000 | 2.47253760 |
| Al | 7.26991000 | 9.78082000 | 8.78548000 | 2.47202160 |
| Al | 9.69003000 | 5.58904000 | 8.78548000 | 2.47183390 |
| Al | 2.42966000 | 1.39726000 | 8.78548000 | 2.47232080 |
| Al | -2.41059000 | 9.78082000 | 8.78548000 | 2.47116070 |

| | | | |
|---|---|---|---|
| Al | 7.26991000 | 1.39726000 | 8.78548000 | 2.47202520 |
| Al | 0.00953000 | 5.58904000 | 8.78548000 | 2.47522050 |
| Al | 2.42966000 | 9.78082000 | 8.78548000 | 2.47596360 |
| Al | 4.84978000 | 5.58904000 | 8.78548000 | 2.47695590 |
| O | 8.01661000 | 2.88194000 | 9.68989000 | -1.64440720 |
| O | 3.17636000 | 11.26550000 | 9.68989000 | -1.63911330 |
| O | 10.45104000 | 1.30159000 | 9.68989000 | -1.65577450 |
| O | 12.85686000 | 2.88194000 | 9.68989000 | -1.64182460 |
| O | 5.59648000 | 7.07372000 | 9.68989000 | -1.63649280 |
| O | 5.61079000 | 9.68514000 | 9.68989000 | -1.65779750 |
| O | 13.02257000 | 0.00826000 | 9.68989000 | -1.66693320 |
| O | 8.01661000 | 11.26550000 | 9.68989000 | -1.64473360 |
| O | 0.77054000 | 1.30159000 | 9.68989000 | -1.66063170 |
| O | 8.03091000 | 5.49337000 | 9.68989000 | -1.65966130 |
| O | 8.18232000 | 8.39182000 | 9.68989000 | -1.66652510 |
| O | 3.17636000 | 2.88194000 | 9.68989000 | -1.64372940 |
| O | 10.43673000 | 7.07372000 | 9.68989000 | -1.64571330 |
| O | -4.06971000 | 9.68514000 | 9.68989000 | -1.66093320 |
| O | 10.60245000 | 4.20004000 | 9.68989000 | -1.66773040 |
| O | 3.34207000 | 0.00826000 | 9.68989000 | -1.66868280 |
| O | -1.66389000 | 11.26550000 | 9.68989000 | -1.64293290 |
| O | -1.64958000 | 5.49336000 | 9.68989000 | -1.64717430 |
| O | 5.61079000 | 1.30159000 | 9.68989000 | -1.66044740 |
| O | -1.49818000 | 8.39182000 | 9.68989000 | -1.66314160 |
| O | 0.75623000 | 7.07372000 | 9.68989000 | -1.63729920 |
| O | 0.77054000 | 9.68514000 | 9.68989000 | -1.64718930 |
| O | 0.92195000 | 4.20004000 | 9.68989000 | -1.65665570 |
| O | 8.18232000 | 0.00827000 | 9.68989000 | -1.66431030 |
| O | 3.19067000 | 5.49337000 | 9.68989000 | -1.64970320 |
| O | 3.34207000 | 8.39182000 | 9.68989000 | -1.65309410 |
| O | 5.76220000 | 4.20004000 | 9.68989000 | -1.65675010 |
| Al | 9.69003000 | 2.79452000 | 10.71830000 | 2.47805830 |
| Al | 4.84978000 | 11.17808000 | 10.71830000 | 2.48145050 |
| Al | 0.00953000 | 2.79452000 | 10.71830000 | 2.47718040 |
| Al | 7.26991000 | 6.98630000 | 10.71830000 | 2.47283240 |
| Al | -4.83072000 | 11.17808000 | 10.71830000 | 2.47690580 |
| Al | -2.41059000 | 6.98630000 | 10.71830000 | 2.47554150 |
| Al | 4.84978000 | 2.79452000 | 10.71830000 | 2.47952890 |
| Al | 0.00953000 | 11.17808000 | 10.71830000 | 2.47557890 |
| Al | 2.42966000 | 6.98630000 | 10.71830000 | 2.47665850 |
| Al | 9.69003000 | 0.00000000 | 10.98601000 | 2.47230230 |
| Al | 4.84978000 | 8.38356000 | 10.98601000 | 2.46871180 |
| Al | 0.00954000 | 0.00000000 | 10.98601000 | 2.46945380 |

| | | | |
|---|---|---|---|
| Al | 7.26991000 | 4.19178000 | 10.98601000 | 2.47070770 |
| Al | -4.83072000 | 8.38356000 | 10.98601000 | 2.47067490 |
| Al | -2.41059000 | 4.19178000 | 10.98601000 | 2.47223780 |
| Al | 4.84978000 | 0.00000000 | 10.98601000 | 2.47071380 |
| Al | 0.00953000 | 8.38356000 | 10.98601000 | 2.47008030 |
| Al | 2.42966000 | 4.19178000 | 10.98601000 | 2.47318860 |
| O | 8.86970000 | 4.09141000 | 11.85311000 | -1.62262620 |
| O | 8.97676000 | 1.43895000 | 11.85276000 | -1.62289780 |
| O | 11.21824000 | 2.86002000 | 11.85342000 | -1.62441380 |
| O | 4.03688000 | 12.47866000 | 11.85889000 | -1.62322670 |
| O | 4.13482000 | 9.80564000 | 11.86298000 | -1.61456640 |
| O | 6.37759000 | 11.25083000 | 11.85663000 | -1.62601140 |
| O | -0.80606000 | 4.10886000 | 11.84753000 | -1.61667370 |
| O | 6.43875000 | 8.29823000 | 11.94190000 | -1.60035790 |
| O | -0.69816000 | 1.43753000 | 11.85179000 | -1.62457870 |
| O | 6.56729000 | 5.64457000 | 11.83039000 | -1.61013950 |
| O | 8.79263000 | 7.05981000 | 11.85864000 | -1.62464870 |
| O | 1.54969000 | 2.85676000 | 11.87136000 | -1.62557210 |
| O | -5.65043000 | 12.46695000 | 11.85041000 | -1.62021730 |
| O | -5.53375000 | 9.81932000 | 11.85151000 | -1.62295380 |
| O | -3.30863000 | 11.24648000 | 11.84872000 | -1.62479090 |
| O | -3.23656000 | 8.27701000 | 11.85448000 | -1.62668710 |
| O | 4.03050000 | 4.11566000 | 11.85478000 | -1.61447590 |
| O | -3.12838000 | 5.63786000 | 11.85728000 | -1.62496850 |
| O | 4.13822000 | 1.43514000 | 11.85315000 | -1.62348520 |
| O | 6.38128000 | 2.85351000 | 11.86081000 | -1.62360950 |
| O | -0.86843000 | 7.05479000 | 11.85576000 | -1.60035440 |
| O | -0.81375000 | 12.46382000 | 11.85299000 | -1.62209110 |
| O | -0.72383000 | 9.80212000 | 11.89477000 | -1.61581370 |
| O | 1.55387000 | 11.21073000 | 11.83318000 | -1.60737820 |
| O | 1.57548000 | 8.30897000 | 11.93256000 | -1.58044610 |
| O | 1.69878000 | 5.60197000 | 11.97330000 | -1.55484860 |
| O | 3.98290000 | 7.00531000 | 11.81788000 | -1.60972330 |
| Al | 12.11187000 | 1.40442000 | 11.89349000 | 2.41599800 |
| Al | 7.28749000 | 9.82051000 | 11.87565000 | 2.40771290 |
| Al | 9.68425000 | 5.59070000 | 11.87037000 | 2.41545980 |
| Al | 2.43094000 | 1.39212000 | 11.87656000 | 2.41466140 |
| Al | -2.44206000 | 9.78323000 | 11.83294000 | 2.41114100 |
| Al | 7.27049000 | 1.39121000 | 11.87727000 | 2.41381840 |
| Al | -0.07181000 | 5.58261000 | 12.40301000 | 2.38394460 |
| Al | 2.45835000 | 9.80724000 | 12.40385000 | 2.38404220 |
| Al | 4.89369000 | 5.57635000 | 12.29765000 | 2.37736470 |
| Pt | 2.09521000 | 9.85435000 | 14.84548000 | -0.74734980 |

| | | | |
|---|---|---|---|
| Pt | 3.45147000 | 7.76779000 | 14.93777000 | -0.40237310 |
| Pt | 5.05299000 | 5.72232000 | 14.77545000 | -0.77597270 |
| Pt | 0.08749000 | 6.01712000 | 14.83133000 | -0.94579510 |
| Sn | 2.43612000 | 5.24433000 | 14.22320000 | 0.95677460 |
| Sn | 5.94877000 | 8.13991000 | 14.34715000 | 0.87119180 |
| Sn | 0.00926000 | 8.53457000 | 14.18874000 | 0.97334610 |

Figure S5

Coordinates for ethylene binding $Pt_4Sn_3$ on alumina, the format is

Number of atom

Comment line: ethylene-cluster-support-isomer

symbol   x   y   z   bader_charge

```
283
c2h4-pt4sn3-al2o3-C
 Al  -4.83072000  11.17808000   0.00000000   2.16835670
 Al  -2.41059000   6.98630000   0.00000000   2.16836610
 Al   0.00953000  11.17808000   0.00000000   2.16828320
 Al   0.00953000   2.79452000   0.00000000   2.16835890
 Al   2.42966000   6.98630000   0.00000000   2.16821110
 Al   4.84978000   2.79452000   0.00000000   2.16810560
 Al   4.84978000  11.17808000   0.00000000   2.16840470
 Al   7.26991000   6.98630000   0.00000000   2.16822290
 Al   9.69003000   2.79452000   0.00000000   2.16838310
 O   -3.89334000   9.78082000   0.85241000  -1.55250690
 O   -1.66922000   8.49672000   0.85241000  -1.56094690
 O   -1.66922000  11.06492000   0.85241000  -1.55851860
 O   -1.47322000   5.58904000   0.85241000  -1.55261790
 O    0.75091000   4.30494000   0.85241000  -1.56100260
 O    0.75091000   6.87314000   0.85241000  -1.55860270
 O    0.94691000   9.78082000   0.85241000  -1.55254850
 O    0.94691000   1.39726000   0.85241000  -1.55256960
 O    3.17104000   0.11316000   0.85241000  -1.56089930
 O    3.17103000   8.49672000   0.85241000  -1.56066150
 O    3.17103000   2.68136000   0.85241000  -1.55857680
 O    3.17103000  11.06492000   0.85241000  -1.55852860
 O    3.36703000   5.58904000   0.85241000  -1.55248500
 O    5.59116000   4.30494000   0.85241000  -1.56091090
 O    5.59116000   6.87314000   0.85241000  -1.55844860
 O    5.78716000   1.39726000   0.85241000  -1.55253210
 O    5.78716000   9.78082000   0.85241000  -1.55243210
 O    8.01128000   8.49672000   0.85241000  -1.56089220
 O    8.01128000   0.11316000   0.85241000  -1.56091470
 O    8.01128000   2.68136000   0.85241000  -1.55859830
 O    8.01128000  11.06492000   0.85241000  -1.55856600
 O    8.20728000   5.58904000   0.85241000  -1.55249750
 O   10.43141000   4.30494000   0.85241000  -1.56108960
 O   10.43141000   6.87314000   0.85241000  -1.55861420
```

| | | | | |
|---|---|---|---|---|
| O | 10.62741000 | 1.39726000 | 0.85241000 | -1.55259440 |
| O | 12.85153000 | 0.11316000 | 0.85241000 | -1.56087530 |
| O | 12.85153000 | 2.68136000 | 0.85241000 | -1.55877710 |
| Al | -4.83072000 | 8.38356000 | 1.70482000 | 2.46345100 |
| Al | -2.41059000 | 4.19178000 | 1.70482000 | 2.46338040 |
| Al | 0.00953000 | 8.38356000 | 1.70482000 | 2.46357470 |
| Al | 0.00954000 | 0.00000000 | 1.70482000 | 2.46338380 |
| Al | 2.42966000 | 4.19178000 | 1.70482000 | 2.46356770 |
| Al | 4.84978000 | 0.00000000 | 1.70482000 | 2.46343960 |
| Al | 4.84978000 | 8.38356000 | 1.70482000 | 2.46322890 |
| Al | 7.26991000 | 4.19178000 | 1.70482000 | 2.46343730 |
| Al | 9.69003000 | 0.00000000 | 1.70482000 | 2.46339910 |
| Al | -2.41059000 | 9.78082000 | 2.20274000 | 2.48276420 |
| Al | 0.00953000 | 5.58904000 | 2.20274000 | 2.48269530 |
| Al | 2.42966000 | 1.39726000 | 2.20274000 | 2.48278940 |
| Al | 2.42966000 | 9.78082000 | 2.20274000 | 2.48280320 |
| Al | 4.84978000 | 5.58904000 | 2.20274000 | 2.48284570 |
| Al | 7.26991000 | 9.78082000 | 2.20274000 | 2.48284850 |
| Al | 7.26991000 | 1.39726000 | 2.20274000 | 2.48276560 |
| Al | 9.69003000 | 5.58904000 | 2.20274000 | 2.48270170 |
| Al | 12.11016000 | 1.39726000 | 2.20274000 | 2.48272260 |
| O | -5.57210000 | 9.66766000 | 3.05515000 | -1.64219710 |
| O | -3.34796000 | 8.38356000 | 3.05515000 | -1.62429430 |
| O | -3.15197000 | 5.47588000 | 3.05515000 | -1.64221710 |
| O | -3.15197000 | 11.29124000 | 3.05515000 | -1.65010920 |
| O | -0.92784000 | 4.19178000 | 3.05515000 | -1.62439080 |
| O | -0.73185000 | 1.28410000 | 3.05515000 | -1.64241950 |
| O | -0.73185000 | 9.66766000 | 3.05515000 | -1.64230560 |
| O | -0.73184000 | 7.09945000 | 3.05515000 | -1.65023570 |
| O | 1.49229000 | 8.38356000 | 3.05515000 | -1.62428300 |
| O | 1.49229000 | 0.00001000 | 3.05515000 | -1.62422990 |
| O | 1.68828000 | 5.47588000 | 3.05515000 | -1.64229550 |
| O | 1.68828000 | 2.90768000 | 3.05515000 | -1.65008330 |
| O | 1.68828000 | 11.29123000 | 3.05515000 | -1.65007540 |
| O | 3.91241000 | 4.19178000 | 3.05515000 | -1.62432390 |
| O | 4.10840000 | 9.66766000 | 3.05515000 | -1.64210320 |
| O | 4.10840000 | 1.28410000 | 3.05515000 | -1.64230500 |
| O | 4.10841000 | 7.09946000 | 3.05515000 | -1.64997940 |
| O | 6.33254000 | 0.00001000 | 3.05515000 | -1.62434370 |
| O | 6.33254000 | 8.38356000 | 3.05515000 | -1.62410060 |
| O | 6.52853000 | 11.29123000 | 3.05515000 | -1.64991690 |
| O | 6.52853000 | 5.47588000 | 3.05515000 | -1.64202000 |
| O | 6.52853000 | 2.90768000 | 3.05515000 | -1.65023850 |
| O | 8.75266000 | 4.19178000 | 3.05515000 | -1.62432400 |
| O | 8.94866000 | 7.09945000 | 3.05515000 | -1.65014920 |
| O | 8.94865000 | 1.28410000 | 3.05515000 | -1.64230910 |
| O | 11.17277000 | 0.00000000 | 3.05515000 | -1.62419920 |
| O | 11.36878000 | 2.90767000 | 3.05515000 | -1.65025400 |
| Al | -4.83072000 | 11.17808000 | 3.90756000 | 2.47791210 |
| Al | -2.41059000 | 6.98630000 | 3.90756000 | 2.47798410 |
| Al | 0.00953000 | 11.17808000 | 3.90756000 | 2.47797380 |
| Al | 0.00953000 | 2.79452000 | 3.90756000 | 2.47801970 |
| Al | 2.42966000 | 6.98630000 | 3.90756000 | 2.47792060 |
| Al | 4.84978000 | 2.79452000 | 3.90756000 | 2.47799510 |
| Al | 4.84978000 | 11.17808000 | 3.90756000 | 2.47795600 |
| Al | 7.26991000 | 6.98630000 | 3.90756000 | 2.47779550 |
| Al | 9.69003000 | 2.79452000 | 3.90756000 | 2.47805510 |
| Al | -4.83072000 | 8.38356000 | 4.40548000 | 2.47237860 |
| Al | -2.41059000 | 4.19178000 | 4.40548000 | 2.47222220 |
| Al | 0.00953000 | 8.38356000 | 4.40548000 | 2.47238060 |
| Al | 0.00954000 | 0.00000000 | 4.40548000 | 2.47235200 |
| Al | 2.42966000 | 4.19178000 | 4.40548000 | 2.47229410 |
| Al | 4.84978000 | 0.00000000 | 4.40548000 | 2.47249800 |
| Al | 4.84978000 | 8.38356000 | 4.40548000 | 2.47230640 |
| Al | 7.26991000 | 4.19178000 | 4.40548000 | 2.47237240 |
| Al | 9.69003000 | 0.00000000 | 4.40548000 | 2.47221490 |
| O | -6.31347000 | 11.17808000 | 5.25789000 | -1.64346670 |
| O | -4.08934000 | 12.46219000 | 5.25789000 | -1.65402070 |
| O | -4.08934000 | 9.89398000 | 5.25789000 | -1.64905620 |
| O | -3.89335000 | 6.98630000 | 5.25789000 | -1.64353540 |
| O | -1.66922000 | 8.27040000 | 5.25789000 | -1.65377440 |
| O | -1.66921000 | 5.70219000 | 5.25789000 | -1.64895890 |
| O | -1.47322000 | 11.17808000 | 5.25789000 | -1.64356100 |
| O | -1.47323000 | 2.79452000 | 5.25789000 | -1.64332600 |
| O | 0.75091000 | 4.07862000 | 5.25789000 | -1.65391280 |
| O | 0.75091000 | 12.46219000 | 5.25789000 | -1.65394680 |
| O | 0.75091000 | 9.89397000 | 5.25789000 | -1.64894110 |
| O | 0.75091000 | 1.51042000 | 5.25789000 | -1.64888330 |
| O | 0.94690000 | 6.98630000 | 5.25789000 | -1.64348210 |
| O | 3.17103000 | 8.27040000 | 5.25789000 | -1.65375600 |
| O | 3.17104000 | 5.70220000 | 5.25789000 | -1.64899350 |
| O | 3.36703000 | 2.79452000 | 5.25789000 | -1.64346170 |
| O | 3.36702000 | 11.17808000 | 5.25789000 | -1.64350340 |
| O | 5.59116000 | 12.46219000 | 5.25789000 | -1.65415290 |
| O | 5.59116000 | 4.07863000 | 5.25789000 | -1.65393650 |
| O | 5.59116000 | 9.89397000 | 5.25789000 | -1.64917700 |

| | | | |
|---|---|---|---|
| O | 5.59116000 | 1.51042000 | 5.25789000 | -1.64897800 |
| O | 5.78715000 | 6.98630000 | 5.25789000 | -1.64332560 |
| O | 8.01128000 | 8.27040000 | 5.25789000 | -1.65398360 |
| O | 8.01129000 | 5.70219000 | 5.25789000 | -1.64920210 |
| O | 8.20727000 | 2.79452000 | 5.25789000 | -1.64321350 |
| O | 10.43141000 | 4.07862000 | 5.25789000 | -1.65398640 |
| O | 10.43141000 | 1.51041000 | 5.25789000 | -1.64891360 |
| Al | -2.41059000 | 9.78082000 | 6.11030000 | 2.47602860 |
| Al | 0.00953000 | 5.58904000 | 6.11030000 | 2.47587930 |
| Al | 2.42966000 | 1.39726000 | 6.11030000 | 2.47605400 |
| Al | 2.42966000 | 9.78082000 | 6.11030000 | 2.47611990 |
| Al | 4.84978000 | 5.58904000 | 6.11030000 | 2.47614150 |
| Al | 7.26991000 | 9.78082000 | 6.11030000 | 2.47625480 |
| Al | 7.26991000 | 1.39726000 | 6.11030000 | 2.47605260 |
| Al | 9.69003000 | 5.58904000 | 6.11030000 | 2.47623650 |
| Al | 12.11016000 | 1.39726000 | 6.11030000 | 2.47597870 |
| Al | -4.83072000 | 11.17808000 | 6.60910000 | 2.47666500 |
| Al | -2.41059000 | 6.98630000 | 6.60910000 | 2.47689680 |
| Al | 0.00953000 | 11.17808000 | 6.60910000 | 2.47667310 |
| Al | 0.00953000 | 2.79452000 | 6.60910000 | 2.47665370 |
| Al | 2.42966000 | 6.98630000 | 6.60910000 | 2.47697370 |
| Al | 4.84978000 | 2.79452000 | 6.60910000 | 2.47678850 |
| Al | 4.84978000 | 11.17808000 | 6.60910000 | 2.47683960 |
| Al | 7.26991000 | 6.98630000 | 6.60910000 | 2.47681010 |
| Al | 9.69003000 | 2.79452000 | 6.60910000 | 2.47677220 |
| O | -3.15680000 | 8.49706000 | 7.46570000 | -1.64971450 |
| O | -3.14925000 | 11.06893000 | 7.46570000 | -1.64973520 |
| O | -0.92572000 | 9.77647000 | 7.46570000 | -1.65382810 |
| O | -0.73667000 | 4.30528000 | 7.46570000 | -1.64992110 |
| O | -0.72913000 | 6.87715000 | 7.46570000 | -1.64998140 |
| O | 1.49440000 | 5.58469000 | 7.46570000 | -1.65360050 |
| O | 1.68345000 | 8.49706000 | 7.46570000 | -1.64926620 |
| O | 1.68345000 | 0.11351000 | 7.46570000 | -1.64901840 |
| O | 1.69100000 | 2.68537000 | 7.46570000 | -1.65041430 |
| O | 1.69100000 | 11.06893000 | 7.46570000 | -1.64922490 |
| O | 3.91453000 | 1.39291000 | 7.46570000 | -1.65400640 |
| O | 3.91452000 | 9.77646000 | 7.46570000 | -1.65342680 |
| O | 4.10358000 | 4.30528000 | 7.46570000 | -1.64899290 |
| O | 4.11112000 | 6.87715000 | 7.46570000 | -1.64912730 |
| O | 6.33465000 | 5.58468000 | 7.46570000 | -1.65365090 |
| O | 6.52370000 | 0.11351000 | 7.46570000 | -1.64895180 |
| O | 6.52370000 | 8.49706000 | 7.46570000 | -1.64844190 |
| O | 6.53125000 | 2.68537000 | 7.46570000 | -1.64971300 |

| | | | |
|---|---|---|---|
| O | 6.53125000 | 11.06893000 | 7.46570000 | -1.64944500 |
| O | 8.75477000 | 9.77646000 | 7.46570000 | -1.65344530 |
| O | 8.75477000 | 1.39291000 | 7.46570000 | -1.65457100 |
| O | 8.94383000 | 4.30528000 | 7.46570000 | -1.64980480 |
| O | 8.95137000 | 6.87715000 | 7.46570000 | -1.64943930 |
| O | 11.17490000 | 5.58468000 | 7.46570000 | -1.65457140 |
| O | 11.36395000 | 0.11351000 | 7.46570000 | -1.64899970 |
| O | 11.37149000 | 2.68537000 | 7.46570000 | -1.65022040 |
| O | 13.59502000 | 1.39291000 | 7.46570000 | -1.65374160 |
| Al | -4.83072000 | 8.38356000 | 8.32833000 | 2.47237380 |
| Al | -2.41059000 | 4.19178000 | 8.32833000 | 2.47238080 |
| Al | 0.00953000 | 8.38356000 | 8.32833000 | 2.47085040 |
| Al | 0.00954000 | 0.00000000 | 8.32833000 | 2.47114120 |
| Al | 2.42966000 | 4.19178000 | 8.32833000 | 2.47076310 |
| Al | 4.84978000 | 0.00000000 | 8.32833000 | 2.47163270 |
| Al | 4.84978000 | 8.38356000 | 8.32833000 | 2.47063110 |
| Al | 7.26991000 | 4.19178000 | 8.32833000 | 2.47180120 |
| Al | 9.69003000 | 0.00000000 | 8.32833000 | 2.47133460 |
| Al | -2.41059000 | 9.78082000 | 8.78548000 | 2.47248270 |
| Al | 0.00953000 | 5.58904000 | 8.78548000 | 2.47209060 |
| Al | 2.42966000 | 1.39726000 | 8.78548000 | 2.47248330 |
| Al | 2.42966000 | 9.78082000 | 8.78548000 | 2.47616250 |
| Al | 4.84978000 | 5.58904000 | 8.78548000 | 2.47657030 |
| Al | 7.26991000 | 9.78082000 | 8.78548000 | 2.47640750 |
| Al | 7.26991000 | 1.39726000 | 8.78548000 | 2.47196760 |
| Al | 9.69003000 | 5.58904000 | 8.78548000 | 2.47218070 |
| Al | 12.11016000 | 1.39726000 | 8.78548000 | 2.47150020 |
| O | 8.01661000 | 11.26550000 | 9.68989000 | -1.64659600 |
| O | 10.43673000 | 7.07371000 | 9.68989000 | -1.64291100 |
| O | -4.06971000 | 9.68514000 | 9.68989000 | -1.65987410 |
| O | -1.66389000 | 11.26550000 | 9.68989000 | -1.64305830 |
| O | 12.85686000 | 2.88194000 | 9.68989000 | -1.64463070 |
| O | -1.64959000 | 5.49336000 | 9.68989000 | -1.66085470 |
| O | -1.49818000 | 8.39182000 | 9.68989000 | -1.66822640 |
| O | 0.75624000 | 7.07372000 | 9.68989000 | -1.64323410 |
| O | 0.77054000 | 9.68514000 | 9.68989000 | -1.64582140 |
| O | 0.77054000 | 1.30159000 | 9.68989000 | -1.65671930 |
| O | 0.92195000 | 4.20004000 | 9.68989000 | -1.66715650 |
| O | 3.17636000 | 11.26550000 | 9.68989000 | -1.63903260 |
| O | 3.17636000 | 2.88194000 | 9.68989000 | -1.64407760 |
| O | 3.19067000 | 5.49336000 | 9.68989000 | -1.64740840 |
| O | 3.34208000 | 0.00826000 | 9.68989000 | -1.66818120 |
| O | 3.34207000 | 8.39182000 | 9.68989000 | -1.65511170 |
| O | 5.59648000 | 7.07372000 | 9.68989000 | -1.63288930 |
| O | 5.61079000 | 1.30159000 | 9.68989000 | -1.65978650 |
| O | 5.61079000 | 9.68514000 | 9.68989000 | -1.64568960 |
| O | 5.76220000 | 4.20004000 | 9.68989000 | -1.65404800 |
| O | 8.01661000 | 2.88194000 | 9.68989000 | -1.64564400 |
| O | 8.03091000 | 5.49336000 | 9.68989000 | -1.66111320 |
| O | 8.18232000 | 0.00826000 | 9.68989000 | -1.66560390 |
| O | 8.18232000 | 8.39182000 | 9.68989000 | -1.65666420 |
| O | 10.45104000 | 1.30158000 | 9.68989000 | -1.65710140 |
| O | 10.60245000 | 4.20004000 | 9.68989000 | -1.66991860 |
| O | 13.02257000 | 0.00826000 | 9.68989000 | -1.65728950 |
| Al | -4.83072000 | 11.17808000 | 10.71830000 | 2.47969740 |
| Al | -2.41059000 | 6.98630000 | 10.71830000 | 2.47478550 |
| Al | 0.00953000 | 11.17808000 | 10.71830000 | 2.48099460 |
| Al | 0.00953000 | 2.79452000 | 10.71830000 | 2.47774100 |
| Al | 2.42966000 | 6.98630000 | 10.71830000 | 2.47890690 |
| Al | 4.84978000 | 2.79452000 | 10.71830000 | 2.47547900 |
| Al | 4.84978000 | 11.17808000 | 10.71830000 | 2.47861060 |
| Al | 7.26991000 | 6.98630000 | 10.71830000 | 2.48052590 |
| Al | 9.69003000 | 2.79452000 | 10.71830000 | 2.47565990 |
| Al | -4.83072000 | 8.38356000 | 10.98601000 | 2.47167240 |
| Al | -2.41059000 | 4.19178000 | 10.98601000 | 2.46743840 |
| Al | 0.00953000 | 8.38356000 | 10.98601000 | 2.47072880 |
| Al | 0.00954000 | 0.00000000 | 10.98601000 | 2.47277390 |
| Al | 2.42966000 | 4.19178000 | 10.98601000 | 2.47221560 |
| Al | 4.84978000 | 0.00000000 | 10.98601000 | 2.47181200 |
| Al | 4.84978000 | 8.38356000 | 10.98601000 | 2.46989020 |
| Al | 7.26991000 | 4.19178000 | 10.98601000 | 2.47218640 |
| Al | 9.69003000 | 0.00000000 | 10.98601000 | 2.47026670 |
| O | -5.65708000 | 12.47535000 | 11.86092000 | -1.62264360 |
| O | -5.54687000 | 9.81548000 | 11.84040000 | -1.60880570 |
| O | -3.29833000 | 11.24129000 | 11.85212000 | -1.62372760 |
| O | -3.22984000 | 8.27324000 | 11.85308000 | -1.61855390 |
| O | -3.10866000 | 5.62900000 | 11.85419000 | -1.62401510 |
| O | -0.87496000 | 7.04916000 | 11.85520000 | -1.62368180 |
| O | -0.81041000 | 12.47104000 | 11.85077000 | -1.62133230 |
| O | -0.81648000 | 4.07611000 | 11.85449000 | -1.61889040 |
| O | -0.70401000 | 9.81287000 | 11.86357000 | -1.62563940 |
| O | -0.70250000 | 1.43440000 | 11.85304000 | -1.62789340 |
| O | 1.53628000 | 2.85967000 | 11.85462000 | -1.62651960 |
| O | 1.54890000 | 11.23861000 | 11.83464000 | -1.61327460 |
| O | 1.64003000 | 8.29058000 | 11.88626000 | -1.59516590 |
| O | 1.70290000 | 5.63586000 | 11.87160000 | -1.63102160 |

| | | | |
|---|---|---|---|
| O | 3.94739000 | 7.01597000 | 11.89191000 | -1.59181880 |
| O | 4.02953000 | 4.09648000 | 11.83595000 | -1.61340990 |
| O | 4.04643000 | 12.48151000 | 11.87298000 | -1.62826510 |
| O | 4.13768000 | 1.43793000 | 11.85256000 | -1.62439800 |
| O | 4.12383000 | 9.85216000 | 11.92694000 | -1.56767550 |
| O | 6.37038000 | 11.21375000 | 11.86612000 | -1.60269560 |
| O | 6.39214000 | 2.85261000 | 11.86249000 | -1.62260160 |
| O | 6.48820000 | 8.29027000 | 11.90907000 | -1.57455600 |
| O | 6.52735000 | 5.64950000 | 11.88677000 | -1.59977110 |
| O | 8.80035000 | 7.03071000 | 11.86746000 | -1.63565210 |
| O | 8.86740000 | 4.09177000 | 11.85487000 | -1.62554170 |
| O | 8.98908000 | 1.44102000 | 11.85132000 | -1.62715450 |
| O | 11.21333000 | 2.86847000 | 11.85194000 | -1.61923240 |
| Al | -2.41538000 | 9.77930000 | 11.89546000 | 2.41549240 |
| Al | -0.01517000 | 5.57866000 | 11.87206000 | 2.41200010 |
| Al | 2.42902000 | 1.39721000 | 11.86723000 | 2.41369920 |
| Al | 2.37142000 | 9.82397000 | 12.40895000 | 2.37532860 |
| Al | 4.82661000 | 5.53641000 | 12.39825000 | 2.38017330 |
| Al | 7.33108000 | 9.81773000 | 12.39833000 | 2.36868150 |
| Al | 7.27868000 | 1.38720000 | 11.86297000 | 2.41665720 |
| Al | 9.70992000 | 5.57329000 | 11.86142000 | 2.41382800 |
| Al | 12.11508000 | 1.42487000 | 11.83988000 | 2.41769370 |
| Pt | 4.76158000 | 8.07055000 | 14.93271000 | -0.55349260 |
| Pt | 4.79470000 | 5.50286000 | 14.88388000 | -0.52205090 |
| Pt | 2.57765000 | 9.73618000 | 14.84512000 | -0.91560440 |
| Pt | 7.03413000 | 9.71819000 | 14.82525000 | -0.91664860 |
| Sn | 4.81761000 | 10.94647000 | 14.02216000 | 1.00310220 |
| Sn | 7.08006000 | 7.14920000 | 14.07472000 | 0.94199810 |
| Sn | 2.44739000 | 7.12950000 | 14.12284000 | 0.93257980 |
| C | 4.07140000 | 3.51059000 | 15.21064000 | -0.14807250 |
| C | 5.47984000 | 3.48308000 | 15.13302000 | -0.22368760 |
| H | 3.56872000 | 3.48284000 | 16.18342000 | 0.07277120 |
| H | 3.47162000 | 3.18856000 | 14.35712000 | 0.09498470 |
| H | 6.08244000 | 3.44064000 | 16.04702000 | 0.09217280 |
| H | 5.97012000 | 3.13490000 | 14.22049000 | 0.13846670 |

283
c2h4-pt4sn3-al2o3-C

| | | | |
|---|---|---|---|
| Al | -4.83072000 | 11.17808000 | 0.00000000 | 2.16822810 |
| Al | -2.41059000 | 6.98630000 | 0.00000000 | 2.16828060 |
| Al | 0.00953000 | 11.17808000 | 0.00000000 | 2.16842880 |
| Al | 0.00953000 | 2.79452000 | 0.00000000 | 2.16826950 |
| Al | 2.42966000 | 6.98630000 | 0.00000000 | 2.16837170 |
| Al | 4.84978000 | 2.79452000 | 0.00000000 | 2.16822870 |

| | | | |
|---|---|---|---|
| Al | 4.84978000 | 11.17808000 | 0.00000000 | 2.16837980 |
| Al | 7.26991000 | 6.98630000 | 0.00000000 | 2.16823950 |
| Al | 9.69003000 | 2.79452000 | 0.00000000 | 2.16832420 |
| O | -3.89334000 | 9.78082000 | 0.85241000 | -1.55253020 |
| O | -1.66922000 | 8.49672000 | 0.85241000 | -1.56083440 |
| O | -1.66922000 | 11.06492000 | 0.85241000 | -1.55858250 |
| O | -1.47322000 | 5.58904000 | 0.85241000 | -1.55255680 |
| O | 0.75091000 | 4.30494000 | 0.85241000 | -1.56065360 |
| O | 0.75091000 | 6.87314000 | 0.85241000 | -1.55852380 |
| O | 0.94691000 | 9.78082000 | 0.85241000 | -1.55258240 |
| O | 0.94691000 | 1.39726000 | 0.85241000 | -1.55248400 |
| O | 3.17104000 | 0.11316000 | 0.85241000 | -1.56091430 |
| O | 3.17103000 | 8.49672000 | 0.85241000 | -1.56087800 |
| O | 3.17103000 | 2.68136000 | 0.85241000 | -1.55845150 |
| O | 3.17103000 | 11.06492000 | 0.85241000 | -1.55872130 |
| O | 3.36703000 | 5.58904000 | 0.85241000 | -1.55243840 |
| O | 5.59116000 | 4.30494000 | 0.85241000 | -1.56089550 |
| O | 5.59116000 | 6.87314000 | 0.85241000 | -1.55853420 |
| O | 5.78716000 | 1.39726000 | 0.85241000 | -1.55248240 |
| O | 5.78716000 | 9.78082000 | 0.85241000 | -1.55256880 |
| O | 8.01128000 | 8.49672000 | 0.85241000 | -1.56089500 |
| O | 8.01128000 | 0.11316000 | 0.85241000 | -1.56106270 |
| O | 8.01128000 | 2.68136000 | 0.85241000 | -1.55862140 |
| O | 8.01128000 | 11.06492000 | 0.85241000 | -1.55859300 |
| O | 8.20728000 | 5.58904000 | 0.85241000 | -1.55254830 |
| O | 10.43141000 | 4.30494000 | 0.85241000 | -1.56095010 |
| O | 10.43141000 | 6.87314000 | 0.85241000 | -1.55852010 |
| O | 10.62741000 | 1.39726000 | 0.85241000 | -1.55261570 |
| O | 12.85153000 | 0.11316000 | 0.85241000 | -1.56101330 |
| O | 12.85153000 | 2.68136000 | 0.85241000 | -1.55862060 |
| Al | -4.83072000 | 8.38356000 | 1.70482000 | 2.46344230 |
| Al | -2.41059000 | 4.19178000 | 1.70482000 | 2.46357340 |
| Al | 0.00953000 | 8.38356000 | 1.70482000 | 2.46338390 |
| Al | -7.25084000 | 12.57534000 | 1.70482000 | 2.46350310 |
| Al | 2.42966000 | 4.19178000 | 1.70482000 | 2.46322730 |
| Al | -2.41060000 | 12.57534000 | 1.70482000 | 2.46343730 |
| Al | 4.84978000 | 8.38356000 | 1.70482000 | 2.46338320 |
| Al | 7.26991000 | 4.19178000 | 1.70482000 | 2.46348430 |
| Al | 2.42965000 | 12.57534000 | 1.70482000 | 2.46337970 |
| Al | -2.41059000 | 9.78082000 | 2.20274000 | 2.48277160 |
| Al | 0.00953000 | 5.58904000 | 2.20274000 | 2.48283600 |
| Al | 2.42966000 | 1.39726000 | 2.20274000 | 2.48284090 |
| Al | 2.42966000 | 9.78082000 | 2.20274000 | 2.48270000 |

| | | | |
|---|---|---|---|
| Al | 4.84978000 | 5.58904000 | 2.20274000 | 2.48285270 |
| Al | 7.26991000 | 9.78082000 | 2.20274000 | 2.48278690 |
| Al | 7.26991000 | 1.39726000 | 2.20274000 | 2.48273010 |
| Al | 9.69003000 | 5.58904000 | 2.20274000 | 2.48276220 |
| Al | 12.11016000 | 1.39726000 | 2.20274000 | 2.48269210 |
| O | -5.57210000 | 9.66766000 | 3.05515000 | -1.64233010 |
| O | -3.34796000 | 8.38356000 | 3.05515000 | -1.62426410 |
| O | -3.15197000 | 5.47588000 | 3.05515000 | -1.64226330 |
| O | -3.15197000 | 11.29124000 | 3.05515000 | -1.65012130 |
| O | -0.92784000 | 4.19178000 | 3.05515000 | -1.62428420 |
| O | -0.73185000 | 1.28410000 | 3.05515000 | -1.64224500 |
| O | -0.73185000 | 9.66766000 | 3.05515000 | -1.64225000 |
| O | -0.73184000 | 7.09945000 | 3.05515000 | -1.65009840 |
| O | 1.49229000 | 8.38356000 | 3.05515000 | -1.62417970 |
| O | 1.49229000 | 0.00001000 | 3.05515000 | -1.62434760 |
| O | 1.68828000 | 5.47588000 | 3.05515000 | -1.64206290 |
| O | 1.68828000 | 2.90768000 | 3.05515000 | -1.64998650 |
| O | 1.68828000 | 11.29123000 | 3.05515000 | -1.65022790 |
| O | 3.91241000 | 4.19178000 | 3.05515000 | -1.62410150 |
| O | 4.10840000 | 9.66766000 | 3.05515000 | -1.64236220 |
| O | 4.10840000 | 1.28410000 | 3.05515000 | -1.64207560 |
| O | 4.10841000 | 7.09946000 | 3.05515000 | -1.64991780 |
| O | 6.33254000 | 0.00001000 | 3.05515000 | -1.62437390 |
| O | 6.33254000 | 8.38356000 | 3.05515000 | -1.62422670 |
| O | 6.52853000 | 11.29123000 | 3.05515000 | -1.65008750 |
| O | 6.52853000 | 5.47588000 | 3.05515000 | -1.64219070 |
| O | 6.52853000 | 2.90768000 | 3.05515000 | -1.65014320 |
| O | 8.75266000 | 4.19178000 | 3.05515000 | -1.62436880 |
| O | 8.94866000 | 7.09945000 | 3.05515000 | -1.65011610 |
| O | 8.94865000 | 1.28410000 | 3.05515000 | -1.64222460 |
| O | 3.91239000 | 12.57534000 | 3.05515000 | -1.62439520 |
| O | 11.36878000 | 2.90767000 | 3.05515000 | -1.65027140 |
| Al | -4.83072000 | 11.17808000 | 3.90756000 | 2.47792450 |
| Al | -2.41059000 | 6.98630000 | 3.90756000 | 2.47790830 |
| Al | 0.00953000 | 11.17808000 | 3.90756000 | 2.47795190 |
| Al | 0.00953000 | 2.79452000 | 3.90756000 | 2.47789250 |
| Al | 2.42966000 | 6.98630000 | 3.90756000 | 2.47795320 |
| Al | 4.84978000 | 2.79452000 | 3.90756000 | 2.47779620 |
| Al | 4.84978000 | 11.17808000 | 3.90756000 | 2.47802310 |
| Al | 7.26991000 | 6.98630000 | 3.90756000 | 2.47791120 |
| Al | 9.69003000 | 2.79452000 | 3.90756000 | 2.47804990 |
| Al | -4.83072000 | 8.38356000 | 4.40548000 | 2.47243510 |
| Al | -2.41059000 | 4.19178000 | 4.40548000 | 2.47233940 |

| | | | |
|---|---|---|---|
| Al | 0.00953000 | 8.38356000 | 4.40548000 | 2.47224870 |
| Al | -7.25084000 | 12.57534000 | 4.40548000 | 2.47227580 |
| Al | 2.42966000 | 4.19178000 | 4.40548000 | 2.47225930 |
| Al | -2.41060000 | 12.57534000 | 4.40548000 | 2.47247270 |
| Al | 4.84978000 | 8.38356000 | 4.40548000 | 2.47229330 |
| Al | 7.26991000 | 4.19178000 | 4.40548000 | 2.47241580 |
| Al | 2.42965000 | 12.57534000 | 4.40548000 | 2.47226860 |
| O | -6.31347000 | 11.17808000 | 5.25789000 | -1.64342680 |
| O | -4.08934000 | 12.46219000 | 5.25789000 | -1.65389250 |
| O | -4.08934000 | 9.89398000 | 5.25789000 | -1.64897640 |
| O | -3.89335000 | 6.98630000 | 5.25789000 | -1.64356200 |
| O | -1.66922000 | 8.27040000 | 5.25789000 | -1.65396360 |
| O | -1.66921000 | 5.70219000 | 5.25789000 | -1.64908080 |
| O | -1.47322000 | 11.17808000 | 5.25789000 | -1.64335090 |
| O | -1.47323000 | 2.79452000 | 5.25789000 | -1.64331670 |
| O | 0.75091000 | 4.07862000 | 5.25789000 | -1.65367200 |
| O | 0.75091000 | 12.46219000 | 5.25789000 | -1.65408680 |
| O | 0.75091000 | 9.89397000 | 5.25789000 | -1.64887850 |
| O | 0.75091000 | 1.51042000 | 5.25789000 | -1.64897560 |
| O | 0.94690000 | 6.98630000 | 5.25789000 | -1.64351030 |
| O | 3.17103000 | 8.27040000 | 5.25789000 | -1.65414900 |
| O | 3.17104000 | 5.70220000 | 5.25789000 | -1.64918630 |
| O | 3.36703000 | 2.79452000 | 5.25789000 | -1.64340940 |
| O | 3.36702000 | 11.17808000 | 5.25789000 | -1.64330410 |
| O | 5.59116000 | 12.46219000 | 5.25789000 | -1.65391070 |
| O | 5.59116000 | 4.07863000 | 5.25789000 | -1.65400400 |
| O | 5.59116000 | 9.89397000 | 5.25789000 | -1.64881690 |
| O | 5.59116000 | 1.51042000 | 5.25789000 | -1.64921400 |
| O | 5.78715000 | 6.98630000 | 5.25789000 | -1.64343430 |
| O | 8.01128000 | 8.27040000 | 5.25789000 | -1.65405380 |
| O | 8.01129000 | 5.70219000 | 5.25789000 | -1.64912020 |
| O | 8.20727000 | 2.79452000 | 5.25789000 | -1.64346020 |
| O | 10.43141000 | 4.07862000 | 5.25789000 | -1.65380940 |
| O | 10.43141000 | 1.51041000 | 5.25789000 | -1.64906390 |
| Al | -2.41059000 | 9.78082000 | 6.11030000 | 2.47610240 |
| Al | 0.00953000 | 5.58904000 | 6.11030000 | 2.47622090 |
| Al | 2.42966000 | 1.39726000 | 6.11030000 | 2.47615970 |
| Al | 2.42966000 | 9.78082000 | 6.11030000 | 2.47591190 |
| Al | 4.84978000 | 5.58904000 | 6.11030000 | 2.47626760 |
| Al | 7.26991000 | 9.78082000 | 6.11030000 | 2.47603740 |
| Al | 7.26991000 | 1.39726000 | 6.11030000 | 2.47620870 |
| Al | 9.69003000 | 5.58904000 | 6.11030000 | 2.47602870 |
| Al | 12.11016000 | 1.39726000 | 6.11030000 | 2.47569370 |

| | | | |
|---|---|---|---|
| Al | -4.83072000 | 11.17808000 | 6.60910000 | 2.47674510 |
| Al | -2.41059000 | 6.98630000 | 6.60910000 | 2.47675240 |
| Al | 0.00953000 | 11.17808000 | 6.60910000 | 2.47673470 |
| Al | 0.00953000 | 2.79452000 | 6.60910000 | 2.47678260 |
| Al | 2.42966000 | 6.98630000 | 6.60910000 | 2.47688290 |
| Al | 4.84978000 | 2.79452000 | 6.60910000 | 2.47673230 |
| Al | 4.84978000 | 11.17808000 | 6.60910000 | 2.47648200 |
| Al | 7.26991000 | 6.98630000 | 6.60910000 | 2.47686170 |
| Al | 9.69003000 | 2.79452000 | 6.60910000 | 2.47685280 |
| O | -3.15680000 | 8.49706000 | 7.46570000 | -1.64901310 |
| O | -3.14925000 | 11.06893000 | 7.46570000 | -1.64972700 |
| O | -0.92572000 | 9.77647000 | 7.46570000 | -1.65444250 |
| O | -0.73667000 | 4.30528000 | 7.46570000 | -1.64932940 |
| O | -0.72913000 | 6.87715000 | 7.46570000 | -1.64904330 |
| O | 1.49440000 | 5.58469000 | 7.46570000 | -1.65346630 |
| O | 1.68345000 | 8.49706000 | 7.46570000 | -1.64887040 |
| O | 1.68345000 | 0.11351000 | 7.46570000 | -1.64871040 |
| O | 1.69100000 | 2.68537000 | 7.46570000 | -1.64909680 |
| O | 1.69100000 | 11.06893000 | 7.46570000 | -1.64986280 |
| O | 3.91453000 | 1.39291000 | 7.46570000 | -1.65355880 |
| O | 3.91452000 | 9.77646000 | 7.46570000 | -1.65377610 |
| O | 4.10358000 | 4.30528000 | 7.46570000 | -1.64863130 |
| O | 4.11112000 | 6.87715000 | 7.46570000 | -1.64935720 |
| O | 6.33465000 | 5.58468000 | 7.46570000 | -1.65345680 |
| O | 6.52370000 | 0.11351000 | 7.46570000 | -1.64961210 |
| O | 6.52370000 | 8.49706000 | 7.46570000 | -1.64907270 |
| O | 6.53125000 | 2.68537000 | 7.46570000 | -1.64954960 |
| O | 6.53125000 | 11.06893000 | 7.46570000 | -1.65056370 |
| O | 8.75477000 | 9.77646000 | 7.46570000 | -1.65408790 |
| O | 8.75477000 | 1.39291000 | 7.46570000 | -1.65474070 |
| O | 8.94383000 | 4.30528000 | 7.46570000 | -1.64974450 |
| O | 8.95137000 | 6.87715000 | 7.46570000 | -1.64981880 |
| O | 11.17490000 | 5.58468000 | 7.46570000 | -1.65388120 |
| O | 11.36395000 | 0.11351000 | 7.46570000 | -1.64965530 |
| O | 11.37149000 | 2.68537000 | 7.46570000 | -1.65006510 |
| O | 13.59502000 | 1.39291000 | 7.46570000 | -1.65363840 |
| Al | -4.83072000 | 8.38356000 | 8.32833000 | 2.47198460 |
| Al | -2.41059000 | 4.19178000 | 8.32833000 | 2.47112980 |
| Al | 0.00953000 | 8.38356000 | 8.32833000 | 2.47130070 |
| Al | -7.25084000 | 12.57534000 | 8.32833000 | 2.47124780 |
| Al | 2.42966000 | 4.19178000 | 8.32833000 | 2.47038420 |
| Al | -2.41060000 | 12.57534000 | 8.32833000 | 2.47167860 |
| Al | 4.84978000 | 8.38356000 | 8.32833000 | 2.47092470 |
| Al | 7.26991000 | 4.19178000 | 8.32833000 | 2.47226690 |
| Al | 2.42965000 | 12.57534000 | 8.32833000 | 2.47197250 |
| Al | -2.41059000 | 9.78082000 | 8.78548000 | 2.47219950 |
| Al | 0.00953000 | 5.58904000 | 8.78548000 | 2.47625090 |
| Al | 2.42966000 | 1.39726000 | 8.78548000 | 2.47640840 |
| Al | 2.42966000 | 9.78082000 | 8.78548000 | 2.47194270 |
| Al | 4.84978000 | 5.58904000 | 8.78548000 | 2.47679740 |
| Al | 7.26991000 | 9.78082000 | 8.78548000 | 2.47249940 |
| Al | 7.26991000 | 1.39726000 | 8.78548000 | 2.47244900 |
| Al | 9.69003000 | 5.58904000 | 8.78548000 | 2.47195780 |
| Al | 12.11016000 | 1.39726000 | 8.78548000 | 2.47153360 |
| O | 8.01661000 | 11.26550000 | 9.68989000 | -1.64437140 |
| O | 10.43673000 | 7.07371000 | 9.68989000 | -1.64280570 |
| O | -4.06971000 | 9.68514000 | 9.68989000 | -1.65984980 |
| O | -1.66389000 | 11.26550000 | 9.68989000 | -1.64490770 |
| O | 12.85686000 | 2.88194000 | 9.68989000 | -1.64356240 |
| O | -1.64959000 | 5.49336000 | 9.68989000 | -1.64558050 |
| O | -1.49818000 | 8.39182000 | 9.68989000 | -1.66442150 |
| O | 0.75624000 | 7.07372000 | 9.68989000 | -1.63734260 |
| O | 0.77054000 | 9.68514000 | 9.68989000 | -1.65676890 |
| O | 0.77054000 | 1.30159000 | 9.68989000 | -1.64806310 |
| O | 0.92195000 | 4.20004000 | 9.68989000 | -1.65603580 |
| O | 3.17636000 | 11.26550000 | 9.68989000 | -1.64541210 |
| O | 3.17636000 | 2.88194000 | 9.68989000 | -1.63233310 |
| O | 3.19067000 | 5.49336000 | 9.68989000 | -1.64584510 |
| O | 3.34208000 | 0.00826000 | 9.68989000 | -1.65427810 |
| O | 3.34207000 | 8.39182000 | 9.68989000 | -1.66027020 |
| O | 5.59648000 | 7.07372000 | 9.68989000 | -1.64235140 |
| O | 5.61079000 | 1.30159000 | 9.68989000 | -1.66013440 |
| O | 5.61079000 | 9.68514000 | 9.68989000 | -1.65684430 |
| O | 5.76220000 | 4.20004000 | 9.68989000 | -1.65701740 |
| O | 8.01661000 | 2.88194000 | 9.68989000 | -1.64296360 |
| O | 8.03091000 | 5.49336000 | 9.68989000 | -1.66094150 |
| O | 8.18232000 | 0.00826000 | 9.68989000 | -1.66878640 |
| O | 8.18232000 | 8.39182000 | 9.68989000 | -1.66845090 |
| O | 10.45104000 | 1.30158000 | 9.68989000 | -1.66087830 |
| O | 10.60245000 | 4.20004000 | 9.68989000 | -1.66953850 |
| O | 13.02257000 | 0.00826000 | 9.68989000 | -1.66795360 |
| Al | -4.83072000 | 11.17808000 | 10.71830000 | 2.47562190 |
| Al | -2.41059000 | 6.98630000 | 10.71830000 | 2.48018870 |
| Al | 0.00953000 | 11.17808000 | 10.71830000 | 2.47565310 |
| Al | 0.00953000 | 2.79452000 | 10.71830000 | 2.47712740 |
| Al | 2.42966000 | 6.98630000 | 10.71830000 | 2.47827230 |

| | | | |
|---|---|---|---|
| Al | 4.84978000 | 2.79452000 | 10.71830000 | 2.47996500 |
| Al | 4.84978000 | 11.17808000 | 10.71830000 | 2.47784450 |
| Al | 7.26991000 | 6.98630000 | 10.71830000 | 2.48093340 |
| Al | 9.69003000 | 2.79452000 | 10.71830000 | 2.47509870 |
| Al | -4.83072000 | 8.38356000 | 10.98601000 | 2.47262440 |
| Al | -2.41059000 | 4.19178000 | 10.98601000 | 2.47103660 |
| Al | 0.00953000 | 8.38356000 | 10.98601000 | 2.46922110 |
| Al | -7.25084000 | 12.57534000 | 10.98601000 | 2.47134900 |
| Al | 2.42966000 | 4.19178000 | 10.98601000 | 2.47322500 |
| Al | -2.41060000 | 12.57534000 | 10.98601000 | 2.47158490 |
| Al | 4.84978000 | 8.38356000 | 10.98601000 | 2.47355290 |
| Al | 7.26991000 | 4.19178000 | 10.98601000 | 2.47153260 |
| Al | 2.42965000 | 12.57534000 | 10.98601000 | 2.46745840 |
| O | -5.65256000 | 12.48065000 | 11.83427000 | -1.61491200 |
| O | -5.54019000 | 9.82115000 | 11.85095000 | -1.62501650 |
| O | -3.29146000 | 11.24285000 | 11.86349000 | -1.62056970 |
| O | -3.23115000 | 8.28187000 | 11.85187000 | -1.62149860 |
| O | -3.12068000 | 5.62174000 | 11.86081000 | -1.62637500 |
| O | -0.87219000 | 7.04726000 | 11.84077000 | -1.61224510 |
| O | -0.81246000 | 12.47635000 | 11.85518000 | -1.62757780 |
| O | -0.78135000 | 4.09375000 | 11.86592000 | -1.59937990 |
| O | -0.69547000 | 9.82534000 | 11.85309000 | -1.62593820 |
| O | -0.71741000 | 1.44687000 | 11.87309000 | -1.62876680 |
| O | 1.52146000 | 2.82784000 | 11.92729000 | -1.57069520 |
| O | 1.53416000 | 11.25209000 | 11.85424000 | -1.62000890 |
| O | 1.62610000 | 8.28937000 | 11.86752000 | -1.62987450 |
| O | 1.69175000 | 5.65702000 | 11.90958000 | -1.57361250 |
| O | 3.95872000 | 7.01158000 | 11.88626000 | -1.60317560 |
| O | 4.06504000 | 4.09386000 | 11.89074000 | -1.58964970 |
| O | 4.02440000 | 12.46026000 | 11.85205000 | -1.61851030 |
| O | 4.11628000 | 1.45856000 | 11.88695000 | -1.59433400 |
| O | 4.13792000 | 9.81682000 | 11.85517000 | -1.62675100 |
| O | 6.37270000 | 11.24772000 | 11.85143000 | -1.62479530 |
| O | 6.38254000 | 2.84021000 | 11.87155000 | -1.63663280 |
| O | 6.44826000 | 8.29281000 | 11.86302000 | -1.62558350 |
| O | 6.55252000 | 5.62464000 | 11.83511000 | -1.60967290 |
| O | 8.80105000 | 7.04765000 | 11.85236000 | -1.62719480 |
| O | 8.87071000 | 4.08389000 | 11.85431000 | -1.62117700 |
| O | 8.99321000 | 1.43842000 | 11.85451000 | -1.62436400 |
| O | 11.24402000 | 2.85762000 | 11.85318000 | -1.62216720 |
| Al | -2.40699000 | 9.77755000 | 11.89448000 | 2.41772700 |
| Al | -0.05268000 | 5.62346000 | 12.39905000 | 2.37122980 |
| Al | 2.42227000 | 1.32434000 | 12.40902000 | 2.37523920 |

| | | | |
|---|---|---|---|
| Al | 2.43332000 | 9.80555000 | 11.86199000 | 2.41741520 |
| Al | 4.90688000 | 5.59485000 | 12.39686000 | 2.37865630 |
| Al | 7.27433000 | 9.79305000 | 11.86418000 | 2.41745340 |
| Al | 7.29155000 | 1.38098000 | 11.87181000 | 2.41346090 |
| Al | 9.69078000 | 5.58824000 | 11.86558000 | 2.41431480 |
| Al | 12.08395000 | 1.38767000 | 11.84053000 | 2.41337560 |
| Pt | 2.75080000 | 4.26630000 | 14.93358000 | -0.55100550 |
| Pt | 4.95319000 | 5.58695000 | 14.88103000 | -0.52815990 |
| Pt | 2.39805000 | 1.54299000 | 14.84469000 | -0.91345590 |
| Pt | 0.18471000 | 5.41271000 | 14.82605000 | -0.91661320 |
| Sn | 0.23144000 | 2.88085000 | 14.02187000 | 1.00005760 |
| Sn | 2.38643000 | 6.73530000 | 14.07624000 | 0.94558260 |
| Sn | 4.72056000 | 2.73209000 | 14.12246000 | 0.93080860 |
| C | 7.04475000 | 5.97873000 | 15.16949000 | -0.19234150 |
| C | 6.34041000 | 7.20083000 | 15.14689000 | -0.21185610 |
| H | 7.35445000 | 5.53311000 | 16.12066000 | 0.09224840 |
| H | 7.60483000 | 5.65889000 | 14.28881000 | 0.12388430 |
| H | 6.08872000 | 7.71093000 | 16.08316000 | 0.07750810 |
| H | 6.36342000 | 7.82796000 | 14.25232000 | 0.13892980 |

283

c2h4-pt4sn3-al2o3-A

| | | | |
|---|---|---|---|
| Al | -4.83072000 | 11.17808000 | 0.00000000 | 2.16869390 |
| Al | -2.41059000 | 6.98630000 | 0.00000000 | 2.16876050 |
| Al | 0.00953000 | 11.17808000 | 0.00000000 | 2.16885340 |
| Al | 0.00953000 | 2.79452000 | 0.00000000 | 2.16872940 |
| Al | 2.42966000 | 6.98630000 | 0.00000000 | 2.16882890 |
| Al | 4.84978000 | 2.79452000 | 0.00000000 | 2.16869260 |
| Al | 4.84978000 | 11.17808000 | 0.00000000 | 2.16882330 |
| Al | 7.26991000 | 6.98630000 | 0.00000000 | 2.16867320 |
| Al | 9.69003000 | 2.79452000 | 0.00000000 | 2.16871550 |
| O | -3.89334000 | 9.78082000 | 0.85241000 | -1.55219080 |
| O | -1.66922000 | 8.49672000 | 0.85241000 | -1.56046290 |
| O | -1.66922000 | 11.06492000 | 0.85241000 | -1.55826660 |
| O | -1.47322000 | 5.58904000 | 0.85241000 | -1.55208510 |
| O | 0.75091000 | 4.30494000 | 0.85241000 | -1.56036100 |
| O | 0.75091000 | 6.87314000 | 0.85241000 | -1.55822260 |
| O | 0.94691000 | 9.78082000 | 0.85241000 | -1.55218610 |
| O | 0.94691000 | 1.39726000 | 0.85241000 | -1.55214990 |
| O | 3.17104000 | 0.11316000 | 0.85241000 | -1.56045610 |
| O | 3.17103000 | 8.49672000 | 0.85241000 | -1.56061540 |
| O | 3.17103000 | 2.68136000 | 0.85241000 | -1.55819880 |
| O | 3.17103000 | 11.06492000 | 0.85241000 | -1.55830060 |
| O | 3.36703000 | 5.58904000 | 0.85241000 | -1.55209470 |

| | | | |
|---|---|---|---|
| O | 5.59116000 | 4.30494000 | 0.85241000 | -1.56050000 |
| O | 5.59116000 | 6.87314000 | 0.85241000 | -1.55823580 |
| O | 5.78716000 | 1.39726000 | 0.85241000 | -1.55229560 |
| O | 5.78716000 | 9.78082000 | 0.85241000 | -1.55223540 |
| O | 8.01128000 | 8.49672000 | 0.85241000 | -1.56053890 |
| O | 8.01128000 | 0.11316000 | 0.85241000 | -1.56074250 |
| O | 8.01128000 | 2.68136000 | 0.85241000 | -1.55831670 |
| O | 8.01128000 | 11.06492000 | 0.85241000 | -1.55831450 |
| O | 8.20728000 | 5.58904000 | 0.85241000 | -1.55225990 |
| O | 10.43141000 | 4.30494000 | 0.85241000 | -1.56056280 |
| O | 10.43141000 | 6.87314000 | 0.85241000 | -1.55829250 |
| O | 10.62741000 | 1.39726000 | 0.85241000 | -1.55234540 |
| O | 12.85153000 | 0.11316000 | 0.85241000 | -1.56062970 |
| O | 12.85153000 | 2.68136000 | 0.85241000 | -1.55831640 |
| Al | -4.83072000 | 8.38356000 | 1.70482000 | 2.46326840 |
| Al | -2.41059000 | 4.19178000 | 1.70482000 | 2.46323440 |
| Al | 0.00953000 | 8.38356000 | 1.70482000 | 2.46320750 |
| Al | 0.00954000 | 0.00000000 | 1.70482000 | 2.46330380 |
| Al | 2.42966000 | 4.19178000 | 1.70482000 | 2.46324550 |
| Al | 4.84978000 | 0.00000000 | 1.70482000 | 2.46333930 |
| Al | 4.84978000 | 8.38356000 | 1.70482000 | 2.46329870 |
| Al | 7.26991000 | 4.19178000 | 1.70482000 | 2.46337030 |
| Al | 9.69003000 | 0.00000000 | 1.70482000 | 2.46335740 |
| Al | -2.41059000 | 9.78082000 | 2.20274000 | 2.48269990 |
| Al | 0.00953000 | 5.58904000 | 2.20274000 | 2.48276710 |
| Al | 2.42966000 | 1.39726000 | 2.20274000 | 2.48271280 |
| Al | 2.42966000 | 9.78082000 | 2.20274000 | 2.48266040 |
| Al | 4.84978000 | 5.58904000 | 2.20274000 | 2.48267110 |
| Al | 7.26991000 | 9.78082000 | 2.20274000 | 2.48271200 |
| Al | 7.26991000 | 1.39726000 | 2.20274000 | 2.48275500 |
| Al | 9.69003000 | 5.58904000 | 2.20274000 | 2.48271480 |
| Al | 12.11016000 | 1.39726000 | 2.20274000 | 2.48272160 |
| O | -5.57210000 | 9.66766000 | 3.05515000 | -1.64209670 |
| O | -3.34796000 | 8.38356000 | 3.05515000 | -1.62409690 |
| O | -3.15197000 | 5.47588000 | 3.05515000 | -1.64210690 |
| O | -3.15197000 | 11.29124000 | 3.05515000 | -1.64977420 |
| O | -0.92784000 | 4.19178000 | 3.05515000 | -1.62417610 |
| O | -0.73185000 | 1.28410000 | 3.05515000 | -1.64198190 |
| O | -0.73185000 | 9.66766000 | 3.05515000 | -1.64215820 |
| O | -0.73184000 | 7.09945000 | 3.05515000 | -1.64976100 |
| O | 1.49229000 | 8.38356000 | 3.05515000 | -1.62406160 |
| O | 1.49229000 | 0.00001000 | 3.05515000 | -1.62412420 |
| O | 1.68828000 | 5.47588000 | 3.05515000 | -1.64164980 |
| O | 1.68828000 | 2.90768000 | 3.05515000 | -1.64968250 |
| O | 1.68828000 | 11.29123000 | 3.05515000 | -1.65007860 |
| O | 3.91241000 | 4.19178000 | 3.05515000 | -1.62408720 |
| O | 4.10840000 | 9.66766000 | 3.05515000 | -1.64218290 |
| O | 4.10840000 | 1.28410000 | 3.05515000 | -1.64190620 |
| O | 4.10841000 | 7.09946000 | 3.05515000 | -1.64991110 |
| O | 6.33254000 | 0.00001000 | 3.05515000 | -1.62416460 |
| O | 6.33254000 | 8.38356000 | 3.05515000 | -1.62410960 |
| O | 6.52853000 | 11.29123000 | 3.05515000 | -1.65008550 |
| O | 6.52853000 | 5.47588000 | 3.05515000 | -1.64210560 |
| O | 6.52853000 | 2.90768000 | 3.05515000 | -1.64986830 |
| O | 8.75266000 | 4.19178000 | 3.05515000 | -1.62425440 |
| O | 8.94866000 | 7.09945000 | 3.05515000 | -1.65012460 |
| O | 8.94865000 | 1.28410000 | 3.05515000 | -1.64228580 |
| O | 11.17277000 | 0.00000000 | 3.05515000 | -1.62437780 |
| O | 11.36878000 | 2.90767000 | 3.05515000 | -1.65020330 |
| Al | -4.83072000 | 11.17808000 | 3.90756000 | 2.47792070 |
| Al | -2.41059000 | 6.98630000 | 3.90756000 | 2.47790070 |
| Al | 0.00953000 | 11.17808000 | 3.90756000 | 2.47802710 |
| Al | 0.00953000 | 2.79452000 | 3.90756000 | 2.47779630 |
| Al | 2.42966000 | 6.98630000 | 3.90756000 | 2.47773510 |
| Al | 4.84978000 | 2.79452000 | 3.90756000 | 2.47777930 |
| Al | 4.84978000 | 11.17808000 | 3.90756000 | 2.47801150 |
| Al | 7.26991000 | 6.98630000 | 3.90756000 | 2.47794440 |
| Al | 9.69003000 | 2.79452000 | 3.90756000 | 2.47792820 |
| Al | -4.83072000 | 8.38356000 | 4.40548000 | 2.47241220 |
| Al | -2.41059000 | 4.19178000 | 4.40548000 | 2.47244860 |
| Al | 0.00953000 | 8.38356000 | 4.40548000 | 2.47231910 |
| Al | 0.00954000 | 0.00000000 | 4.40548000 | 2.47240780 |
| Al | 2.42966000 | 4.19178000 | 4.40548000 | 2.47220560 |
| Al | 4.84978000 | 0.00000000 | 4.40548000 | 2.47235080 |
| Al | 4.84978000 | 8.38356000 | 4.40548000 | 2.47224640 |
| Al | 7.26991000 | 4.19178000 | 4.40548000 | 2.47217040 |
| Al | 9.69003000 | 0.00000000 | 4.40548000 | 2.47223010 |
| O | -6.31347000 | 11.17808000 | 5.25789000 | -1.64347500 |
| O | -4.08934000 | 12.46219000 | 5.25789000 | -1.65401430 |
| O | -4.08934000 | 9.89398000 | 5.25789000 | -1.64895150 |
| O | -3.89335000 | 6.98630000 | 5.25789000 | -1.64358600 |
| O | -1.66922000 | 8.27040000 | 5.25789000 | -1.65398770 |
| O | -1.66921000 | 5.70219000 | 5.25789000 | -1.64912760 |
| O | -1.47322000 | 11.17808000 | 5.25789000 | -1.64343760 |
| O | -1.47323000 | 2.79452000 | 5.25789000 | -1.64345470 |
| O | 0.75091000 | 4.07862000 | 5.25789000 | -1.65381740 |

| | | | |
|---|---|---|---|
| O | 0.75091000 | 12.46219000 | 5.25789000 | -1.65414810 |
| O | 0.75091000 | 9.89397000 | 5.25789000 | -1.64898740 |
| O | 0.75091000 | 1.51042000 | 5.25789000 | -1.64905320 |
| O | 0.94690000 | 6.98630000 | 5.25789000 | -1.64354280 |
| O | 3.17103000 | 8.27040000 | 5.25789000 | -1.65424240 |
| O | 3.17104000 | 5.70220000 | 5.25789000 | -1.64895180 |
| O | 3.36703000 | 2.79452000 | 5.25789000 | -1.64313630 |
| O | 3.36702000 | 11.17808000 | 5.25789000 | -1.64328050 |
| O | 5.59116000 | 12.46219000 | 5.25789000 | -1.65396390 |
| O | 5.59116000 | 4.07863000 | 5.25789000 | -1.65418690 |
| O | 5.59116000 | 9.89397000 | 5.25789000 | -1.64889580 |
| O | 5.59116000 | 1.51042000 | 5.25789000 | -1.64928290 |
| O | 5.78715000 | 6.98630000 | 5.25789000 | -1.64336560 |
| O | 8.01128000 | 8.27040000 | 5.25789000 | -1.65404700 |
| O | 8.01129000 | 5.70219000 | 5.25789000 | -1.64888510 |
| O | 8.20727000 | 2.79452000 | 5.25789000 | -1.64351660 |
| O | 10.43141000 | 4.07862000 | 5.25789000 | -1.65382630 |
| O | 10.43141000 | 1.51041000 | 5.25789000 | -1.64891220 |
| Al | -2.41059000 | 9.78082000 | 6.11030000 | 2.47601570 |
| Al | 0.00953000 | 5.58904000 | 6.11030000 | 2.47631740 |
| Al | 2.42966000 | 1.39726000 | 6.11030000 | 2.47629230 |
| Al | 2.42966000 | 9.78082000 | 6.11030000 | 2.47595400 |
| Al | 4.84978000 | 5.58904000 | 6.11030000 | 2.47629030 |
| Al | 7.26991000 | 9.78082000 | 6.11030000 | 2.47611490 |
| Al | 7.26991000 | 1.39726000 | 6.11030000 | 2.47616890 |
| Al | 9.69003000 | 5.58904000 | 6.11030000 | 2.47596200 |
| Al | 12.11016000 | 1.39726000 | 6.11030000 | 2.47563290 |
| Al | -4.83072000 | 11.17808000 | 6.60910000 | 2.47684370 |
| Al | -2.41059000 | 6.98630000 | 6.60910000 | 2.47689250 |
| Al | 0.00953000 | 11.17808000 | 6.60910000 | 2.47695960 |
| Al | 0.00953000 | 2.79452000 | 6.60910000 | 2.47706260 |
| Al | 2.42966000 | 6.98630000 | 6.60910000 | 2.47687470 |
| Al | 4.84978000 | 2.79452000 | 6.60910000 | 2.47694380 |
| Al | 4.84978000 | 11.17808000 | 6.60910000 | 2.47657760 |
| Al | 7.26991000 | 6.98630000 | 6.60910000 | 2.47679620 |
| Al | 9.69003000 | 2.79452000 | 6.60910000 | 2.47690460 |
| O | -3.15680000 | 8.49706000 | 7.46570000 | -1.64891810 |
| O | -3.14925000 | 11.06893000 | 7.46570000 | -1.64953670 |
| O | -0.92572000 | 9.77647000 | 7.46570000 | -1.65427490 |
| O | -0.73667000 | 4.30528000 | 7.46570000 | -1.64901500 |
| O | -0.72913000 | 6.87715000 | 7.46570000 | -1.64888570 |
| O | 1.49440000 | 5.58469000 | 7.46570000 | -1.65322420 |
| O | 1.68345000 | 8.49706000 | 7.46570000 | -1.64840790 |
| O | 1.68345000 | 0.11351000 | 7.46570000 | -1.64900090 |
| O | 1.69100000 | 2.68537000 | 7.46570000 | -1.64882790 |
| O | 1.69100000 | 11.06893000 | 7.46570000 | -1.64997540 |
| O | 3.91453000 | 1.39291000 | 7.46570000 | -1.65364040 |
| O | 3.91452000 | 9.77646000 | 7.46570000 | -1.65403180 |
| O | 4.10358000 | 4.30528000 | 7.46570000 | -1.64858320 |
| O | 4.11112000 | 6.87715000 | 7.46570000 | -1.64978900 |
| O | 6.33465000 | 5.58468000 | 7.46570000 | -1.65361810 |
| O | 6.52370000 | 0.11351000 | 7.46570000 | -1.64940360 |
| O | 6.52370000 | 8.49706000 | 7.46570000 | -1.64930500 |
| O | 6.53125000 | 2.68537000 | 7.46570000 | -1.64932890 |
| O | 6.53125000 | 11.06893000 | 7.46570000 | -1.65037640 |
| O | 8.75477000 | 9.77646000 | 7.46570000 | -1.65366860 |
| O | 8.75477000 | 1.39291000 | 7.46570000 | -1.65468190 |
| O | 8.94383000 | 4.30528000 | 7.46570000 | -1.64967740 |
| O | 8.95137000 | 6.87715000 | 7.46570000 | -1.64991130 |
| O | 11.17490000 | 5.58468000 | 7.46570000 | -1.65369500 |
| O | 11.36395000 | 0.11351000 | 7.46570000 | -1.64969340 |
| O | 11.37149000 | 2.68537000 | 7.46570000 | -1.65014450 |
| O | 13.59502000 | 1.39291000 | 7.46570000 | -1.65360460 |
| Al | -4.83072000 | 8.38356000 | 8.32833000 | 2.47144110 |
| Al | -2.41059000 | 4.19178000 | 8.32833000 | 2.47112070 |
| Al | 0.00953000 | 8.38356000 | 8.32833000 | 2.47086700 |
| Al | 0.00954000 | 0.00000000 | 8.32833000 | 2.47100310 |
| Al | 2.42966000 | 4.19178000 | 8.32833000 | 2.47106620 |
| Al | 4.84978000 | 0.00000000 | 8.32833000 | 2.47147210 |
| Al | 4.84978000 | 8.38356000 | 8.32833000 | 2.47115200 |
| Al | 7.26991000 | 4.19178000 | 8.32833000 | 2.47215790 |
| Al | 9.69003000 | 0.00000000 | 8.32833000 | 2.47187580 |
| Al | -2.41059000 | 9.78082000 | 8.78548000 | 2.47230660 |
| Al | 0.00953000 | 5.58904000 | 8.78548000 | 2.47631920 |
| Al | 2.42966000 | 1.39726000 | 8.78548000 | 2.47641370 |
| Al | 2.42966000 | 9.78082000 | 8.78548000 | 2.47287530 |
| Al | 4.84978000 | 5.58904000 | 8.78548000 | 2.47579780 |
| Al | 7.26991000 | 9.78082000 | 8.78548000 | 2.47232030 |
| Al | 7.26991000 | 1.39726000 | 8.78548000 | 2.47224400 |
| Al | 9.69003000 | 5.58904000 | 8.78548000 | 2.47201320 |
| Al | 12.11016000 | 1.39726000 | 8.78548000 | 2.47111670 |
| O | 8.01661000 | 11.26550000 | 9.68989000 | -1.64624020 |
| O | 10.43673000 | 7.07371000 | 9.68989000 | -1.64334880 |
| O | -4.06971000 | 9.68514000 | 9.68989000 | -1.65942400 |
| O | -1.66389000 | 11.26550000 | 9.68989000 | -1.64488810 |
| O | 12.85686000 | 2.88194000 | 9.68989000 | -1.64570140 |

| | | | |
|---|---|---|---|
| O | -1.64959000 | 5.49336000 | 9.68989000 | -1.64342240 |
| O | -1.49818000 | 8.39182000 | 9.68989000 | -1.66391670 |
| O | 0.75624000 | 7.07372000 | 9.68989000 | -1.63763860 |
| O | 0.77054000 | 9.68514000 | 9.68989000 | -1.65435610 |
| O | 0.77054000 | 1.30159000 | 9.68989000 | -1.64769560 |
| O | 0.92195000 | 4.20004000 | 9.68989000 | -1.65613110 |
| O | 3.17636000 | 11.26550000 | 9.68989000 | -1.64386830 |
| O | 3.17636000 | 2.88194000 | 9.68989000 | -1.63386150 |
| O | 3.19067000 | 5.49336000 | 9.68989000 | -1.64545600 |
| O | 3.34208000 | 0.00826000 | 9.68989000 | -1.65473740 |
| O | 3.34207000 | 8.39182000 | 9.68989000 | -1.66318640 |
| O | 5.59648000 | 7.07372000 | 9.68989000 | -1.63985730 |
| O | 5.61079000 | 1.30159000 | 9.68989000 | -1.65990020 |
| O | 5.61079000 | 9.68514000 | 9.68989000 | -1.65789290 |
| O | 5.76220000 | 4.20004000 | 9.68989000 | -1.65733020 |
| O | 8.01661000 | 2.88194000 | 9.68989000 | -1.64409490 |
| O | 8.03091000 | 5.49336000 | 9.68989000 | -1.66041940 |
| O | 8.18232000 | 0.00826000 | 9.68989000 | -1.66864930 |
| O | 8.18232000 | 8.39182000 | 9.68989000 | -1.66837010 |
| O | 10.45104000 | 1.30158000 | 9.68989000 | -1.66226990 |
| O | 10.60245000 | 4.20004000 | 9.68989000 | -1.66936850 |
| O | 13.02257000 | 0.00826000 | 9.68989000 | -1.66778970 |
| Al | -4.83072000 | 11.17808000 | 10.71830000 | 2.47772660 |
| Al | -2.41059000 | 6.98630000 | 10.71830000 | 2.47953330 |
| Al | 0.00953000 | 11.17808000 | 10.71830000 | 2.47588610 |
| Al | 0.00953000 | 2.79452000 | 10.71830000 | 2.47726800 |
| Al | 2.42966000 | 6.98630000 | 10.71830000 | 2.47684890 |
| Al | 4.84978000 | 2.79452000 | 10.71830000 | 2.47772760 |
| Al | 4.84978000 | 11.17808000 | 10.71830000 | 2.47650890 |
| Al | 7.26991000 | 6.98630000 | 10.71830000 | 2.48016510 |
| Al | 9.69003000 | 2.79452000 | 10.71830000 | 2.47320640 |
| Al | -4.83072000 | 8.38356000 | 10.98601000 | 2.47273080 |
| Al | -2.41059000 | 4.19178000 | 10.98601000 | 2.47150890 |
| Al | 0.00953000 | 8.38356000 | 10.98601000 | 2.46879500 |
| Al | 0.00954000 | 0.00000000 | 10.98601000 | 2.47118240 |
| Al | 2.42966000 | 4.19178000 | 10.98601000 | 2.47488600 |
| Al | 4.84978000 | 0.00000000 | 10.98601000 | 2.47290330 |
| Al | 4.84978000 | 8.38356000 | 10.98601000 | 2.47268060 |
| Al | 7.26991000 | 4.19178000 | 10.98601000 | 2.47365880 |
| Al | 9.69003000 | 0.00000000 | 10.98601000 | 2.46538560 |
| O | -5.65639000 | 12.48009000 | 11.84080000 | -1.61050340 |
| O | -5.54233000 | 9.81909000 | 11.85090000 | -1.62372710 |
| O | -3.29207000 | 11.23743000 | 11.86086000 | -1.62300270 |

| | | | |
|---|---|---|---|
| O | -3.23156000 | 8.28113000 | 11.85216000 | -1.62054320 |
| O | -3.12396000 | 5.62208000 | 11.85906000 | -1.62512750 |
| O | -0.85987000 | 7.03902000 | 11.84533000 | -1.61355950 |
| O | -0.80917000 | 12.47612000 | 11.85186000 | -1.62605130 |
| O | -0.77639000 | 4.09302000 | 11.87698000 | -1.58344710 |
| O | -0.69076000 | 9.82462000 | 11.85605000 | -1.63028980 |
| O | -0.72493000 | 1.44664000 | 11.87635000 | -1.62722160 |
| O | 1.52289000 | 2.82574000 | 11.90477000 | -1.58025960 |
| O | 1.53167000 | 11.26265000 | 11.86370000 | -1.61514600 |
| O | 1.64129000 | 8.29203000 | 11.87912000 | -1.62361340 |
| O | 1.69419000 | 5.64297000 | 11.81585000 | -1.61049270 |
| O | 3.97657000 | 7.04880000 | 11.95594000 | -1.56078860 |
| O | 4.05256000 | 4.09787000 | 11.94853000 | -1.57194050 |
| O | 4.02270000 | 12.45541000 | 11.85330000 | -1.62099540 |
| O | 4.11386000 | 1.45402000 | 11.84997000 | -1.60889770 |
| O | 4.13442000 | 9.81384000 | 11.85434000 | -1.63363940 |
| O | 6.36880000 | 11.24667000 | 11.84980000 | -1.62567060 |
| O | 6.37837000 | 2.85369000 | 11.87809000 | -1.63235500 |
| O | 6.44409000 | 8.29428000 | 11.86745000 | -1.63211530 |
| O | 6.54794000 | 5.63048000 | 11.82392000 | -1.61154340 |
| O | 8.79661000 | 7.04973000 | 11.85388000 | -1.62510720 |
| O | 8.86963000 | 4.08869000 | 11.85586000 | -1.62427070 |
| O | 8.98502000 | 1.44068000 | 11.84836000 | -1.61815820 |
| O | 11.22105000 | 2.85791000 | 11.85222000 | -1.62515170 |
| Al | -2.40335000 | 9.77583000 | 11.89531000 | 2.41658710 |
| Al | 0.02188000 | 5.63313000 | 12.40472000 | 2.37539190 |
| Al | 2.43569000 | 1.33428000 | 12.38103000 | 2.37232740 |
| Al | 2.42564000 | 9.82154000 | 11.93280000 | 2.40442170 |
| Al | 4.92674000 | 5.59531000 | 12.42698000 | 2.38107920 |
| Al | 7.27410000 | 9.79462000 | 11.85211000 | 2.41782220 |
| Al | 7.27730000 | 1.39209000 | 11.86201000 | 2.41386010 |
| Al | 9.69031000 | 5.59165000 | 11.86612000 | 2.41363780 |
| Al | 12.07507000 | 1.38607000 | 11.81506000 | 2.41381190 |
| Pt | 2.53514000 | 5.26091000 | 14.80715000 | -0.49233900 |
| Pt | -0.00000000 | 5.27245000 | 14.83359000 | -0.83106980 |
| Pt | 5.11706000 | 5.73427000 | 14.89636000 | -0.63811450 |
| Pt | 2.30375000 | 1.73533000 | 14.78581000 | -0.93952940 |
| Sn | 4.29525000 | 3.22784000 | 14.21843000 | 0.92317650 |
| Sn | 0.06262000 | 2.85085000 | 14.00287000 | 1.03031570 |
| Sn | 3.24820000 | 7.59679000 | 14.09641000 | 0.90955380 |
| C | 7.26523000 | 5.63627000 | 15.24941000 | -0.16773360 |
| C | 6.87769000 | 6.98582000 | 15.22049000 | -0.18965580 |
| H | 7.42623000 | 5.12708000 | 16.20499000 | 0.05187030 |

| | | | |
|---|---|---|---|
| H | 7.73716000 | 5.17276000 | 14.38067000 | 0.11484340 |
| H | 6.74195000 | 7.54110000 | 16.15361000 | 0.08183400 |
| H | 7.05308000 | 7.58325000 | 14.32341000 | 0.12463380 |

283

c2h4-pt4sn3-al2o3-A

| | | | | |
|---|---|---|---|---|
| Al | -4.83072000 | 11.17808000 | 0.00000000 | 2.16888640 |
| Al | -2.41059000 | 6.98630000 | 0.00000000 | 2.16887110 |
| Al | 0.00953000 | 11.17808000 | 0.00000000 | 2.16884510 |
| Al | 0.00953000 | 2.79452000 | 0.00000000 | 2.16888520 |
| Al | 2.42966000 | 6.98630000 | 0.00000000 | 2.16876990 |
| Al | 4.84978000 | 2.79452000 | 0.00000000 | 2.16875040 |
| Al | 4.84978000 | 11.17808000 | 0.00000000 | 2.16888560 |
| Al | 7.26991000 | 6.98630000 | 0.00000000 | 2.16884390 |
| Al | 9.69003000 | 2.79452000 | 0.00000000 | 2.16896990 |
| O | -3.89334000 | 9.78082000 | 0.85241000 | -1.55216600 |
| O | -1.66922000 | 8.49672000 | 0.85241000 | -1.56047710 |
| O | -1.66922000 | 11.06492000 | 0.85241000 | -1.55823520 |
| O | -1.47322000 | 5.58904000 | 0.85241000 | -1.55225630 |
| O | 0.75091000 | 4.30494000 | 0.85241000 | -1.56049640 |
| O | 0.75091000 | 6.87314000 | 0.85241000 | -1.55826520 |
| O | 0.94691000 | 9.78082000 | 0.85241000 | -1.55205760 |
| O | 0.94691000 | 1.39726000 | 0.85241000 | -1.55209250 |
| O | 3.17104000 | 0.11316000 | 0.85241000 | -1.56048740 |
| O | 3.17103000 | 8.49672000 | 0.85241000 | -1.56035520 |
| O | 3.17103000 | 2.68136000 | 0.85241000 | -1.55826440 |
| O | 3.17103000 | 11.06492000 | 0.85241000 | -1.55823870 |
| O | 3.36703000 | 5.58904000 | 0.85241000 | -1.55217710 |
| O | 5.59116000 | 4.30494000 | 0.85241000 | -1.56052820 |
| O | 5.59116000 | 6.87314000 | 0.85241000 | -1.55816870 |
| O | 5.78716000 | 1.39726000 | 0.85241000 | -1.55218750 |
| O | 5.78716000 | 9.78082000 | 0.85241000 | -1.55203550 |
| O | 8.01128000 | 8.49672000 | 0.85241000 | -1.56040850 |
| O | 8.01128000 | 0.11316000 | 0.85241000 | -1.56054540 |
| O | 8.01128000 | 2.68136000 | 0.85241000 | -1.55828360 |
| O | 8.01128000 | 11.06492000 | 0.85241000 | -1.55817520 |
| O | 8.20728000 | 5.58904000 | 0.85241000 | -1.55226060 |
| O | 10.43141000 | 4.30494000 | 0.85241000 | -1.56078790 |
| O | 10.43141000 | 6.87314000 | 0.85241000 | -1.55829900 |
| O | 10.62741000 | 1.39726000 | 0.85241000 | -1.55230750 |
| O | 12.85153000 | 0.11316000 | 0.85241000 | -1.56050780 |
| O | 12.85153000 | 2.68136000 | 0.85241000 | -1.55832330 |
| Al | -4.83072000 | 8.38356000 | 1.70482000 | 2.46327950 |
| Al | -2.41059000 | 4.19178000 | 1.70482000 | 2.46346500 |

| | | | |
|---|---|---|---|
| Al | 0.00953000 | 8.38356000 | 1.70482000 | 2.46320720 |
| Al | 0.00954000 | 0.00000000 | 1.70482000 | 2.46320700 |
| Al | 2.42966000 | 4.19178000 | 1.70482000 | 2.46324200 |
| Al | 4.84978000 | 0.00000000 | 1.70482000 | 2.46326510 |
| Al | 4.84978000 | 8.38356000 | 1.70482000 | 2.46324680 |
| Al | 7.26991000 | 4.19178000 | 1.70482000 | 2.46341020 |
| Al | 9.69003000 | 0.00000000 | 1.70482000 | 2.46326670 |
| Al | -2.41059000 | 9.78082000 | 2.20274000 | 2.48269000 |
| Al | 0.00953000 | 5.58904000 | 2.20274000 | 2.48271310 |
| Al | 2.42966000 | 1.39726000 | 2.20274000 | 2.48263150 |
| Al | 2.42966000 | 9.78082000 | 2.20274000 | 2.48272970 |
| Al | 4.84978000 | 5.58904000 | 2.20274000 | 2.48271180 |
| Al | 7.26991000 | 9.78082000 | 2.20274000 | 2.48270240 |
| Al | 7.26991000 | 1.39726000 | 2.20274000 | 2.48265850 |
| Al | 9.69003000 | 5.58904000 | 2.20274000 | 2.48269320 |
| Al | 12.11016000 | 1.39726000 | 2.20274000 | 2.48275620 |
| O | -5.57210000 | 9.66766000 | 3.05515000 | -1.64171430 |
| O | -3.34796000 | 8.38356000 | 3.05515000 | -1.62413790 |
| O | -3.15197000 | 5.47588000 | 3.05515000 | -1.64214460 |
| O | -3.15197000 | 11.29124000 | 3.05515000 | -1.64992530 |
| O | -0.92784000 | 4.19178000 | 3.05515000 | -1.62434480 |
| O | -0.73185000 | 1.28410000 | 3.05515000 | -1.64216760 |
| O | -0.73185000 | 9.66766000 | 3.05515000 | -1.64182860 |
| O | -0.73184000 | 7.09945000 | 3.05515000 | -1.65002780 |
| O | 1.49229000 | 8.38356000 | 3.05515000 | -1.62411360 |
| O | 1.49229000 | 0.00001000 | 3.05515000 | -1.62406240 |
| O | 1.68828000 | 5.47588000 | 3.05515000 | -1.64186240 |
| O | 1.68828000 | 2.90768000 | 3.05515000 | -1.65005330 |
| O | 1.68828000 | 11.29123000 | 3.05515000 | -1.64985470 |
| O | 3.91241000 | 4.19178000 | 3.05515000 | -1.62408760 |
| O | 4.10840000 | 9.66766000 | 3.05515000 | -1.64160470 |
| O | 4.10840000 | 1.28410000 | 3.05515000 | -1.64216620 |
| O | 4.10841000 | 7.09946000 | 3.05515000 | -1.64966870 |
| O | 6.33254000 | 0.00001000 | 3.05515000 | -1.62406210 |
| O | 6.33254000 | 8.38356000 | 3.05515000 | -1.62400510 |
| O | 6.52853000 | 11.29123000 | 3.05515000 | -1.64984520 |
| O | 6.52853000 | 5.47588000 | 3.05515000 | -1.64190180 |
| O | 6.52853000 | 2.90768000 | 3.05515000 | -1.65004090 |
| O | 8.75266000 | 4.19178000 | 3.05515000 | -1.62430490 |
| O | 8.94866000 | 7.09945000 | 3.05515000 | -1.64993500 |
| O | 8.94865000 | 1.28410000 | 3.05515000 | -1.64213570 |
| O | 11.17277000 | 0.00000000 | 3.05515000 | -1.62404770 |
| O | 11.36878000 | 2.90767000 | 3.05515000 | -1.65012590 |

| | | | |
|---|---|---|---|
| Al | -4.83072000 | 11.17808000 | 3.90756000 | 2.47774530 |
| Al | -2.41059000 | 6.98630000 | 3.90756000 | 2.47776260 |
| Al | 0.00953000 | 11.17808000 | 3.90756000 | 2.47775380 |
| Al | 0.00953000 | 2.79452000 | 3.90756000 | 2.47793840 |
| Al | 2.42966000 | 6.98630000 | 3.90756000 | 2.47777500 |
| Al | 4.84978000 | 2.79452000 | 3.90756000 | 2.47797040 |
| Al | 4.84978000 | 11.17808000 | 3.90756000 | 2.47779290 |
| Al | 7.26991000 | 6.98630000 | 3.90756000 | 2.47777960 |
| Al | 9.69003000 | 2.79452000 | 3.90756000 | 2.47803650 |
| Al | -4.83072000 | 8.38356000 | 4.40548000 | 2.47226360 |
| Al | -2.41059000 | 4.19178000 | 4.40548000 | 2.47223860 |
| Al | 0.00953000 | 8.38356000 | 4.40548000 | 2.47243600 |
| Al | 0.00954000 | 0.00000000 | 4.40548000 | 2.47230400 |
| Al | 2.42966000 | 4.19178000 | 4.40548000 | 2.47238770 |
| Al | 4.84978000 | 0.00000000 | 4.40548000 | 2.47227630 |
| Al | 4.84978000 | 8.38356000 | 4.40548000 | 2.47226160 |
| Al | 7.26991000 | 4.19178000 | 4.40548000 | 2.47223990 |
| Al | 9.69003000 | 0.00000000 | 4.40548000 | 2.47224830 |
| O | -6.31347000 | 11.17808000 | 5.25789000 | -1.64358200 |
| O | -4.08934000 | 12.46219000 | 5.25789000 | -1.65410130 |
| O | -4.08934000 | 9.89398000 | 5.25789000 | -1.64917950 |
| O | -3.89335000 | 6.98630000 | 5.25789000 | -1.64339820 |
| O | -1.66922000 | 8.27040000 | 5.25789000 | -1.65386120 |
| O | -1.66921000 | 5.70219000 | 5.25789000 | -1.64903040 |
| O | -1.47322000 | 11.17808000 | 5.25789000 | -1.64350010 |
| O | -1.47323000 | 2.79452000 | 5.25789000 | -1.64354320 |
| O | 0.75091000 | 4.07862000 | 5.25789000 | -1.65393650 |
| O | 0.75091000 | 12.46219000 | 5.25789000 | -1.65411340 |
| O | 0.75091000 | 9.89397000 | 5.25789000 | -1.64899580 |
| O | 0.75091000 | 1.51042000 | 5.25789000 | -1.64891220 |
| O | 0.94690000 | 6.98630000 | 5.25789000 | -1.64345260 |
| O | 3.17103000 | 8.27040000 | 5.25789000 | -1.65385880 |
| O | 3.17104000 | 5.70220000 | 5.25789000 | -1.64909450 |
| O | 3.36703000 | 2.79452000 | 5.25789000 | -1.64352630 |
| O | 3.36702000 | 11.17808000 | 5.25789000 | -1.64367300 |
| O | 5.59116000 | 12.46219000 | 5.25789000 | -1.65415650 |
| O | 5.59116000 | 4.07863000 | 5.25789000 | -1.65396990 |
| O | 5.59116000 | 9.89397000 | 5.25789000 | -1.64901980 |
| O | 5.59116000 | 1.51042000 | 5.25789000 | -1.64894190 |
| O | 5.78715000 | 6.98630000 | 5.25789000 | -1.64319620 |
| O | 8.01128000 | 8.27040000 | 5.25789000 | -1.65398400 |
| O | 8.01129000 | 5.70219000 | 5.25789000 | -1.64914110 |
| O | 8.20727000 | 2.79452000 | 5.25789000 | -1.64322210 |

| | | | |
|---|---|---|---|
| O | 10.43141000 | 4.07862000 | 5.25789000 | -1.65407140 |
| O | 10.43141000 | 1.51041000 | 5.25789000 | -1.64886180 |
| Al | -2.41059000 | 9.78082000 | 6.11030000 | 2.47610110 |
| Al | 0.00953000 | 5.58904000 | 6.11030000 | 2.47587430 |
| Al | 2.42966000 | 1.39726000 | 6.11030000 | 2.47611470 |
| Al | 2.42966000 | 9.78082000 | 6.11030000 | 2.47639070 |
| Al | 4.84978000 | 5.58904000 | 6.11030000 | 2.47608120 |
| Al | 7.26991000 | 9.78082000 | 6.11030000 | 2.47630350 |
| Al | 7.26991000 | 1.39726000 | 6.11030000 | 2.47619620 |
| Al | 9.69003000 | 5.58904000 | 6.11030000 | 2.47615830 |
| Al | 12.11016000 | 1.39726000 | 6.11030000 | 2.47597570 |
| Al | -4.83072000 | 11.17808000 | 6.60910000 | 2.47682360 |
| Al | -2.41059000 | 6.98630000 | 6.60910000 | 2.47680120 |
| Al | 0.00953000 | 11.17808000 | 6.60910000 | 2.47682020 |
| Al | 0.00953000 | 2.79452000 | 6.60910000 | 2.47663580 |
| Al | 2.42966000 | 6.98630000 | 6.60910000 | 2.47690880 |
| Al | 4.84978000 | 2.79452000 | 6.60910000 | 2.47672660 |
| Al | 4.84978000 | 11.17808000 | 6.60910000 | 2.47717000 |
| Al | 7.26991000 | 6.98630000 | 6.60910000 | 2.47672200 |
| Al | 9.69003000 | 2.79452000 | 6.60910000 | 2.47673010 |
| O | -3.15680000 | 8.49706000 | 7.46570000 | -1.64964190 |
| O | -3.14925000 | 11.06893000 | 7.46570000 | -1.64984710 |
| O | -0.92572000 | 9.77647000 | 7.46570000 | -1.65368200 |
| O | -0.73667000 | 4.30528000 | 7.46570000 | -1.64967120 |
| O | -0.72913000 | 6.87715000 | 7.46570000 | -1.64972620 |
| O | 1.49440000 | 5.58469000 | 7.46570000 | -1.65377490 |
| O | 1.68345000 | 8.49706000 | 7.46570000 | -1.64919460 |
| O | 1.68345000 | 0.11351000 | 7.46570000 | -1.64897740 |
| O | 1.69100000 | 2.68537000 | 7.46570000 | -1.65042330 |
| O | 1.69100000 | 11.06893000 | 7.46570000 | -1.64935960 |
| O | 3.91453000 | 1.39291000 | 7.46570000 | -1.65412070 |
| O | 3.91452000 | 9.77646000 | 7.46570000 | -1.65343820 |
| O | 4.10358000 | 4.30528000 | 7.46570000 | -1.64896580 |
| O | 4.11112000 | 6.87715000 | 7.46570000 | -1.64900530 |
| O | 6.33465000 | 5.58468000 | 7.46570000 | -1.65397260 |
| O | 6.52370000 | 0.11351000 | 7.46570000 | -1.64900160 |
| O | 6.52370000 | 8.49706000 | 7.46570000 | -1.64836580 |
| O | 6.53125000 | 2.68537000 | 7.46570000 | -1.64998190 |
| O | 6.53125000 | 11.06893000 | 7.46570000 | -1.64953910 |
| O | 8.75477000 | 9.77646000 | 7.46570000 | -1.65367250 |
| O | 8.75477000 | 1.39291000 | 7.46570000 | -1.65454730 |
| O | 8.94383000 | 4.30528000 | 7.46570000 | -1.64956580 |
| O | 8.95137000 | 6.87715000 | 7.46570000 | -1.64951040 |

| | | | |
|---|---|---|---|
| O | 11.17490000 | 5.58468000 | 7.46570000 | -1.65431800 |
| O | 11.36395000 | 0.11351000 | 7.46570000 | -1.64896790 |
| O | 11.37149000 | 2.68537000 | 7.46570000 | -1.65040500 |
| O | 13.59502000 | 1.39291000 | 7.46570000 | -1.65363240 |
| Al | -4.83072000 | 8.38356000 | 8.32833000 | 2.47192440 |
| Al | -2.41059000 | 4.19178000 | 8.32833000 | 2.47239220 |
| Al | 0.00953000 | 8.38356000 | 8.32833000 | 2.47089590 |
| Al | 0.00954000 | 0.00000000 | 8.32833000 | 2.47108080 |
| Al | 2.42966000 | 4.19178000 | 8.32833000 | 2.47112660 |
| Al | 4.84978000 | 0.00000000 | 8.32833000 | 2.47149800 |
| Al | 4.84978000 | 8.38356000 | 8.32833000 | 2.47065150 |
| Al | 7.26991000 | 4.19178000 | 8.32833000 | 2.47174070 |
| Al | 9.69003000 | 0.00000000 | 8.32833000 | 2.47159750 |
| Al | -2.41059000 | 9.78082000 | 8.78548000 | 2.47267870 |
| Al | 0.00953000 | 5.58904000 | 8.78548000 | 2.47215640 |
| Al | 2.42966000 | 1.39726000 | 8.78548000 | 2.47260810 |
| Al | 2.42966000 | 9.78082000 | 8.78548000 | 2.47646710 |
| Al | 4.84978000 | 5.58904000 | 8.78548000 | 2.47598440 |
| Al | 7.26991000 | 9.78082000 | 8.78548000 | 2.47672030 |
| Al | 7.26991000 | 1.39726000 | 8.78548000 | 2.47170150 |
| Al | 9.69003000 | 5.58904000 | 8.78548000 | 2.47346310 |
| Al | 12.11016000 | 1.39726000 | 8.78548000 | 2.47109140 |
| O | 8.01661000 | 11.26550000 | 9.68989000 | -1.64732730 |
| O | 10.43673000 | 7.07371000 | 9.68989000 | -1.64172190 |
| O | -4.06971000 | 9.68514000 | 9.68989000 | -1.66057590 |
| O | -1.66389000 | 11.26550000 | 9.68989000 | -1.64386500 |
| O | 12.85686000 | 2.88194000 | 9.68989000 | -1.64606360 |
| O | -1.64959000 | 5.49336000 | 9.68989000 | -1.66087630 |
| O | -1.49818000 | 8.39182000 | 9.68989000 | -1.66835060 |
| O | 0.75624000 | 7.07372000 | 9.68989000 | -1.64352320 |
| O | 0.77054000 | 9.68514000 | 9.68989000 | -1.64657080 |
| O | 0.77054000 | 1.30159000 | 9.68989000 | -1.65625800 |
| O | 0.92195000 | 4.20004000 | 9.68989000 | -1.66853790 |
| O | 3.17636000 | 11.26550000 | 9.68989000 | -1.63973980 |
| O | 3.17636000 | 2.88194000 | 9.68989000 | -1.64287440 |
| O | 3.19067000 | 5.49336000 | 9.68989000 | -1.64775960 |
| O | 3.34208000 | 0.00826000 | 9.68989000 | -1.66835630 |
| O | 3.34207000 | 8.39182000 | 9.68989000 | -1.65614370 |
| O | 5.59648000 | 7.07372000 | 9.68989000 | -1.63055990 |
| O | 5.61079000 | 1.30159000 | 9.68989000 | -1.65943280 |
| O | 5.61079000 | 9.68514000 | 9.68989000 | -1.64768510 |
| O | 5.76220000 | 4.20004000 | 9.68989000 | -1.65553120 |
| O | 8.01661000 | 2.88194000 | 9.68989000 | -1.64506880 |

| | | | |
|---|---|---|---|
| O | 8.03091000 | 5.49336000 | 9.68989000 | -1.65950380 |
| O | 8.18232000 | 0.00826000 | 9.68989000 | -1.66448820 |
| O | 8.18232000 | 8.39182000 | 9.68989000 | -1.65337230 |
| O | 10.45104000 | 1.30158000 | 9.68989000 | -1.65745890 |
| O | 10.60245000 | 4.20004000 | 9.68989000 | -1.66754790 |
| O | 13.02257000 | 0.00826000 | 9.68989000 | -1.65697660 |
| Al | -4.83072000 | 11.17808000 | 10.71830000 | 2.48139360 |
| Al | -2.41059000 | 6.98630000 | 10.71830000 | 2.47494800 |
| Al | 0.00953000 | 11.17808000 | 10.71830000 | 2.47986690 |
| Al | 0.00953000 | 2.79452000 | 10.71830000 | 2.47800300 |
| Al | 2.42966000 | 6.98630000 | 10.71830000 | 2.47716630 |
| Al | 4.84978000 | 2.79452000 | 10.71830000 | 2.47540840 |
| Al | 4.84978000 | 11.17808000 | 10.71830000 | 2.47813800 |
| Al | 7.26991000 | 6.98630000 | 10.71830000 | 2.47658330 |
| Al | 9.69003000 | 2.79452000 | 10.71830000 | 2.47520150 |
| Al | -4.83072000 | 8.38356000 | 10.98601000 | 2.46933270 |
| Al | -2.41059000 | 4.19178000 | 10.98601000 | 2.46863910 |
| Al | 0.00953000 | 8.38356000 | 10.98601000 | 2.47027500 |
| Al | 0.00954000 | 0.00000000 | 10.98601000 | 2.47288930 |
| Al | 2.42966000 | 4.19178000 | 10.98601000 | 2.47257150 |
| Al | 4.84978000 | 0.00000000 | 10.98601000 | 2.47233060 |
| Al | 4.84978000 | 8.38356000 | 10.98601000 | 2.47541090 |
| Al | 7.26991000 | 4.19178000 | 10.98601000 | 2.47226300 |
| Al | 9.69003000 | 0.00000000 | 10.98601000 | 2.46653110 |
| O | -5.65495000 | 12.47796000 | 11.85903000 | -1.62099520 |
| O | -5.56065000 | 9.80906000 | 11.84541000 | -1.60898150 |
| O | -3.29894000 | 11.24183000 | 11.85212000 | -1.62450360 |
| O | -3.23238000 | 8.26970000 | 11.85566000 | -1.61861170 |
| O | -3.09856000 | 5.62574000 | 11.86298000 | -1.62676500 |
| O | -0.87662000 | 7.04641000 | 11.85179000 | -1.62114650 |
| O | -0.81103000 | 12.47376000 | 11.85115000 | -1.62253180 |
| O | -0.81058000 | 4.08206000 | 11.84817000 | -1.61764520 |
| O | -0.70854000 | 9.81592000 | 11.86055000 | -1.62312240 |
| O | -0.70100000 | 1.43683000 | 11.85214000 | -1.62787470 |
| O | 1.54068000 | 2.85837000 | 11.85618000 | -1.62802390 |
| O | 1.54905000 | 11.24317000 | 11.84045000 | -1.61060560 |
| O | 1.63719000 | 8.29467000 | 11.85047000 | -1.60873550 |
| O | 1.71649000 | 5.63309000 | 11.87842000 | -1.63280880 |
| O | 3.95513000 | 7.02647000 | 11.94736000 | -1.57234020 |
| O | 4.03602000 | 4.09678000 | 11.82222000 | -1.61454010 |
| O | 4.04983000 | 12.48788000 | 11.87597000 | -1.62513630 |
| O | 4.14156000 | 1.44123000 | 11.85432000 | -1.62826880 |
| O | 4.12028000 | 9.85177000 | 11.90470000 | -1.57628260 |

| | | | |
|---|---|---|---|
| O | 6.36810000 | 11.20855000 | 11.87785000 | -1.58759120 |
| O | 6.39537000 | 2.85578000 | 11.86770000 | -1.62480890 |
| O | 6.47284000 | 8.29363000 | 11.81490000 | -1.60827210 |
| O | 6.54729000 | 5.61246000 | 11.95720000 | -1.56335110 |
| O | 8.79408000 | 7.01602000 | 11.87877000 | -1.63334520 |
| O | 8.86650000 | 4.09683000 | 11.85396000 | -1.63395140 |
| O | 8.99007000 | 1.44495000 | 11.84976000 | -1.62437820 |
| O | 11.21058000 | 2.87219000 | 11.85296000 | -1.61773410 |
| Al | -2.41865000 | 9.77714000 | 11.89535000 | 2.41554850 |
| Al | 0.00107000 | 5.58516000 | 11.86037000 | 2.41450040 |
| Al | 2.43189000 | 1.39567000 | 11.86733000 | 2.41524080 |
| Al | 2.37222000 | 9.80881000 | 12.37995000 | 2.37330790 |
| Al | 4.81243000 | 5.52021000 | 12.42531000 | 2.38202450 |
| Al | 7.30142000 | 9.74477000 | 12.40438000 | 2.37150090 |
| Al | 7.27985000 | 1.38700000 | 11.85454000 | 2.41505360 |
| Al | 9.72674000 | 5.57357000 | 11.93030000 | 2.40351940 |
| Al | 12.11791000 | 1.43284000 | 11.81646000 | 2.41533080 |
| Pt | 5.72470000 | 7.72984000 | 14.80884000 | -0.49783400 |
| Pt | 6.99033000 | 9.92422000 | 14.83410000 | -0.82554250 |
| Pt | 4.80158000 | 5.27174000 | 14.89087000 | -0.63720580 |
| Pt | 2.78002000 | 9.72901000 | 14.78469000 | -0.94011960 |
| Sn | 3.07569000 | 7.25896000 | 14.21475000 | 0.93115490 |
| Sn | 4.87415000 | 11.10360000 | 14.00482000 | 1.02586230 |
| Sn | 7.37101000 | 5.92380000 | 14.10771000 | 0.90419400 |
| C | 4.90429000 | 3.10667000 | 15.19063000 | -0.20320750 |
| C | 3.56032000 | 3.51129000 | 15.22921000 | -0.15785460 |
| H | 5.44611000 | 2.90811000 | 16.12006000 | 0.06020850 |
| H | 5.30816000 | 2.64671000 | 14.28620000 | 0.14008830 |
| H | 3.04851000 | 3.64101000 | 16.18846000 | 0.06559610 |
| H | 2.90900000 | 3.37517000 | 14.36288000 | 0.10996860 |

283

c2h4-pt4sn3-al2o3-C

| | | | |
|---|---|---|---|
| Al | -4.83072000 | 11.17808000 | 0.00000000 | 2.16613310 |
| Al | -2.41059000 | 6.98630000 | 0.00000000 | 2.16613650 |
| Al | 0.00953000 | 11.17808000 | 0.00000000 | 2.16616300 |
| Al | 0.00953000 | 2.79452000 | 0.00000000 | 2.16635110 |
| Al | 2.42966000 | 6.98630000 | 0.00000000 | 2.16639280 |
| Al | 4.84978000 | 2.79452000 | 0.00000000 | 2.16644420 |
| Al | 4.84978000 | 11.17808000 | 0.00000000 | 2.16611660 |
| Al | 7.26991000 | 6.98630000 | 0.00000000 | 2.16620960 |
| Al | 9.69003000 | 2.79452000 | 0.00000000 | 2.16620300 |
| O | -3.89334000 | 9.78082000 | 0.85241000 | -1.55326970 |
| O | -1.66922000 | 8.49672000 | 0.85241000 | -1.56164780 |

| | | | |
|---|---|---|---|
| O | -1.66922000 | 11.06492000 | 0.85241000 | -1.55919560 |
| O | -1.47322000 | 5.58904000 | 0.85241000 | -1.55326140 |
| O | 0.75091000 | 4.30494000 | 0.85241000 | -1.56158900 |
| O | 0.75091000 | 6.87314000 | 0.85241000 | -1.55919660 |
| O | 0.94691000 | 9.78082000 | 0.85241000 | -1.55331220 |
| O | 0.94691000 | 1.39726000 | 0.85241000 | -1.55325180 |
| O | 3.17104000 | 0.11316000 | 0.85241000 | -1.56160870 |
| O | 3.17103000 | 8.49672000 | 0.85241000 | -1.56176320 |
| O | 3.17103000 | 2.68136000 | 0.85241000 | -1.55915420 |
| O | 3.17103000 | 11.06492000 | 0.85241000 | -1.55935100 |
| O | 3.36703000 | 5.58904000 | 0.85241000 | -1.55315520 |
| O | 5.59116000 | 4.30494000 | 0.85241000 | -1.56164190 |
| O | 5.59116000 | 6.87314000 | 0.85241000 | -1.55917940 |
| O | 5.78716000 | 1.39726000 | 0.85241000 | -1.55333270 |
| O | 5.78716000 | 9.78082000 | 0.85241000 | -1.55328520 |
| O | 8.01128000 | 8.49672000 | 0.85241000 | -1.56162800 |
| O | 8.01128000 | 0.11316000 | 0.85241000 | -1.56187010 |
| O | 8.01128000 | 2.68136000 | 0.85241000 | -1.55929860 |
| O | 8.01128000 | 11.06492000 | 0.85241000 | -1.55916130 |
| O | 8.20728000 | 5.58904000 | 0.85241000 | -1.55330670 |
| O | 10.43141000 | 4.30494000 | 0.85241000 | -1.56163690 |
| O | 10.43141000 | 6.87314000 | 0.85241000 | -1.55916280 |
| O | 10.62741000 | 1.39726000 | 0.85241000 | -1.55338830 |
| O | 12.85153000 | 0.11316000 | 0.85241000 | -1.56169000 |
| O | 12.85153000 | 2.68136000 | 0.85241000 | -1.55923280 |
| Al | -4.83072000 | 8.38356000 | 1.70482000 | 2.46337910 |
| Al | -2.41059000 | 4.19178000 | 1.70482000 | 2.46337890 |
| Al | 0.00953000 | 8.38356000 | 1.70482000 | 2.46342850 |
| Al | -7.25084000 | 12.57534000 | 1.70482000 | 2.46337520 |
| Al | 2.42966000 | 4.19178000 | 1.70482000 | 2.46332140 |
| Al | -2.41060000 | 12.57534000 | 1.70482000 | 2.46342650 |
| Al | 4.84978000 | 8.38356000 | 1.70482000 | 2.46342230 |
| Al | 7.26991000 | 4.19178000 | 1.70482000 | 2.46343780 |
| Al | 2.42965000 | 12.57534000 | 1.70482000 | 2.46350630 |
| Al | -2.41059000 | 9.78082000 | 2.20274000 | 2.48266120 |
| Al | 0.00953000 | 5.58904000 | 2.20274000 | 2.48272740 |
| Al | 2.42966000 | 1.39726000 | 2.20274000 | 2.48271530 |
| Al | 2.42966000 | 9.78082000 | 2.20274000 | 2.48260940 |
| Al | 4.84978000 | 5.58904000 | 2.20274000 | 2.48264750 |
| Al | 7.26991000 | 9.78082000 | 2.20274000 | 2.48265500 |
| Al | 7.26991000 | 1.39726000 | 2.20274000 | 2.48265710 |
| Al | 9.69003000 | 5.58904000 | 2.20274000 | 2.48272580 |
| Al | 12.11016000 | 1.39726000 | 2.20274000 | 2.48268300 |

| | | | |
|---|---|---|---|
| O | -5.57210000 | 9.66766000 | 3.05515000 | -1.64243450 |
| O | -3.34796000 | 8.38356000 | 3.05515000 | -1.62446400 |
| O | -3.15197000 | 5.47588000 | 3.05515000 | -1.64260720 |
| O | -3.15197000 | 11.29124000 | 3.05515000 | -1.65040590 |
| O | -0.92784000 | 4.19178000 | 3.05515000 | -1.62448160 |
| O | -0.73185000 | 1.28410000 | 3.05515000 | -1.64254120 |
| O | -0.73185000 | 9.66766000 | 3.05515000 | -1.64250120 |
| O | -0.73184000 | 7.09945000 | 3.05515000 | -1.65042690 |
| O | 1.49229000 | 8.38356000 | 3.05515000 | -1.62452810 |
| O | 1.49229000 | 0.00001000 | 3.05515000 | -1.62455230 |
| O | 1.68828000 | 5.47588000 | 3.05515000 | -1.64242860 |
| O | 1.68828000 | 2.90768000 | 3.05515000 | -1.65038500 |
| O | 1.68828000 | 11.29123000 | 3.05515000 | -1.65058660 |
| O | 3.91241000 | 4.19178000 | 3.05515000 | -1.62437720 |
| O | 4.10840000 | 9.66766000 | 3.05515000 | -1.64248760 |
| O | 4.10840000 | 1.28410000 | 3.05515000 | -1.64232460 |
| O | 4.10841000 | 7.09946000 | 3.05515000 | -1.65044930 |
| O | 6.33254000 | 0.00001000 | 3.05515000 | -1.62463570 |
| O | 6.33254000 | 8.38356000 | 3.05515000 | -1.62447410 |
| O | 6.52853000 | 11.29123000 | 3.05515000 | -1.65042170 |
| O | 6.52853000 | 5.47588000 | 3.05515000 | -1.64239700 |
| O | 6.52853000 | 2.90768000 | 3.05515000 | -1.65045360 |
| O | 8.75266000 | 4.19178000 | 3.05515000 | -1.62472730 |
| O | 8.94866000 | 7.09945000 | 3.05515000 | -1.65044730 |
| O | 8.94865000 | 1.28410000 | 3.05515000 | -1.64263350 |
| O | 3.91239000 | 12.57534000 | 3.05515000 | -1.62460250 |
| O | 11.36878000 | 2.90767000 | 3.05515000 | -1.65054650 |
| Al | -4.83072000 | 11.17808000 | 3.90756000 | 2.47801860 |
| Al | -2.41059000 | 6.98630000 | 3.90756000 | 2.47803540 |
| Al | 0.00953000 | 11.17808000 | 3.90756000 | 2.47807070 |
| Al | 0.00953000 | 2.79452000 | 3.90756000 | 2.47797720 |
| Al | 2.42966000 | 6.98630000 | 3.90756000 | 2.47806910 |
| Al | 4.84978000 | 2.79452000 | 3.90756000 | 2.47805670 |
| Al | 4.84978000 | 11.17808000 | 3.90756000 | 2.47790520 |
| Al | 7.26991000 | 6.98630000 | 3.90756000 | 2.47807420 |
| Al | 9.69003000 | 2.79452000 | 3.90756000 | 2.47792010 |
| Al | -4.83072000 | 8.38356000 | 4.40548000 | 2.47218140 |
| Al | -2.41059000 | 4.19178000 | 4.40548000 | 2.47223630 |
| Al | 0.00953000 | 8.38356000 | 4.40548000 | 2.47204900 |
| Al | -7.25084000 | 12.57534000 | 4.40548000 | 2.47235340 |
| Al | 2.42966000 | 4.19178000 | 4.40548000 | 2.47239510 |
| Al | -2.41060000 | 12.57534000 | 4.40548000 | 2.47245890 |
| Al | 4.84978000 | 8.38356000 | 4.40548000 | 2.47215970 |
| Al | 7.26991000 | 4.19178000 | 4.40548000 | 2.47246030 |
| Al | 2.42965000 | 12.57534000 | 4.40548000 | 2.47231030 |
| O | -6.31347000 | 11.17808000 | 5.25789000 | -1.64364280 |
| O | -4.08934000 | 12.46219000 | 5.25789000 | -1.65404280 |
| O | -4.08934000 | 9.89398000 | 5.25789000 | -1.64902290 |
| O | -3.89335000 | 6.98630000 | 5.25789000 | -1.64348080 |
| O | -1.66922000 | 8.27040000 | 5.25789000 | -1.65400930 |
| O | -1.66921000 | 5.70219000 | 5.25789000 | -1.64908400 |
| O | -1.47322000 | 11.17808000 | 5.25789000 | -1.64339510 |
| O | -1.47323000 | 2.79452000 | 5.25789000 | -1.64358350 |
| O | 0.75091000 | 4.07862000 | 5.25789000 | -1.65399560 |
| O | 0.75091000 | 12.46219000 | 5.25789000 | -1.65416590 |
| O | 0.75091000 | 9.89397000 | 5.25789000 | -1.64896370 |
| O | 0.75091000 | 1.51042000 | 5.25789000 | -1.64915490 |
| O | 0.94690000 | 6.98630000 | 5.25789000 | -1.64325920 |
| O | 3.17103000 | 8.27040000 | 5.25789000 | -1.65415800 |
| O | 3.17104000 | 5.70220000 | 5.25789000 | -1.64909690 |
| O | 3.36703000 | 2.79452000 | 5.25789000 | -1.64357860 |
| O | 3.36702000 | 11.17808000 | 5.25789000 | -1.64365840 |
| O | 5.59116000 | 12.46219000 | 5.25789000 | -1.65383700 |
| O | 5.59116000 | 4.07863000 | 5.25789000 | -1.65421540 |
| O | 5.59116000 | 9.89397000 | 5.25789000 | -1.64918150 |
| O | 5.59116000 | 1.51042000 | 5.25789000 | -1.64938860 |
| O | 5.78715000 | 6.98630000 | 5.25789000 | -1.64343190 |
| O | 8.01128000 | 8.27040000 | 5.25789000 | -1.65384390 |
| O | 8.01129000 | 5.70219000 | 5.25789000 | -1.64936440 |
| O | 8.20727000 | 2.79452000 | 5.25789000 | -1.64339170 |
| O | 10.43141000 | 4.07862000 | 5.25789000 | -1.65391490 |
| O | 10.43141000 | 1.51041000 | 5.25789000 | -1.64922390 |
| Al | -2.41059000 | 9.78082000 | 6.11030000 | 2.47634410 |
| Al | 0.00953000 | 5.58904000 | 6.11030000 | 2.47609570 |
| Al | 2.42966000 | 1.39726000 | 6.11030000 | 2.47617080 |
| Al | 2.42966000 | 9.78082000 | 6.11030000 | 2.47582930 |
| Al | 4.84978000 | 5.58904000 | 6.11030000 | 2.47621360 |
| Al | 7.26991000 | 9.78082000 | 6.11030000 | 2.47604990 |
| Al | 7.26991000 | 1.39726000 | 6.11030000 | 2.47597780 |
| Al | 9.69003000 | 5.58904000 | 6.11030000 | 2.47603100 |
| Al | 12.11016000 | 1.39726000 | 6.11030000 | 2.47588120 |
| Al | -4.83072000 | 11.17808000 | 6.60910000 | 2.47695260 |
| Al | -2.41059000 | 6.98630000 | 6.60910000 | 2.47673310 |
| Al | 0.00953000 | 11.17808000 | 6.60910000 | 2.47682630 |
| Al | 0.00953000 | 2.79452000 | 6.60910000 | 2.47702520 |
| Al | 2.42966000 | 6.98630000 | 6.60910000 | 2.47685810 |

| | | | |
|---|---|---|---|
| Al | 4.84978000 | 2.79452000 | 6.60910000 | 2.47692590 |
| Al | 4.84978000 | 11.17808000 | 6.60910000 | 2.47664300 |
| Al | 7.26991000 | 6.98630000 | 6.60910000 | 2.47672730 |
| Al | 9.69003000 | 2.79452000 | 6.60910000 | 2.47709230 |
| O | -3.15680000 | 8.49706000 | 7.46570000 | -1.64908040 |
| O | -3.14925000 | 11.06893000 | 7.46570000 | -1.65000730 |
| O | -0.92572000 | 9.77647000 | 7.46570000 | -1.65470180 |
| O | -0.73667000 | 4.30528000 | 7.46570000 | -1.64896800 |
| O | -0.72913000 | 6.87715000 | 7.46570000 | -1.64929670 |
| O | 1.49440000 | 5.58469000 | 7.46570000 | -1.65387350 |
| O | 1.68345000 | 8.49706000 | 7.46570000 | -1.64877270 |
| O | 1.68345000 | 0.11351000 | 7.46570000 | -1.64867250 |
| O | 1.69100000 | 2.68537000 | 7.46570000 | -1.64933070 |
| O | 1.69100000 | 11.06893000 | 7.46570000 | -1.65045870 |
| O | 3.91453000 | 1.39291000 | 7.46570000 | -1.65369310 |
| O | 3.91452000 | 9.77646000 | 7.46570000 | -1.65405750 |
| O | 4.10358000 | 4.30528000 | 7.46570000 | -1.64887470 |
| O | 4.11112000 | 6.87715000 | 7.46570000 | -1.64966690 |
| O | 6.33465000 | 5.58468000 | 7.46570000 | -1.65318300 |
| O | 6.52370000 | 0.11351000 | 7.46570000 | -1.64936620 |
| O | 6.52370000 | 8.49706000 | 7.46570000 | -1.64913510 |
| O | 6.53125000 | 2.68537000 | 7.46570000 | -1.64988020 |
| O | 6.53125000 | 11.06893000 | 7.46570000 | -1.65030180 |
| O | 8.75477000 | 9.77646000 | 7.46570000 | -1.65387720 |
| O | 8.75477000 | 1.39291000 | 7.46570000 | -1.65488160 |
| O | 8.94383000 | 4.30528000 | 7.46570000 | -1.64960240 |
| O | 8.95137000 | 6.87715000 | 7.46570000 | -1.64953440 |
| O | 11.17490000 | 5.58468000 | 7.46570000 | -1.65393670 |
| O | 11.36395000 | 0.11351000 | 7.46570000 | -1.64985130 |
| O | 11.37149000 | 2.68537000 | 7.46570000 | -1.65021220 |
| O | 13.59502000 | 1.39291000 | 7.46570000 | -1.65390410 |
| Al | -4.83072000 | 8.38356000 | 8.32833000 | 2.47180440 |
| Al | -2.41059000 | 4.19178000 | 8.32833000 | 2.47089990 |
| Al | 0.00953000 | 8.38356000 | 8.32833000 | 2.47127500 |
| Al | -7.25084000 | 12.57534000 | 8.32833000 | 2.47084230 |
| Al | 2.42966000 | 4.19178000 | 8.32833000 | 2.47082530 |
| Al | -2.41060000 | 12.57534000 | 8.32833000 | 2.47177370 |
| Al | 4.84978000 | 8.38356000 | 8.32833000 | 2.47094520 |
| Al | 7.26991000 | 4.19178000 | 8.32833000 | 2.47229530 |
| Al | 2.42965000 | 12.57534000 | 8.32833000 | 2.47262720 |
| Al | -2.41059000 | 9.78082000 | 8.78548000 | 2.47247300 |
| Al | 0.00953000 | 5.58904000 | 8.78548000 | 2.47663270 |
| Al | 2.42966000 | 1.39726000 | 8.78548000 | 2.47672220 |

| | | | |
|---|---|---|---|
| Al | 2.42966000 | 9.78082000 | 8.78548000 | 2.47118220 |
| Al | 4.84978000 | 5.58904000 | 8.78548000 | 2.47660100 |
| Al | 7.26991000 | 9.78082000 | 8.78548000 | 2.47289020 |
| Al | 7.26991000 | 1.39726000 | 8.78548000 | 2.47168950 |
| Al | 9.69003000 | 5.58904000 | 8.78548000 | 2.47221590 |
| Al | 12.11016000 | 1.39726000 | 8.78548000 | 2.47110180 |
| O | 8.01661000 | 11.26550000 | 9.68989000 | -1.64483230 |
| O | 10.43673000 | 7.07371000 | 9.68989000 | -1.64278920 |
| O | -4.06971000 | 9.68514000 | 9.68989000 | -1.65907070 |
| O | -1.66389000 | 11.26550000 | 9.68989000 | -1.64566370 |
| O | 12.85686000 | 2.88194000 | 9.68989000 | -1.64447220 |
| O | -1.64959000 | 5.49336000 | 9.68989000 | -1.64491210 |
| O | -1.49818000 | 8.39182000 | 9.68989000 | -1.66472560 |
| O | 0.75624000 | 7.07372000 | 9.68989000 | -1.63756980 |
| O | 0.77054000 | 9.68514000 | 9.68989000 | -1.65813500 |
| O | 0.77054000 | 1.30159000 | 9.68989000 | -1.64802440 |
| O | 0.92195000 | 4.20004000 | 9.68989000 | -1.65678060 |
| O | 3.17636000 | 11.26550000 | 9.68989000 | -1.64520040 |
| O | 3.17636000 | 2.88194000 | 9.68989000 | -1.63269410 |
| O | 3.19067000 | 5.49336000 | 9.68989000 | -1.64598810 |
| O | 3.34208000 | 0.00826000 | 9.68989000 | -1.65287890 |
| O | 3.34207000 | 8.39182000 | 9.68989000 | -1.65784260 |
| O | 5.59648000 | 7.07372000 | 9.68989000 | -1.64639510 |
| O | 5.61079000 | 1.30159000 | 9.68989000 | -1.66112660 |
| O | 5.61079000 | 9.68514000 | 9.68989000 | -1.65598210 |
| O | 5.76220000 | 4.20004000 | 9.68989000 | -1.65695230 |
| O | 8.01661000 | 2.88194000 | 9.68989000 | -1.64414860 |
| O | 8.03091000 | 5.49336000 | 9.68989000 | -1.65971710 |
| O | 8.18232000 | 0.00826000 | 9.68989000 | -1.67090910 |
| O | 8.18232000 | 8.39182000 | 9.68989000 | -1.66707680 |
| O | 10.45104000 | 1.30158000 | 9.68989000 | -1.66211720 |
| O | 10.60245000 | 4.20004000 | 9.68989000 | -1.66857190 |
| O | 13.02257000 | 0.00826000 | 9.68989000 | -1.66613370 |
| Al | -4.83072000 | 11.17808000 | 10.71830000 | 2.47675390 |
| Al | -2.41059000 | 6.98630000 | 10.71830000 | 2.48066230 |
| Al | 0.00953000 | 11.17808000 | 10.71830000 | 2.47588240 |
| Al | 0.00953000 | 2.79452000 | 10.71830000 | 2.47789410 |
| Al | 2.42966000 | 6.98630000 | 10.71830000 | 2.47742340 |
| Al | 4.84978000 | 2.79452000 | 10.71830000 | 2.47783490 |
| Al | 4.84978000 | 11.17808000 | 10.71830000 | 2.47778010 |
| Al | 7.26991000 | 6.98630000 | 10.71830000 | 2.48149050 |
| Al | 9.69003000 | 2.79452000 | 10.71830000 | 2.47493000 |
| Al | -4.83072000 | 8.38356000 | 10.98601000 | 2.47188580 |

| | | | |
|---|---|---|---|
| Al | -2.41059000 | 4.19178000 | 10.98601000 | 2.47076320 |
| Al | 0.00953000 | 8.38356000 | 10.98601000 | 2.46999980 |
| Al | -7.25084000 | 12.57534000 | 10.98601000 | 2.47025610 |
| Al | 2.42966000 | 4.19178000 | 10.98601000 | 2.46980040 |
| Al | -2.41060000 | 12.57534000 | 10.98601000 | 2.47126810 |
| Al | 4.84978000 | 8.38356000 | 10.98601000 | 2.47273370 |
| Al | 7.26991000 | 4.19178000 | 10.98601000 | 2.47121930 |
| Al | 2.42965000 | 12.57534000 | 10.98601000 | 2.46717350 |
| O | -5.64997000 | 12.49618000 | 11.84525000 | -1.62165260 |
| O | -5.54278000 | 9.82157000 | 11.85241000 | -1.62245760 |
| O | -3.29072000 | 11.24326000 | 11.86374000 | -1.61985330 |
| O | -3.22855000 | 8.28053000 | 11.85287000 | -1.61991460 |
| O | -3.12322000 | 5.61951000 | 11.86456000 | -1.62619360 |
| O | -0.86485000 | 7.04157000 | 11.84143000 | -1.61568200 |
| O | -0.81198000 | 12.47542000 | 11.85398000 | -1.62586180 |
| O | -0.77974000 | 4.09522000 | 11.88019000 | -1.59389570 |
| O | -0.69462000 | 9.82374000 | 11.85324000 | -1.62714460 |
| O | -0.72156000 | 1.45151000 | 11.87802000 | -1.62606330 |
| O | 1.51327000 | 2.83796000 | 11.91414000 | -1.57334280 |
| O | 1.53210000 | 11.25270000 | 11.85131000 | -1.61975570 |
| O | 1.62717000 | 8.28730000 | 11.87850000 | -1.62641790 |
| O | 1.71535000 | 5.65577000 | 11.92848000 | -1.57097990 |
| O | 3.95214000 | 7.02053000 | 11.87500000 | -1.60353740 |
| O | 4.06300000 | 4.08154000 | 11.92083000 | -1.56540550 |
| O | 4.02234000 | 12.45723000 | 11.85250000 | -1.61905570 |
| O | 4.11543000 | 1.46188000 | 11.88212000 | -1.59591890 |
| O | 4.13622000 | 9.81753000 | 11.85367000 | -1.62817570 |
| O | 6.37377000 | 11.24637000 | 11.85287000 | -1.62412760 |
| O | 6.37883000 | 2.83608000 | 11.87874000 | -1.62946870 |
| O | 6.44419000 | 8.28739000 | 11.86419000 | -1.62327830 |
| O | 6.54567000 | 5.61965000 | 11.84111000 | -1.61293230 |
| O | 8.80108000 | 7.04760000 | 11.85305000 | -1.62444370 |
| O | 8.86930000 | 4.08050000 | 11.85302000 | -1.61871820 |
| O | 8.99511000 | 1.43914000 | 11.85169000 | -1.62583790 |
| O | 11.22523000 | 2.85812000 | 11.85454000 | -1.62125620 |
| Al | -2.40742000 | 9.77801000 | 11.89565000 | 2.41693110 |
| Al | -0.04119000 | 5.62828000 | 12.42030000 | 2.37111490 |
| Al | 2.42179000 | 1.33440000 | 12.46740000 | 2.36269200 |
| Al | 2.43354000 | 9.80854000 | 11.83752000 | 2.41787280 |
| Al | 4.90894000 | 5.62339000 | 12.42514000 | 2.36163680 |
| Al | 7.27081000 | 9.78657000 | 11.89595000 | 2.41620120 |
| Al | 7.29506000 | 1.37608000 | 11.83358000 | 2.41253630 |
| Al | 9.68578000 | 5.58726000 | 11.89335000 | 2.41513160 |

| | | | |
|---|---|---|---|
| Al | 12.07842000 | 1.38473000 | 11.83419000 | 2.41369270 |
| Pt | 2.40026000 | 4.39388000 | 16.25880000 | -0.07834830 |
| Pt | 0.32559000 | 5.37583000 | 14.83665000 | -0.96371690 |
| Pt | 4.48150000 | 5.38270000 | 14.83547000 | -0.96367390 |
| Pt | 2.43242000 | 2.01410000 | 14.84134000 | -1.00119290 |
| Sn | 4.70002000 | 2.91204000 | 13.95541000 | 1.03345580 |
| Sn | 0.14668000 | 2.90686000 | 13.96427000 | 1.02945340 |
| Sn | 2.41116000 | 6.75825000 | 14.02632000 | 1.02628690 |
| C | 3.14108000 | 4.32108000 | 18.21025000 | -0.17120030 |
| C | 1.71032000 | 4.39180000 | 18.22925000 | -0.17059150 |
| H | 3.73443000 | 5.20620000 | 18.46179000 | 0.09879460 |
| H | 3.64494000 | 3.37036000 | 18.41244000 | 0.06731880 |
| H | 1.21389000 | 5.33055000 | 18.49567000 | 0.06980580 |
| H | 1.12060000 | 3.49511000 | 18.44670000 | 0.08004510 |

283

c2h4-pt4sn3-al2o3-B

| | | | |
|---|---|---|---|
| Al | -4.83072000 | 11.17808000 | 0.00000000 | 2.17034410 |
| Al | -2.41059000 | 6.98630000 | 0.00000000 | 2.17034420 |
| Al | 0.00953000 | 11.17808000 | 0.00000000 | 2.17027320 |
| Al | 0.00953000 | 2.79452000 | 0.00000000 | 2.17041670 |
| Al | 2.42966000 | 6.98630000 | 0.00000000 | 2.17002110 |
| Al | 4.84978000 | 2.79452000 | 0.00000000 | 2.17016310 |
| Al | 4.84978000 | 11.17808000 | 0.00000000 | 2.17032330 |
| Al | 7.26991000 | 6.98630000 | 0.00000000 | 2.17017100 |
| Al | 9.69003000 | 2.79452000 | 0.00000000 | 2.17044090 |
| O | -3.89334000 | 9.78082000 | 0.85241000 | -1.55179730 |
| O | -1.66922000 | 8.49672000 | 0.85241000 | -1.55996640 |
| O | -1.66922000 | 11.06492000 | 0.85241000 | -1.55784220 |
| O | -1.47322000 | 5.58904000 | 0.85241000 | -1.55182930 |
| O | 0.75091000 | 4.30494000 | 0.85241000 | -1.55997140 |
| O | 0.75091000 | 6.87314000 | 0.85241000 | -1.55791450 |
| O | 0.94691000 | 9.78082000 | 0.85241000 | -1.55152770 |
| O | 0.94691000 | 1.39726000 | 0.85241000 | -1.55180380 |
| O | 3.17104000 | 0.11316000 | 0.85241000 | -1.56002610 |
| O | 3.17103000 | 8.49672000 | 0.85241000 | -1.55989390 |
| O | 3.17103000 | 2.68136000 | 0.85241000 | -1.55786640 |
| O | 3.17103000 | 11.06492000 | 0.85241000 | -1.55784650 |
| O | 3.36703000 | 5.58904000 | 0.85241000 | -1.55185950 |
| O | 5.59116000 | 4.30494000 | 0.85241000 | -1.55993940 |
| O | 5.59116000 | 6.87314000 | 0.85241000 | -1.55777010 |
| O | 5.78716000 | 1.39726000 | 0.85241000 | -1.55181210 |
| O | 5.78716000 | 9.78082000 | 0.85241000 | -1.55169220 |
| O | 8.01128000 | 8.49672000 | 0.85241000 | -1.55994500 |

| | | | |
|---|---|---|---|
| O | 8.01128000 | 0.11316000 | 0.85241000 | -1.56015430 |
| O | 8.01128000 | 2.68136000 | 0.85241000 | -1.55780370 |
| O | 8.01128000 | 11.06492000 | 0.85241000 | -1.55787660 |
| O | 8.20728000 | 5.58904000 | 0.85241000 | -1.55180780 |
| O | 10.43141000 | 4.30494000 | 0.85241000 | -1.56009580 |
| O | 10.43141000 | 6.87314000 | 0.85241000 | -1.55778380 |
| O | 10.62741000 | 1.39726000 | 0.85241000 | -1.55186710 |
| O | 12.85153000 | 0.11316000 | 0.85241000 | -1.56009790 |
| O | 12.85153000 | 2.68136000 | 0.85241000 | -1.55796130 |
| Al | -4.83072000 | 8.38356000 | 1.70482000 | 2.46328010 |
| Al | -2.41059000 | 4.19178000 | 1.70482000 | 2.46329240 |
| Al | 0.00953000 | 8.38356000 | 1.70482000 | 2.46333390 |
| Al | 0.00954000 | 0.00000000 | 1.70482000 | 2.46328090 |
| Al | 2.42966000 | 4.19178000 | 1.70482000 | 2.46345620 |
| Al | 4.84978000 | 0.00000000 | 1.70482000 | 2.46327760 |
| Al | 4.84978000 | 8.38356000 | 1.70482000 | 2.46338160 |
| Al | 7.26991000 | 4.19178000 | 1.70482000 | 2.46331960 |
| Al | 9.69003000 | 0.00000000 | 1.70482000 | 2.46330940 |
| Al | -2.41059000 | 9.78082000 | 2.20274000 | 2.48280620 |
| Al | 0.00953000 | 5.58904000 | 2.20274000 | 2.48280960 |
| Al | 2.42966000 | 1.39726000 | 2.20274000 | 2.48270590 |
| Al | 2.42966000 | 9.78082000 | 2.20274000 | 2.48272350 |
| Al | 4.84978000 | 5.58904000 | 2.20274000 | 2.48283710 |
| Al | 7.26991000 | 9.78082000 | 2.20274000 | 2.48263160 |
| Al | 7.26991000 | 1.39726000 | 2.20274000 | 2.48273280 |
| Al | 9.69003000 | 5.58904000 | 2.20274000 | 2.48272750 |
| Al | 12.11016000 | 1.39726000 | 2.20274000 | 2.48264410 |
| O | -5.57210000 | 9.66766000 | 3.05515000 | -1.64146570 |
| O | -3.34796000 | 8.38356000 | 3.05515000 | -1.62386400 |
| O | -3.15197000 | 5.47588000 | 3.05515000 | -1.64182910 |
| O | -3.15197000 | 11.29124000 | 3.05515000 | -1.64960390 |
| O | -0.92784000 | 4.19178000 | 3.05515000 | -1.62404370 |
| O | -0.73185000 | 1.28410000 | 3.05515000 | -1.64202720 |
| O | -0.73185000 | 9.66766000 | 3.05515000 | -1.64145170 |
| O | -0.73184000 | 7.09945000 | 3.05515000 | -1.64961460 |
| O | 1.49229000 | 8.38356000 | 3.05515000 | -1.62389490 |
| O | 1.49229000 | 0.00001000 | 3.05515000 | -1.62391260 |
| O | 1.68828000 | 5.47588000 | 3.05515000 | -1.64163820 |
| O | 1.68828000 | 2.90768000 | 3.05515000 | -1.64945700 |
| O | 1.68828000 | 11.29123000 | 3.05515000 | -1.64937810 |
| O | 3.91241000 | 4.19178000 | 3.05515000 | -1.62377230 |
| O | 4.10840000 | 9.66766000 | 3.05515000 | -1.64132920 |
| O | 4.10840000 | 1.28410000 | 3.05515000 | -1.64149240 |
| O | 4.10841000 | 7.09946000 | 3.05515000 | -1.64941640 |
| O | 6.33254000 | 0.00001000 | 3.05515000 | -1.62389140 |
| O | 6.33254000 | 8.38356000 | 3.05515000 | -1.62385340 |
| O | 6.52853000 | 11.29123000 | 3.05515000 | -1.64949210 |
| O | 6.52853000 | 5.47588000 | 3.05515000 | -1.64152440 |
| O | 6.52853000 | 2.90768000 | 3.05515000 | -1.64950810 |
| O | 8.75266000 | 4.19178000 | 3.05515000 | -1.62397280 |
| O | 8.94866000 | 7.09945000 | 3.05515000 | -1.64960860 |
| O | 8.94865000 | 1.28410000 | 3.05515000 | -1.64163770 |
| O | 11.17277000 | 0.00000000 | 3.05515000 | -1.62404800 |
| O | 11.36878000 | 2.90767000 | 3.05515000 | -1.64987390 |
| Al | -4.83072000 | 11.17808000 | 3.90756000 | 2.47778610 |
| Al | -2.41059000 | 6.98630000 | 3.90756000 | 2.47766280 |
| Al | 0.00953000 | 11.17808000 | 3.90756000 | 2.47763240 |
| Al | 0.00953000 | 2.79452000 | 3.90756000 | 2.47785160 |
| Al | 2.42966000 | 6.98630000 | 3.90756000 | 2.47754170 |
| Al | 4.84978000 | 2.79452000 | 3.90756000 | 2.47769830 |
| Al | 4.84978000 | 11.17808000 | 3.90756000 | 2.47766520 |
| Al | 7.26991000 | 6.98630000 | 3.90756000 | 2.47770470 |
| Al | 9.69003000 | 2.79452000 | 3.90756000 | 2.47779240 |
| Al | -4.83072000 | 8.38356000 | 4.40548000 | 2.47229910 |
| Al | -2.41059000 | 4.19178000 | 4.40548000 | 2.47237020 |
| Al | 0.00953000 | 8.38356000 | 4.40548000 | 2.47228430 |
| Al | 0.00954000 | 0.00000000 | 4.40548000 | 2.47238660 |
| Al | 2.42966000 | 4.19178000 | 4.40548000 | 2.47249090 |
| Al | 4.84978000 | 0.00000000 | 4.40548000 | 2.47234850 |
| Al | 4.84978000 | 8.38356000 | 4.40548000 | 2.47249300 |
| Al | 7.26991000 | 4.19178000 | 4.40548000 | 2.47230910 |
| Al | 9.69003000 | 0.00000000 | 4.40548000 | 2.47219070 |
| O | -6.31347000 | 11.17808000 | 5.25789000 | -1.64349310 |
| O | -4.08934000 | 12.46219000 | 5.25789000 | -1.65407060 |
| O | -4.08934000 | 9.89398000 | 5.25789000 | -1.64918830 |
| O | -3.89335000 | 6.98630000 | 5.25789000 | -1.64347740 |
| O | -1.66222000 | 8.27040000 | 5.25789000 | -1.65378930 |
| O | -1.66921000 | 5.70219000 | 5.25789000 | -1.64912160 |
| O | -1.47322000 | 11.17808000 | 5.25789000 | -1.64367030 |
| O | -1.47323000 | 2.79452000 | 5.25789000 | -1.64357920 |
| O | 0.75091000 | 4.07862000 | 5.25789000 | -1.65369400 |
| O | 0.75091000 | 12.46219000 | 5.25789000 | -1.65417670 |
| O | 0.75091000 | 9.89397000 | 5.25789000 | -1.64901730 |
| O | 0.75091000 | 1.51042000 | 5.25789000 | -1.64905420 |
| O | 0.94690000 | 6.98630000 | 5.25789000 | -1.64326830 |
| O | 3.17103000 | 8.27040000 | 5.25789000 | -1.65409130 |

| | | | |
|---|---|---|---|
| O | 3.17104000 | 5.70220000 | 5.25789000 | -1.64911240 |
| O | 3.36703000 | 2.79452000 | 5.25789000 | -1.64309820 |
| O | 3.36702000 | 11.17808000 | 5.25789000 | -1.64344720 |
| O | 5.59116000 | 12.46219000 | 5.25789000 | -1.65417270 |
| O | 5.59116000 | 4.07863000 | 5.25789000 | -1.65414240 |
| O | 5.59116000 | 9.89397000 | 5.25789000 | -1.64911050 |
| O | 5.59116000 | 1.51042000 | 5.25789000 | -1.64923330 |
| O | 5.78715000 | 6.98630000 | 5.25789000 | -1.64280390 |
| O | 8.01128000 | 8.27040000 | 5.25789000 | -1.65408440 |
| O | 8.01129000 | 5.70219000 | 5.25789000 | -1.64929380 |
| O | 8.20727000 | 2.79452000 | 5.25789000 | -1.64332680 |
| O | 10.43141000 | 4.07862000 | 5.25789000 | -1.65402510 |
| O | 10.43141000 | 1.51041000 | 5.25789000 | -1.64890860 |
| Al | -2.41059000 | 9.78082000 | 6.11030000 | 2.47616830 |
| Al | 0.00953000 | 5.58904000 | 6.11030000 | 2.47612670 |
| Al | 2.42966000 | 1.39726000 | 6.11030000 | 2.47611320 |
| Al | 2.42966000 | 9.78082000 | 6.11030000 | 2.47621890 |
| Al | 4.84978000 | 5.58904000 | 6.11030000 | 2.47637900 |
| Al | 7.26991000 | 9.78082000 | 6.11030000 | 2.47615620 |
| Al | 7.26991000 | 1.39726000 | 6.11030000 | 2.47598580 |
| Al | 9.69003000 | 5.58904000 | 6.11030000 | 2.47598640 |
| Al | 12.11016000 | 1.39726000 | 6.11030000 | 2.47587070 |
| Al | -4.83072000 | 11.17808000 | 6.60910000 | 2.47688980 |
| Al | -2.41059000 | 6.98630000 | 6.60910000 | 2.47685970 |
| Al | 0.00953000 | 11.17808000 | 6.60910000 | 2.47717280 |
| Al | 0.00953000 | 2.79452000 | 6.60910000 | 2.47679130 |
| Al | 2.42966000 | 6.98630000 | 6.60910000 | 2.47711800 |
| Al | 4.84978000 | 2.79452000 | 6.60910000 | 2.46662340 |
| Al | 4.84978000 | 11.17808000 | 6.60910000 | 2.47710410 |
| Al | 7.26991000 | 6.98630000 | 6.60910000 | 2.47679540 |
| Al | 9.69003000 | 2.79452000 | 6.60910000 | 2.47684480 |
| O | -3.15680000 | 8.49706000 | 7.46570000 | -1.64925090 |
| O | -3.14925000 | 11.06893000 | 7.46570000 | -1.64985950 |
| O | -0.92572000 | 9.77647000 | 7.46570000 | -1.65366730 |
| O | -0.73667000 | 4.30528000 | 7.46570000 | -1.64952310 |
| O | -0.72913000 | 6.87715000 | 7.46570000 | -1.64921650 |
| O | 1.49440000 | 5.58469000 | 7.46570000 | -1.65335760 |
| O | 1.68345000 | 8.49706000 | 7.46570000 | -1.64837940 |
| O | 1.68345000 | 0.11351000 | 7.46570000 | -1.64915490 |
| O | 1.69100000 | 2.68537000 | 7.46570000 | -1.64969010 |
| O | 1.69100000 | 11.06893000 | 7.46570000 | -1.64945140 |
| O | 3.91453000 | 1.39291000 | 7.46570000 | -1.65409710 |
| O | 3.91452000 | 9.77646000 | 7.46570000 | -1.65384390 |

| | | | |
|---|---|---|---|
| O | 4.10358000 | 4.30528000 | 7.46570000 | -1.64879230 |
| O | 4.11112000 | 6.87715000 | 7.46570000 | -1.64897990 |
| O | 6.33465000 | 5.58468000 | 7.46570000 | -1.65394390 |
| O | 6.52370000 | 0.11351000 | 7.46570000 | -1.64874980 |
| O | 6.52370000 | 8.49706000 | 7.46570000 | -1.64814570 |
| O | 6.53125000 | 2.68537000 | 7.46570000 | -1.64940720 |
| O | 6.53125000 | 11.06893000 | 7.46570000 | -1.64967220 |
| O | 8.75477000 | 9.77646000 | 7.46570000 | -1.65361670 |
| O | 8.75477000 | 1.39291000 | 7.46570000 | -1.65433800 |
| O | 8.94383000 | 4.30528000 | 7.46570000 | -1.64985090 |
| O | 8.95137000 | 6.87715000 | 7.46570000 | -1.64923830 |
| O | 11.17490000 | 5.58468000 | 7.46570000 | -1.65437390 |
| O | 11.36395000 | 0.11351000 | 7.46570000 | -1.64891590 |
| O | 11.37149000 | 2.68537000 | 7.46570000 | -1.65050340 |
| O | 13.59502000 | 1.39291000 | 7.46570000 | -1.65379160 |
| Al | -4.83072000 | 8.38356000 | 8.32833000 | 2.47210910 |
| Al | -2.41059000 | 4.19178000 | 8.32833000 | 2.47183470 |
| Al | 0.00953000 | 8.38356000 | 8.32833000 | 2.47059100 |
| Al | 0.00954000 | 0.00000000 | 8.32833000 | 2.47060410 |
| Al | 2.42966000 | 4.19178000 | 8.32833000 | 2.47074970 |
| Al | 4.84978000 | 0.00000000 | 8.32833000 | 2.47153020 |
| Al | 4.84978000 | 8.38356000 | 8.32833000 | 2.47039100 |
| Al | 7.26991000 | 4.19178000 | 8.32833000 | 2.47199560 |
| Al | 9.69003000 | 0.00000000 | 8.32833000 | 2.47126890 |
| Al | -2.41059000 | 9.78082000 | 8.78548000 | 2.47245690 |
| Al | 0.00953000 | 5.58904000 | 8.78548000 | 2.47228730 |
| Al | 2.42966000 | 1.39726000 | 8.78548000 | 2.47229800 |
| Al | 2.42966000 | 9.78082000 | 8.78548000 | 2.47618720 |
| Al | 4.84978000 | 5.58904000 | 8.78548000 | 2.47619860 |
| Al | 7.26991000 | 9.78082000 | 8.78548000 | 2.47594250 |
| Al | 7.26991000 | 1.39726000 | 8.78548000 | 2.47231860 |
| Al | 9.69003000 | 5.58904000 | 8.78548000 | 2.47216410 |
| Al | 12.11016000 | 1.39726000 | 8.78548000 | 2.47037930 |
| O | 8.01661000 | 11.26550000 | 9.68989000 | -1.64749400 |
| O | 10.43673000 | 7.07371000 | 9.68989000 | -1.64391050 |
| O | -4.06971000 | 9.68514000 | 9.68989000 | -1.66008410 |
| O | -1.66389000 | 11.26550000 | 9.68989000 | -1.64314040 |
| O | 12.85686000 | 2.88194000 | 9.68989000 | -1.64908410 |
| O | -1.64959000 | 5.49336000 | 9.68989000 | -1.65914170 |
| O | -1.49818000 | 8.39182000 | 9.68989000 | -1.66758990 |
| O | 0.75624000 | 7.07372000 | 9.68989000 | -1.64200590 |
| O | 0.77054000 | 9.68514000 | 9.68989000 | -1.64769450 |
| O | 0.77054000 | 1.30159000 | 9.68989000 | -1.65684170 |

| | | | |
|---|---|---|---|
| O | 0.92195000 | 4.20004000 | 9.68989000 | -1.66803730 |
| O | 3.17636000 | 11.26550000 | 9.68989000 | -1.63894600 |
| O | 3.17636000 | 2.88194000 | 9.68989000 | -1.64233570 |
| O | 3.19067000 | 5.49336000 | 9.68989000 | -1.64604680 |
| O | 3.34208000 | 0.00826000 | 9.68989000 | -1.66770690 |
| O | 3.34207000 | 8.39182000 | 9.68989000 | -1.65487530 |
| O | 5.59648000 | 7.07372000 | 9.68989000 | -1.63462050 |
| O | 5.61079000 | 1.30159000 | 9.68989000 | -1.65902470 |
| O | 5.61079000 | 9.68514000 | 9.68989000 | -1.64533100 |
| O | 5.76220000 | 4.20004000 | 9.68989000 | -1.65371060 |
| O | 8.01661000 | 2.88194000 | 9.68989000 | -1.64572360 |
| O | 8.03091000 | 5.49336000 | 9.68989000 | -1.66141990 |
| O | 8.18232000 | 0.00826000 | 9.68989000 | -1.66373870 |
| O | 8.18232000 | 8.39182000 | 9.68989000 | -1.65749760 |
| O | 10.45104000 | 1.30158000 | 9.68989000 | -1.65839440 |
| O | 10.60245000 | 4.20004000 | 9.68989000 | -1.66960180 |
| O | 13.02257000 | 0.00826000 | 9.68989000 | -1.65788460 |
| Al | -4.83072000 | 11.17808000 | 10.71830000 | 2.48140070 |
| Al | -2.41059000 | 6.98630000 | 10.71830000 | 2.47827290 |
| Al | 0.00953000 | 11.17808000 | 10.71830000 | 2.47255730 |
| Al | 0.00953000 | 2.79452000 | 10.71830000 | 2.47832740 |
| Al | 2.42966000 | 6.98630000 | 10.71830000 | 2.47487510 |
| Al | 4.84978000 | 2.79452000 | 10.71830000 | 2.46989710 |
| Al | 4.84978000 | 11.17808000 | 10.71830000 | 2.47450310 |
| Al | 7.26991000 | 6.98630000 | 10.71830000 | 2.47852860 |
| Al | 9.69003000 | 2.79452000 | 10.71830000 | 2.47553410 |
| Al | -4.83072000 | 8.38356000 | 10.98601000 | 2.47410800 |
| Al | -2.41059000 | 4.19178000 | 10.98601000 | 2.46965420 |
| Al | 0.00953000 | 8.38356000 | 10.98601000 | 2.47288720 |
| Al | 0.00954000 | 0.00000000 | 10.98601000 | 2.47298950 |
| Al | 2.42966000 | 4.19178000 | 10.98601000 | 2.46999710 |
| Al | 4.84978000 | 0.00000000 | 10.98601000 | 2.47010600 |
| Al | 4.84978000 | 8.38356000 | 10.98601000 | 2.47008920 |
| Al | 7.26991000 | 4.19178000 | 10.98601000 | 2.47265680 |
| Al | 9.69003000 | 0.00000000 | 10.98601000 | 2.46728390 |
| O | -5.65925000 | 12.47506000 | 11.86245000 | -1.62451710 |
| O | -5.56522000 | 9.81837000 | 11.84694000 | -1.61135930 |
| O | -3.30263000 | 11.23859000 | 11.85090000 | -1.62404240 |
| O | -3.23680000 | 8.27694000 | 11.85442000 | -1.62071410 |
| O | -3.11953000 | 5.62861000 | 11.84965000 | -1.62032020 |
| O | -0.88348000 | 7.04352000 | 11.85082000 | -1.62423900 |
| O | -0.80428000 | 12.47122000 | 11.85016000 | -1.61750450 |
| O | -0.81382000 | 4.08619000 | 11.84891000 | -1.61914510 |
| O | -0.70849000 | 9.82082000 | 11.85960000 | -1.62149870 |
| O | -0.70283000 | 1.43269000 | 11.85030000 | -1.62998620 |
| O | 1.53134000 | 2.85074000 | 11.84890000 | -1.62521600 |
| O | 1.58151000 | 11.21513000 | 11.85580000 | -1.59833770 |
| O | 1.62694000 | 8.29724000 | 11.84497000 | -1.61286370 |
| O | 1.71846000 | 5.62660000 | 11.86448000 | -1.62435030 |
| O | 3.99097000 | 7.01869000 | 11.86876000 | -1.55231090 |
| O | 4.02845000 | 4.07471000 | 11.82639000 | -1.60740290 |
| O | 4.06390000 | 12.49443000 | 11.87750000 | -1.61086340 |
| O | 4.14035000 | 1.43040000 | 11.85954000 | -1.62002200 |
| O | 4.15177000 | 9.85313000 | 11.86715000 | -1.59853300 |
| O | 6.36151000 | 11.24192000 | 11.89795000 | -1.58182270 |
| O | 6.39361000 | 2.85286000 | 11.86392000 | -1.62314480 |
| O | 6.48351000 | 8.29912000 | 11.92456000 | -1.56598290 |
| O | 6.52445000 | 5.65488000 | 11.87138000 | -1.59462150 |
| O | 8.79173000 | 7.03871000 | 11.87975000 | -1.63171990 |
| O | 8.86646000 | 4.09159000 | 11.85599000 | -1.62603360 |
| O | 8.99154000 | 1.43729000 | 11.85156000 | -1.63152340 |
| O | 11.20518000 | 2.87605000 | 11.84918000 | -1.61656660 |
| Al | -2.41432000 | 9.77790000 | 11.89494000 | 2.41334380 |
| Al | 0.01210000 | 5.58894000 | 11.89982000 | 2.41446520 |
| Al | 2.43110000 | 1.39897000 | 11.87477000 | 2.41466830 |
| Al | 2.43593000 | 9.75753000 | 12.34950000 | 2.37657330 |
| Al | 4.81340000 | 5.51367000 | 12.38896000 | 2.38147290 |
| Al | 7.31255000 | 9.82418000 | 12.39669000 | 2.37191960 |
| Al | 7.27867000 | 1.39028000 | 11.87622000 | 2.41641430 |
| Al | 9.69969000 | 5.58093000 | 11.85394000 | 2.41306580 |
| Al | 12.11950000 | 1.45076000 | 11.76971000 | 2.41230140 |
| Pt | 4.98789000 | 5.32925000 | 14.84513000 | -0.49271220 |
| Pt | 6.96320000 | 9.85595000 | 14.87645000 | -1.03540850 |
| Pt | 4.06077000 | 7.66209000 | 14.26330000 | -0.42301140 |
| Pt | 2.88311000 | 10.16813000 | 14.81478000 | -0.86703900 |
| Sn | 5.00403000 | 11.48248000 | 14.02141000 | 1.00703640 |
| Sn | 6.76183000 | 7.26880000 | 14.06107000 | 0.98838930 |
| Sn | 4.77932000 | 8.98835000 | 16.25955000 | 0.68932200 |
| C | 6.25534000 | 3.65997000 | 15.34553000 | -0.18048050 |
| C | 4.92084000 | 3.22710000 | 15.21821000 | -0.21367240 |
| H | 6.97428000 | 3.49142000 | 14.54061000 | 0.11917480 |
| H | 6.68955000 | 3.83055000 | 16.33638000 | 0.09265060 |
| H | 4.59324000 | 2.72400000 | 14.30423000 | 0.12653450 |
| H | 4.31202000 | 3.04636000 | 16.11061000 | 0.07910090 |

283

c2h4-pt4sn3-al2o3-B

| | | | |
|---|---|---|---|
| Al | -4.83072000 | 11.17808000 | 0.00000000 | 2.17002530 |
| Al | -2.41059000 | 6.98630000 | 0.00000000 | 2.17013970 |
| Al | 0.00953000 | 11.17808000 | 0.00000000 | 2.17015090 |
| Al | 0.00953000 | 2.79452000 | 0.00000000 | 2.17006390 |
| Al | 2.42966000 | 6.98630000 | 0.00000000 | 2.16989150 |
| Al | 4.84978000 | 2.79452000 | 0.00000000 | 2.16978230 |
| Al | 4.84978000 | 11.17808000 | 0.00000000 | 2.17022660 |
| Al | 7.26991000 | 6.98630000 | 0.00000000 | 2.16991570 |
| Al | 9.69003000 | 2.79452000 | 0.00000000 | 2.17027120 |
| O | -3.89334000 | 9.78082000 | 0.85241000 | -1.55184310 |
| O | -1.66922000 | 8.49672000 | 0.85241000 | -1.56006640 |
| O | -1.66922000 | 11.06492000 | 0.85241000 | -1.55780910 |
| O | -1.47322000 | 5.58904000 | 0.85241000 | -1.55182580 |
| O | 0.75091000 | 4.30494000 | 0.85241000 | -1.55995660 |
| O | 0.75091000 | 6.87314000 | 0.85241000 | -1.55795300 |
| O | 0.94691000 | 9.78082000 | 0.85241000 | -1.55187830 |
| O | 0.94691000 | 1.39726000 | 0.85241000 | -1.55178260 |
| O | 3.17104000 | 0.11316000 | 0.85241000 | -1.55997470 |
| O | 3.17103000 | 8.49672000 | 0.85241000 | -1.56018480 |
| O | 3.17103000 | 2.68136000 | 0.85241000 | -1.55785810 |
| O | 3.17103000 | 11.06492000 | 0.85241000 | -1.55800010 |
| O | 3.36703000 | 5.58904000 | 0.85241000 | -1.55151590 |
| O | 5.59116000 | 4.30494000 | 0.85241000 | -1.55994410 |
| O | 5.59116000 | 6.87314000 | 0.85241000 | -1.55781220 |
| O | 5.78716000 | 1.39726000 | 0.85241000 | -1.55178950 |
| O | 5.78716000 | 9.78082000 | 0.85241000 | -1.55186100 |
| O | 8.01128000 | 8.49672000 | 0.85241000 | -1.56005760 |
| O | 8.01128000 | 0.11316000 | 0.85241000 | -1.56015830 |
| O | 8.01128000 | 2.68136000 | 0.85241000 | -1.55783590 |
| O | 8.01128000 | 11.06492000 | 0.85241000 | -1.55807710 |
| O | 8.20728000 | 5.58904000 | 0.85241000 | -1.55178950 |
| O | 10.43141000 | 4.30494000 | 0.85241000 | -1.56008320 |
| O | 10.43141000 | 6.87314000 | 0.85241000 | -1.55792790 |
| O | 10.62741000 | 1.39726000 | 0.85241000 | -1.55193720 |
| O | 12.85153000 | 0.11316000 | 0.85241000 | -1.56010570 |
| O | 12.85153000 | 2.68136000 | 0.85241000 | -1.55805340 |
| Al | -4.83072000 | 8.38356000 | 1.70482000 | 2.46328110 |
| Al | -2.41059000 | 4.19178000 | 1.70482000 | 2.46328110 |
| Al | 0.00953000 | 8.38356000 | 1.70482000 | 2.46328090 |
| Al | 0.00954000 | 0.00000000 | 1.70482000 | 2.46334010 |
| Al | 2.42966000 | 4.19178000 | 1.70482000 | 2.46333080 |
| Al | 4.84978000 | 0.00000000 | 1.70482000 | 2.46328910 |
| Al | 4.84978000 | 8.38356000 | 1.70482000 | 2.46331060 |

| | | | |
|---|---|---|---|
| Al | 7.26991000 | 4.19178000 | 1.70482000 | 2.46331950 |
| Al | 9.69003000 | 0.00000000 | 1.70482000 | 2.46324030 |
| Al | -2.41059000 | 9.78082000 | 2.20274000 | 2.48273890 |
| Al | 0.00953000 | 5.58904000 | 2.20274000 | 2.48281380 |
| Al | 2.42966000 | 1.39726000 | 2.20274000 | 2.48282120 |
| Al | 2.42966000 | 9.78082000 | 2.20274000 | 2.48265320 |
| Al | 4.84978000 | 5.58904000 | 2.20274000 | 2.48265680 |
| Al | 7.26991000 | 9.78082000 | 2.20274000 | 2.48262970 |
| Al | 7.26991000 | 1.39726000 | 2.20274000 | 2.48273750 |
| Al | 9.69003000 | 5.58904000 | 2.20274000 | 2.48273570 |
| Al | 12.11016000 | 1.39726000 | 2.20274000 | 2.48269490 |
| O | -5.57210000 | 9.66766000 | 3.05515000 | -1.64157290 |
| O | -3.34796000 | 8.38356000 | 3.05515000 | -1.62389890 |
| O | -3.15197000 | 5.47588000 | 3.05515000 | -1.64172430 |
| O | -3.15197000 | 11.29124000 | 3.05515000 | -1.64957690 |
| O | -0.92784000 | 4.19178000 | 3.05515000 | -1.62387830 |
| O | -0.73185000 | 1.28410000 | 3.05515000 | -1.64173850 |
| O | -0.73185000 | 9.66766000 | 3.05515000 | -1.64157590 |
| O | -0.73184000 | 7.09945000 | 3.05515000 | -1.64957390 |
| O | 1.49229000 | 8.38356000 | 3.05515000 | -1.62398740 |
| O | 1.49229000 | 0.00001000 | 3.05515000 | -1.62379570 |
| O | 1.68828000 | 5.47588000 | 3.05515000 | -1.64146210 |
| O | 1.68828000 | 2.90768000 | 3.05515000 | -1.64958660 |
| O | 1.68828000 | 11.29123000 | 3.05515000 | -1.64961890 |
| O | 3.91241000 | 4.19178000 | 3.05515000 | -1.62392620 |
| O | 4.10840000 | 9.66766000 | 3.05515000 | -1.64180620 |
| O | 4.10840000 | 1.28410000 | 3.05515000 | -1.64139950 |
| O | 4.10841000 | 7.09946000 | 3.05515000 | -1.64942460 |
| O | 6.33254000 | 0.00001000 | 3.05515000 | -1.62393660 |
| O | 6.33254000 | 8.38356000 | 3.05515000 | -1.62396910 |
| O | 6.52853000 | 11.29123000 | 3.05515000 | -1.64976350 |
| O | 6.52853000 | 5.47588000 | 3.05515000 | -1.64142180 |
| O | 6.52853000 | 2.90768000 | 3.05515000 | -1.64933980 |
| O | 8.75266000 | 4.19178000 | 3.05515000 | -1.62391670 |
| O | 8.94866000 | 7.09945000 | 3.05515000 | -1.64961480 |
| O | 8.94865000 | 1.28410000 | 3.05515000 | -1.64181820 |
| O | 11.17277000 | 0.00000000 | 3.05515000 | -1.62411290 |
| O | 11.36878000 | 2.90767000 | 3.05515000 | -1.64975110 |
| Al | -4.83072000 | 11.17808000 | 3.90756000 | 2.47772530 |
| Al | -2.41059000 | 6.98630000 | 3.90756000 | 2.47777090 |
| Al | 0.00953000 | 11.17808000 | 3.90756000 | 2.47770300 |
| Al | 0.00953000 | 2.79452000 | 3.90756000 | 2.47774030 |
| Al | 2.42966000 | 6.98630000 | 3.90756000 | 2.47761930 |

| | | | |
|---|---|---|---|
| Al | 4.84978000 | 2.79452000 | 3.90756000 | 2.47755960 |
| Al | 4.84978000 | 11.17808000 | 3.90756000 | 2.47796000 |
| Al | 7.26991000 | 6.98630000 | 3.90756000 | 2.47777740 |
| Al | 9.69003000 | 2.79452000 | 3.90756000 | 2.47777220 |
| Al | -4.83072000 | 8.38356000 | 4.40548000 | 2.47226090 |
| Al | -2.41059000 | 4.19178000 | 4.40548000 | 2.47226940 |
| Al | 0.00953000 | 8.38356000 | 4.40548000 | 2.47215930 |
| Al | 0.00954000 | 0.00000000 | 4.40548000 | 2.47270140 |
| Al | 2.42966000 | 4.19178000 | 4.40548000 | 2.47229530 |
| Al | 4.84978000 | 0.00000000 | 4.40548000 | 2.47228690 |
| Al | 4.84978000 | 8.38356000 | 4.40548000 | 2.47220450 |
| Al | 7.26991000 | 4.19178000 | 4.40548000 | 2.47242540 |
| Al | 9.69003000 | 0.00000000 | 4.40548000 | 2.47237260 |
| O | -6.31347000 | 11.17808000 | 5.25789000 | -1.64349930 |
| O | -4.08934000 | 12.46219000 | 5.25789000 | -1.65390010 |
| O | -4.08934000 | 9.89398000 | 5.25789000 | -1.64917460 |
| O | -3.89335000 | 6.98630000 | 5.25789000 | -1.64346200 |
| O | -1.66922000 | 8.27040000 | 5.25789000 | -1.65404320 |
| O | -1.66921000 | 5.70219000 | 5.25789000 | -1.64907130 |
| O | -1.47322000 | 11.17808000 | 5.25789000 | -1.64319880 |
| O | -1.47323000 | 2.79452000 | 5.25789000 | -1.64342950 |
| O | 0.75091000 | 4.07862000 | 5.25789000 | -1.65362950 |
| O | 0.75091000 | 12.46219000 | 5.25789000 | -1.65418150 |
| O | 0.75091000 | 9.89397000 | 5.25789000 | -1.64899940 |
| O | 0.75091000 | 1.51042000 | 5.25789000 | -1.64924900 |
| O | 0.94690000 | 6.98630000 | 5.25789000 | -1.64365620 |
| O | 3.17103000 | 8.27040000 | 5.25789000 | -1.65424870 |
| O | 3.17104000 | 5.70220000 | 5.25789000 | -1.64905210 |
| O | 3.36703000 | 2.79452000 | 5.25789000 | -1.64296530 |
| O | 3.36702000 | 11.17808000 | 5.25789000 | -1.64346990 |
| O | 5.59116000 | 12.46219000 | 5.25789000 | -1.65407860 |
| O | 5.59116000 | 4.07863000 | 5.25789000 | -1.65417250 |
| O | 5.59116000 | 9.89397000 | 5.25789000 | -1.64882830 |
| O | 5.59116000 | 1.51042000 | 5.25789000 | -1.64927530 |
| O | 5.78715000 | 6.98630000 | 5.25789000 | -1.64338060 |
| O | 8.01128000 | 8.27040000 | 5.25789000 | -1.65417680 |
| O | 8.01129000 | 5.70219000 | 5.25789000 | -1.64914780 |
| O | 8.20727000 | 2.79452000 | 5.25789000 | -1.64320220 |
| O | 10.43141000 | 4.07862000 | 5.25789000 | -1.65408940 |
| O | 10.43141000 | 1.51041000 | 5.25789000 | -1.64904560 |
| Al | -2.41059000 | 9.78082000 | 6.11030000 | 2.47623540 |
| Al | 0.00953000 | 5.58904000 | 6.11030000 | 2.47629350 |
| Al | 2.42966000 | 1.39726000 | 6.11030000 | 2.47628880 |

| | | | |
|---|---|---|---|
| Al | 2.42966000 | 9.78082000 | 6.11030000 | 2.47621870 |
| Al | 4.84978000 | 5.58904000 | 6.11030000 | 2.47604260 |
| Al | 7.26991000 | 9.78082000 | 6.11030000 | 2.47616410 |
| Al | 7.26991000 | 1.39726000 | 6.11030000 | 2.47622440 |
| Al | 9.69003000 | 5.58904000 | 6.11030000 | 2.47602360 |
| Al | 12.11016000 | 1.39726000 | 6.11030000 | 2.47562000 |
| Al | -4.83072000 | 11.17808000 | 6.60910000 | 2.47690760 |
| Al | -2.41059000 | 6.98630000 | 6.60910000 | 2.47678090 |
| Al | 0.00953000 | 11.17808000 | 6.60910000 | 2.47659980 |
| Al | 0.00953000 | 2.79452000 | 6.60910000 | 2.47691130 |
| Al | 2.42966000 | 6.98630000 | 6.60910000 | 2.47722250 |
| Al | 4.84978000 | 2.79452000 | 6.60910000 | 2.47720160 |
| Al | 4.84978000 | 11.17808000 | 6.60910000 | 2.47668270 |
| Al | 7.26991000 | 6.98630000 | 6.60910000 | 2.47703410 |
| Al | 9.69003000 | 2.79452000 | 6.60910000 | 2.47666280 |
| O | -3.15680000 | 8.49706000 | 7.46570000 | -1.64927990 |
| O | -3.14925000 | 11.06893000 | 7.46570000 | -1.64964380 |
| O | -0.92572000 | 9.77647000 | 7.46570000 | -1.65434580 |
| O | -0.73667000 | 4.30528000 | 7.46570000 | -1.64924640 |
| O | -0.72913000 | 6.87715000 | 7.46570000 | -1.64883550 |
| O | 1.49440000 | 5.58469000 | 7.46570000 | -1.65369530 |
| O | 1.68345000 | 8.49706000 | 7.46570000 | -1.64862770 |
| O | 1.68345000 | 0.11351000 | 7.46570000 | -1.64897150 |
| O | 1.69100000 | 2.68537000 | 7.46570000 | -1.64905750 |
| O | 1.69100000 | 11.06893000 | 7.46570000 | -1.64999630 |
| O | 3.91453000 | 1.39291000 | 7.46570000 | -1.65338420 |
| O | 3.91452000 | 9.77646000 | 7.46570000 | -1.65421140 |
| O | 4.10358000 | 4.30528000 | 7.46570000 | -1.64778710 |
| O | 4.11112000 | 6.87715000 | 7.46570000 | -1.64951060 |
| O | 6.33465000 | 5.58468000 | 7.46570000 | -1.65366290 |
| O | 6.52370000 | 0.11351000 | 7.46570000 | -1.64947190 |
| O | 6.52370000 | 8.49706000 | 7.46570000 | -1.64869270 |
| O | 6.53125000 | 2.68537000 | 7.46570000 | -1.64944000 |
| O | 6.53125000 | 11.06893000 | 7.46570000 | -1.65053560 |
| O | 8.75477000 | 9.77646000 | 7.46570000 | -1.65377790 |
| O | 8.75477000 | 1.39291000 | 7.46570000 | -1.65464580 |
| O | 8.94383000 | 4.30528000 | 7.46570000 | -1.64965570 |
| O | 8.95137000 | 6.87715000 | 7.46570000 | -1.64947510 |
| O | 11.17490000 | 5.58468000 | 7.46570000 | -1.65408330 |
| O | 11.36395000 | 0.11351000 | 7.46570000 | -1.64971450 |
| O | 11.37149000 | 2.68537000 | 7.46570000 | -1.64984030 |
| O | 13.59502000 | 1.39291000 | 7.46570000 | -1.65370400 |
| Al | -4.83072000 | 8.38356000 | 8.32833000 | 2.47154780 |

| | | | |
|---|---|---|---|
| Al | -2.41059000 | 4.19178000 | 8.32833000 | 2.47084870 |
| Al | 0.00953000 | 8.38356000 | 8.32833000 | 2.47140840 |
| Al | 0.00954000 | 0.00000000 | 8.32833000 | 2.47050690 |
| Al | 2.42966000 | 4.19178000 | 8.32833000 | 2.46974770 |
| Al | 4.84978000 | 0.00000000 | 8.32833000 | 2.47145200 |
| Al | 4.84978000 | 8.38356000 | 8.32833000 | 2.47125680 |
| Al | 7.26991000 | 4.19178000 | 8.32833000 | 2.47183720 |
| Al | 9.69003000 | 0.00000000 | 8.32833000 | 2.47152000 |
| Al | -2.41059000 | 9.78082000 | 8.78548000 | 2.47280610 |
| Al | 0.00953000 | 5.58904000 | 8.78548000 | 2.47593300 |
| Al | 2.42966000 | 1.39726000 | 8.78548000 | 2.47588010 |
| Al | 2.42966000 | 9.78082000 | 8.78548000 | 2.47173900 |
| Al | 4.84978000 | 5.58904000 | 8.78548000 | 2.47640160 |
| Al | 7.26991000 | 9.78082000 | 8.78548000 | 2.47220410 |
| Al | 7.26991000 | 1.39726000 | 8.78548000 | 2.47263380 |
| Al | 9.69003000 | 5.58904000 | 8.78548000 | 2.47234850 |
| Al | 12.11016000 | 1.39726000 | 8.78548000 | 2.47002330 |
| O | 8.01661000 | 11.26550000 | 9.68989000 | -1.64631050 |
| O | 10.43673000 | 7.07371000 | 9.68989000 | -1.64301290 |
| O | -4.06971000 | 9.68514000 | 9.68989000 | -1.65907750 |
| O | -1.66389000 | 11.26550000 | 9.68989000 | -1.64383900 |
| O | 12.85686000 | 2.88194000 | 9.68989000 | -1.64550370 |
| O | -1.64959000 | 5.49336000 | 9.68989000 | -1.64613860 |
| O | -1.49818000 | 8.39182000 | 9.68989000 | -1.66441330 |
| O | 0.75624000 | 7.07372000 | 9.68989000 | -1.63787870 |
| O | 0.77054000 | 9.68514000 | 9.68989000 | -1.65716240 |
| O | 0.77054000 | 1.30159000 | 9.68989000 | -1.64750570 |
| O | 0.92195000 | 4.20004000 | 9.68989000 | -1.65624890 |
| O | 3.17636000 | 11.26550000 | 9.68989000 | -1.64594470 |
| O | 3.17636000 | 2.88194000 | 9.68989000 | -1.63300300 |
| O | 3.19067000 | 5.49336000 | 9.68989000 | -1.64510620 |
| O | 3.34208000 | 0.00826000 | 9.68989000 | -1.65560010 |
| O | 3.34207000 | 8.39182000 | 9.68989000 | -1.65661380 |
| O | 5.59648000 | 7.07372000 | 9.68989000 | -1.64636530 |
| O | 5.61079000 | 1.30159000 | 9.68989000 | -1.65941590 |
| O | 5.61079000 | 9.68514000 | 9.68989000 | -1.65659980 |
| O | 5.76220000 | 4.20004000 | 9.68989000 | -1.65463030 |
| O | 8.01661000 | 2.88194000 | 9.68989000 | -1.64408590 |
| O | 8.03091000 | 5.49336000 | 9.68989000 | -1.65967520 |
| O | 8.18232000 | 0.00826000 | 9.68989000 | -1.66628340 |
| O | 8.18232000 | 8.39182000 | 9.68989000 | -1.66803720 |
| O | 10.45104000 | 1.30158000 | 9.68989000 | -1.66390420 |
| O | 10.60245000 | 4.20004000 | 9.68989000 | -1.66844750 |

| | | | |
|---|---|---|---|
| O | 13.02257000 | 0.00826000 | 9.68989000 | -1.66680270 |
| Al | -4.83072000 | 11.17808000 | 10.71830000 | 2.47868740 |
| Al | -2.41059000 | 6.98630000 | 10.71830000 | 2.47996860 |
| Al | 0.00953000 | 11.17808000 | 10.71830000 | 2.47691720 |
| Al | 0.00953000 | 2.79452000 | 10.71830000 | 2.47658540 |
| Al | 2.42966000 | 6.98630000 | 10.71830000 | 2.47820730 |
| Al | 4.84978000 | 2.79452000 | 10.71830000 | 2.47891200 |
| Al | 4.84978000 | 11.17808000 | 10.71830000 | 2.47382300 |
| Al | 7.26991000 | 6.98630000 | 10.71830000 | 2.47919020 |
| Al | 9.69003000 | 2.79452000 | 10.71830000 | 2.47365720 |
| Al | -4.83072000 | 8.38356000 | 10.98601000 | 2.47090780 |
| Al | -2.41059000 | 4.19178000 | 10.98601000 | 2.47154750 |
| Al | 0.00953000 | 8.38356000 | 10.98601000 | 2.47207660 |
| Al | 0.00954000 | 0.00000000 | 10.98601000 | 2.46890800 |
| Al | 2.42966000 | 4.19178000 | 10.98601000 | 2.47293080 |
| Al | 4.84978000 | 0.00000000 | 10.98601000 | 2.47409460 |
| Al | 4.84978000 | 8.38356000 | 10.98601000 | 2.47315260 |
| Al | 7.26991000 | 4.19178000 | 10.98601000 | 2.47195380 |
| Al | 9.69003000 | 0.00000000 | 10.98601000 | 2.46937450 |
| O | -5.64813000 | 12.52131000 | 11.85477000 | -1.60511270 |
| O | -5.54237000 | 9.82726000 | 11.85026000 | -1.61958540 |
| O | -3.29619000 | 11.23575000 | 11.86025000 | -1.62344890 |
| O | -3.22667000 | 8.27983000 | 11.85074000 | -1.62010290 |
| O | -3.12033000 | 5.62040000 | 11.86272000 | -1.62739110 |
| O | -0.86548000 | 7.02977000 | 11.84785000 | -1.61422000 |
| O | -0.80337000 | 12.47171000 | 11.85084000 | -1.62112810 |
| O | -0.80053000 | 4.07159000 | 11.89115000 | -1.58941180 |
| O | -0.69488000 | 9.81740000 | 11.85426000 | -1.62763100 |
| O | -0.73577000 | 1.45415000 | 11.87809000 | -1.61653030 |
| O | 1.50393000 | 2.85030000 | 11.88068000 | -1.59091220 |
| O | 1.53990000 | 11.24300000 | 11.84986000 | -1.62204200 |
| O | 1.62349000 | 8.27826000 | 11.88053000 | -1.62607760 |
| O | 1.68413000 | 5.64839000 | 11.92195000 | -1.56852910 |
| O | 3.95407000 | 7.00850000 | 11.86851000 | -1.59777650 |
| O | 4.04226000 | 4.13365000 | 11.86954000 | -1.55782670 |
| O | 4.02273000 | 12.45018000 | 11.84879000 | -1.61675140 |
| O | 4.11895000 | 1.44435000 | 11.84221000 | -1.61048040 |
| O | 4.13879000 | 9.81635000 | 11.85688000 | -1.62383300 |
| O | 6.37528000 | 11.25177000 | 11.85182000 | -1.62596740 |
| O | 6.38275000 | 2.85838000 | 11.86467000 | -1.62436930 |
| O | 6.44744000 | 8.29521000 | 11.86259000 | -1.62659030 |
| O | 6.56971000 | 5.63415000 | 11.82929000 | -1.61699650 |
| O | 8.80584000 | 7.05503000 | 11.85934000 | -1.61863590 |

| | | | |
|---|---|---|---|
| O | 8.88037000 | 4.08435000 | 11.84881000 | -1.61673950 |
| O | 8.98278000 | 1.43539000 | 11.84895000 | -1.62266840 |
| O | 11.22500000 | 2.85878000 | 11.85040000 | -1.62853360 |
| Al | -2.40618000 | 9.77938000 | 11.89425000 | 2.41724650 |
| Al | -0.05169000 | 5.60041000 | 12.39381000 | 2.37413320 |
| Al | 2.45394000 | 1.40888000 | 12.35254000 | 2.37359260 |
| Al | 2.43232000 | 9.79371000 | 11.85173000 | 2.41487800 |
| Al | 4.93195000 | 5.60076000 | 12.39705000 | 2.38195210 |
| Al | 7.27291000 | 9.79219000 | 11.87383000 | 2.41714590 |
| Al | 7.26861000 | 1.39947000 | 11.89934000 | 2.41649890 |
| Al | 9.68644000 | 5.58990000 | 11.87986000 | 2.41565700 |
| Al | 12.06002000 | 1.38004000 | 11.77006000 | 2.41173920 |
| Pt | 4.97955000 | 5.93342000 | 14.85547000 | -0.49382420 |
| Pt | 0.11167000 | 5.26328000 | 14.87516000 | -1.03554350 |
| Pt | 3.49027000 | 3.90517000 | 14.26507000 | -0.41967510 |
| Pt | 1.97360000 | 1.60712000 | 14.82969000 | -0.87104670 |
| Sn | -0.24200000 | 2.74296000 | 14.01239000 | 1.01231360 |
| Sn | 2.42296000 | 6.42960000 | 14.05290000 | 0.99107710 |
| Sn | 1.98638000 | 3.85866000 | 16.26220000 | 0.69186780 |
| C | 6.72840000 | 7.04435000 | 15.37191000 | -0.16980050 |
| C | 5.65275000 | 7.94426000 | 15.22913000 | -0.22745690 |
| H | 7.45909000 | 6.93313000 | 14.56617000 | 0.10969180 |
| H | 7.06677000 | 6.74651000 | 16.37021000 | 0.08829170 |
| H | 5.53815000 | 8.53494000 | 14.31735000 | 0.12820320 |
| H | 5.14188000 | 8.34098000 | 16.11257000 | 0.08768260 |

283
c2h4-pt4sn3-al2o3-B

| | | | |
|---|---|---|---|
| Al | -4.83072000 | 11.17808000 | 0.00000000 | 2.17023470 |
| Al | -2.41059000 | 6.98630000 | 0.00000000 | 2.17035270 |
| Al | 0.00953000 | 11.17808000 | 0.00000000 | 2.17037350 |
| Al | 0.00953000 | 2.79452000 | 0.00000000 | 2.17025540 |
| Al | 2.42966000 | 6.98630000 | 0.00000000 | 2.17013200 |
| Al | 4.84978000 | 2.79452000 | 0.00000000 | 2.17001160 |
| Al | 4.84978000 | 11.17808000 | 0.00000000 | 2.17039330 |
| Al | 7.26991000 | 6.98630000 | 0.00000000 | 2.17018300 |
| Al | 9.69003000 | 2.79452000 | 0.00000000 | 2.17042750 |
| O | -3.89334000 | 9.78082000 | 0.85241000 | -1.55183100 |
| O | -1.66922000 | 8.49672000 | 0.85241000 | -1.56005000 |
| O | -1.66922000 | 11.06492000 | 0.85241000 | -1.55779900 |
| O | -1.47322000 | 5.58904000 | 0.85241000 | -1.55182130 |
| O | 0.75091000 | 4.30494000 | 0.85241000 | -1.55997200 |
| O | 0.75091000 | 6.87314000 | 0.85241000 | -1.55794420 |
| O | 0.94691000 | 9.78082000 | 0.85241000 | -1.55188100 |
| O | 0.94691000 | 1.39726000 | 0.85241000 | -1.55177160 |
| O | 3.17104000 | 0.11316000 | 0.85241000 | -1.55999290 |
| O | 3.17103000 | 8.49672000 | 0.85241000 | -1.56019000 |
| O | 3.17103000 | 2.68136000 | 0.85241000 | -1.55783460 |
| O | 3.17103000 | 11.06492000 | 0.85241000 | -1.55795690 |
| O | 3.36703000 | 5.58904000 | 0.85241000 | -1.55151190 |
| O | 5.59116000 | 4.30494000 | 0.85241000 | -1.55996320 |
| O | 5.59116000 | 6.87314000 | 0.85241000 | -1.55780110 |
| O | 5.78716000 | 1.39726000 | 0.85241000 | -1.55178200 |
| O | 5.78716000 | 9.78082000 | 0.85241000 | -1.55183470 |
| O | 8.01128000 | 8.49672000 | 0.85241000 | -1.56007180 |
| O | 8.01128000 | 0.11316000 | 0.85241000 | -1.56014040 |
| O | 8.01128000 | 2.68136000 | 0.85241000 | -1.55779960 |
| O | 8.01128000 | 11.06492000 | 0.85241000 | -1.55802970 |
| O | 8.20728000 | 5.58904000 | 0.85241000 | -1.55179730 |
| O | 10.43141000 | 4.30494000 | 0.85241000 | -1.56006080 |
| O | 10.43141000 | 6.87314000 | 0.85241000 | -1.55787700 |
| O | 10.62741000 | 1.39726000 | 0.85241000 | -1.55191370 |
| O | 12.85153000 | 0.11316000 | 0.85241000 | -1.56008760 |
| O | 12.85153000 | 2.68136000 | 0.85241000 | -1.55800710 |
| Al | -4.83072000 | 8.38356000 | 1.70482000 | 2.46328170 |
| Al | -2.41059000 | 4.19178000 | 1.70482000 | 2.46328150 |
| Al | 0.00953000 | 8.38356000 | 1.70482000 | 2.46328150 |
| Al | 0.00954000 | 0.00000000 | 1.70482000 | 2.46334070 |
| Al | 2.42966000 | 4.19178000 | 1.70482000 | 2.46333680 |
| Al | 4.84978000 | 0.00000000 | 1.70482000 | 2.46328950 |
| Al | 4.84978000 | 8.38356000 | 1.70482000 | 2.46331120 |
| Al | 7.26991000 | 4.19178000 | 1.70482000 | 2.46332000 |
| Al | 9.69003000 | 0.00000000 | 1.70482000 | 2.46324090 |
| Al | -2.41059000 | 9.78082000 | 2.20274000 | 2.48274050 |
| Al | 0.00953000 | 5.58904000 | 2.20274000 | 2.48281570 |
| Al | 2.42966000 | 1.39726000 | 2.20274000 | 2.48282250 |
| Al | 2.42966000 | 9.78082000 | 2.20274000 | 2.48265430 |
| Al | 4.84978000 | 5.58904000 | 2.20274000 | 2.48272840 |
| Al | 7.26991000 | 9.78082000 | 2.20274000 | 2.48263070 |
| Al | 7.26991000 | 1.39726000 | 2.20274000 | 2.48273820 |
| Al | 9.69003000 | 5.58904000 | 2.20274000 | 2.48273680 |
| Al | 12.11016000 | 1.39726000 | 2.20274000 | 2.48269720 |
| O | -5.57210000 | 9.66766000 | 3.05515000 | -1.64157090 |
| O | -3.34796000 | 8.38356000 | 3.05515000 | -1.62389620 |
| O | -3.15197000 | 5.47588000 | 3.05515000 | -1.64172170 |
| O | -3.15197000 | 11.29124000 | 3.05515000 | -1.64957470 |
| O | -0.92784000 | 4.19178000 | 3.05515000 | -1.62387330 |
| O | -0.73185000 | 1.28410000 | 3.05515000 | -1.64173290 |
| O | -0.73185000 | 9.66766000 | 3.05515000 | -1.64157330 |
| O | -0.73184000 | 7.09945000 | 3.05515000 | -1.64956900 |
| O | 1.49229000 | 8.38356000 | 3.05515000 | -1.62398630 |
| O | 1.49229000 | 0.00001000 | 3.05515000 | -1.62379330 |
| O | 1.68828000 | 5.47588000 | 3.05515000 | -1.64145710 |
| O | 1.68828000 | 2.90768000 | 3.05515000 | -1.64958310 |
| O | 1.68828000 | 11.29123000 | 3.05515000 | -1.64961680 |
| O | 3.91241000 | 4.19178000 | 3.05515000 | -1.62392470 |
| O | 4.10840000 | 9.66766000 | 3.05515000 | -1.64180520 |
| O | 4.10840000 | 1.28410000 | 3.05515000 | -1.64138480 |
| O | 4.10841000 | 7.09946000 | 3.05515000 | -1.64942310 |
| O | 6.33254000 | 0.00001000 | 3.05515000 | -1.62393540 |
| O | 6.33254000 | 8.38356000 | 3.05515000 | -1.62396720 |
| O | 6.52853000 | 11.29123000 | 3.05515000 | -1.64976160 |
| O | 6.52853000 | 5.47588000 | 3.05515000 | -1.64149160 |
| O | 6.52853000 | 2.90768000 | 3.05515000 | -1.64933860 |
| O | 8.75266000 | 4.19178000 | 3.05515000 | -1.62391450 |
| O | 8.94866000 | 7.09945000 | 3.05515000 | -1.64964770 |
| O | 8.94865000 | 1.28410000 | 3.05515000 | -1.64181660 |
| O | 11.17277000 | 0.00000000 | 3.05515000 | -1.62410750 |
| O | 11.36878000 | 2.90767000 | 3.05515000 | -1.64974600 |
| Al | -4.83072000 | 11.17808000 | 3.90756000 | 2.47772460 |
| Al | -2.41059000 | 6.98630000 | 3.90756000 | 2.47776990 |
| Al | 0.00953000 | 11.17808000 | 3.90756000 | 2.47770270 |
| Al | 0.00953000 | 2.79452000 | 3.90756000 | 2.47773930 |
| Al | 2.42966000 | 6.98630000 | 3.90756000 | 2.47761910 |
| Al | 4.84978000 | 2.79452000 | 3.90756000 | 2.47755920 |
| Al | 4.84978000 | 11.17808000 | 3.90756000 | 2.47795940 |
| Al | 7.26991000 | 6.98630000 | 3.90756000 | 2.47777740 |
| Al | 9.69003000 | 2.79452000 | 3.90756000 | 2.47777160 |
| Al | -4.83072000 | 8.38356000 | 4.40548000 | 2.47226150 |
| Al | -2.41059000 | 4.19178000 | 4.40548000 | 2.47233680 |
| Al | 0.00953000 | 8.38356000 | 4.40548000 | 2.47215970 |
| Al | 0.00954000 | 0.00000000 | 4.40548000 | 2.47270200 |
| Al | 2.42966000 | 4.19178000 | 4.40548000 | 2.47229780 |
| Al | 4.84978000 | 0.00000000 | 4.40548000 | 2.47228760 |
| Al | 4.84978000 | 8.38356000 | 4.40548000 | 2.47217260 |
| Al | 7.26991000 | 4.19178000 | 4.40548000 | 2.47242260 |
| Al | 9.69003000 | 0.00000000 | 4.40548000 | 2.47237530 |
| O | -6.31347000 | 11.17808000 | 5.25789000 | -1.64350200 |
| O | -4.08934000 | 12.46219000 | 5.25789000 | -1.65389540 |
| O | -4.08934000 | 9.89398000 | 5.25789000 | -1.64917370 |

| | | | |
|---|---|---|---|
| O | -3.89335000 | 6.98630000 | 5.25789000 | -1.64346190 |
| O | -1.66922000 | 8.27040000 | 5.25789000 | -1.65404490 |
| O | -1.66921000 | 5.70219000 | 5.25789000 | -1.64913590 |
| O | -1.47322000 | 11.17808000 | 5.25789000 | -1.64319940 |
| O | -1.47323000 | 2.79452000 | 5.25789000 | -1.64343020 |
| O | 0.75091000 | 4.07862000 | 5.25789000 | -1.65362810 |
| O | 0.75091000 | 12.46219000 | 5.25789000 | -1.65412950 |
| O | 0.75091000 | 9.89397000 | 5.25789000 | -1.64899580 |
| O | 0.75091000 | 1.51042000 | 5.25789000 | -1.64925040 |
| O | 0.94690000 | 6.98630000 | 5.25789000 | -1.64363430 |
| O | 3.17103000 | 8.27040000 | 5.25789000 | -1.65422670 |
| O | 3.17104000 | 5.70220000 | 5.25789000 | -1.64902710 |
| O | 3.36703000 | 2.79452000 | 5.25789000 | -1.64293010 |
| O | 3.36702000 | 11.17808000 | 5.25789000 | -1.64346220 |
| O | 5.59116000 | 12.46219000 | 5.25789000 | -1.65405880 |
| O | 5.59116000 | 4.07863000 | 5.25789000 | -1.65417650 |
| O | 5.59116000 | 9.89397000 | 5.25789000 | -1.64882430 |
| O | 5.59116000 | 1.51042000 | 5.25789000 | -1.64927210 |
| O | 5.78715000 | 6.98630000 | 5.25789000 | -1.64337900 |
| O | 8.01128000 | 8.27040000 | 5.25789000 | -1.65417210 |
| O | 8.01129000 | 5.70219000 | 5.25789000 | -1.64915160 |
| O | 8.20727000 | 2.79452000 | 5.25789000 | -1.64320430 |
| O | 10.43141000 | 4.07862000 | 5.25789000 | -1.65405130 |
| O | 10.43141000 | 1.51041000 | 5.25789000 | -1.64904530 |
| Al | -2.41059000 | 9.78082000 | 6.11030000 | 2.47623550 |
| Al | 0.00953000 | 5.58904000 | 6.11030000 | 2.47629080 |
| Al | 2.42966000 | 1.39726000 | 6.11030000 | 2.47635950 |
| Al | 2.42966000 | 9.78082000 | 6.11030000 | 2.47619830 |
| Al | 4.84978000 | 5.58904000 | 6.11030000 | 2.47601120 |
| Al | 7.26991000 | 9.78082000 | 6.11030000 | 2.47616050 |
| Al | 7.26991000 | 1.39726000 | 6.11030000 | 2.47617410 |
| Al | 9.69003000 | 5.58904000 | 6.11030000 | 2.47608400 |
| Al | 12.11016000 | 1.39726000 | 6.11030000 | 2.47562140 |
| Al | -4.83072000 | 11.17808000 | 6.60910000 | 2.47688890 |
| Al | -2.41059000 | 6.98630000 | 6.60910000 | 2.47683240 |
| Al | 0.00953000 | 11.17808000 | 6.60910000 | 2.47662160 |
| Al | 0.00953000 | 2.79452000 | 6.60910000 | 2.47691900 |
| Al | 2.42966000 | 6.98630000 | 6.60910000 | 2.47715960 |
| Al | 4.84978000 | 2.79452000 | 6.60910000 | 2.47714480 |
| Al | 4.84978000 | 11.17808000 | 6.60910000 | 2.47678510 |
| Al | 7.26991000 | 6.98630000 | 6.60910000 | 2.47702870 |
| Al | 9.69003000 | 2.79452000 | 6.60910000 | 2.47658600 |
| O | -3.15680000 | 8.49706000 | 7.46570000 | -1.64928060 |

| | | | |
|---|---|---|---|
| O | -3.14925000 | 11.06893000 | 7.46570000 | -1.64966140 |
| O | -0.92572000 | 9.77647000 | 7.46570000 | -1.65433460 |
| O | -0.73667000 | 4.30528000 | 7.46570000 | -1.64927130 |
| O | -0.72913000 | 6.87715000 | 7.46570000 | -1.64888260 |
| O | 1.49440000 | 5.58469000 | 7.46570000 | -1.65367130 |
| O | 1.68345000 | 8.49706000 | 7.46570000 | -1.64872620 |
| O | 1.68345000 | 0.11351000 | 7.46570000 | -1.64882130 |
| O | 1.69100000 | 2.68537000 | 7.46570000 | -1.64917640 |
| O | 1.69100000 | 11.06893000 | 7.46570000 | -1.64994450 |
| O | 3.91453000 | 1.39291000 | 7.46570000 | -1.65339420 |
| O | 3.91452000 | 9.77646000 | 7.46570000 | -1.65387900 |
| O | 4.10358000 | 4.30528000 | 7.46570000 | -1.64770920 |
| O | 4.11112000 | 6.87715000 | 7.46570000 | -1.64962140 |
| O | 6.33465000 | 5.58468000 | 7.46570000 | -1.65359040 |
| O | 6.52370000 | 0.11351000 | 7.46570000 | -1.64945070 |
| O | 6.52370000 | 8.49706000 | 7.46570000 | -1.64865900 |
| O | 6.53125000 | 2.68537000 | 7.46570000 | -1.64937950 |
| O | 6.53125000 | 11.06893000 | 7.46570000 | -1.65048730 |
| O | 8.75477000 | 9.77646000 | 7.46570000 | -1.65381250 |
| O | 8.75477000 | 1.39291000 | 7.46570000 | -1.65459360 |
| O | 8.94383000 | 4.30528000 | 7.46570000 | -1.64954340 |
| O | 8.95137000 | 6.87715000 | 7.46570000 | -1.64938860 |
| O | 11.17490000 | 5.58468000 | 7.46570000 | -1.65413020 |
| O | 11.36395000 | 0.11351000 | 7.46570000 | -1.64984380 |
| O | 11.37149000 | 2.68537000 | 7.46570000 | -1.64980420 |
| O | 13.59502000 | 1.39291000 | 7.46570000 | -1.65369260 |
| Al | -4.83072000 | 8.38356000 | 8.32833000 | 2.47149570 |
| Al | -2.41059000 | 4.19178000 | 8.32833000 | 2.47082780 |
| Al | 0.00953000 | 8.38356000 | 8.32833000 | 2.47140790 |
| Al | 0.00954000 | 0.00000000 | 8.32833000 | 2.47060010 |
| Al | 2.42966000 | 4.19178000 | 8.32833000 | 2.46989740 |
| Al | 4.84978000 | 0.00000000 | 8.32833000 | 2.47143960 |
| Al | 4.84978000 | 8.38356000 | 8.32833000 | 2.47102220 |
| Al | 7.26991000 | 4.19178000 | 8.32833000 | 2.47176430 |
| Al | 9.69003000 | 0.00000000 | 8.32833000 | 2.47154020 |
| Al | -2.41059000 | 9.78082000 | 8.78548000 | 2.47267640 |
| Al | 0.00953000 | 5.58904000 | 8.78548000 | 2.47589940 |
| Al | 2.42966000 | 1.39726000 | 8.78548000 | 2.47582620 |
| Al | 2.42966000 | 9.78082000 | 8.78548000 | 2.47178130 |
| Al | 4.84978000 | 5.58904000 | 8.78548000 | 2.47631590 |
| Al | 7.26991000 | 9.78082000 | 8.78548000 | 2.47203680 |
| Al | 7.26991000 | 1.39726000 | 8.78548000 | 2.47264300 |
| Al | 9.69003000 | 5.58904000 | 8.78548000 | 2.47240900 |
| Al | 12.11016000 | 1.39726000 | 8.78548000 | 2.47006440 |
| O | 8.01661000 | 11.26550000 | 9.68989000 | -1.64653990 |
| O | 10.43673000 | 7.07371000 | 9.68989000 | -1.64314000 |
| O | -4.06971000 | 9.68514000 | 9.68989000 | -1.65907820 |
| O | -1.66389000 | 11.26550000 | 9.68989000 | -1.64378970 |
| O | 12.85686000 | 2.88194000 | 9.68989000 | -1.64555080 |
| O | -1.64959000 | 5.49336000 | 9.68989000 | -1.64595750 |
| O | -1.49818000 | 8.39182000 | 9.68989000 | -1.66435650 |
| O | 0.75624000 | 7.07372000 | 9.68989000 | -1.63791210 |
| O | 0.77054000 | 9.68514000 | 9.68989000 | -1.65742510 |
| O | 0.77054000 | 1.30159000 | 9.68989000 | -1.64767330 |
| O | 0.92195000 | 4.20004000 | 9.68989000 | -1.65630790 |
| O | 3.17636000 | 11.26550000 | 9.68989000 | -1.64596440 |
| O | 3.17636000 | 2.88194000 | 9.68989000 | -1.63294660 |
| O | 3.19067000 | 5.49336000 | 9.68989000 | -1.64420850 |
| O | 3.34208000 | 0.00826000 | 9.68989000 | -1.65573120 |
| O | 3.34207000 | 8.39182000 | 9.68989000 | -1.65748060 |
| O | 5.59648000 | 7.07372000 | 9.68989000 | -1.64648980 |
| O | 5.61079000 | 1.30159000 | 9.68989000 | -1.65926180 |
| O | 5.61079000 | 9.68514000 | 9.68989000 | -1.65617580 |
| O | 5.76220000 | 4.20004000 | 9.68989000 | -1.65399410 |
| O | 8.01661000 | 2.88194000 | 9.68989000 | -1.64430390 |
| O | 8.03091000 | 5.49336000 | 9.68989000 | -1.66005370 |
| O | 8.18232000 | 0.00826000 | 9.68989000 | -1.66635330 |
| O | 8.18232000 | 8.39182000 | 9.68989000 | -1.66776830 |
| O | 10.45104000 | 1.30158000 | 9.68989000 | -1.66341380 |
| O | 10.60245000 | 4.20004000 | 9.68989000 | -1.66838220 |
| O | 13.02257000 | 0.00826000 | 9.68989000 | -1.66671270 |
| Al | -4.83072000 | 11.17808000 | 10.71830000 | 2.47886740 |
| Al | -2.41059000 | 6.98630000 | 10.71830000 | 2.47989170 |
| Al | 0.00953000 | 11.17808000 | 10.71830000 | 2.47709580 |
| Al | 0.00953000 | 2.79452000 | 10.71830000 | 2.47736150 |
| Al | 2.42966000 | 6.98630000 | 10.71830000 | 2.47802490 |
| Al | 4.84978000 | 2.79452000 | 10.71830000 | 2.47909610 |
| Al | 4.84978000 | 11.17808000 | 10.71830000 | 2.47474310 |
| Al | 7.26991000 | 6.98630000 | 10.71830000 | 2.47802820 |
| Al | 9.69003000 | 2.79452000 | 10.71830000 | 2.47409980 |
| Al | -4.83072000 | 8.38356000 | 10.98601000 | 2.47133320 |
| Al | -2.41059000 | 4.19178000 | 10.98601000 | 2.47157220 |
| Al | 0.00953000 | 8.38356000 | 10.98601000 | 2.47158440 |
| Al | 0.00954000 | 0.00000000 | 10.98601000 | 2.46951800 |
| Al | 2.42966000 | 4.19178000 | 10.98601000 | 2.47302060 |
| Al | 4.84978000 | 0.00000000 | 10.98601000 | 2.47444990 |

| | | | |
|---|---|---|---|
| Al | 4.84978000 | 8.38356000 | 10.98601000 | 2.47364560 |
| Al | 7.26991000 | 4.19178000 | 10.98601000 | 2.47197000 |
| Al | 9.69003000 | 0.00000000 | 10.98601000 | 2.46934790 |
| O | -5.64864000 | 12.51869000 | 11.85483000 | -1.60590650 |
| O | -5.54291000 | 9.82707000 | 11.85020000 | -1.62025300 |
| O | -3.29594000 | 11.23563000 | 11.85999000 | -1.62417530 |
| O | -3.22696000 | 8.27989000 | 11.85098000 | -1.62020680 |
| O | -3.12079000 | 5.62029000 | 11.86288000 | -1.62704730 |
| O | -0.86472000 | 7.03081000 | 11.84837000 | -1.61406880 |
| O | -0.80372000 | 12.47188000 | 11.85084000 | -1.62134740 |
| O | -0.79652000 | 4.07607000 | 11.89604000 | -1.58800400 |
| O | -0.69471000 | 9.81760000 | 11.85422000 | -1.62790370 |
| O | -0.73430000 | 1.45509000 | 11.87745000 | -1.61816560 |
| O | 1.50632000 | 2.85056000 | 11.87917000 | -1.59055370 |
| O | 1.53923000 | 11.24433000 | 11.85026000 | -1.62153300 |
| O | 1.62626000 | 8.27994000 | 11.87715000 | -1.62681460 |
| O | 1.68968000 | 5.65077000 | 11.91109000 | -1.57215630 |
| O | 3.96195000 | 7.00491000 | 11.88678000 | -1.59602180 |
| O | 4.04328000 | 4.12976000 | 11.86479000 | -1.54774500 |
| O | 4.02222000 | 12.45057000 | 11.84940000 | -1.61743180 |
| O | 4.11812000 | 1.44434000 | 11.84180000 | -1.61107680 |
| O | 4.13875000 | 9.81632000 | 11.85486000 | -1.62444340 |
| O | 6.37391000 | 11.25336000 | 11.85177000 | -1.62616860 |
| O | 6.38234000 | 2.85852000 | 11.86446000 | -1.62478430 |
| O | 6.45105000 | 8.29663000 | 11.85043000 | -1.62660780 |
| O | 6.56712000 | 5.63293000 | 11.83940000 | -1.61579700 |
| O | 8.80488000 | 7.05482000 | 11.85962000 | -1.61837110 |
| O | 8.88011000 | 4.08479000 | 11.84920000 | -1.61718800 |
| O | 8.98276000 | 1.43592000 | 11.84918000 | -1.62330850 |
| O | 11.22617000 | 2.85877000 | 11.85125000 | -1.62767700 |
| Al | -2.40595000 | 9.77925000 | 11.89433000 | 2.41769200 |
| Al | -0.03776000 | 5.60771000 | 12.39303000 | 2.37452380 |
| Al | 2.45242000 | 1.40531000 | 12.34868000 | 2.37364980 |
| Al | 2.43251000 | 9.79620000 | 11.84896000 | 2.41555210 |
| Al | 4.93520000 | 5.59136000 | 12.42329000 | 2.37432010 |
| Al | 7.27210000 | 9.79464000 | 11.86550000 | 2.41879270 |
| Al | 7.26852000 | 1.39950000 | 11.89939000 | 2.41645600 |
| Al | 9.68651000 | 5.59032000 | 11.88504000 | 2.41589570 |
| Al | 12.06180000 | 1.37996000 | 11.77906000 | 2.41208570 |
| Pt | 4.96973000 | 5.84754000 | 14.97159000 | -0.47832600 |
| Pt | 0.13524000 | 5.29933000 | 14.87532000 | -1.03624530 |
| Pt | 3.47594000 | 3.92494000 | 14.20130000 | -0.42712480 |
| Pt | 1.97858000 | 1.63361000 | 14.82353000 | -0.86452420 |

| | | | |
|---|---|---|---|
| Sn | -0.24338000 | 2.77950000 | 14.01325000 | 1.01257110 |
| Sn | 2.46890000 | 6.48448000 | 14.05986000 | 0.98060180 |
| Sn | 2.00551000 | 3.89369000 | 16.23653000 | 0.69257130 |
| C | 6.26124000 | 7.57450000 | 14.96670000 | -0.13207260 |
| C | 6.41023000 | 6.86554000 | 16.17401000 | -0.22452410 |
| H | 5.65759000 | 8.48485000 | 14.92493000 | 0.06118590 |
| H | 6.98683000 | 7.46199000 | 14.15832000 | 0.13978340 |
| H | 5.92020000 | 7.21621000 | 17.08866000 | 0.07660560 |
| H | 7.27300000 | 6.21015000 | 16.32754000 | 0.08612840 |

283

c2h4-pt4sn3-al2o3-B

| | | | |
|---|---|---|---|
| Al | -4.83072000 | 11.17808000 | 0.00000000 | 2.16966430 |
| Al | -2.41059000 | 6.98630000 | 0.00000000 | 2.16989210 |
| Al | 0.00953000 | 11.17808000 | 0.00000000 | 2.16979640 |
| Al | 0.00953000 | 2.79452000 | 0.00000000 | 2.16984760 |
| Al | 2.42966000 | 6.98630000 | 0.00000000 | 2.16947640 |
| Al | 4.84978000 | 2.79452000 | 0.00000000 | 2.16957860 |
| Al | 4.84978000 | 11.17808000 | 0.00000000 | 2.16968620 |
| Al | 7.26991000 | 6.98630000 | 0.00000000 | 2.16943270 |
| Al | 9.69003000 | 2.79452000 | 0.00000000 | 2.16994900 |
| O | -3.89334000 | 9.78082000 | 0.85241000 | -1.55173490 |
| O | -1.66922000 | 8.49672000 | 0.85241000 | -1.55997570 |
| O | -1.66922000 | 11.06492000 | 0.85241000 | -1.55785590 |
| O | -1.47322000 | 5.58904000 | 0.85241000 | -1.55185450 |
| O | 0.75091000 | 4.30494000 | 0.85241000 | -1.55999720 |
| O | 0.75091000 | 6.87314000 | 0.85241000 | -1.55794320 |
| O | 0.94691000 | 9.78082000 | 0.85241000 | -1.55177950 |
| O | 0.94691000 | 1.39726000 | 0.85241000 | -1.55180600 |
| O | 3.17104000 | 0.11316000 | 0.85241000 | -1.56002930 |
| O | 3.17103000 | 8.49672000 | 0.85241000 | -1.55989630 |
| O | 3.17103000 | 2.68136000 | 0.85241000 | -1.55800030 |
| O | 3.17103000 | 11.06492000 | 0.85241000 | -1.55792480 |
| O | 3.36703000 | 5.58904000 | 0.85241000 | -1.55181330 |
| O | 5.59116000 | 4.30494000 | 0.85241000 | -1.55990120 |
| O | 5.59116000 | 6.87314000 | 0.85241000 | -1.55773600 |
| O | 5.78716000 | 1.39726000 | 0.85241000 | -1.55180260 |
| O | 5.78716000 | 9.78082000 | 0.85241000 | -1.55147250 |
| O | 8.01128000 | 8.49672000 | 0.85241000 | -1.55993520 |
| O | 8.01128000 | 0.11316000 | 0.85241000 | -1.56005820 |
| O | 8.01128000 | 2.68136000 | 0.85241000 | -1.55778900 |
| O | 8.01128000 | 11.06492000 | 0.85241000 | -1.55787250 |
| O | 8.20728000 | 5.58904000 | 0.85241000 | -1.55184180 |
| O | 10.43141000 | 4.30494000 | 0.85241000 | -1.56007180 |
| O | 10.43141000 | 6.87314000 | 0.85241000 | -1.55783630 |
| O | 10.62741000 | 1.39726000 | 0.85241000 | -1.55190080 |
| O | 12.85153000 | 0.11316000 | 0.85241000 | -1.56014770 |
| O | 12.85153000 | 2.68136000 | 0.85241000 | -1.55799680 |
| Al | -4.83072000 | 8.38356000 | 1.70482000 | 2.46330720 |
| Al | -2.41059000 | 4.19178000 | 1.70482000 | 2.46324930 |
| Al | 0.00953000 | 8.38356000 | 1.70482000 | 2.46334530 |
| Al | 0.00954000 | 0.00000000 | 1.70482000 | 2.46327820 |
| Al | 2.42966000 | 4.19178000 | 1.70482000 | 2.46351200 |
| Al | 4.84978000 | 0.00000000 | 1.70482000 | 2.46327530 |
| Al | 4.84978000 | 8.38356000 | 1.70482000 | 2.46334520 |
| Al | 7.26991000 | 4.19178000 | 1.70482000 | 2.46339050 |
| Al | 9.69003000 | 0.00000000 | 1.70482000 | 2.46330750 |
| Al | -2.41059000 | 9.78082000 | 2.20274000 | 2.48270800 |
| Al | 0.00953000 | 5.58904000 | 2.20274000 | 2.48279190 |
| Al | 2.42966000 | 1.39726000 | 2.20274000 | 2.48272580 |
| Al | 2.42966000 | 9.78082000 | 2.20274000 | 2.48278100 |
| Al | 4.84978000 | 5.58904000 | 2.20274000 | 2.48280750 |
| Al | 7.26991000 | 9.78082000 | 2.20274000 | 2.48268820 |
| Al | 7.26991000 | 1.39726000 | 2.20274000 | 2.48275860 |
| Al | 9.69003000 | 5.58904000 | 2.20274000 | 2.48276550 |
| Al | 12.11016000 | 1.39726000 | 2.20274000 | 2.48266800 |
| O | -5.57210000 | 9.66766000 | 3.05515000 | -1.64147310 |
| O | -3.34796000 | 8.38356000 | 3.05515000 | -1.62382830 |
| O | -3.15197000 | 5.47588000 | 3.05515000 | -1.64172500 |
| O | -3.15197000 | 11.29124000 | 3.05515000 | -1.64949210 |
| O | -0.92784000 | 4.19178000 | 3.05515000 | -1.62410780 |
| O | -0.73185000 | 1.28410000 | 3.05515000 | -1.64195980 |
| O | -0.73185000 | 9.66766000 | 3.05515000 | -1.64151450 |
| O | -0.73184000 | 7.09945000 | 3.05515000 | -1.64965870 |
| O | 1.49229000 | 8.38356000 | 3.05515000 | -1.62387610 |
| O | 1.49229000 | 0.00001000 | 3.05515000 | -1.62399010 |
| O | 1.68828000 | 5.47588000 | 3.05515000 | -1.64153350 |
| O | 1.68828000 | 2.90768000 | 3.05515000 | -1.64975870 |
| O | 1.68828000 | 11.29123000 | 3.05515000 | -1.64951610 |
| O | 3.91241000 | 4.19178000 | 3.05515000 | -1.62376310 |
| O | 4.10840000 | 9.66766000 | 3.05515000 | -1.64140760 |
| O | 4.10840000 | 1.28410000 | 3.05515000 | -1.64143060 |
| O | 4.10841000 | 7.09946000 | 3.05515000 | -1.64950430 |
| O | 6.33254000 | 0.00001000 | 3.05515000 | -1.62387750 |
| O | 6.33254000 | 8.38356000 | 3.05515000 | -1.62395550 |
| O | 6.52853000 | 11.29123000 | 3.05515000 | -1.64945650 |
| O | 6.52853000 | 5.47588000 | 3.05515000 | -1.64144060 |
| O | 6.52853000 | 2.90768000 | 3.05515000 | -1.64929560 |
| O | 8.75266000 | 4.19178000 | 3.05515000 | -1.62391260 |
| O | 8.94866000 | 7.09945000 | 3.05515000 | -1.64935160 |
| O | 8.94865000 | 1.28410000 | 3.05515000 | -1.64164890 |
| O | 11.17277000 | 0.00000000 | 3.05515000 | -1.62394690 |
| O | 11.36878000 | 2.90767000 | 3.05515000 | -1.64979710 |
| Al | -4.83072000 | 11.17808000 | 3.90756000 | 2.47768700 |
| Al | -2.41059000 | 6.98630000 | 3.90756000 | 2.47769810 |
| Al | 0.00953000 | 11.17808000 | 3.90756000 | 2.47778840 |
| Al | 0.00953000 | 2.79452000 | 3.90756000 | 2.47785270 |
| Al | 2.42966000 | 6.98630000 | 3.90756000 | 2.47764360 |
| Al | 4.84978000 | 2.79452000 | 3.90756000 | 2.47763460 |
| Al | 4.84978000 | 11.17808000 | 3.90756000 | 2.47754690 |
| Al | 7.26991000 | 6.98630000 | 3.90756000 | 2.47749680 |
| Al | 9.69003000 | 2.79452000 | 3.90756000 | 2.47776770 |
| Al | -4.83072000 | 8.38356000 | 4.40548000 | 2.47245790 |
| Al | -2.41059000 | 4.19178000 | 4.40548000 | 2.47224690 |
| Al | 0.00953000 | 8.38356000 | 4.40548000 | 2.47228050 |
| Al | 0.00954000 | 0.00000000 | 4.40548000 | 2.47240320 |
| Al | 2.42966000 | 4.19178000 | 4.40548000 | 2.47245040 |
| Al | 4.84978000 | 0.00000000 | 4.40548000 | 2.47228190 |
| Al | 4.84978000 | 8.38356000 | 4.40548000 | 2.47226460 |
| Al | 7.26991000 | 4.19178000 | 4.40548000 | 2.47243210 |
| Al | 9.69003000 | 0.00000000 | 4.40548000 | 2.47207330 |
| O | -6.31347000 | 11.17808000 | 5.25789000 | -1.64361880 |
| O | -4.08934000 | 12.46219000 | 5.25789000 | -1.65411370 |
| O | -4.08934000 | 9.89398000 | 5.25789000 | -1.64916140 |
| O | -3.89335000 | 6.98630000 | 5.25789000 | -1.64322020 |
| O | -1.66922000 | 8.27040000 | 5.25789000 | -1.65389140 |
| O | -1.66921000 | 5.70219000 | 5.25789000 | -1.64932260 |
| O | -1.47322000 | 11.17808000 | 5.25789000 | -1.64350300 |
| O | -1.47323000 | 2.79452000 | 5.25789000 | -1.64322510 |
| O | 0.75091000 | 4.07862000 | 5.25789000 | -1.65393350 |
| O | 0.75091000 | 12.46219000 | 5.25789000 | -1.65409500 |
| O | 0.75091000 | 9.89397000 | 5.25789000 | -1.64906760 |
| O | 0.75091000 | 1.51042000 | 5.25789000 | -1.64885310 |
| O | 0.94690000 | 6.98630000 | 5.25789000 | -1.64338250 |
| O | 3.17103000 | 8.27040000 | 5.25789000 | -1.65398680 |
| O | 3.17104000 | 5.70220000 | 5.25789000 | -1.64903400 |
| O | 3.36703000 | 2.79452000 | 5.25789000 | -1.64315800 |
| O | 3.36702000 | 11.17808000 | 5.25789000 | -1.64353130 |
| O | 5.59116000 | 12.46219000 | 5.25789000 | -1.65393490 |
| O | 5.59116000 | 4.07863000 | 5.25789000 | -1.65397420 |

| | | | |
|---|---|---|---|
| O | 5.59116000 | 9.89397000 | 5.25789000 | -1.64866470 |
| O | 5.59116000 | 1.51042000 | 5.25789000 | -1.64928060 |
| O | 5.78715000 | 6.98630000 | 5.25789000 | -1.64279620 |
| O | 8.01128000 | 8.27040000 | 5.25789000 | -1.65412070 |
| O | 8.01129000 | 5.70219000 | 5.25789000 | -1.64927280 |
| O | 8.20727000 | 2.79452000 | 5.25789000 | -1.64314970 |
| O | 10.43141000 | 4.07862000 | 5.25789000 | -1.65422140 |
| O | 10.43141000 | 1.51041000 | 5.25789000 | -1.64894860 |
| Al | -2.41059000 | 9.78082000 | 6.11030000 | 2.47621980 |
| Al | 0.00953000 | 5.58904000 | 6.11030000 | 2.47592830 |
| Al | 2.42966000 | 1.39726000 | 6.11030000 | 2.47594880 |
| Al | 2.42966000 | 9.78082000 | 6.11030000 | 2.47650700 |
| Al | 4.84978000 | 5.58904000 | 6.11030000 | 2.47645860 |
| Al | 7.26991000 | 9.78082000 | 6.11030000 | 2.47602220 |
| Al | 7.26991000 | 1.39726000 | 6.11030000 | 2.47623370 |
| Al | 9.69003000 | 5.58904000 | 6.11030000 | 2.47617990 |
| Al | 12.11016000 | 1.39726000 | 6.11030000 | 2.47576850 |
| Al | -4.83072000 | 11.17808000 | 6.60910000 | 2.47702400 |
| Al | -2.41059000 | 6.98630000 | 6.60910000 | 2.47684340 |
| Al | 0.00953000 | 11.17808000 | 6.60910000 | 2.47684170 |
| Al | 0.00953000 | 2.79452000 | 6.60910000 | 2.47680730 |
| Al | 2.42966000 | 6.98630000 | 6.60910000 | 2.47713950 |
| Al | 4.84978000 | 2.79452000 | 6.60910000 | 2.47691500 |
| Al | 4.84978000 | 11.17808000 | 6.60910000 | 2.47701870 |
| Al | 7.26991000 | 6.98630000 | 6.60910000 | 2.47706270 |
| Al | 9.69003000 | 2.79452000 | 6.60910000 | 2.47670390 |
| O | -3.15680000 | 8.49706000 | 7.46570000 | -1.64929080 |
| O | -3.14925000 | 11.06893000 | 7.46570000 | -1.64982720 |
| O | -0.92572000 | 9.77647000 | 7.46570000 | -1.65395010 |
| O | -0.73667000 | 4.30528000 | 7.46570000 | -1.64920890 |
| O | -0.72913000 | 6.87715000 | 7.46570000 | -1.64955120 |
| O | 1.49440000 | 5.58469000 | 7.46570000 | -1.65333440 |
| O | 1.68345000 | 8.49706000 | 7.46570000 | -1.64901500 |
| O | 1.68345000 | 0.11351000 | 7.46570000 | -1.64887630 |
| O | 1.69100000 | 2.68537000 | 7.46570000 | -1.65021660 |
| O | 1.69100000 | 11.06893000 | 7.46570000 | -1.64927040 |
| O | 3.91453000 | 1.39291000 | 7.46570000 | -1.65378310 |
| O | 3.91452000 | 9.77646000 | 7.46570000 | -1.65349520 |
| O | 4.10358000 | 4.30528000 | 7.46570000 | -1.64911500 |
| O | 4.11112000 | 6.87715000 | 7.46570000 | -1.64938000 |
| O | 6.33465000 | 5.58468000 | 7.46570000 | -1.65344220 |
| O | 6.52370000 | 0.11351000 | 7.46570000 | -1.64900760 |
| O | 6.52370000 | 8.49706000 | 7.46570000 | -1.64776280 |

| | | | |
|---|---|---|---|
| O | 6.53125000 | 2.68537000 | 7.46570000 | -1.64948510 |
| O | 6.53125000 | 11.06893000 | 7.46570000 | -1.64984400 |
| O | 8.75477000 | 9.77646000 | 7.46570000 | -1.65369000 |
| O | 8.75477000 | 1.39291000 | 7.46570000 | -1.65455870 |
| O | 8.94383000 | 4.30528000 | 7.46570000 | -1.64981580 |
| O | 8.95137000 | 6.87715000 | 7.46570000 | -1.64936840 |
| O | 11.17490000 | 5.58468000 | 7.46570000 | -1.65448650 |
| O | 11.36395000 | 0.11351000 | 7.46570000 | -1.64920760 |
| O | 11.37149000 | 2.68537000 | 7.46570000 | -1.65006950 |
| O | 13.59502000 | 1.39291000 | 7.46570000 | -1.65405460 |
| Al | -4.83072000 | 8.38356000 | 8.32833000 | 2.47155740 |
| Al | -2.41059000 | 4.19178000 | 8.32833000 | 2.47221530 |
| Al | 0.00953000 | 8.38356000 | 8.32833000 | 2.47104450 |
| Al | 0.00954000 | 0.00000000 | 8.32833000 | 2.47064630 |
| Al | 2.42966000 | 4.19178000 | 8.32833000 | 2.47081890 |
| Al | 4.84978000 | 0.00000000 | 8.32833000 | 2.47138570 |
| Al | 4.84978000 | 8.38356000 | 8.32833000 | 2.47017070 |
| Al | 7.26991000 | 4.19178000 | 8.32833000 | 2.47145870 |
| Al | 9.69003000 | 0.00000000 | 8.32833000 | 2.47087720 |
| Al | -2.41059000 | 9.78082000 | 8.78548000 | 2.47278230 |
| Al | 0.00953000 | 5.58904000 | 8.78548000 | 2.47166880 |
| Al | 2.42966000 | 1.39726000 | 8.78548000 | 2.47232130 |
| Al | 2.42966000 | 9.78082000 | 8.78548000 | 2.47630240 |
| Al | 4.84978000 | 5.58904000 | 8.78548000 | 2.47594440 |
| Al | 7.26991000 | 9.78082000 | 8.78548000 | 2.47569010 |
| Al | 7.26991000 | 1.39726000 | 8.78548000 | 2.47247740 |
| Al | 9.69003000 | 5.58904000 | 8.78548000 | 2.47256950 |
| Al | 12.11016000 | 1.39726000 | 8.78548000 | 2.46997850 |
| O | 8.01661000 | 11.26550000 | 9.68989000 | -1.64793110 |
| O | 10.43673000 | 7.07371000 | 9.68989000 | -1.64420160 |
| O | -4.06971000 | 9.68514000 | 9.68989000 | -1.65939600 |
| O | -1.66389000 | 11.26550000 | 9.68989000 | -1.64347110 |
| O | 12.85686000 | 2.88194000 | 9.68989000 | -1.64975850 |
| O | -1.64959000 | 5.49336000 | 9.68989000 | -1.61164460 |
| O | -1.49818000 | 8.39182000 | 9.68989000 | -1.66789930 |
| O | 0.75624000 | 7.07372000 | 9.68989000 | -1.64346940 |
| O | 0.77054000 | 9.68514000 | 9.68989000 | -1.64737940 |
| O | 0.77054000 | 1.30159000 | 9.68989000 | -1.65651940 |
| O | 0.92195000 | 4.20004000 | 9.68989000 | -1.66755410 |
| O | 3.17636000 | 11.26550000 | 9.68989000 | -1.63766310 |
| O | 3.17636000 | 2.88194000 | 9.68989000 | -1.64387550 |
| O | 3.19067000 | 5.49336000 | 9.68989000 | -1.64704080 |
| O | 3.34208000 | 0.00826000 | 9.68989000 | -1.66789480 |
| O | 3.34207000 | 8.39182000 | 9.68989000 | -1.65735320 |
| O | 5.59648000 | 7.07372000 | 9.68989000 | -1.63273100 |
| O | 5.61079000 | 1.30159000 | 9.68989000 | -1.65875820 |
| O | 5.61079000 | 9.68514000 | 9.68989000 | -1.64636130 |
| O | 5.76220000 | 4.20004000 | 9.68989000 | -1.65478100 |
| O | 8.01661000 | 2.88194000 | 9.68989000 | -1.64468550 |
| O | 8.03091000 | 5.49336000 | 9.68989000 | -1.65942100 |
| O | 8.18232000 | 0.00826000 | 9.68989000 | -1.66502810 |
| O | 8.18232000 | 8.39182000 | 9.68989000 | -1.65542270 |
| O | 10.45104000 | 1.30158000 | 9.68989000 | -1.66003000 |
| O | 10.60245000 | 4.20004000 | 9.68989000 | -1.66675830 |
| O | 13.02257000 | 0.00826000 | 9.68989000 | -1.65796510 |
| Al | -4.83072000 | 11.17808000 | 10.71830000 | 2.48222290 |
| Al | -2.41059000 | 6.98630000 | 10.71830000 | 2.47661400 |
| Al | 0.00953000 | 11.17808000 | 10.71830000 | 2.47685560 |
| Al | 0.00953000 | 2.79452000 | 10.71830000 | 2.47750050 |
| Al | 2.42966000 | 6.98630000 | 10.71830000 | 2.47590650 |
| Al | 4.84978000 | 2.79452000 | 10.71830000 | 2.47821860 |
| Al | 4.84978000 | 11.17808000 | 10.71830000 | 2.47564910 |
| Al | 7.26991000 | 6.98630000 | 10.71830000 | 2.47839570 |
| Al | 9.69003000 | 2.79452000 | 10.71830000 | 2.47599240 |
| Al | -4.83072000 | 8.38356000 | 10.98601000 | 2.47437000 |
| Al | -2.41059000 | 4.19178000 | 10.98601000 | 2.46971750 |
| Al | 0.00953000 | 8.38356000 | 10.98601000 | 2.46851050 |
| Al | 0.00954000 | 0.00000000 | 10.98601000 | 2.47300270 |
| Al | 2.42966000 | 4.19178000 | 10.98601000 | 2.47302570 |
| Al | 4.84978000 | 0.00000000 | 10.98601000 | 2.47088000 |
| Al | 4.84978000 | 8.38356000 | 10.98601000 | 2.47426890 |
| Al | 7.26991000 | 4.19178000 | 10.98601000 | 2.47108640 |
| Al | 9.69003000 | 0.00000000 | 10.98601000 | 2.46922640 |
| O | -5.65882000 | 12.48046000 | 11.85966000 | -1.62466740 |
| O | -5.57244000 | 9.82071000 | 11.85170000 | -1.60662390 |
| O | -3.30049000 | 11.24152000 | 11.85261000 | -1.62437010 |
| O | -3.22502000 | 8.28052000 | 11.85124000 | -1.62170630 |
| O | -3.12205000 | 5.62706000 | 11.85002000 | -1.62111310 |
| O | -0.87406000 | 7.04790000 | 11.85337000 | -1.62430310 |
| O | -0.80522000 | 12.47269000 | 11.84972000 | -1.62005440 |
| O | -0.81243000 | 4.07774000 | 11.84814000 | -1.61784540 |
| O | -0.70048000 | 9.81819000 | 11.86198000 | -1.62640480 |
| O | -0.70221000 | 1.43096000 | 11.84973000 | -1.63004310 |
| O | 1.53489000 | 2.85796000 | 11.85581000 | -1.62761050 |
| O | 1.55882000 | 11.24298000 | 11.85356000 | -1.60317960 |
| O | 1.64752000 | 8.30765000 | 11.86697000 | -1.60723510 |
| O | 1.71028000 | 5.63911000 | 11.87951000 | -1.62868170 |
| O | 3.96221000 | 7.01151000 | 11.94247000 | -1.55227940 |
| O | 4.03556000 | 4.10109000 | 11.82419000 | -1.61140480 |
| O | 4.06228000 | 12.48930000 | 11.87706000 | -1.61263220 |
| O | 4.13950000 | 1.43454000 | 11.85657000 | -1.62442690 |
| O | 4.13202000 | 9.85089000 | 11.86898000 | -1.59745590 |
| O | 6.36125000 | 11.21925000 | 11.90905000 | -1.57463340 |
| O | 6.38852000 | 2.84802000 | 11.86433000 | -1.62916270 |
| O | 6.44568000 | 8.29298000 | 11.85537000 | -1.58918400 |
| O | 6.55570000 | 5.62803000 | 11.83547000 | -1.60866550 |
| O | 8.80394000 | 7.05083000 | 11.86616000 | -1.62300320 |
| O | 8.87757000 | 4.08397000 | 11.84881000 | -1.61970020 |
| O | 8.99478000 | 1.43365000 | 11.85186000 | -1.63212440 |
| O | 11.21113000 | 2.87128000 | 11.84815000 | -1.61770010 |
| Al | -2.41160000 | 9.78208000 | 11.89381000 | 2.41514360 |
| Al | -0.00365000 | 5.58233000 | 11.85883000 | 2.41318500 |
| Al | 2.43262000 | 1.39926000 | 11.87344000 | 2.41528430 |
| Al | 2.42156000 | 9.82020000 | 12.33662000 | 2.37407050 |
| Al | 4.91926000 | 5.49391000 | 12.40564000 | 2.37889090 |
| Al | 7.29110000 | 9.73489000 | 12.37924000 | 2.37664120 |
| Al | 7.28162000 | 1.39302000 | 11.87590000 | 2.41591420 |
| Al | 9.68592000 | 5.58970000 | 11.89910000 | 2.41576890 |
| Al | 12.12008000 | 1.44712000 | 11.75463000 | 2.41287380 |
| Pt | 7.03841000 | 10.04620000 | 14.87148000 | -0.87828130 |
| Pt | 2.93307000 | 9.93295000 | 14.82655000 | -0.99975760 |
| Pt | 5.87370000 | 7.61300000 | 14.47749000 | -0.43455970 |
| Pt | 4.78874000 | 5.31282000 | 14.86424000 | -0.51951050 |
| Sn | 3.15302000 | 7.31299000 | 14.10287000 | 0.98694310 |
| Sn | 4.99893000 | 11.46104000 | 13.98885000 | 1.01415240 |
| Sn | 5.00483000 | 9.06951000 | 16.31280000 | 0.72914500 |
| C | 4.71305000 | 3.20271000 | 15.25614000 | -0.16229990 |
| C | 3.39180000 | 3.68044000 | 15.19875000 | -0.22193530 |
| H | 5.18855000 | 3.01619000 | 16.22535000 | 0.04326770 |
| H | 5.14461000 | 2.67970000 | 14.39909000 | 0.13691760 |
| H | 2.82999000 | 3.87662000 | 16.11739000 | 0.10041980 |
| H | 2.78804000 | 3.54353000 | 14.29904000 | 0.12524280 |

283
c2h4-pt4sn3-al2o3-B

| | | | | |
|---|---|---|---|---|
| Al | -4.83072000 | 11.17808000 | 0.00000000 | 2.16988470 |
| Al | -2.41059000 | 6.98630000 | 0.00000000 | 2.16977920 |
| Al | 0.00953000 | 11.17808000 | 0.00000000 | 2.16995420 |
| Al | 0.00953000 | 2.79452000 | 0.00000000 | 2.16978950 |
| Al | 2.42966000 | 6.98630000 | 0.00000000 | 2.16951550 |

| | | | |
|---|---|---|---|
| Al | 4.84978000 | 2.79452000 | 0.00000000 | 2.16960250 |
| Al | 4.84978000 | 11.17808000 | 0.00000000 | 2.17000360 |
| Al | 7.26991000 | 6.98630000 | 0.00000000 | 2.16973570 |
| Al | 9.69003000 | 2.79452000 | 0.00000000 | 2.17000360 |
| O | -3.89334000 | 9.78082000 | 0.85241000 | -1.55180060 |
| O | -1.66922000 | 8.49672000 | 0.85241000 | -1.55998280 |
| O | -1.66922000 | 11.06492000 | 0.85241000 | -1.55787470 |
| O | -1.47322000 | 5.58904000 | 0.85241000 | -1.55186470 |
| O | 0.75091000 | 4.30494000 | 0.85241000 | -1.55993020 |
| O | 0.75091000 | 6.87314000 | 0.85241000 | -1.55781760 |
| O | 0.94691000 | 9.78082000 | 0.85241000 | -1.55164330 |
| O | 0.94691000 | 1.39726000 | 0.85241000 | -1.55176770 |
| O | 3.17104000 | 0.11316000 | 0.85241000 | -1.56000450 |
| O | 3.17103000 | 8.49672000 | 0.85241000 | -1.56001940 |
| O | 3.17103000 | 2.68136000 | 0.85241000 | -1.55776810 |
| O | 3.17103000 | 11.06492000 | 0.85241000 | -1.55793020 |
| O | 3.36703000 | 5.58904000 | 0.85241000 | -1.55172690 |
| O | 5.59116000 | 4.30494000 | 0.85241000 | -1.55993320 |
| O | 5.59116000 | 6.87314000 | 0.85241000 | -1.55779970 |
| O | 5.78716000 | 1.39726000 | 0.85241000 | -1.55183110 |
| O | 5.78716000 | 9.78082000 | 0.85241000 | -1.55173180 |
| O | 8.01128000 | 8.49672000 | 0.85241000 | -1.55999930 |
| O | 8.01128000 | 0.11316000 | 0.85241000 | -1.56021620 |
| O | 8.01128000 | 2.68136000 | 0.85241000 | -1.55783140 |
| O | 8.01128000 | 11.06492000 | 0.85241000 | -1.55795770 |
| O | 8.20728000 | 5.58904000 | 0.85241000 | -1.55180810 |
| O | 10.43141000 | 4.30494000 | 0.85241000 | -1.56006980 |
| O | 10.43141000 | 6.87314000 | 0.85241000 | -1.55786320 |
| O | 10.62741000 | 1.39726000 | 0.85241000 | -1.55188420 |
| O | 12.85153000 | 0.11316000 | 0.85241000 | -1.56011210 |
| O | 12.85153000 | 2.68136000 | 0.85241000 | -1.55803870 |
| Al | -4.83072000 | 8.38356000 | 1.70482000 | 2.46327900 |
| Al | -2.41059000 | 4.19178000 | 1.70482000 | 2.46346190 |
| Al | 0.00953000 | 8.38356000 | 1.70482000 | 2.46320180 |
| Al | 0.00954000 | 0.00000000 | 1.70482000 | 2.46327520 |
| Al | 2.42966000 | 4.19178000 | 1.70482000 | 2.46345710 |
| Al | 4.84978000 | 0.00000000 | 1.70482000 | 2.46327640 |
| Al | 4.84978000 | 8.38356000 | 1.70482000 | 2.46338120 |
| Al | 7.26991000 | 4.19178000 | 1.70482000 | 2.46331780 |
| Al | 9.69003000 | 0.00000000 | 1.70482000 | 2.46326880 |
| Al | -2.41059000 | 9.78082000 | 2.20274000 | 2.48273810 |
| Al | 0.00953000 | 5.58904000 | 2.20274000 | 2.48282180 |
| Al | 2.42966000 | 1.39726000 | 2.20274000 | 2.48271650 |

| | | | |
|---|---|---|---|
| Al | 2.42966000 | 9.78082000 | 2.20274000 | 2.48263220 |
| Al | 4.84978000 | 5.58904000 | 2.20274000 | 2.48278320 |
| Al | 7.26991000 | 9.78082000 | 2.20274000 | 2.48261620 |
| Al | 7.26991000 | 1.39726000 | 2.20274000 | 2.48272380 |
| Al | 9.69003000 | 5.58904000 | 2.20274000 | 2.48270140 |
| Al | 12.11016000 | 1.39726000 | 2.20274000 | 2.48269200 |
| O | -5.57210000 | 9.66766000 | 3.05515000 | -1.64150600 |
| O | -3.34796000 | 8.38356000 | 3.05515000 | -1.62385660 |
| O | -3.15197000 | 5.47588000 | 3.05515000 | -1.64168630 |
| O | -3.15197000 | 11.29124000 | 3.05515000 | -1.64956870 |
| O | -0.92784000 | 4.19178000 | 3.05515000 | -1.62394450 |
| O | -0.73185000 | 1.28410000 | 3.05515000 | -1.64168270 |
| O | -0.73185000 | 9.66766000 | 3.05515000 | -1.64151010 |
| O | -0.73184000 | 7.09945000 | 3.05515000 | -1.64960780 |
| O | 1.49229000 | 8.38356000 | 3.05515000 | -1.62387670 |
| O | 1.49229000 | 0.00001000 | 3.05515000 | -1.62384750 |
| O | 1.68828000 | 5.47588000 | 3.05515000 | -1.64130970 |
| O | 1.68828000 | 2.90768000 | 3.05515000 | -1.64950820 |
| O | 1.68828000 | 11.29123000 | 3.05515000 | -1.64967800 |
| O | 3.91241000 | 4.19178000 | 3.05515000 | -1.62384940 |
| O | 4.10840000 | 9.66766000 | 3.05515000 | -1.64162300 |
| O | 4.10840000 | 1.28410000 | 3.05515000 | -1.64141010 |
| O | 4.10841000 | 7.09946000 | 3.05515000 | -1.64939300 |
| O | 6.33254000 | 0.00001000 | 3.05515000 | -1.62388470 |
| O | 6.33254000 | 8.38356000 | 3.05515000 | -1.62395930 |
| O | 6.52853000 | 11.29123000 | 3.05515000 | -1.64972370 |
| O | 6.52853000 | 5.47588000 | 3.05515000 | -1.64145750 |
| O | 6.52853000 | 2.90768000 | 3.05515000 | -1.64947500 |
| O | 8.75266000 | 4.19178000 | 3.05515000 | -1.62388380 |
| O | 8.94866000 | 7.09945000 | 3.05515000 | -1.64957850 |
| O | 8.94865000 | 1.28410000 | 3.05515000 | -1.64184210 |
| O | 11.17277000 | 0.00000000 | 3.05515000 | -1.62407330 |
| O | 11.36878000 | 2.90767000 | 3.05515000 | -1.64966570 |
| Al | -4.83072000 | 11.17808000 | 3.90756000 | 2.47773150 |
| Al | -2.41059000 | 6.98630000 | 3.90756000 | 2.47770430 |
| Al | 0.00953000 | 11.17808000 | 3.90756000 | 2.47771680 |
| Al | 0.00953000 | 2.79452000 | 3.90756000 | 2.47774190 |
| Al | 2.42966000 | 6.98630000 | 3.90756000 | 2.47749950 |
| Al | 4.84978000 | 2.79452000 | 3.90756000 | 2.47764860 |
| Al | 4.84978000 | 11.17808000 | 3.90756000 | 2.47782490 |
| Al | 7.26991000 | 6.98630000 | 3.90756000 | 2.47777110 |
| Al | 9.69003000 | 2.79452000 | 3.90756000 | 2.47777640 |
| Al | -4.83072000 | 8.38356000 | 4.40548000 | 2.47213280 |

| | | | |
|---|---|---|---|
| Al | -2.41059000 | 4.19178000 | 4.40548000 | 2.47243860 |
| Al | 0.00953000 | 8.38356000 | 4.40548000 | 2.47216720 |
| Al | 0.00954000 | 0.00000000 | 4.40548000 | 2.47249580 |
| Al | 2.42966000 | 4.19178000 | 4.40548000 | 2.47248000 |
| Al | 4.84978000 | 0.00000000 | 4.40548000 | 2.47226780 |
| Al | 4.84978000 | 8.38356000 | 4.40548000 | 2.47229260 |
| Al | 7.26991000 | 4.19178000 | 4.40548000 | 2.47236440 |
| Al | 9.69003000 | 0.00000000 | 4.40548000 | 2.47234230 |
| O | -6.31347000 | 11.17808000 | 5.25789000 | -1.64330620 |
| O | -4.08934000 | 12.46219000 | 5.25789000 | -1.65395210 |
| O | -4.08934000 | 9.89398000 | 5.25789000 | -1.64897210 |
| O | -3.89335000 | 6.98630000 | 5.25789000 | -1.64342840 |
| O | -1.66922000 | 8.27040000 | 5.25789000 | -1.65393610 |
| O | -1.66921000 | 5.70219000 | 5.25789000 | -1.64916280 |
| O | -1.47322000 | 11.17808000 | 5.25789000 | -1.64331750 |
| O | -1.47323000 | 2.79452000 | 5.25789000 | -1.64340990 |
| O | 0.75091000 | 4.07862000 | 5.25789000 | -1.65374380 |
| O | 0.75091000 | 12.46219000 | 5.25789000 | -1.65424410 |
| O | 0.75091000 | 9.89397000 | 5.25789000 | -1.64885440 |
| O | 0.75091000 | 1.51042000 | 5.25789000 | -1.64908500 |
| O | 0.94690000 | 6.98630000 | 5.25789000 | -1.64343940 |
| O | 3.17103000 | 8.27040000 | 5.25789000 | -1.65423980 |
| O | 3.17104000 | 5.70220000 | 5.25789000 | -1.64898390 |
| O | 3.36703000 | 2.79452000 | 5.25789000 | -1.64289100 |
| O | 3.36702000 | 11.17808000 | 5.25789000 | -1.64343460 |
| O | 5.59116000 | 12.46219000 | 5.25789000 | -1.65417220 |
| O | 5.59116000 | 4.07863000 | 5.25789000 | -1.65400650 |
| O | 5.59116000 | 9.89397000 | 5.25789000 | -1.64879040 |
| O | 5.59116000 | 1.51042000 | 5.25789000 | -1.64944870 |
| O | 5.78715000 | 6.98630000 | 5.25789000 | -1.64303710 |
| O | 8.01128000 | 8.27040000 | 5.25789000 | -1.65418790 |
| O | 8.01129000 | 5.70219000 | 5.25789000 | -1.64922410 |
| O | 8.20727000 | 2.79452000 | 5.25789000 | -1.64325990 |
| O | 10.43141000 | 4.07862000 | 5.25789000 | -1.65409660 |
| O | 10.43141000 | 1.51041000 | 5.25789000 | -1.64881180 |
| Al | -2.41059000 | 9.78082000 | 6.11030000 | 2.47610010 |
| Al | 0.00953000 | 5.58904000 | 6.11030000 | 2.47637130 |
| Al | 2.42966000 | 1.39726000 | 6.11030000 | 2.47626810 |
| Al | 2.42966000 | 9.78082000 | 6.11030000 | 2.47605400 |
| Al | 4.84978000 | 5.58904000 | 6.11030000 | 2.47619020 |
| Al | 7.26991000 | 9.78082000 | 6.11030000 | 2.47605270 |
| Al | 7.26991000 | 1.39726000 | 6.11030000 | 2.47598710 |
| Al | 9.69003000 | 5.58904000 | 6.11030000 | 2.47591410 |
| Al | 12.11016000 | 1.39726000 | 6.11030000 | 2.47565740 |
| Al | -4.83072000 | 11.17808000 | 6.60910000 | 2.47672530 |
| Al | -2.41059000 | 6.98630000 | 6.60910000 | 2.47688960 |
| Al | 0.00953000 | 11.17808000 | 6.60910000 | 2.47680720 |
| Al | 0.00953000 | 2.79452000 | 6.60910000 | 2.47682130 |
| Al | 2.42966000 | 6.98630000 | 6.60910000 | 2.47759320 |
| Al | 4.84978000 | 2.79452000 | 6.60910000 | 2.47676590 |
| Al | 4.84978000 | 11.17808000 | 6.60910000 | 2.47686880 |
| Al | 7.26991000 | 6.98630000 | 6.60910000 | 2.47724910 |
| Al | 9.69003000 | 2.79452000 | 6.60910000 | 2.47673750 |
| O | -3.15680000 | 8.49706000 | 7.46570000 | -1.64897390 |
| O | -3.14925000 | 11.06893000 | 7.46570000 | -1.64975440 |
| O | -0.92572000 | 9.77647000 | 7.46570000 | -1.65386720 |
| O | -0.73667000 | 4.30528000 | 7.46570000 | -1.64923980 |
| O | -0.72913000 | 6.87715000 | 7.46570000 | -1.64904180 |
| O | 1.49440000 | 5.58469000 | 7.46570000 | -1.65374750 |
| O | 1.68345000 | 8.49706000 | 7.46570000 | -1.64820280 |
| O | 1.68345000 | 0.11351000 | 7.46570000 | -1.64902700 |
| O | 1.69100000 | 2.68537000 | 7.46570000 | -1.64892220 |
| O | 1.69100000 | 11.06893000 | 7.46570000 | -1.65026050 |
| O | 3.91453000 | 1.39291000 | 7.46570000 | -1.65348000 |
| O | 3.91452000 | 9.77646000 | 7.46570000 | -1.65413330 |
| O | 4.10358000 | 4.30528000 | 7.46570000 | -1.64883440 |
| O | 4.11112000 | 6.87715000 | 7.46570000 | -1.64894060 |
| O | 6.33465000 | 5.58468000 | 7.46570000 | -1.65381220 |
| O | 6.52370000 | 0.11351000 | 7.46570000 | -1.64933940 |
| O | 6.52370000 | 8.49706000 | 7.46570000 | -1.64844540 |
| O | 6.53125000 | 2.68537000 | 7.46570000 | -1.64918990 |
| O | 6.53125000 | 11.06893000 | 7.46570000 | -1.65027390 |
| O | 8.75477000 | 9.77646000 | 7.46570000 | -1.65395610 |
| O | 8.75477000 | 1.39291000 | 7.46570000 | -1.65483860 |
| O | 8.94383000 | 4.30528000 | 7.46570000 | -1.64984560 |
| O | 8.95137000 | 6.87715000 | 7.46570000 | -1.64933910 |
| O | 11.17490000 | 5.58468000 | 7.46570000 | -1.65393580 |
| O | 11.36395000 | 0.11351000 | 7.46570000 | -1.64923680 |
| O | 11.37149000 | 2.68537000 | 7.46570000 | -1.65019880 |
| O | 13.59502000 | 1.39291000 | 7.46570000 | -1.65413250 |
| Al | -4.83072000 | 8.38356000 | 8.32833000 | 2.47156960 |
| Al | -2.41059000 | 4.19178000 | 8.32833000 | 2.47106160 |
| Al | 0.00953000 | 8.38356000 | 8.32833000 | 2.47050070 |
| Al | 0.00954000 | 0.00000000 | 8.32833000 | 2.47042770 |
| Al | 2.42966000 | 4.19178000 | 8.32833000 | 2.46995500 |
| Al | 4.84978000 | 0.00000000 | 8.32833000 | 2.47150090 |
| Al | 4.84978000 | 8.38356000 | 8.32833000 | 2.47052060 |
| Al | 7.26991000 | 4.19178000 | 8.32833000 | 2.47221430 |
| Al | 9.69003000 | 0.00000000 | 8.32833000 | 2.47104360 |
| Al | -2.41059000 | 9.78082000 | 8.78548000 | 2.47264210 |
| Al | 0.00953000 | 5.58904000 | 8.78548000 | 2.47591410 |
| Al | 2.42966000 | 1.39726000 | 8.78548000 | 2.47652310 |
| Al | 2.42966000 | 9.78082000 | 8.78548000 | 2.47218900 |
| Al | 4.84978000 | 5.58904000 | 8.78548000 | 2.47649970 |
| Al | 7.26991000 | 9.78082000 | 8.78548000 | 2.47242540 |
| Al | 7.26991000 | 1.39726000 | 8.78548000 | 2.47246960 |
| Al | 9.69003000 | 5.58904000 | 8.78548000 | 2.47234280 |
| Al | 12.11016000 | 1.39726000 | 8.78548000 | 2.46996290 |
| O | 8.01661000 | 11.26550000 | 9.68989000 | -1.64801640 |
| O | 10.43673000 | 7.07371000 | 9.68989000 | -1.64299260 |
| O | -4.06971000 | 9.68514000 | 9.68989000 | -1.66007680 |
| O | -1.66389000 | 11.26550000 | 9.68989000 | -1.64465020 |
| O | 12.85686000 | 2.88194000 | 9.68989000 | -1.64724420 |
| O | -1.64959000 | 5.49336000 | 9.68989000 | -1.64411490 |
| O | -1.49818000 | 8.39182000 | 9.68989000 | -1.66276290 |
| O | 0.75624000 | 7.07372000 | 9.68989000 | -1.63936470 |
| O | 0.77054000 | 9.68514000 | 9.68989000 | -1.65634610 |
| O | 0.77054000 | 1.30159000 | 9.68989000 | -1.64907110 |
| O | 0.92195000 | 4.20004000 | 9.68989000 | -1.65773630 |
| O | 3.17636000 | 11.26550000 | 9.68989000 | -1.64481550 |
| O | 3.17636000 | 2.88194000 | 9.68989000 | -1.63292260 |
| O | 3.19067000 | 5.49336000 | 9.68989000 | -1.64313180 |
| O | 3.34208000 | 0.00826000 | 9.68989000 | -1.65477530 |
| O | 3.34207000 | 8.39182000 | 9.68989000 | -1.66322990 |
| O | 5.59648000 | 7.07372000 | 9.68989000 | -1.63875760 |
| O | 5.61079000 | 1.30159000 | 9.68989000 | -1.65998220 |
| O | 5.61079000 | 9.68514000 | 9.68989000 | -1.65772100 |
| O | 5.76220000 | 4.20004000 | 9.68989000 | -1.65539370 |
| O | 8.01661000 | 2.88194000 | 9.68989000 | -1.64464810 |
| O | 8.03091000 | 5.49336000 | 9.68989000 | -1.66112990 |
| O | 8.18232000 | 0.00826000 | 9.68989000 | -1.66767850 |
| O | 8.18232000 | 8.39182000 | 9.68989000 | -1.66737350 |
| O | 10.45104000 | 1.30158000 | 9.68989000 | -1.66352910 |
| O | 10.60245000 | 4.20004000 | 9.68989000 | -1.66939340 |
| O | 13.02257000 | 0.00826000 | 9.68989000 | -1.66384020 |
| Al | -4.83072000 | 11.17808000 | 10.71830000 | 2.47837650 |
| Al | -2.41059000 | 6.98630000 | 10.71830000 | 2.48049800 |
| Al | 0.00953000 | 11.17808000 | 10.71830000 | 2.47817660 |
| Al | 0.00953000 | 2.79452000 | 10.71830000 | 2.47738210 |
| Al | 2.42966000 | 6.98630000 | 10.71830000 | 2.48170280 |
| Al | 4.84978000 | 2.79452000 | 10.71830000 | 2.47912610 |
| Al | 4.84978000 | 11.17808000 | 10.71830000 | 2.47606050 |
| Al | 7.26991000 | 6.98630000 | 10.71830000 | 2.48062110 |
| Al | 9.69003000 | 2.79452000 | 10.71830000 | 2.47425250 |
| Al | -4.83072000 | 8.38356000 | 10.98601000 | 2.47285770 |
| Al | -2.41059000 | 4.19178000 | 10.98601000 | 2.47051490 |
| Al | 0.00953000 | 8.38356000 | 10.98601000 | 2.47328890 |
| Al | 0.00954000 | 0.00000000 | 10.98601000 | 2.46973560 |
| Al | 2.42966000 | 4.19178000 | 10.98601000 | 2.47185660 |
| Al | 4.84978000 | 0.00000000 | 10.98601000 | 2.47245680 |
| Al | 4.84978000 | 8.38356000 | 10.98601000 | 2.47098010 |
| Al | 7.26991000 | 4.19178000 | 10.98601000 | 2.47358120 |
| Al | 9.69003000 | 0.00000000 | 10.98601000 | 2.46961150 |
| O | -5.66017000 | 12.49122000 | 11.85448000 | -1.60832860 |
| O | -5.54449000 | 9.82488000 | 11.85001000 | -1.62220250 |
| O | -3.29794000 | 11.24310000 | 11.86221000 | -1.62054050 |
| O | -3.23060000 | 8.27982000 | 11.85241000 | -1.62004630 |
| O | -3.12234000 | 5.61830000 | 11.85981000 | -1.62622970 |
| O | -0.86620000 | 7.02463000 | 11.84939000 | -1.61288390 |
| O | -0.81127000 | 12.47731000 | 11.85342000 | -1.62341290 |
| O | -0.78263000 | 4.08184000 | 11.90189000 | -1.57264790 |
| O | -0.70380000 | 9.82503000 | 11.85128000 | -1.62248450 |
| O | -0.73240000 | 1.45631000 | 11.87910000 | -1.61497780 |
| O | 1.51584000 | 2.83463000 | 11.87512000 | -1.59858770 |
| O | 1.54282000 | 11.24062000 | 11.84974000 | -1.62396480 |
| O | 1.60646000 | 8.28022000 | 11.86595000 | -1.62432590 |
| O | 1.70738000 | 5.61508000 | 11.87108000 | -1.57621370 |
| O | 3.95932000 | 7.04852000 | 11.83190000 | -1.60813790 |
| O | 4.06850000 | 4.11282000 | 11.93074000 | -1.55539460 |
| O | 4.02363000 | 12.45736000 | 11.84787000 | -1.61962920 |
| O | 4.10008000 | 1.45659000 | 11.86576000 | -1.60965730 |
| O | 4.13918000 | 9.82946000 | 11.84965000 | -1.62435110 |
| O | 6.37715000 | 11.25551000 | 11.85283000 | -1.62328290 |
| O | 6.37673000 | 2.84469000 | 11.88011000 | -1.63055800 |
| O | 6.45291000 | 8.29052000 | 11.86508000 | -1.62717050 |
| O | 6.53930000 | 5.62800000 | 11.82756000 | -1.60656740 |
| O | 8.80313000 | 7.04812000 | 11.85576000 | -1.62755650 |
| O | 8.87318000 | 4.08286000 | 11.85176000 | -1.62118600 |
| O | 8.99010000 | 1.43992000 | 11.84874000 | -1.62582420 |
| O | 11.22650000 | 2.86031000 | 11.84956000 | -1.62516460 |
| Al | -2.41100000 | 9.77947000 | 11.89461000 | 2.41448620 |
| Al | 0.02256000 | 5.62807000 | 12.38124000 | 2.38050520 |

| | | | |
|---|---|---|---|
| Al | 2.40824000 | 1.36954000 | 12.34758000 | 2.36834320 |
| Al | 2.43099000 | 9.77566000 | 11.89821000 | 2.41557010 |
| Al | 4.89276000 | 5.70366000 | 12.41758000 | 2.37230500 |
| Al | 7.26777000 | 9.79066000 | 11.88207000 | 2.41688210 |
| Al | 7.28287000 | 1.38906000 | 11.86357000 | 2.41460880 |
| Al | 9.68909000 | 5.58747000 | 11.86717000 | 2.41364980 |
| Al | 12.06163000 | 1.38212000 | 11.75403000 | 2.41299530 |
| Pt | -0.14550000 | 5.23187000 | 14.86140000 | -0.87751190 |
| Pt | 2.10771000 | 1.77919000 | 14.83826000 | -0.99938470 |
| Pt | 2.53270000 | 5.44776000 | 14.33639000 | -0.41277700 |
| Pt | 5.00620000 | 5.74700000 | 14.92069000 | -0.52788820 |
| Sn | 4.25373000 | 3.27692000 | 14.10385000 | 0.97230950 |
| Sn | -0.25816000 | 2.75167000 | 13.98813000 | 1.02148130 |
| Sn | 1.79463000 | 4.05167000 | 16.27404000 | 0.72255190 |
| C | 7.18090000 | 5.77736000 | 15.05666000 | -0.13033990 |
| C | 6.61992000 | 6.79130000 | 15.85302000 | -0.22560160 |
| H | 7.53051000 | 4.84496000 | 15.50706000 | 0.04528940 |
| H | 7.57768000 | 6.01236000 | 14.06720000 | 0.12939470 |
| H | 6.52967000 | 6.66337000 | 16.93747000 | 0.07069770 |
| H | 6.60526000 | 7.82652000 | 15.49974000 | 0.12228310 |